%% file: IMoG.tex
\documentclass[figures=plain,11pt]{scrreprt}
\usepackage[T1]{fontenc}
\usepackage{lmodern}
\pdfoutput=1

\usepackage{csquotes} 
\usepackage[hyphens,spaces,obeyspaces]{url}
\usepackage[hidelinks]{hyperref}
\usepackage{todonotes}
\usepackage{tikz}
\usepackage{float}
\usepackage{wrapfig} 
\usepackage{caption}
\usepackage{subcaption}
\usepackage{rotating} 
\usepackage{placeins} 

\usepackage{bbding} 
\usepackage{setspace}
\usepackage{pdfpages}

\title       {Innovation Modeling Grid}
\subtitle    {Technical Documentation}
\author      {Oliver Klemp}

\input{content/tikzgraphics.tex}
\begin{document}
	\maketitle
	\newpage
	DLR - Institute for Systems Engineering for future mobility
	\newpage
	\input{content/abstract.tex}
	\tableofcontents
	\part{Overview over IMoG}
	\input{content/motivation.tex}
	\input{content/delimitation.tex}
	\input{content/process.tex}
	\input{content/methodology.tex}
	\part{Details on the IMoG Methodology}
	\input{content/strategy_perspective.tex}
	\input{content/functional_perspective.tex}
	\input{content/quality_perspective.tex}
	\input{content/structural_perspective.tex}
	\input{content/knowledge_perspective.tex}
	\input{content/roadmapping.tex}
	\input{content/updating_the_roadmap.tex}
	\input{content/inter_perspectives_relations.tex}
	\part{Tooling, Evaluation and Closing}
	\input{content/tooling.tex}
	\input{content/evaluation_results.tex}
	\input{content/closing.tex}
	\part{Appendix}
	\input{content/appendix.tex}
	\bibliographystyle{splncs04}
	\bibliography{bibliography}
\end{document}

%% file: content/tikzgraphics.tex
\usetikzlibrary{shadows,patterns.meta}

\definecolor{carbongray}{rgb}{0.35, 0.35, 0.35}
\newcommand{\valuechain}{
	\node[fill, bottom color=carbongray!10!white, minimum width=8cm, minimum height=0.95cm] at (4,2.5) {\textbf{Original Equipment Manufacturers (OEMs)}};
	\node[fill, bottom color=green!70!white, top color=green!70!black, minimum width=3.5cm, minimum height=0.95cm] at (3.5/2,1.5) {\small \textbf{System Supplier 1}};
	\node[fill, bottom color=green!20!white, top color=green!70!black, minimum width=0.9cm, minimum height=0.95cm] at (4,1.5) {\small \textbf{\ldots}};
	\node[fill, bottom color=green!70!white, top color=green!70!black, minimum width=3.5cm, minimum height=0.95cm] at (8-3.5/2,1.5) {\small \textbf{System Supplier n}};
	\node[fill, bottom color=blue!20!white, top color=blue!70!black!50!white, minimum width=2.3cm, minimum height=0.95cm,align=center] at (2.3/2,0.5) {\scriptsize \textbf{Semiconductor}\\\scriptsize \textbf{Supplier 1}};
	\node[fill, bottom color=blue!20!white, top color=blue!70!black!50!white, minimum width=2.3cm, minimum height=0.95cm,align=center] at (2.35+2.3/2,0.5) {\scriptsize \textbf{Semiconductor}\\\scriptsize \textbf{Supplier 2}};
	\node[fill, bottom color=blue!10!white, top color=blue!70!black!40!white, minimum width=0.95cm, minimum height=0.95cm] at (5.175,0.5) {\small \textbf{\ldots}};
	\node[fill, bottom color=blue!20!white, top color=blue!70!black!50!white, minimum width=2.3cm, minimum height=0.95cm,align=center] at (8-2.3/2,0.5) {\scriptsize \textbf{Semiconductor}\\\scriptsize \textbf{Supplier n}};
	
	\begin{scope}[xshift=0.3cm]
		\node at (8.7,3.35) {\scriptsize common invisible};
		\node at (8.7,3.1) {\scriptsize Guideline};
		\draw[ultra thick]  plot[smooth, tension=.7] coordinates {(8.3,2.8) (8.2,2.4) (8.4,1.8) (8.3,1.2) (8.3,0.7) (8.5,0.4) (8.3,0.1)};
		\node at (8.3,-0.2) {\scriptsize 2000};
		\node at (9.3,-0.2) {\scriptsize nowadays};	
	\end{scope}
	\begin{scope}[xshift=1.2cm]
		\draw[ultra thick]  plot[smooth, tension=.7] coordinates {(8.3,2.8) (8.2,2.4) (8.4,1.8) (8.3,1.2) (8.3,0.7) (8.5,0.4) (8.3,0.1)};
		\filldraw[white] (8.3,2)rectangle (8.4,2.2);
		\draw[thick, red] (8.3356,2.1275) -- (8.4181,2.1405);
		\draw[thick, red] (8.3261,2.0935) -- (8.426,2.0522);
		\draw[thick, red] (8.287,2.0674) -- (8.2522,1.9979);
		\filldraw[white] (8.2,1.5)rectangle (8.5,1.7);
		\draw[thick, red] (8.4441,1.5785) -- (8.4256,1.6561);
		\draw[thick, red] (8.3023,1.641) -- (8.3556,1.6763);
		\draw[thick, red] (8.3175,1.5584) -- (8.3338,1.5199);
		\draw[thick, red] (8.4072,1.5416) -- (8.3772,1.5047);
		\draw[thick, red] (8.3587,1.5606) -- (8.3562,1.5111);
		\filldraw[white] (8.2,1)rectangle (8.4,1.2);
		\draw[thick, red] (8.3159,1.1693) -- (8.3434,1.1217);
		\draw[thick, red] (8.2702,1.1802) -- (8.2379,1.1317);
		\draw[thick, red] (8.2776,1.0155) -- (8.293,1.0616);
		\filldraw[white] (8.2,0.5)rectangle (8.4,0.7);
		\draw[thick, red] (8.2581,0.6181) -- (8.2787,0.6777);
		\draw[thick, red] (8.3136,0.6048) -- (8.2978,0.6749);
		\draw[thick, red] (8.371,0.6157) -- (8.3243,0.6681);
		\draw[thick, red] (8.3761,0.5471) -- (8.3002,0.5428);
		\draw[thick, red] (8.3686,0.522) -- (8.3532,0.4783);
	\end{scope}
}

\newcommand{\processarrow}[2][1]{
	\begin{scope}[xshift=#1*2.8cm]
		\draw (0,0) -- (2.8,0) -- (3.2,0.7) -- (2.8,1.4) -- (0,1.4) -- (0.4,0.7) -- (0,0);
		\node[text width=7em, align=center] at (1.59,0.7) {#2};
	\end{scope}
}

\newcommand{\stone}{
	\draw[ultra thick, fill=gray!70, shade, shading angle=135,drop shadow={opacity=0.25, shadow xshift=-1ex, shadow yshift=1ex}]  plot[smooth, tension=0.5] coordinates {(-1.5,0.85) (-0.5,0.5) (0,0.5) (0.5,1) (1,1) (1,1.5) (-0.5,2) (-1.5,1.5) (-1.55,0.95)}--cycle;
	\draw[fill=black] (-1.04,0.68)  -- (-0.97,0.65) -- (-0.79,1.1) -- (-0.84,1.13);
	\draw[fill=black]  (-0.81,1.05) -- (-0.8,1.1)  -- (-1.05,1.3);
	\draw[fill=black] (-0.81,1.05) -- (-0.81,1.1) -- (-0.4,1.3);
}
\newcommand{\stonetwo}{
	\draw[ultra thick, fill=gray!70, shade, shading angle=135,drop shadow={opacity=0.25, shadow xshift=-1ex, shadow yshift=1ex}]  plot[smooth, tension=.7] coordinates {(0,1.2)(-0.23,0.8) (0,0) (2,1) (3.5,1) (4,2) (3.25,3.15) (2,3) (1,2.5) (0.5,1.7)}--cycle;
}
\newcommand{\signstick}{
	\draw[ultra thick,fill, left color=brown!50!black, right color=brown!70!white,drop shadow={opacity=0.25, shadow xshift=-1ex, shadow yshift=1ex}]   plot[smooth, tension=.7] coordinates {(0,3.5) (0,1) (0.8,0.8) (1,2.1) (1,7.5) (1.2,10.2) (1,12) (0.6,12.3) (0.2,12) (0.1,11.2) (0.1,10.2) (0,9.5) (0,7.5) (-0.2,5.7) (0,4)}--cycle;
	\draw[ultra thick]  plot[smooth, tension=.7] coordinates {(0.3,12) (0.6,11.9) (0.9,12)};
}

\newcommand{\signnatural}{
	\draw[ultra thick,fill, left color=brown!50!black, right color=brown!70!white]
	(0,0) -- (0,1) -- (-0.1,1.1) -- (0,2) -- (3,2.5) -- (4.3,1.9) -- (3.4,0.9) -- cycle;
	\draw[ultra thick,fill, left color=brown!30!black, right color=brown!70!black]
	(0,0) -- (0,1) -- (-0.1,1.1) -- (0,2) -- (-0.49,2.4) -- (-0.59,1.1) -- (-0.49, 1) -- (-0.5,0.35) -- cycle;
	\draw[ultra thick,fill, left color=brown!30!black, right color=brown!70!white]
	(0,2) -- (-0.49,2.4) -- (2.8,2.59) --  (3,2.5) -- cycle;
	\draw[ultra thick] (0.2414,1.6631) -- (0.7335,1.2413) -- (0.9444,1.7334) -- (1.2607,1.4873) -- (1.706,1.8423) -- (2.1708,1.639) -- (2.6647,1.9876) -- (3.2166,1.8133);
	\draw[ultra thick] (0.2826,0.9127) -- (0.5441,0.506) -- (0.7474,0.9418) -- (1.096,0.6222) -- (1.3865,0.9999) -- (1.5898,0.9127) -- (1.706,1.087);
	\draw[ultra thick] (2.2289,1.2323) -- (2.4904,1.1161) -- (2.6356,1.4937) --(2.9261,1.2613) ;
}

\newcommand{\signnaturalleft}{
	\draw[ultra thick,fill, left color=brown!70!white, right color=brown!50!black]
	(0,0) -- (0,0.8) -- (0.1,0.9) -- (0,2) -- (-3,1.9) -- (-4.3,1.2) -- (-3.0,0.2) -- cycle;
	\draw[ultra thick,fill, left color=brown!30!black, right color=brown!70!white]
	(0,2)  -- (-3,1.9) -- (-4.3,1.2) -- (-2.9,2.05) -- (-0.1,2.2) -- cycle;
	\draw[ultra thick] (-2.8182,1.5103) -- (-2.4581,1.0542) -- (-2.29,1.5103) -- (-1.9299,1.2223) -- (-1.6658,1.5584) -- (-1.5218,1.2223) -- (-1.2817,1.3423) -- (-1.0656,1.6304) -- (-0.8015,1.0542) -- (-0.8015,1.3423) -- (-0.4414,1.3423);
	\draw[ultra thick] (-2.7221,0.8141) -- (-2.4341,0.502) -- (-2.0739,0.9342) -- (-1.6898,0.478) -- (-1.2097,0.6941);
}

\newcommand{\signnaturalleftfront}{
	\draw[ultra thick,fill, left color=brown!70!white, right color=brown!50!black]
	(0,0) -- (0,0.8) -- (0.1,0.9) -- (0,2) -- (-2.5,0.1) -- (-3,-1.55) -- (-2.2,-2.3) -- cycle;
	\draw[ultra thick,fill, left color=brown!70!white, right color=brown!50!black]
	(0,2)  -- (-2.5,0.1) -- (-3,-1.55) -- (-2.2,-2.3) -- (-3.2,-1.5) -- (-2.79,0.15)-- (-0.3,2.05)   -- cycle;
	\draw[ultra thick] (-3,-1.55) -- (-3.2,-1.5) ;
	\draw[ultra thick] (-2.3435,-0.2381) -- (-1.8929,-0.4288) -- (-1.9102,0.0045) -- (-1.4769,-0.3248) -- (-1.5462,0.3685) -- (-1.3556,0.1952) -- (-1.0089,0.5418) -- (-0.9223,0.3338) -- (-0.8356,0.7325) -- (-0.5583,0.5765) -- (-0.4196,0.9925) -- (-0.2117,0.9925);
	\draw[ultra thick] (-1.9969,-1.2087) -- (-1.5809,-1.4167) -- (-1.6502,-0.9314) -- (-1.4423,-0.9141) -- (-1.3036,-0.6194) -- (-1.1476,-0.7061) -- (-1.0783,-0.4635) -- (-0.7836,-0.4115);
	\draw[ultra thick] (-0.801,-0.0128) -- (-0.5583,-0.2381) -- (-0.5583,0.2645) -- (-0.2983,0.1432) -- (-0.2463,0.2992);
}

\newcommand{\signnaturalrightfront}{
	\draw[ultra thick,fill, left color=brown!50!black, right color=brown!70!white]
	(0,0) -- (0,0.8) -- (-0.1,0.9) -- (0,2) -- (3,0.3) -- (3.5,-1.25) -- (2.5,-1.65) -- cycle;
	\draw[ultra thick,fill, left color=brown!50!black, right color=brown!70!white]
	(0,2)  -- (3,0.3) -- (3.5,-1.25) -- (2.5,-1.65) -- (3.65,-1.25) -- (3.2,0.3)-- (0.2,2.05)   -- cycle;
	\draw[ultra thick] (0.3253,1.5132) -- (0.4827,0.8521) -- (0.7135,1.1774) -- (0.8709,0.7576) -- (1.2067,1.0095) -- (1.2802,0.3589) -- (1.6894,0.4953) -- (1.9306,0.2855) -- (2.34,0.3065) -- (2.382,-0.0714) -- (2.8122,-0.1447);
	\draw[ultra thick] (0.2833,0.4009) -- (0.2833,0.0021) -- (0.5771,0.1805) -- (0.5771,-0.1658) -- (0.8604,-0.0503) -- (1.1018,-0.2287) -- (1.1438,-0.4281) -- (1.3326,-0.2812);
	\draw[ultra thick] (1.6789,-0.4701) -- (1.6894,-0.7849) -- (2.0252,-0.7324) -- (2.382,-1.0367);
}

\newcommand{\complexitysign}{
	\begin{scope}[yshift =0cm]
		\stonetwo
	\end{scope}
	\begin{scope}[xshift =2cm, yshift=8.9cm]
		\signnatural
	\end{scope}
	\begin{scope}[xshift=2cm, yshift=8.8cm]
		\signnaturalrightfront
	\end{scope}
	\begin{scope}[xshift =1cm, yshift=-0.3cm]
		\signstick
	\end{scope}
	\begin{scope}[xshift=1.2cm, yshift=9.1cm]
		\signnaturalleft
	\end{scope}
	\begin{scope}[xshift=1.4cm, yshift=8.8cm]
		\signnaturalleftfront
	\end{scope}
	\begin{scope}[xshift=3cm,yshift =-0.5cm]
		\stone
	\end{scope}
}

%% file: content/abstract.tex
\chapter*{Abstract}
This technical document presents the committee driven innovation modeling methodology \enquote{Innovation Modeling Grid} in detail.
This document is the successor of three publications on IMoG \cite{Fakih2021, klemp2023imog, shakeri2023shaping} and focuses on presenting all details of the methodology.

\subsubsection*{Acknowledgments}
This work has been supported by the GENIAL! project as funded by the German Federal Ministry of Education and Research (BMBF) under the funding code 16ES0865-16ES0876 in the ICT 2020 funding programme.

%% file: content/motivation.tex
\chapter{Introduction}
\label{chap:Introduction}

This document presents the modeling methodology Innovation Modeling Grid.
The Innovation Modeling Grid (IMoG) targets the discussion and modeling of innovations in a committee.
The methodology shall reduce the start-up time for innovation modeling by pre-structuring the innovation in the sense of advising what type of elements exist and how they relate to each other.
The modeling methodology originates from a project within the context of the automobile industry, which is used here to motivate the methodology.


The automobile industry is undergoing a major transformation and is facing the following situation:
First, there is the huge demand for autonomous and highly automated driving.
Autonomous driving shall provide a safer and more efficient transportation, while allowing passengers to focus on other things.
If the driver wants to enjoy driving, then highly automated systems shall support the driver with several assistance systems to ensure a safe journey.
This demand is on a different complexity level than the typical innovations known in the automotive industry and will shape the future development.

Secondly, there is also the huge demand of more sustainability due to the climate change.
The electrification of the transport sector and the limited amount of rare resources require new technologies and design principles.

Individualization is another demand. The passenger demands more comfort and custom functionality in vehicles.
Individualization requires a rethinking towards data driven and software defined vehicles.
Software defined vehicles also relate to high complexity and high loads of external communication.
Additionally, data driven in-vehicle applications represent potential for new business models for software companies.

Referencing business models, mobility as a service is an uprising trend as a business model for car manufacturers.
Not only conventional vehicles are considered, but also the whole transportation sector including trains, the aerospace and the last mile.
This trend requires a rethinking of the structure of the automotive industry as a whole.
The question raises of what is holding the automotive industry back of just addressing these demands today?
Well, each of the demands represent a special challenge.

\begin{wrapfigure}{l}{0.3\textwidth}
	\includegraphics[width=\linewidth]{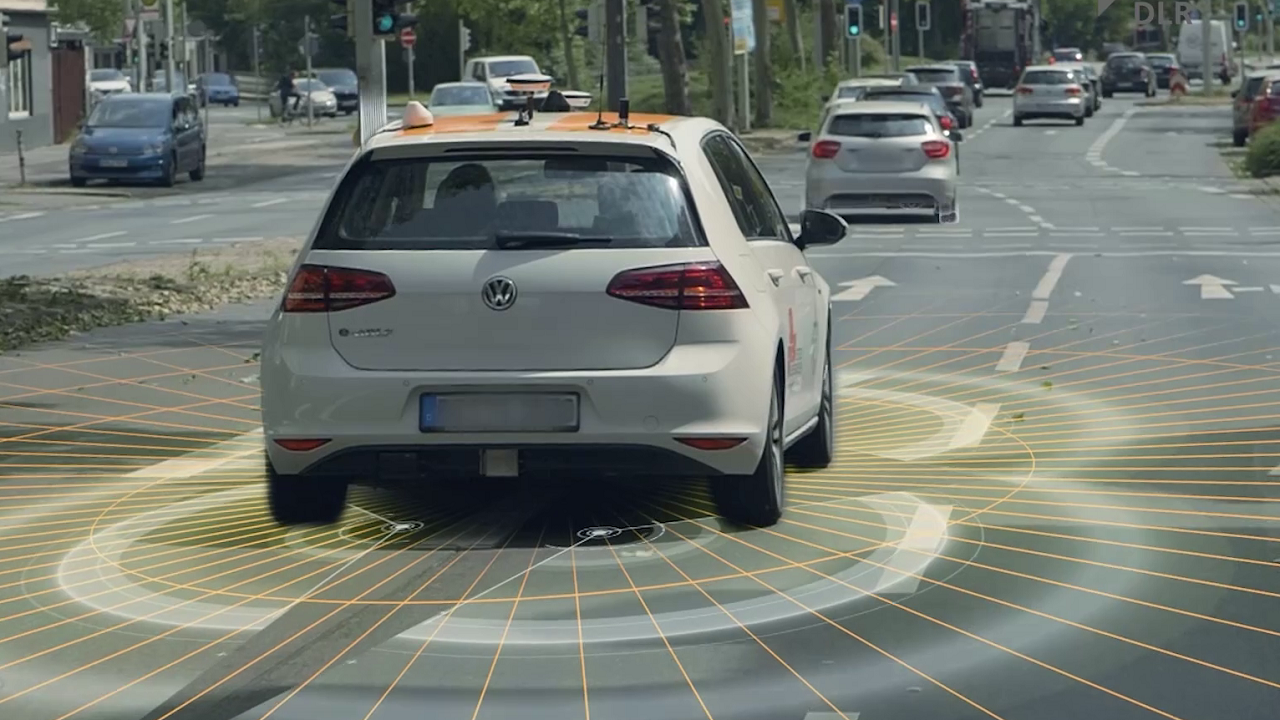}
	\footnotesize \textbf{Autonomous driving / highly automated driving}\newline \hspace*{\fill}{\scriptsize Source: DLR (CC BY-NC-ND 3.0)}
	\label{fig:demands:autonomous}
\end{wrapfigure}
The autonomous driving functionality relies on an accurate perception of the environment, an accurate localization of the car's position, an accurate prediction of the behavior of the other traffic participants – optionally by sharing the intends by communication with the other participants or with the infrastructure –, a sophisticated control of its own behavior, including a computation of trajectories, monitoring its own behavior and securing the vehicle from unauthorized manipulation and the driving functionality relies on an accurate planning and navigating of routes from point A to point B.
This functionality poses a major challenge to the industry with a high level of complexity, including high safety and security requirements.
The computation of this functionality is expected to be mainly implemented in software.
To stay competitive the automotive value chain needs to adjust to this new software focus.

\begin{wrapfigure}{l}{0.3\textwidth}
	\includegraphics[width=\linewidth]{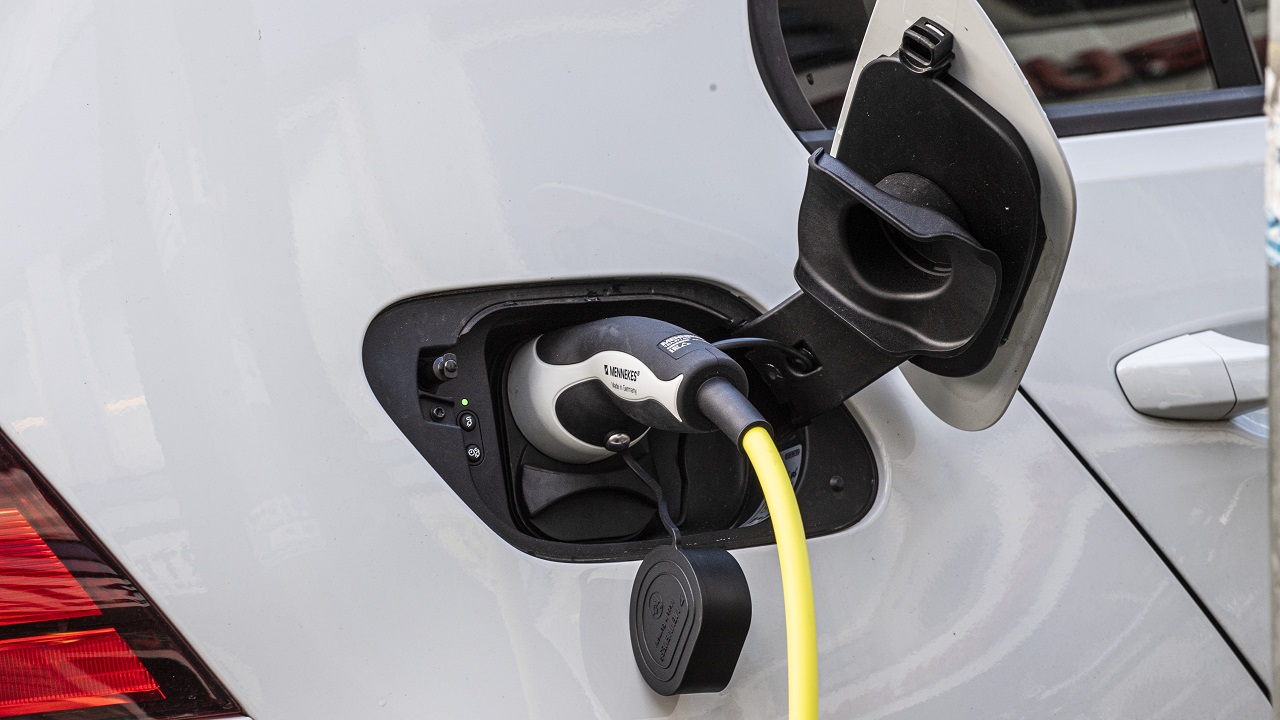}
	\footnotesize \textbf{Electrification / sustainability}\\ \hspace*{\fill}{\scriptsize Source: DLR (CC BY 3.0)}
	\label{fig:demands:sustainability}
\end{wrapfigure}
The climate change and the implied departure from combustion engines to electric engines is a another challenge.
Combustion engines are decried as not acceptable anymore as a future transport solution for the masses.
The expected technology shift goes towards the electrification of cars, which represents a far easier technology than highly optimized combustion engines for market newcomers.
On the other hand, the electrification requires a lot of rare resources - like Lithium – and requires huge accumulators.
Up-scaling the power net in the vehicle itself proposes a challenge of its own by considering electromagnetic conductivity, cable weight and so on.
These factors increase the market pressure for all members of the automotive value chain.

\begin{wrapfigure}{l}{0.3\textwidth}
	\includegraphics[width=\linewidth]{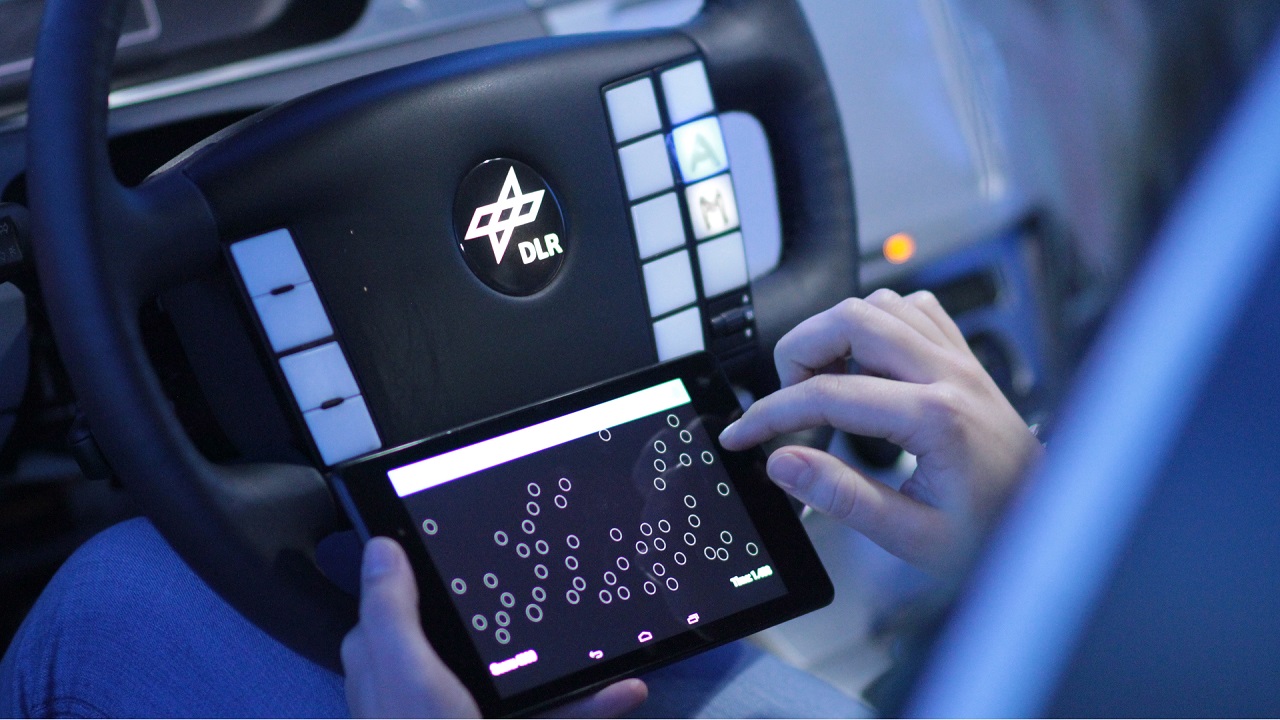}
	\footnotesize \textbf{Individualization / software defined vehicles} \hspace*{\fill}{\scriptsize Source: DLR (CC BY 3.0)}
	\label{fig:demands:individualization}
\end{wrapfigure}
The next challenge is proposed by the individualization of software defined vehicles \cite{softwareAsInnovationDriverCariad, softwareAsInnovationDriverBosch, softwareAsInnovationDriverInfineon}.
The deciding factor for autonomous vehicles lies in the software centered complexity as mentioned in the autonomous challenge.
Individualization requires on top of that short market times.
For example a user likes to connect the newest smartphone generation with the vehicle.
These vehicles can not be simply called into the next workshop to update the software, so over the air updates are required.
These demands further amplify the need for a software focus.
The value chain has to face this software focus, the implied complexity and furthermore the implied safety and security requirements.

\begin{wrapfigure}{l}{0.3\textwidth}
	\includegraphics[width=\linewidth]{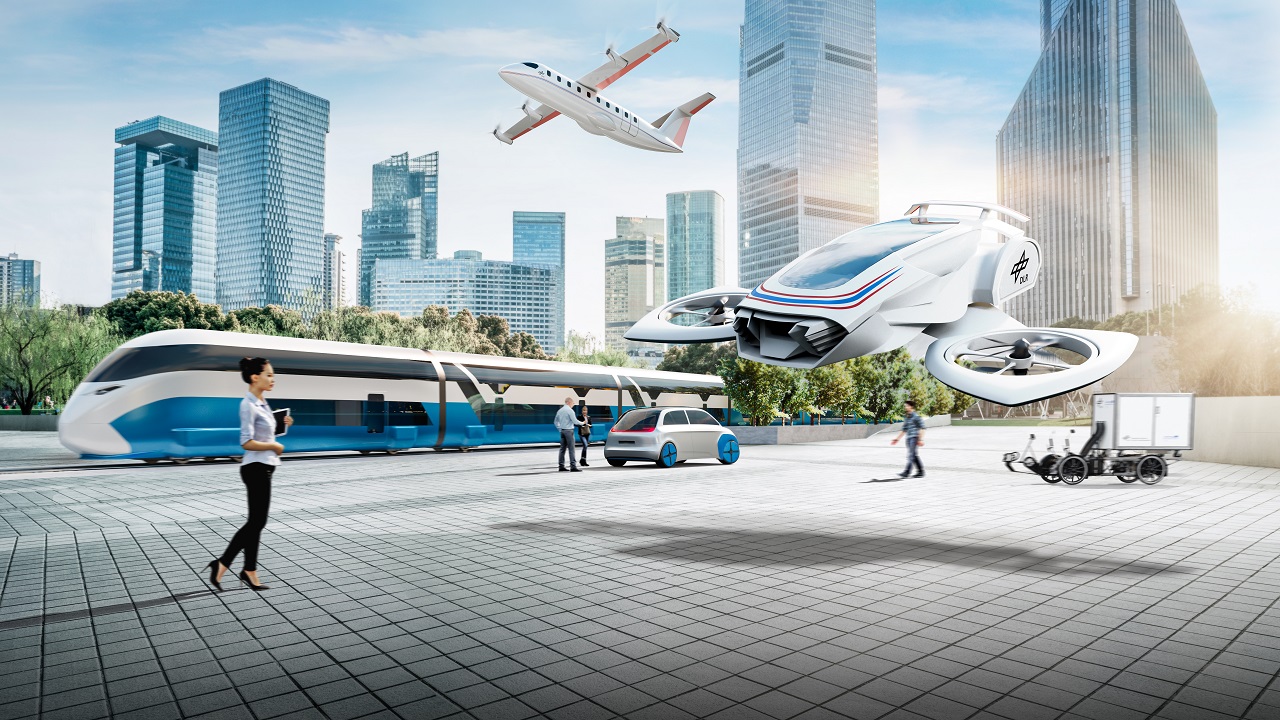}
	\footnotesize \textbf{Mobility as a service }\\ \hspace*{\fill}{\scriptsize Source: DLR (CC BY-NC-ND 3.0)}
	\label{fig:demands:maas}
\end{wrapfigure}
Finally, the general trend of car manufacturers is their move towards new business models that focus on mobility as a service.
Cause for the shift is the globalization and climate change, which demand a rethinking of mobility.
Cities get more and more interconnected with various optimized solutions for mobility: from (underground and intercity) trains, to various models of bus systems, over motorcycles, e-scooters and e-bikes for short trips.
Owning cars is thus not mandatory in larger citizens anymore and therefore, a lower demand for owning vehicles can be expected.
Additionally, improvements in the drive train of new cars tend to be insignificant from the view of the end user equalizing the quality of cars in the market and decreasing the importance of the brand of the car.
The car manufacturers have to think about how they can bind their current customers if they do not want to loose market stakes.
The general trend towards retaining current customers lies in adapting to their new demands.
Future business models are expected to be focused around the aspect of mobility as a service.
This restructuring poses a major challenge to the whole value chain.

\begin{figure}[h!]
	\centering
	\begin{tikzpicture}
		\valuechain
	\end{tikzpicture}
	\caption{A broad view on the structure of the current automotive value chain. Its understanding of the common future is not as easy as in the last decades. Thus their common invisible guideline can not be simply assumed to exist anymore. Instead the automotive value chain has to collaborate and explicitly design it to achieve their maximum efficiency.}
	\label{fig:valuechain}
\end{figure}
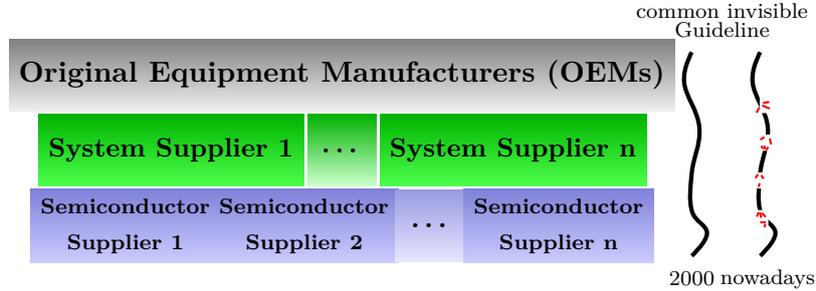
There is another challenge that is not induced by customer demands.
That is the current structure of the value chain (see Figure \ref{fig:valuechain}).
The long established automotive value chain between the Original Equipment Manufacturers (OEMs), chip manufacturers (Tier 1) and semiconductor suppliers (Tier 2) is very fragmented and optimized for producing vehicles with long product- and lifecycles \cite{pwc2018}.
This structure works well for modular design with hardware elements that have a long cycle times.
However, following this principle of modular design leads to a sequential working process, resulting in long communication times and slow innovation speed with no effective use of horizontal connections between the suppliers.
Additionally, the structure does not address the new complexity, the software focus and a service oriented business model in a suitable manner.
The demanded fast and safe realizations represent an enormous technological and methodical challenge.
On the one side, the car manufacturer has to anticipate the very rapidly changing possibilities of future microelectronic platforms, sensors and semiconductor technologies already at the time of product definition to include them in the next generation.
On the other side, the suppliers have to know early enough the requirements of future functionality to strategically invest into technology developments on a quantitative and reliable basis.
The missing communication between the value chain makes it hard to understand and predict the future and slows the value chain significantly down.

\begin{wrapfigure}{l}{0.17\textwidth}
	\centering
	\resizebox{\linewidth}{!}{
		\begin{tikzpicture}
			\complexitysign
		\end{tikzpicture}
	}
	\textbf{Uncertainty}
\end{wrapfigure}
These challenges introduce together a lot of uncertainty - a well known challenge in requirements engineering:
The complexity, including safety and security, adds uncertainty due to the vast exploration space, as not every solution can be explored.
Additionally, this complexity makes it hard to predict the future market trends and new directions of technology.
The limited resources, competition and time to market challenges add uncertainty in the sense of limited time with the pressure to find good solutions in time.
The new business models and new demands challenge the partners of the value chain by being uncertain on how they will cooperate together in the future.
Also directly related is the fact that the value chain has been indirectly guided by a common understanding of future technology over the past decade.
With the raising complexity and uncertainty this invisible guideline more and more disappears.
Overall, the uncertainty and missing of knowledge makes it hard to predict the future, plan innovations and do the right investment decisions.
The pressuring question is therefore how the value chain can sustain the new business models, autonomisation, electrification and individualization with their high requirements.

\begin{wrapfigure}{l}{0.255\textwidth}
	\includegraphics[width=\linewidth]{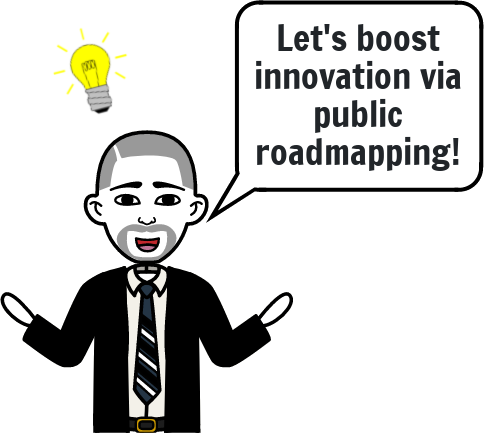}
	\tiny (Storyboard figures are owned property of \\ http://storyboardthat.com)
\end{wrapfigure}
One way to cope with these challenges is by trying to boost innovation with a public roadmapping approach.
This road mapping approach focuses on shaping the understanding of the innovation, which is in essence a requirements engineering problem.
The goal of the roadmap is to better understand and communicate future innovations, the required future technologies and the decisions about their future directions of the other partners of the value chain.
The expected gain of this synchronizing of the strategies across the value chain is an acceleration of the development of future innovative applications.

\begin{figure}[h]
	\centering
	\includegraphics[width=0.75\textwidth]{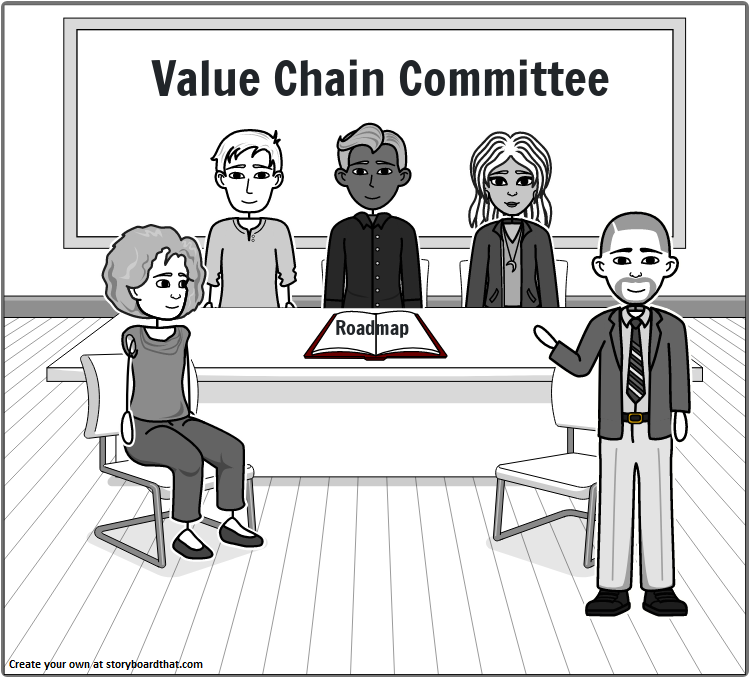}
	\caption{An example committee that aims to boost innovation with a public roadmapping approach.}
	\label{fig:committee}
\end{figure}
The approach can be understood as followed:
The automotive value chain forms a committee for creating a public roadmap on a specific innovation (see Figure \ref{fig:committee}).
This committee may include several car manufacturers (Original Equipment Manufacturers, also called OEMs), several software and hardware component suppliers (Tier 1) as well as several semiconductor suppliers (Tier 2).
The committee is open for the public and for new members to comply with the compliance laws.
The committee meets and discusses the innovation by focusing on the strategies of the stakeholders, the features and functions of the innovation and by exploring the possible solutions of the innovation.
It is crucial for the success to discuss on an appropriate level, which however varies from innovation to innovation.
An appropriate level includes the understanding of the problems and its technical constraints, but it does not include too many details about the development of the innovation as the innovation in itself is not implemented by the committee.

The immediate question appears as for every approach: “How does a consistent public road maps based information transfer in the value chain tackle the challenges?”.
By understanding the future of the innovation and the value chain, the innovation becomes plannable, which reduces the uncertainty about the future for each partner and reduces the involved risks in investment decisions.
The manufacturer can reliably plan with the discussed chip technology long before it is available.
The suppliers gain an early insight into forthcoming requirements with the certainty that their newly developed chip technologies and components will suit an existing demand.
The exploration of solutions directly helps with handling the complexity and boosting quality management.
The competition aspect is covered by using the roadmap to prepare early for the future innovation.
The discussion of the strategies helps with adjusting the value chain to a software defined focus and addressing the new business models.
Overall the public roadmap enables to adjust the value chain to the new demands.
This holistic road mapping is new to the value chain and requires to be open to the public to play within the rules of compliance.
It also enables to communicate in a horizontal manner between suppliers, that boosts synergies in the innovations design.

\begin{figure}[h]
	\centering
	\includegraphics[width=0.45\textwidth]{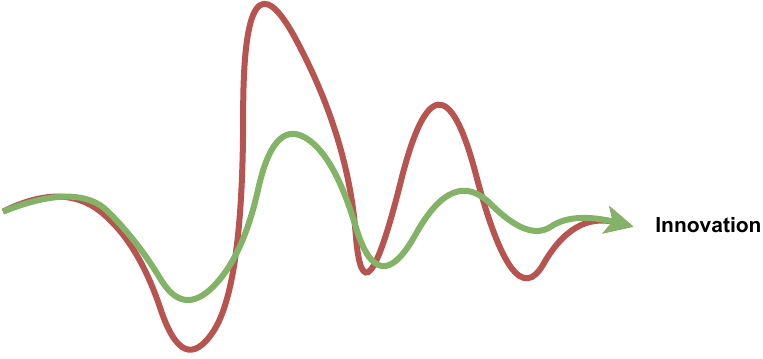}
	\includegraphics[width=0.53\textwidth]{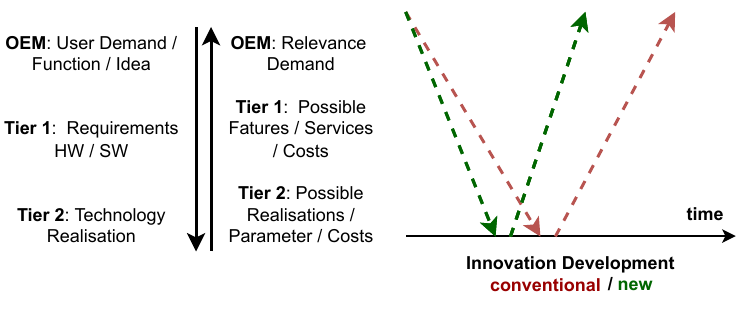}
	\caption{Expected gain of the approach.}
	\label{fig:path to innovation}
\end{figure}
One major gain is the expected speed up in the innovation cycles that is crucial to meet the short time to market demand known from software development.
This speed up can be imagined as followed (see Figure \ref{fig:path to innovation}):
A company without a roadmap explores an innovation by starting with a seemingly feasible direction.
They may spend some time exploring, explaining and discussing with other suppliers and adjusting the direction as needed.
They may find a detail that is not satisfiable by the chosen direction and thus try out the next feasible direction.
They proceed this way - with some more minor adjustments here and there - until the innovation is sufficiently explored.
A committee with a roadmap may be faster by discarding unsatisfiable directions earlier, which smoothens the path taken.
This speed is achieved for several reasons:
First, by discussing and sharing expectations with the committee, technological possibilities are better understood.
This directly leads to an earlier discarding of unsatisfiable directions.
Secondary, the value chain can parallelize the development.
This parallelization is achieved through the already mentioned reduction of uncertainties and risks, which leads to the confidence to initiate investments at an early stage.
Lastly, the committee has the opportunity to standardize common terms to significantly reduce the communication overhead via misunderstandings across the whole value chain.
Finally, a short note about the limitations of this road mapping approach.
The development and engineering tasks are unaffected by this approach.
That means that the development still has to cope with the increased complexity and the technologies have to be developed as well.

Given this road mapping approach, we investigated the research question of what an appropriate process, methodology and tool for this approach would be.
In our opinion a dedicated methodology supported by a process and a tailored tooling is required to efficiently handle this specific context.
We developed the Innovation Modeling Grid (also called IMoG) \cite{Fakih2021,imog2023} as a methodology with a process and a tooling prototype to enable an open, fair and compliant communication along the value chain.
The methodology’s goal is to efficiently represent and model early microelectronic innovations to enable a consistent information transfer along the value chain.
The process recommends who is doing what with which tool to produce which artifacts and the tooling supports as good as possible the above mentioned process and methodology.
The process, the methodology and the tooling are the main focus of this document, which are described in detail in the further sections.
The document will finish with the preliminary evaluation results and a closing.

%% file: content/delimitation.tex
\chapter{Delimitation of IMoG to relevant thematic fields}
\label{chap:scope}

IMoG relates to the umbrella term of innovation management, IMoG shares similarities with the general roadmapping approach and IMoG shares similarities with other well known fields like requirements engineering and systems engineering.

Innovation management refers to the systematic approach of planning, controlling and executing activities related to innovations.
It includes the generation of ideas, evaluation of the feasibility of the ideas, managing development prototypes and guiding the whole process up to the product.
Innovation management is used in companies to drive growth, competitiveness and sustainability.
Furthermore, innovation management also refers to the management of innovation outside of companies.
It provides means to communicate with the stakeholders of the corporation and supports the synchronization and harmonization of agreements between the corporation and external business units.

IMoG can be considered an innovation management technique, although it does not specifically target any particular company or emphasize decision-making activities.
Thus, IMoG is only applicable for the specific class of committee applications in innovation management, where the focus does not lie on presenting the decisions of a company to their stakeholder.

Roadmapping is a technique used in innovation management that relates to IMoG.
Roadmapping is a creative analysis procedure used to analyse, forecast and visualize the development paths of products, services and technologies \cite{roadmappingdef}.
Roadmapping is widely recognized as a strategic management tool to forecast the future development.
Roadmaps serve various purposes depending on the involved stakeholders.
They support achieving a robust and market-oriented technological positioning as well as for enhancing, protecting an utilizing the competence of the organization \cite{Moehrle2017}.
Furthermore, roadmaps play a crucial role in providing orientation for employees, for external stakeholder and for marketing when published.
IMoG shares many similarities with roadmapping, however, IMoG's context slightly differs from the typical roadmapping context.
The typical roadmapping approach requires perfect or quite complete knowledge about the scope of the roadmap, topics of interest and future directions of the object under consideration.
However, this knowledge is often not given in a microelectronic value chain with a huge number of different stakeholder and varying expertise.
Therefore, the IMoG methodology does not build on the assumption of perfect knowledge and includes a sophisticated investigation of the problem space before investigating the future possibilities.
Furthermore, we assume, that the topic under investigation is quite complex and that the solution space is not yet fully understood by the committee.
To tackle this complexity, the IMoG methodology recommends a model-based supported investigation of possible future solutions with dedicated tools.
To better understand the solution space and decision making, the IMoG methodology recommends to parameterize and decompose the solutions until they are sufficiently understood.
This parameterization and decomposition requires dedicated tools to handle the complexity.
Nonetheless, many typical roadmapping techniques can be applied at the later states of the IMoG methodology when the problem is better understood and when the possible solutions are collected.
The difference in imperfect knowledge and complexity also requires to handle the workshops with the committee differently to the typical roadmapping workshops.
This document gives in the later chapters recommendations on how the IMoG methodology may be applied in these workshops and how the roadmap for the microelectronic value chain as a whole can be addressed.

IMoG also shares similarities with requirements engineering.
IMoG divides the innovation modeling into the problem and solution space, which is a common approach in requirements engineering.
Furthermore the alignment and understanding of the innovation is a crucial part of IMoG, which relates to the goal of requirements engineering to foster a better understanding between all stakeholder.
IMoG distinguishes itself from requirements engineering by focusing on the class of innovation modeling in committees while requirements engineering covers the more general and abstract guidelines for stakeholder and system investigation.

Systems engineering focuses on how a system can be systematically developed and systems engineering does not specifically consider committees or abstract concepts.
IMoG also investigates the system decomposition and shares similar concepts.
However, IMoG does not require the level of detail known from systems engineering models, because the innovation is not developed by the committee members.
The development of the innovation happens after the public committee phase internally in the corporations.

%% file: content/process.tex
\chapter{Process for IMoG}
\label{chap:process}

This chapter covers the process that is recommended for the committee to create a roadmap for their innovation.
This section presents the recommended process for the committee to create an innovation roadmap, referred to as IMoG's process.
The description of IMoG's process commences by introducing the various roles involved in Section \ref{sec:process:roles}.
Subsequently, it outlines the process activities, the produced artifacts, and the tools involved in \ref{sec:process:parts}.
Notably, IMoG's process does not propose any template for milestones.
The decision to exclude such a template is based on the assumption that it would vary significantly for each specific innovation.
The process description is finally illustrated with a \enquote{Mobility with an e-scooter} innovation from the time before e-scooters got popular in cities in Section \ref{sec:process:escooter}.

\section{Roles and Responsibility}
\label{sec:process:roles}

The roles of the members of the committee are presented first.
IMoG defines three disjunctive sets of roles.
Each member of the committee may take zero, one or more roles from each role set.
This implies that each member of the committee may represent several roles and that their roles may differ depending on the task.

The first set of roles defines roles of the corporation each member may represent.
The corporation roles include the role of the OEM (Original Equipment Manufacturer), the role of the Tier 1 supplier and the role of the Tier 2 supplier.
The three roles are defined as follows (inspired by Knauf \cite{knauf}):
\begin{itemize}
	\item The \textbf{OEM} (Original Equipment Manufacturer) is the manufacturer of the end product, which deals with the market launch of the vehicle.
	\item The \textbf{Tier 1} suppliers develop system solutions that are tailored to the end product without major changes.
	\item The \textbf{Tier 2} creates the components needed to be integrated into systems. This includes the production of semiconductors and microcontrollers.
\end{itemize}
The corporation roles of the automotive value chain are often more differentiated than in OEM, Tier 1 and Tier 2.
However, the roles defined were evaluated as sufficient enough for automotive committees discussing microelectronic  innovations.

The second set of roles defines the roles of the members of the committee. The roles are described in Table \ref{tab:roles:committee}.
The third set of roles are the roles of the corporation employees, which specialize the role of the \enquote{Corporation Representative} to execute the specific activities of IMoG.
These (in-house) employees help the committee by providing and compiling information.
These roles are described in Table \ref{tab:roles:interns}.

\begin{table}[h]
	\caption{The involved roles in the automotive value chain committee}
	\begin{tabular}{p{0.22\textwidth}|p{0.75\textwidth}}
		\textbf{Roles} & \textbf{Description}\\\hline
		Committee Leader & The responsible person leading the roadmap committee.\\
		Corporation Representative & The responsible person of a corporation to coordinate the corporation internal tasks to produce the needed inputs for the roadmap.\\
		IMoG responsible Model Expert & The responsible person of creating and maintaining the IMoG model on the command of the committee members. The IMoG responsible Model Expert is also called IMoG Modeler.\\
		Roadmap Manager of the Committee & The roadmap manager of the committee is responsible for the creation and maintenance of the roadmap.
	\end{tabular}
	\label{tab:roles:committee}
\end{table}

\begin{table}[h]
	\caption{The involved roles executing the required activities of the recommended process for IMoG}
	\begin{tabular}{p{0.22\textwidth}|p{0.75\textwidth}}
		\textbf{Roles} & \textbf{Description}\\\hline
		Roadmap Manager & The roadmap manager monitors the innovation status, reports to top management on the feasibility of the innovation, surveys new technologies from other partners, and updates the roadmap. The roadmap manager investigates trends and innovations. During innovation modeling, the roadmap manager performs the initial tasks and writes the roadmap after consulting with the other domain experts, requirements engineers, and system architects.\\
		Requirements Engineer & The requirements engineer creates initial top-level requirements for the innovation and captures them uniformly (formally or in natural language). The requirements engineer leverages the expertise of the domain experts and system architects to uniformly refine the requirements in the system models.\\
		System Architect & The system architect has the role of an interdisciplinary expert who designs systems by using modeling techniques. The system architect has know-how in the area of software-hardware design. In innovation modeling, the system architect takes on the role of the innovation modeler and its decomposition into subsystems.\\
		Domain Expert & The domain expert represents a specialist of a particular discipline covering subdomains of development. The domain expert supports the innovation modeling and evaluates its influences and dependencies of certain domain elements on other domain elements.
	\end{tabular}
	\label{tab:roles:interns}
\end{table}

\rule{\textwidth}{1pt}\\%
{\centering \textit{Examples}\\}%
\rule{\textwidth}{1pt}

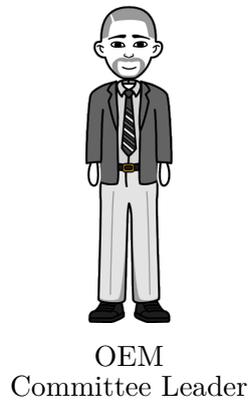
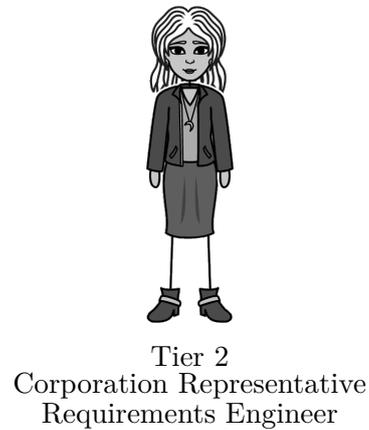
\begin{figure}[H]
	\centering
	\begin{subfigure}[t]{0.49\textwidth}
		\vspace{0pt}
		\centering
		\begin{tikzpicture}
			\node[anchor=south] (img) at (0,0) {\includegraphics[height=0.2\textheight]{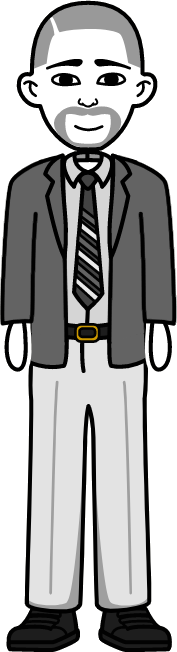}};
			\node at (0,-0.3) {OEM};
			\node at (0,-0.7) {Committee Leader};
			\node at (0,-1.1) {\phantom{Roadmap Manager}};
		\end{tikzpicture}
		\caption{The committee leader.}
		\label{fig:roles:committeeLeader}
	\end{subfigure}
	\hfill
	\begin{subfigure}[t]{0.49\textwidth}
		\vspace{0pt}
		\centering
		\begin{tikzpicture}
			\node[anchor=south] (img) at (0,0) {\includegraphics[height=0.2\textheight]{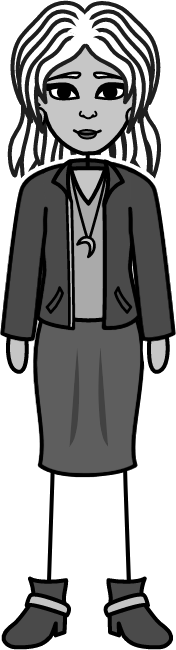}};
			\node at (0,-0.3) {Tier 2};
			\node at (0,-0.7) {Corporation Representative};
			\node at (0,-1.1) {Requirements Engineer};
		\end{tikzpicture}
		\caption{The invited Tier 2 representative.}
		\label{fig:roles:tier2}
	\end{subfigure}
	\caption[Two role examples]{Two role examples. Figures by StoryboardThat (\copyright), \url{www.storyboardthat.com}, used by permission.}
	\label{fig:roles}
\end{figure}

Two examples for a set of chosen roles are shown in the following:
A corporation member of an Original Equipment Manufacturer (see Figure \ref{fig:roles:committeeLeader}) has an idea for a new innovation he likes to discuss.
He founds a committee for discussing the new innovation and takes the role of the committee leader.
He thus have two roles assigned: The role of an OEM representative and the role of the committee leader.

The committee leader invites a member of a Tier 2 supplier to join the committee see Figure \ref{fig:roles:tier2}).
She decides to represent her corporation and play an active role in the committee.
She additionally brings her expertise as a roadmap manager.
She thus have three roles assigned: The role of a Tier 2 representative, the role of the corporation representative for the corporation she works for and the role of the roadmap manager of her corporation.

\section{Process parts: The Activities, Artifacts and Tools}
\label{sec:process:parts}
This section presents the recommended activities, the target artifacts and the recommended tools of IMoG's process.
The section starts with the overview in Section \ref{sec:process:parts:overview} of the activities, the produced artifacts  and the involved tools.
Based on this overview the process details are described from the side of the activities in Section \ref{sec:process:mnemonic}.

In chapter \ref{chap:methodology}, the IMoG methodology is introduced.
The methodology shows \textit{what} is captured in the roadmap model, in \textit{which} way these elements relate and \textit{how} details shall be processed.
However, the process description encompasses the artifacts, which represent the results of IMoG's methodology.
Because of this dependency, it is recommended to read the methodology chapter first before reading the artifact description.
Similarly, it is recommended to read the description about the involved tools after the artifacts.

\subsection{An abstract overview over the activities, artifacts and tools}
\label{sec:process:parts:overview}

\subsubsection{Activities}
\label{sec:process:activities}
\begin{figure}[H]
	\includegraphics[width=\linewidth]{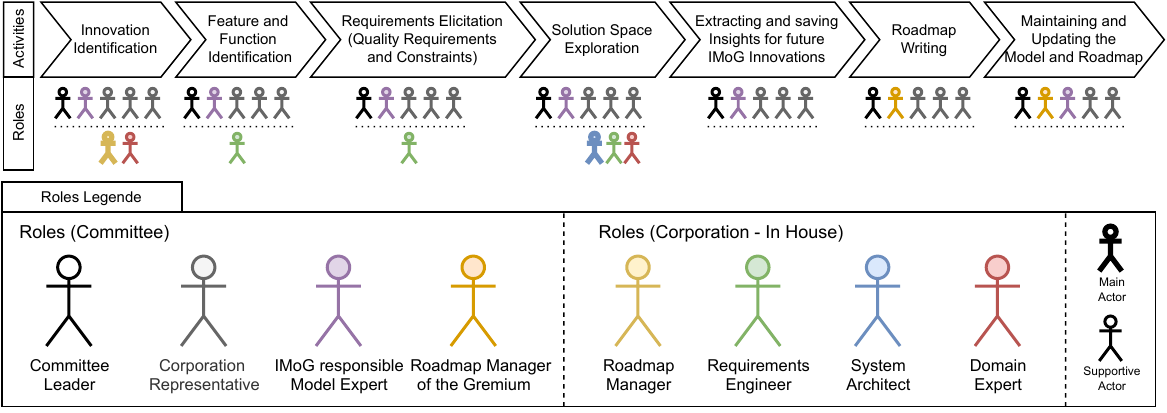}
	\caption{Activities (arrows) and roles of the working process.
		The space between the activities represents nothing special and is for the sake of the graphical representation only.
		The roles are described in Section \ref{sec:process:roles}.}
	\label{fig:process:activities}
\end{figure}

IMoG recommends seven activities for modeling the innovation.
Every of these activities is processed by people taking the recommended roles of IMoG's process, which (the roles) were presented in Section \ref{sec:process:roles}.
The mapping of which activity is processed by which roles is presented in Figure \ref{fig:process:activities}.
Note that the roles are now depicted as colored stick figures.
Despise their graphical depiction their meaning remains the same as before.
The activities are described in the following.

The first activity is called \textbf{Innovation Identification}.
The innovation identification includes creative methods as well as market segment analysis to develop a new innovation idea and create an initial description of the innovation.
The involved roles include the committee leader, the IMoG modeler and the corporation representatives.
The committee leader sets up and coordinates the meetings.
The IMoG modeler is responsible for creating the models and the corporation representatives are responsible for proposing their interests.
The in-house roles include the roadmap manager and the domain experts that help the representatives to identify and describe their interests.

The second activity is called \textbf{Feature and Function Identification}.
The purpose of this activity is to refine the problem understanding and create a feature hierarchy  based on the description of the innovation.
As a further refinement, the feature hierarchy may include user stories and use cases.
The involved roles include the same committee members of the innovation identification activity:  the committee leader, the IMoG modeler and the corporation representatives.
Requirements engineers of the corporations support the creation of the feature hierarchy.

The third activity is called \textbf{Requirements Elicitation}, which adds quality requirements and constraints to the feature hierarchy and refines the problem space further.
It is the last activity focusing on the problem space.
The roles that are involved in this activity are the same as in feature and function identification activity.

The solution space of the innovation is examined after the problem is sufficiently understood.
The corresponding activity is called \textbf{Solution Space Exploration}.
It consists of modeling the possible solutions of the innovation with (sufficient) technical details.
The involved committee roles include the committee leader, the IMoG modeler and the corporation representatives.
The corporation internal leader of this activity is the system architect to examine and analyze the possible solutions.
The system architect gets support from the requirements engineer and the domain expert, however, their help is of supportive nature.

After the solutions are examined, the committee extracts the insights gained by the generated model and saves them in their database for further innovations.
This activity is called \textbf{Extraction and Saving of the Insights}.
No in-house corporation roles are needed.

The \textbf{roadmap writing} is the next activity building upon the insights from the last activity.
The committee members meets again to discuss the roadmap together.
The modeling activities are finished and thus the IMoG modeler does not take part in this activity.
The roadmap manager takes responsibility for the roadmap writing, structures the document, and assigns tasks.
After this activity, the main roadmapping activies are done.

Based on this roadmap, reoccurring meetings are established to \textbf{maintain and update} the roadmap.
The same roles are involved as in the writing of the roadmap.

It is not required to complete each of the seven activities before starting the next one (as usual).
Instead, it is sufficient to draft each model of each activity and refine them when necessary, similarly to what was proposed with the twin peaks model \cite{nuseibeh2001weaving}.

\subsubsection{Artifacts}
\label{sec:process:artifacts}
\vspace{-0.5cm}
\textit{(It is recommended to read Chapter \ref{chap:methodology} before this Section.)}\\

\begin{figure}[H]
	\includegraphics[width=\linewidth]{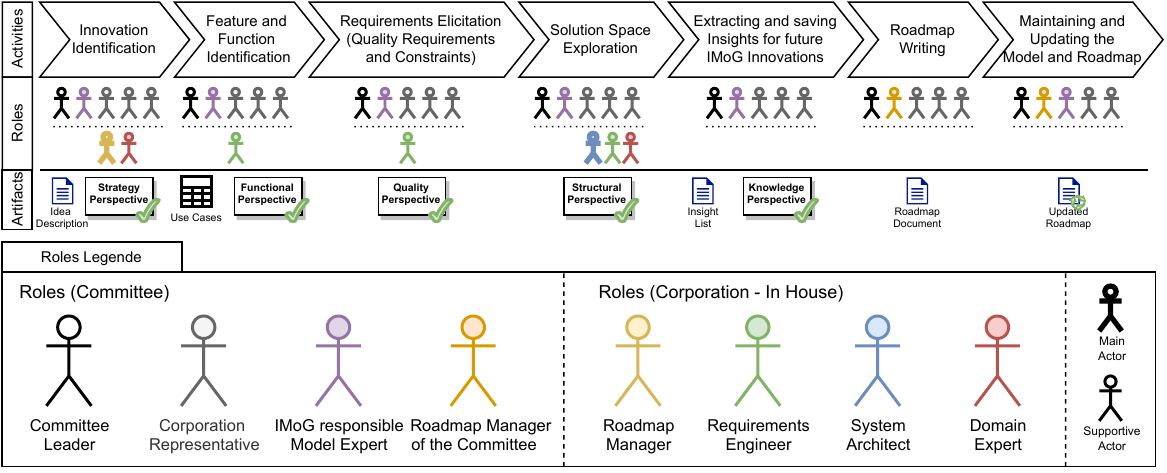}
	\caption{IMoG process with artifacts appended}
	\label{fig:process:artifacts}
\end{figure}

The artifacts are also added to the process image in Figure \ref{fig:process:artifacts}.
The idea description and the filled Strategy Perspective constitute the artifacts of the \enquote{Innovation Identification} activity.
The details of the perspective are presented in Chapter \ref{chap:methodology}.
The artifacts of the \enquote{Feature and Function Identification} activity are the user stories, use cases and the filled Functional Perspective.
The filled Quality Perspective constitutes the artifact of the \enquote{Requirements Elicitation (Quality Requirements and Constraints)} activity.
The filled Structural Perspective constitutes the artifact of the \enquote{Solution Space Exploration} activity.
The (Domain) Knowledge Perspective and the list of insights constitute the artifacts of the \enquote{Extracting and saving Insights for future IMoG Innovations} activity.
Finally, the roadmap is the artifact of the \enquote{roadmap writing} activity, which is updated within the \enquote{maintaining and updating the model and the roadmap} activity.
The artifacts are illustrated with the presentation of the perspectives in Chapter \ref{chap:methodology}

\subsubsection{Tools}
\label{sec:process:tools}
\vspace{-0.5cm}
\textit{(It is recommended to read Chapter \ref{chap:methodology} before this Section.)}\\

\begin{figure}[H]
	\includegraphics[width=\linewidth]{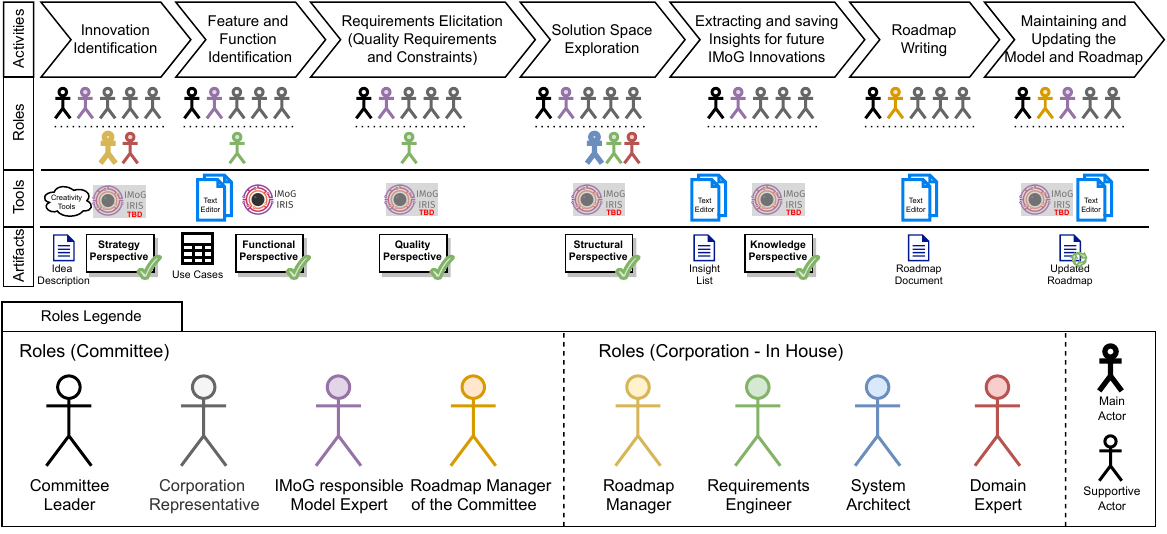}
	\caption{IMoG's process description appended with the proposed tooling (final representation).}
	\label{fig:imog-process}
\end{figure}

The proposed tools are added to the process image in Figure \ref{fig:imog-process}.
Overall, we think that a dedicated tooling for IMoG is required and thus the recommended tooling for IMoG shown in the figure is such dedicated tooling.
A tooling prototype for the Functional Perspective is already implemented and called \enquote{IMoG IRIS} prototype.
Unfortunately, the resources for implementing were not enough to extend the dedicated prototype for the remaining activities.
These left open implementations are marked in the figure with \enquote{To Be Done}.

Next to a dedicated tooling, some activities are best supported by using extra tools:
The \enquote{Innovation Identification} activity would be best supported by a creativity tool that the committee is well versed with.
This, for example, may be a mind mapping tool, some whiteboards or something else.
The \enquote{Feature and Function Identification} activity would be best supported with a dedicated tool to create and manage use cases and user stories.
This could be a common text and table manipulation tool or a more sophisticated requirements engineering tool that supports user stories and use cases well.
The \enquote{Extracting and saving Insights for future IMoG Innovations} activity would be best supported by a text editor to write the insights down.
A text editor is helpful for the \enquote{Roadmap Writing} activity and the roadmap updating activity.

\subsection{Detailed activities description}
\label{sec:process:mnemonic}
This section describes each activity in detail.

\begin{tabular}{p{\textwidth}}
	{(Activity) \Large \textbf{Innovation Identification}}\\
	\begin{center}
		\begin{tikzpicture}
			\processarrow{Innovation Identification};
		\end{tikzpicture}
	\end{center}\\
	\vspace{-1cm}
	\begin{center}
		\includegraphics{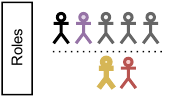}
	\end{center}\\
	{(Roles) The involved roles include the committee leader to set up and manage the meetings, the IMoG modeler responsible for creating the models and the corporation representatives in the roles of the roadmap manager for proposing their interests in the innovations as well as some domain experts for supporting the roadmap managers.}\\\\
	{(Short activity description) This activity uses creative methods as well as market segment analysis to develop a new innovation idea and create an initial description (see figure \ref{fig:mnemo:process:strat-diagram}).}\\
	\vspace{0.25cm}(Sub-activities)\vspace{-0.25cm}
	\begin{center}
		\begin{tikzpicture}
			\processarrow{Creative methods};
			\processarrow[2]{Innovation description};
			\processarrow[3]{\small Strategy Perspective modeling};
			\processarrow[4]{Refinement};
		\end{tikzpicture}
	\end{center}\\
	\vspace{-0.5cm}
	{(Detailed description) The committee leader invites the committee members with the above mentioned roles to a meeting to develop a new innovation idea.
	The committee decides on a creative method or decides on a market analysis technique and carries out the creative method.
	Which creative method or market analysis to choose is not defined nor restricted.
	Creative methods include for example \enquote{Brainstorming}, \enquote{Mind maps}, \enquote{Zwicky Boxes}, \enquote{Walt Disney method}, \enquote{Scenario projection}, etc.
	Market analysis include for example \enquote{User needs projection}, \enquote{User stories}, \enquote{Time to market analysis} or \enquote{Business model analysis}.
	The method that fits the committee members and their idea the best is the one to choose.
	Once the committee finished the creative method and closes the meeting, the committee leader writes down a description of the result of the creative method.
	Based on this description the IMoG responsible person translates this description into a draft of the Strategy Perspective (see Section \ref{sec:imog:methodology:strategy}).
	Now, the committee refines this model of the Strategy Perspective in a few more meetings or by assigning personal tasks.
	The refinement process can also include refinement and review processes in the corporations itself by ask internals (probably people with the roadmap manager role or domain expert role) to give their inputs.
	The input may include more information about the innovation, refined descriptions, goals or identify elements that shall be traced.
	This refinement process goes on until they are sufficiently satisfied with the result (Strategy Perspective).}\\
\end{tabular}

\begin{tabular}{p{\textwidth}}
	\begin{center}
		\includegraphics{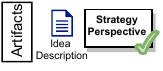}
	\end{center}\\
	\vspace{-0.5cm}
	{(Artifacts) The innovation description (including the common vision and possibly some diagrams) and the filled Strategy Perspective (presenting the vision, the diagrams as well as the stakeholders interests, concerns and strategy and textual goals) constitute the artifacts of the \enquote{Innovation Identification} activity.}\\
	\vspace{-0.5cm}
	\begin{center}
		\includegraphics{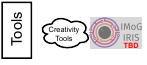}
	\end{center}\\
	\vspace{-0.5cm}
	{(Tools) Overall, we think that a dedicated tooling for IMoG is needed and thus the proposed tool for IMoG would be such dedicated tooling.
	We already implemented a tooling prototype for the Functional Perspective, called \enquote{IMoG IRIS} prototype.
	Unfortunately, we do not have enough resources to extend the dedicated prototype (IMoG IRIS) for the Strategy Perspective activities.
	Additionally, this activity is best supported by a creativity tool that the individual committee that they efficiently use already.
	This creativity tool may be any tool that supports the creative techniques and analysis (paper, mindmaps, documents, scratchboards, drawio, etc...).}\\
\end{tabular}

\begin{figure}[h]
	\includegraphics[width=\linewidth]{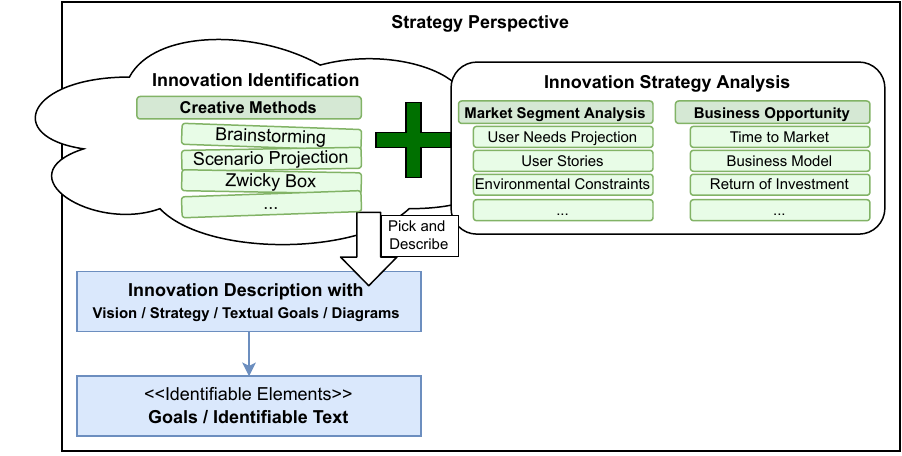}
	\caption{Possible methods for the Innovation Identification activity.}
	\label{fig:mnemo:process:strat-diagram}
\end{figure}

\newpage
\begin{tabular}{p{\textwidth}}
	{(Activity) \Large \textbf{Feature and Function Identification}}\\
	\begin{center}
		\begin{tikzpicture}
			\processarrow{Feature and Function Identification};
		\end{tikzpicture}
	\end{center}\\
	\vspace{-1cm}
	\begin{center}
		\includegraphics{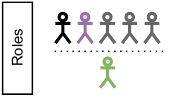}
	\end{center}\\
	{(Roles) The involved roles include the committee leader, the IMoG modeler and the corporation representatives.
	In-house requirements engineers are also involved in this activity.}\\\\
	{(Short activity description) The goal of this activity and the Functional Perspective is to refine the problem space and create a feature hierarchy including optional user stories and use cases based on the description of the innovation.}\\
	\vspace{0.25cm}(Sub-activities)\vspace{-0.25cm}
	\begin{center}
		\begin{tikzpicture}
			\processarrow{User Stories and Use Cases};
			\processarrow[2]{Functional Perspective modeling};
		\end{tikzpicture}
	\end{center}\\
	\vspace{-0.5cm}
	{(Detailed description) After the identification of the innovation in the \enquote{Innovation Identification} activity, the next activity focuses on the identification of the features and functions needed to fulfill the innovation.
	The features and functions shall represent a refining of the problem space of the innovation.
	The committee leader invites the committee members with the above mentioned roles to a meeting.
	First they decide if they want to create user stories and use cases for understanding the general conditions of their innovation or if the general conditions of the innovation are sufficiently understood without user stories and use cases.
	If they decide to create user stories and use cases, they use the meeting to identify the user stories and use cases.
	The corporation representatives are responsible for giving and checking the input for the user stories and use cases.
	They may request their in-house requirements engineers for supporting this task.
	The IMoG modeler supports the corporation representatives by creating templates and giving advice for formulation.
	The committee leader moderates the meetings.
	The outcome of the meeting is a draft of these user stories and use cases.
	The members of the committee then distribute tasks to refine the user stories and use cases to a sufficient degree.
	They meet and refine again until they are sufficiently satisfied with the result.}\\
\end{tabular}

\begin{tabular}{p{\textwidth}}
	{(Detailed description continued) Then the committee leader invites the committee members for another meeting to create a draft of the feature model in the Functional Perspective.
	The IMoG modeler creates a draft of the Functional Perspective based on the inputs of the corporation representatives.
	The committee refines the model by giving input via in-house meetings of the corporations and by additional committee meetings.
	The in-house requirements engineers help the corporation representative and checks the validity and consistency of the Functional Perspective.
	This refinement process goes on until they are sufficiently happy with the result (Strategy Perspective).}\\
	\begin{center}
		\includegraphics{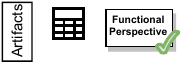}
	\end{center}\\
	\vspace{-0.5cm}
	{(Artifacts) The artifacts of the \enquote{Feature and Function Identification} activity are the features and functions in the form of the Functional Perspective and the optional user stories or use cases that are mapped on the features and functions.
	All features, functions, user stories, use cases as well as all other information are represented in the Functional Perspective.}\\
	\vspace{-0.5cm}
	\begin{center}
		\includegraphics{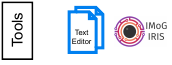}
	\end{center}\\
	\vspace{-0.5cm}
	{(Tools) Overall, we think that a dedicated tooling for IMoG is needed and thus the proposed tool for IMoG would be such dedicated tooling.
	We implemented a tooling prototype for the Functional Perspective, called \enquote{IMoG IRIS} prototype.
	Additionally, this activity is best supported with a dedicated tool to create and manage the use cases and user stories.
	This could be a common text and table manipulation tool or a more sophisticated requirements engineering tool that supports well user stories and use cases.}\\
\end{tabular}

\newpage
\begin{tabular}{p{\textwidth}}
	{(Activity) \Large \textbf{Requirements Elicitation} \normalsize(Quality Requirements and Constraints)}\\
	\begin{center}
		\begin{tikzpicture}
			\processarrow{Requirements Elicitation};
		\end{tikzpicture}
	\end{center}\\
	\vspace{-1cm}
	\begin{center}
		\includegraphics{figures/fp_process_roles.pdf}
	\end{center}\\
	{(Roles) The involved roles include the committee leader, the IMoG modeler and the corporation representatives.
		In-house requirements engineers are involved in this activity.}\\\\
	{(Short activity description) The goal of this activity and the Quality Perspective is to refine the problem space by adding quality requirements and constraints to the feature hierarchy and finishing the refinement of the problem space.}\\
	\vspace{0.25cm}(Sub-activities)\vspace{-0.25cm}
	\begin{center}
		\begin{tikzpicture}
			\processarrow{\small Adding already recorded Requirements};
			\processarrow[2]{\small Requirement Elicitation};
			\processarrow[3]{\small Requirement Analysis};
			\processarrow[4]{\small Requirement Documentation};
			\processarrow[5]{\small Requirement Validation and Verification};
		\end{tikzpicture}
	\end{center}\\
	\vspace{-0.5cm}
	{(Detailed description) The committee leader invites once again the committee members with the above mentioned roles to a meeting.
	Every quality requirement and constraint that came up during the identification of the features and functions is now placed into the Quality Perspective of IMoG and mapped on the features and functions of the Functional Perspective.
	(Note: Process Requirements are not relevant, because the innovation is not built in the committee!).
	Afterwards, a dedicated round of meetings moderated by the committee leader and focusing on the structured elicitation of missing quality requirements and constraints is started.
	These meetings follow the typical steps of requirement engineering (requirements elicitation, requirements analysis requirements documentation, requirements verification and validation \cite{iso29148}).
	The corporation representatives are responsible for eliciting the requirements and checking the consistency of the requirements.
	They may request their in-house requirements engineers for supporting this task.
	The IMoG modeler supports the corporation representatives by filling the requirements into the Quality Perspective.
	They stop the rounds of meetings once they are sufficiently satisfied with the result.
	After the meetings, the modeling of requirements and the problem space is done.}\\
\end{tabular}

\begin{tabular}{p{\textwidth}}
	\begin{center}
		\includegraphics{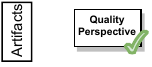}
	\end{center}\\
	\vspace{-0.5cm}
	{(Artifacts) The artifacts of the \enquote{Requirements Elicitation} activity are the added quality requirements and constraints in the Quality Perspective, which are mapped on the features and functions of the Functional Perspective.
	With the requirements elicited the modeling of the problem space is finished (or at least interpreted as a draft like in agile work / twin peaks \cite{nuseibeh2001weaving}).}\\
	\vspace{-0.5cm}
	\begin{center}
		\includegraphics{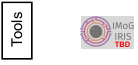}
	\end{center}\\
	\vspace{-0.5cm}
	{(Tools) Overall, we think that a dedicated tooling for IMoG is needed and thus the proposed tool for IMoG would be such dedicated tooling.
	We already implemented a tooling prototype for the Functional Perspective, called \enquote{IMoG IRIS} prototype.
	Unfortunately, we do not have enough resources to extend the dedicated prototype (IMoG IRIS) for the Quality Perspective activities.
	Thus standard requirements managing tools like IBM Rational DOORS or Jama are recommended.}\\
\end{tabular}

\newpage
\begin{tabular}{p{\textwidth}}
	{(Activity) \Large \textbf{Solution Space Exploration}}\\
	\begin{center}
		\begin{tikzpicture}
			\processarrow{Solution Space Exploration};
		\end{tikzpicture}
	\end{center}\\
	\vspace{-1cm}
	\begin{center}
		\includegraphics{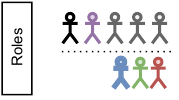}
	\end{center}\\
	{(Roles) The involved roles include the committee leader, the IMoG modeler and the corporation representatives.
		In-house requirements engineers are involved in this activity.
		The leader of this task is the system architect to examine and analyze the possible solutions.
		The system architect gets support from the requirements engineer and the domain expert, however, their help is of supportive nature.}\\\\
	{(Short activity description) The goal of this activity and the goal of the Structural Perspective is to model the solution space of the innovation  with (sufficient) technical details.}\\
	\vspace{0.25cm}(Sub-activities)\vspace{-0.25cm}
	\begin{center}
		\begin{tikzpicture}
			\processarrow{\small Context Level modeling};
			\processarrow[2]{\small System \\Decomposition and FP mapping};
			\processarrow[3]{\small Effect Chain \\and Impact Analysis};
			\processarrow[4]{\small Requirements Elicitation for solutions};
			\processarrow[5]{\small Alternatives Exploration};
		\end{tikzpicture}
	\end{center}\\
	\vspace{-0.5cm}
	{(Detailed description) The modeling of the solution space is the next step.
	The committee leader invites the committee members with the above mentioned roles to explore and discuss the possible solutions.
	The exploration of the solutions may include the following steps (the order does not have to be strictly followed):
	Starting with the context level, a model describing the environment of the intended innovation solution and the innovation itself is designed.
	In this context model, the innovation can be understood and represented as a black box (meaning that the innovation is not decomposed or any of its parts further described).
	The next step may include the system decomposition focusing on how the innovation can be constructed.
	The innovation is represented in detail (white box).
	The general decomposition concepts of using logical components, \enquote{solution principles} (e.g. combustion or electric) and actual solutions (e.g. specific engines) as well as hardware and software mappings are part of the system decomposition step.
	This step may also include the mapping of the features and functions of the Functional Perspective on the components of the system decomposition to achieve traceability to the problem space.
	The third step may include an effect chain modeling to depict and analyze the connections between the innovation components (system decomposition parts) and its environment.}\\
\end{tabular}

\begin{tabular}{p{\textwidth}}
	{(Detailed description continued)
	The analysis allows to understand the dependencies between the innovation components and its environment and their impact on changes.
	The fourth step may include the structured elicitation of missing quality requirements and constraints for the solutions.
	This step uses the same steps already mentioned in the Quality Perspective (Adding solution requirements to QP, requirements elicitation, requirements analysis requirements documentation, requirements verification and validation \cite{iso29148}).
	The last step may include an alternatives exploration including the use of Key Performance Indicators (KPIs) to describe the possible alternatives of the system components and their advantages and limitations.

	These steps can be divided over a series of meetings with internal discussions in the corporations, where the committee leader manages the formalities.
	The IMoG modeler and the corporation representative including their internal roles of the system architect, requirements engineer and domain expert, are responsible for exploring the solution space using the five steps.
	Additionally, the IMoG modeler is responsible for the creation of the Structural Perspective.
	The importance of each internal role varies between the steps:
	The system architect is most important during the context level modeling, the system decomposition and FP mapping, the effect chain analysis and the alternatives exploration.
	The system architect however plays a smaller role in the requirements elicitation for solutions step, where the requirements engineer has the responsibility.
	On the other side the requirements engineer plays a less important role in the other steps.
	The domain experts gives their expertise in all steps.
	However, their input is especially in the system decomposition, the effect chain modeling and in the alternatives exploration needed.

	After the steps, the modeling of requirements and the solution space is done.}\\
	\vspace{-0.5cm}
	\begin{center}
		\includegraphics{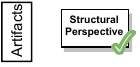}
	\end{center}\\
	\vspace{-0.5cm}
	{(Artifacts) The artifacts of the \enquote{Solution Space Exploration} activity are the model of the solutions covered in the Structural Perspective and its dependencies to the other perspectives.
	This includes the added quality requirements and constraints to the Quality Perspective and the mapping on the features and functions of the Functional Perspective.
	The solution space exploration may include a context model and the decomposition of the innovation including effects and alternatives.}\\
	\vspace{-0.5cm}
	\begin{center}
		\includegraphics{figures/qp_process_tools.pdf}
	\end{center}\\
	\vspace{-0.5cm}
	{(Tools) Overall, we think that a dedicated tooling for IMoG is needed and thus the proposed tool for IMoG would be such dedicated tooling.
	We already implemented a tooling prototype for the Functional Perspective, called \enquote{IMoG IRIS} prototype.
	Unfortunately, we do not have enough resources to extend the dedicated prototype (IMoG IRIS) for the Structural Perspective activities.
	Thus standard system modeling tools like any UML tool or SysML tool are recommended.}\\
\end{tabular}

\newpage
\begin{tabular}{p{\textwidth}}
	{(Activity) \Large \textbf{Extracting and saving insights} \normalsize(for future IMoG innovations)}\\
	\begin{center}
		\begin{tikzpicture}
			\processarrow{Extracting and saving Insights};
		\end{tikzpicture}
	\end{center}\\
	\vspace{-1cm}
	\begin{center}
		\includegraphics{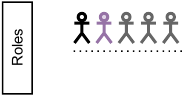}
	\end{center}\\
	{(Roles) The involved roles include the committee leader, the IMoG modeler and the corporation representatives.
		In-house roles are not needed in this activity.}\\\\
	{(Short activity description) The goal of this activity and the (Domain) Knowledge Perspective is to extract the insights of the innovation to use them as a basis for the roadmap and to save the insights of the innovation for future IMoG innovations.}\\
	\vspace{0.25cm}(Sub-activities)\vspace{-0.25cm}
	\begin{center}
		\begin{tikzpicture}
			\processarrow{Extracting Insights};
			\processarrow[2]{\small Saving IMoG elements into the database};
		\end{tikzpicture}
	\end{center}\\
	\vspace{-0.5cm}
	{(Detailed description)
	The modeling of the problem space and solution space is finished.
	The committee leader invites the committee members with the above mentioned roles to extract the insights of the model to use them as a basis for the roadmap.
	The insight extraction is done by the committee members.
	This includes the committee leader, the IMoG modeler and the corporation representatives.
	The outcome of this first activity is the list of insights written down into a document.
	Afterwards the IMoG modeler exports the IMoG elements into a publicly available database.
	For this activity, the IMoG modeler asks the committee members which elements to save and which not.
	The IMoG modeler suggests dependencies to draw between IMoG elements and already available information and discusses these suggestions with the committee members.
	The committee members may also suggest dependencies.
	The outcome of this second activity is the publicly available database enhanced by elements of the regarded innovation.}\\
\end{tabular}

\begin{tabular}{p{\textwidth}}
	\begin{center}
		\includegraphics{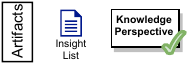}
	\end{center}\\
	\vspace{-0.5cm}
	{(Artifacts) The artifacts of the \enquote{Extracting and saving insight} activity are a list of insights as a basis for the roadmap activities and the publicly available database enhanced by elements of the regarded innovation.}\\
	\vspace{-0.5cm}
	\begin{center}
		\includegraphics{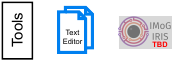}
	\end{center}\\
	\vspace{-0.5cm}
	{(Tools) Overall, we think that a dedicated tooling for IMoG including a publicly available cloud service containing the IMoG model is needed and thus the proposed tool for IMoG would be such dedicated tooling.
	We already implemented a tooling prototype for the Functional Perspective, called \enquote{IMoG IRIS} prototype.
	Unfortunately, we do not have enough resources to extend the dedicated prototype (IMoG IRIS) for the Domain Knowledge Perspective activities.
	Thus standard text editors and databases are recommended.}\\
\end{tabular}

\newpage
\begin{tabular}{p{\textwidth}}
	{(Activity) \Large \textbf{Roadmap Writing}}\\
	\begin{center}
		\begin{tikzpicture}
			\processarrow{Roadmap Writing};
		\end{tikzpicture}
	\end{center}\\
	\vspace{-1cm}
	\begin{center}
		\includegraphics{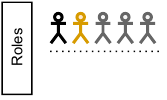}
	\end{center}\\
	{(Roles) The involved roles include the committee leader, the roadmap manager of the committee and the corporation representatives.
	The roadmap manager takes responsibility for the roadmap writing, structures the document, and assigns tasks.
	The corporation representatives are responsible for giving sufficient help in the roadmap creation, reviewing and refinement.
	In-house roles are not needed in this activity.}\\\\
	{(Short activity description) The goal of this activity is to use the extracted insights of the innovation to write the roadmap.}\\\\
	{(Detailed description)
	The modeling activities are finished and the insights of the innovation are extracted.
	The committee leader invites the roadmap manager of the committee and the committee members to write the roadmap based on the insights of the innovation.
	The roadmap manager of the committee creates a first draft of the document structure with sufficient support of the committee members and then refines the roadmap together with the committee members.
	This refinement process goes on until they are sufficiently satisfied with the roadmap.}\\
	\vspace{-0.5cm}
	\begin{center}
		\includegraphics{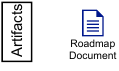}
	\end{center}\\
	\vspace{-0.5cm}
	{(Artifacts) The artifact of the \enquote{Roadmap Writing} activity is the roadmap.}\\
	\vspace{-0.5cm}
	\begin{center}
		\includegraphics{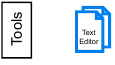}
	\end{center}\\
	\vspace{-0.5cm}
	{(Tools) Standard text editors (e.g., \LaTeX, word, etc.) are recommended.}\\
\end{tabular}

\newpage
\begin{tabular}{p{\textwidth}}
	{(Activity) \Large \textbf{Maintain and update the roadmap}}\\
	\begin{center}
		\begin{tikzpicture}
			\processarrow{Maintain and Update the Roadmap};
		\end{tikzpicture}
	\end{center}\\
	\vspace{-1cm}
	\begin{center}
		\includegraphics{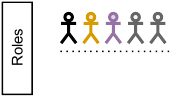}
	\end{center}\\
	{(Roles) The involved roles include the committee leader, committee leader, the IMoG modeler, the roadmap manager of the committee and the corporation representatives.
	In-house roles are not needed in this activity.}\\\\
	{(Short activity description) Reoccurring meetings are established to maintain and update the roadmap.}\\\\
	\vspace{-0.5cm}
	{(Detailed description)
	The roadmap is written!
	Now the committee meets once in a defined time frame to maintain and update the IMoG model and roadmap.
	The committee leader manages the formalities, the IMoG modeler is responsible for updating of the model and the roadmap manager of the committee is responsible for updating the roadmap.
	The corporation representatives are the most important members here as they decide in which direction the roadmap should be point.
	The outcome of this activity is the updated model and roadmap.}\\
	\vspace{-0.5cm}
	\begin{center}
		\includegraphics{figures/rw_process_artifacts.pdf}
	\end{center}\\
	\vspace{-0.5cm}
	{(Artifacts) The artifacts of the \enquote{Maintain and Update the Roadmap} activity are an updated model and an updated roadmap.}\\
	\vspace{-0.5cm}
	\begin{center}
		\includegraphics{figures/rw_process_tools.pdf}
	\end{center}\\
	\vspace{-0.5cm}
	{(Tools) All of the tools mentioned in the other activities are required to maintain the model and the roadmap.}\\
\end{tabular}

\section{Example -- Mobility with an e-scooter}
\label{sec:process:escooter}

\begin{figure}
	\begin{minipage}{0.475\textwidth}
		\includegraphics[width=\textwidth]{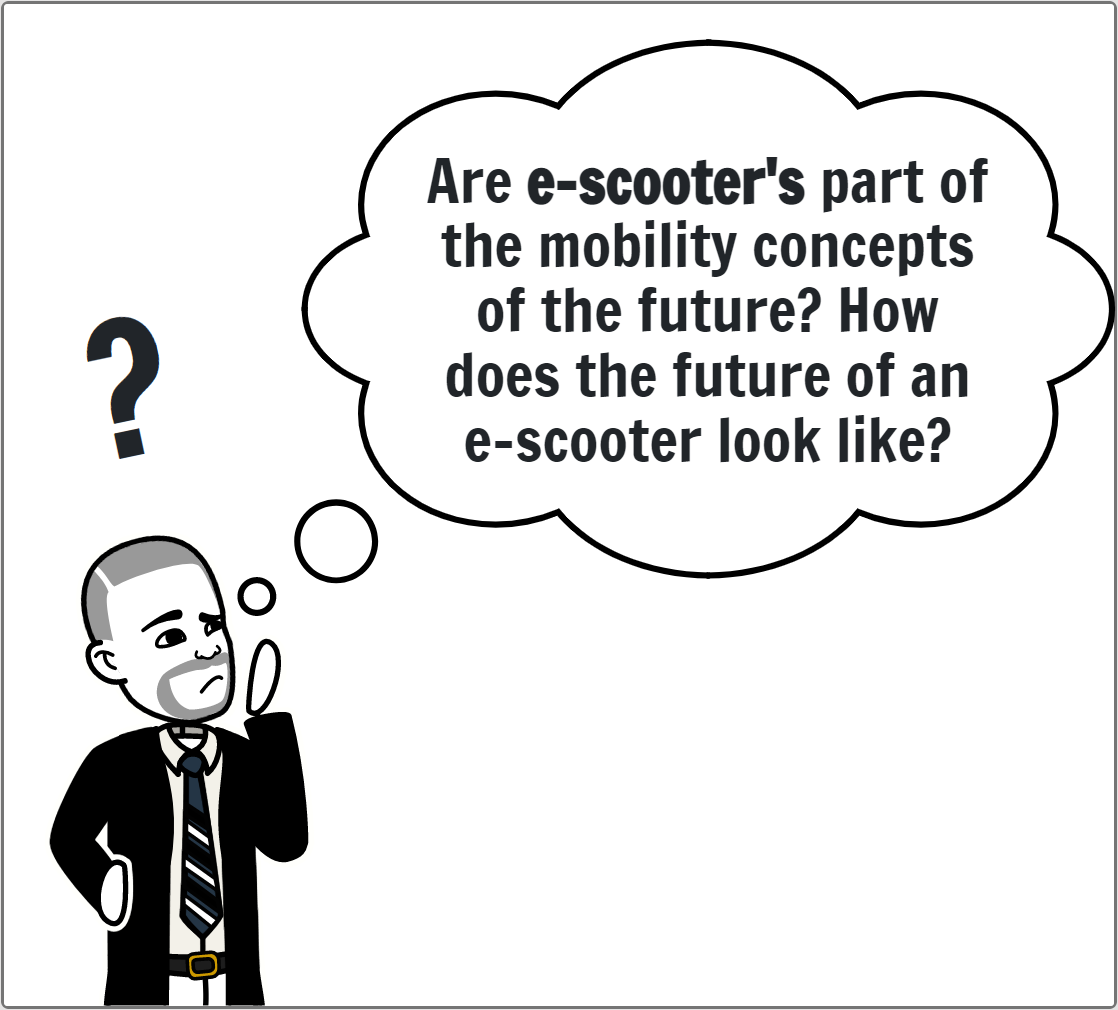}
	\end{minipage}
	\hfill
	\begin{minipage}{0.475\textwidth}
		\includegraphics[width=\textwidth]{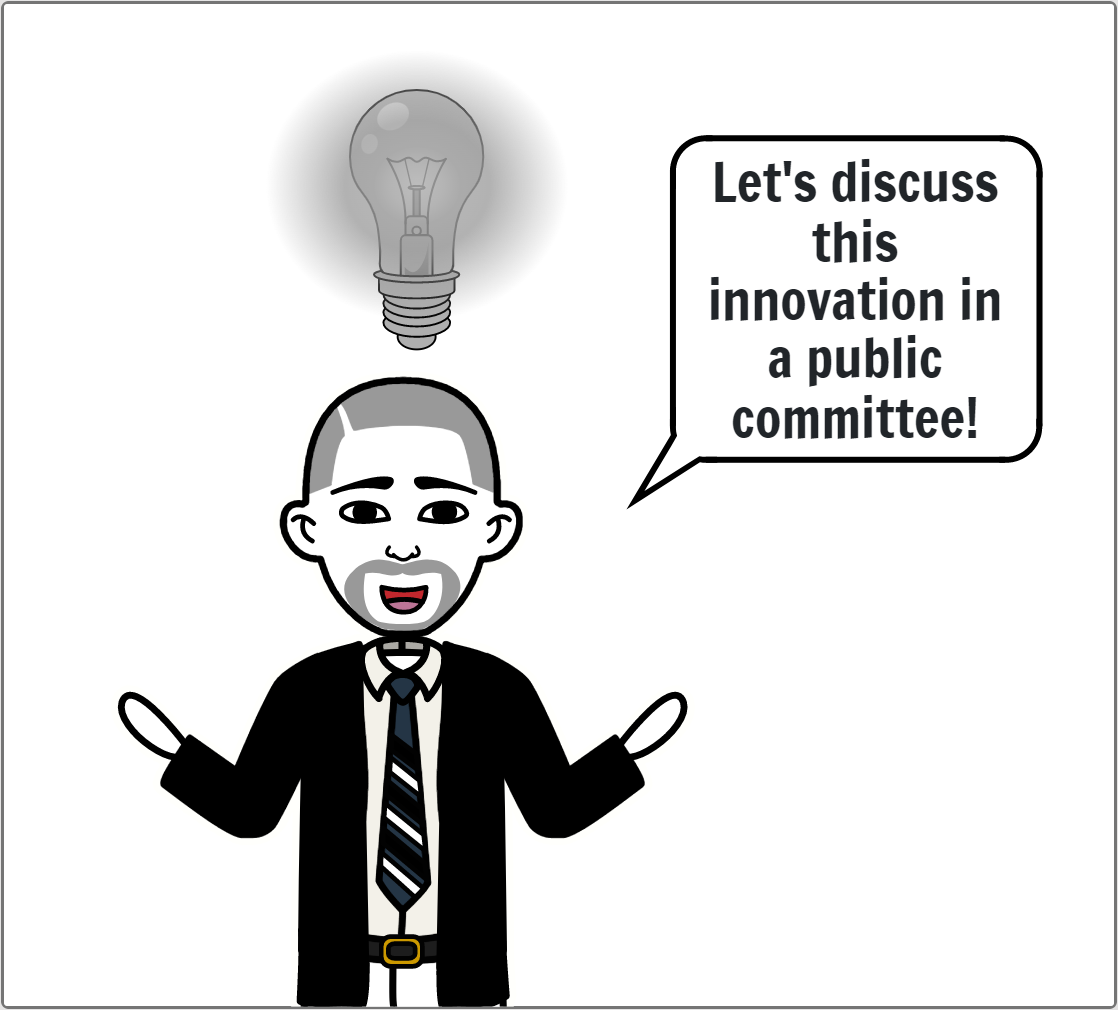}
	\end{minipage}

	\vspace{0.45cm}
	\begin{minipage}{0.475\textwidth}
		\includegraphics[width=\textwidth]{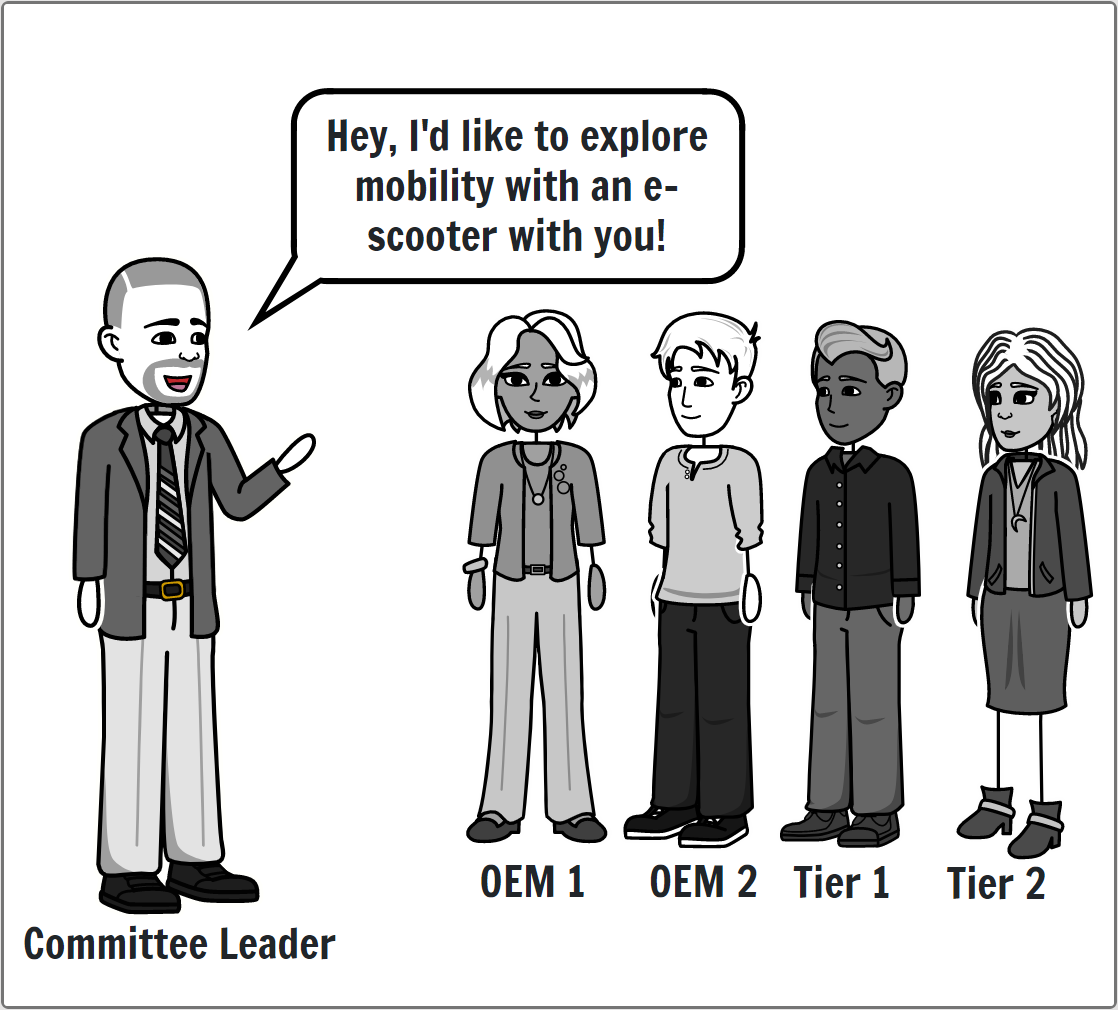}
	\end{minipage}
	\hfill
	\begin{minipage}{0.475\textwidth}
		\includegraphics[width=\textwidth]{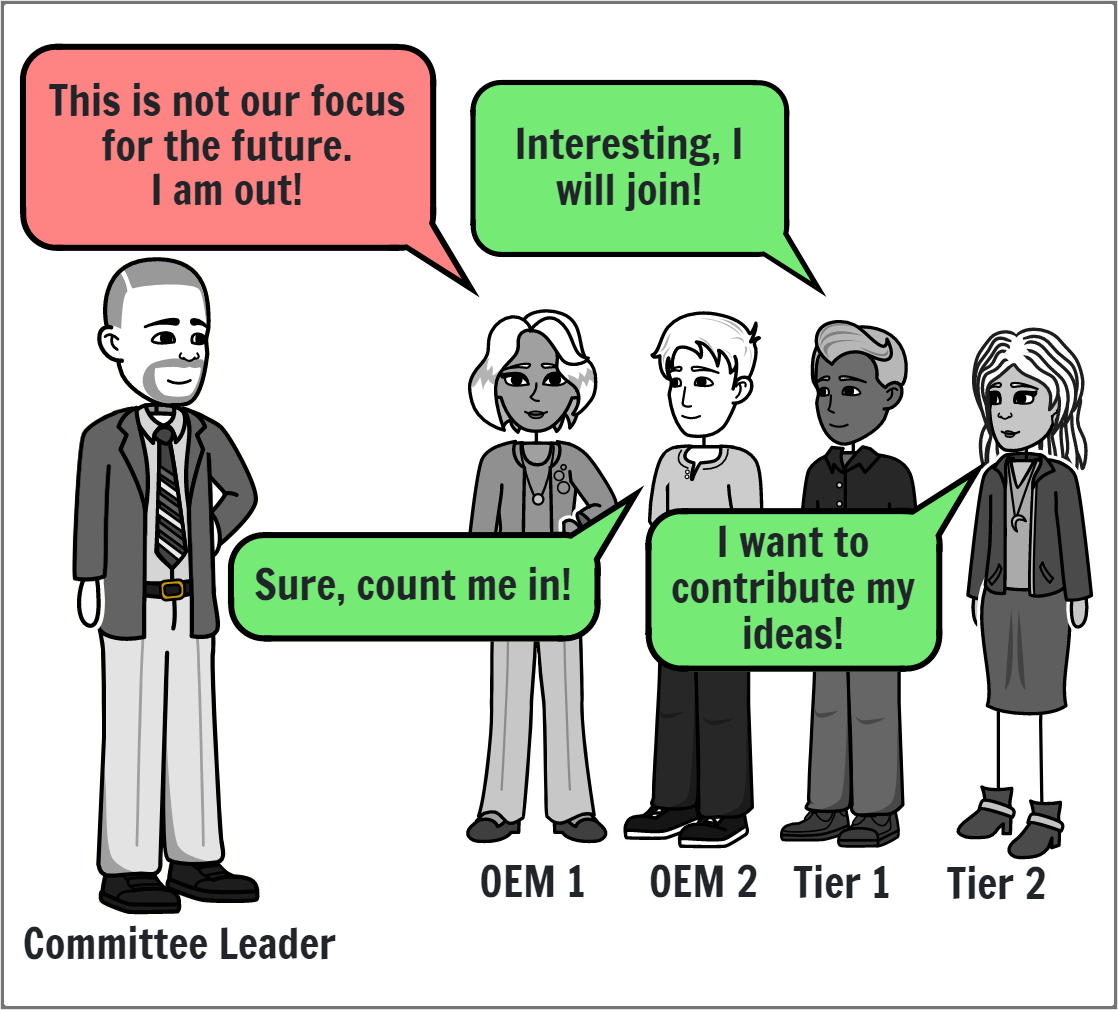}
	\end{minipage}

	\vspace{0.45cm}
	\begin{minipage}{0.475\textwidth}
		\includegraphics[width=\textwidth]{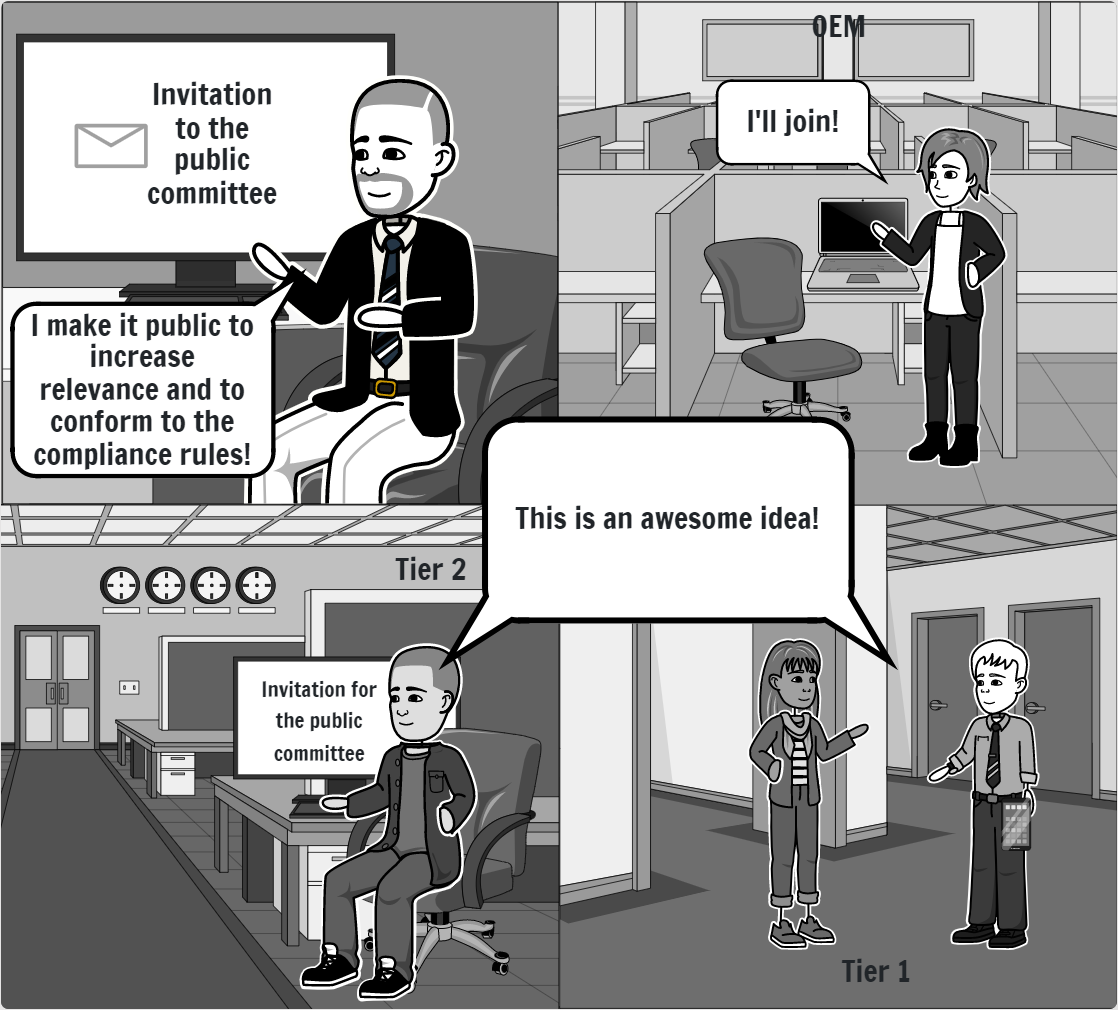}
	\end{minipage}
	\hfill
	\begin{minipage}{0.475\textwidth}
		\resizebox{\textwidth}{!}{%
			\begin{tikzpicture}[every node/.style={inner sep=0,outer sep=0}]
				\node[anchor=south west] at (0.12,0.12) {\includegraphics{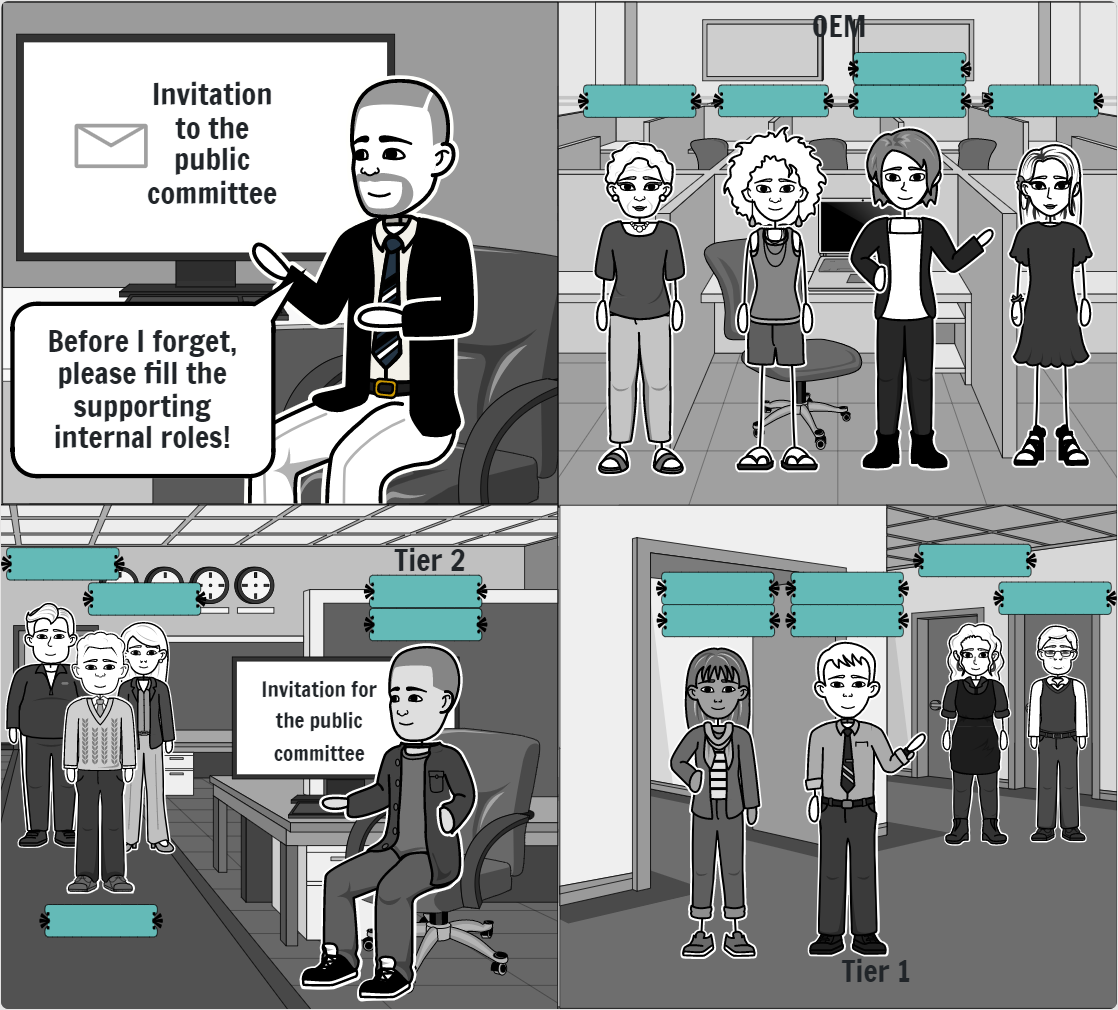}};
				\node at (17.05,24.35) {\small \textbf{Roadmap}};
				\node at (17.05,24) {\small \textbf{Manager}};
				\node at (20.6,24.35) {\small \textbf{Requirements}};
				\node at (20.6,24) {\small \textbf{Engineer}};
				\node at (24.2,24.35) {\small \textbf{System}};
				\node at (24.2,24) {\small \textbf{Architect}};
				\node at (27.7,24.4) {\small \textbf{Domain}};
				\node at (27.7,24) {\small \textbf{Expert}};
				\node at (24.2,25.2) {\small \textbf{Corporation}};
				\node at (24.2,24.85) {\small \textbf{Representative}};
				\node at (19.1,10.6) {\small \textbf{Roadmap}};
				\node at (19.1,10.2) {\small \textbf{Manager}};
				\node at (22.6,10.6) {\small \textbf{Requirements}};
				\node at (22.6,10.2) {\small \textbf{Engineer}};
				\node at (26,12.2) {\small \textbf{System}};
				\node at (26,11.85) {\small \textbf{Architect}};
				\node at (28.1,11.2) {\small \textbf{Domain}};
				\node at (28.1,10.8) {\small \textbf{Expert}};
				\node at (19.1,11.45) {\small \textbf{Corporation}};
				\node at (19.1,11.1) {\small \textbf{Representative}};
				\node at (22.6,11.45) {\small \textbf{Corporation}};
				\node at (22.6,11.1) {\small \textbf{Representative}};
				\node at (1.8,12.1) {\small \textbf{Roadmap}};
				\node at (1.8,11.75) {\small \textbf{Manager}};
				\node at (2.8,2.65) {\small \textbf{Requirements}};
				\node at (2.8,2.25) {\small \textbf{Engineer}};
				\node at (4,11.2) {\small \textbf{System}};
				\node at (4,10.85) {\small \textbf{Architect}};
				\node at (11.3,10.5) {\small \textbf{Domain}};
				\node at (11.3,10.1) {\small \textbf{Expert}};
				\node at (11.4,11.35) {\small \textbf{Corporation}};
				\node at (11.4,11) {\small \textbf{Representative}};
			\end{tikzpicture}
		}
	\end{minipage}
	\caption{Roles storyboard.}
	\label{fig:storyboard:roles:1}
\end{figure}

\begin{figure}
	\begin{minipage}{0.475\textwidth}
		\includegraphics[width=\textwidth]{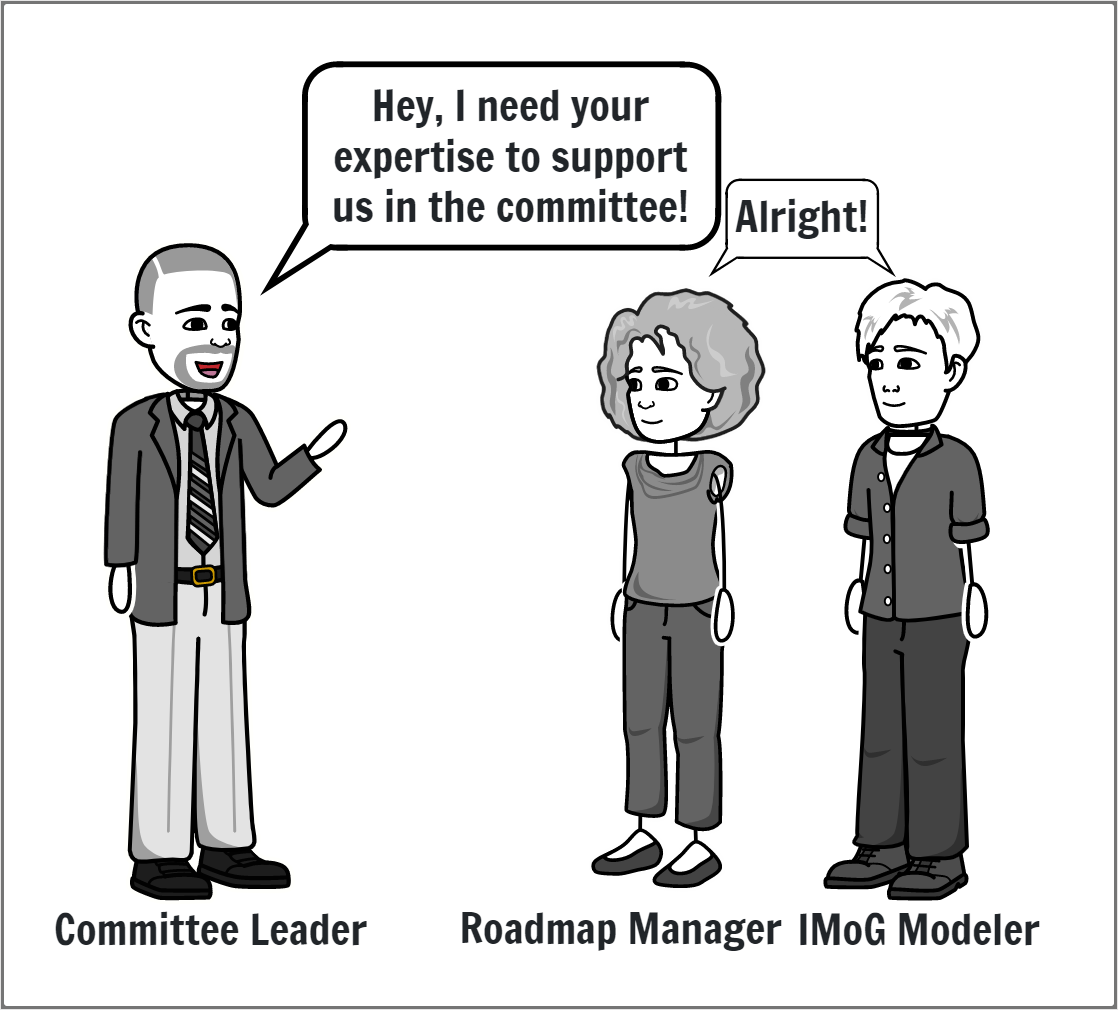}
	\end{minipage}
	\hfill
	\begin{minipage}{0.475\textwidth}
		\resizebox{\textwidth}{!}{%
			\begin{tikzpicture}[every node/.style={inner sep=0,outer sep=0}]
				\node[anchor=south west] at (0.12,0.12) {\includegraphics{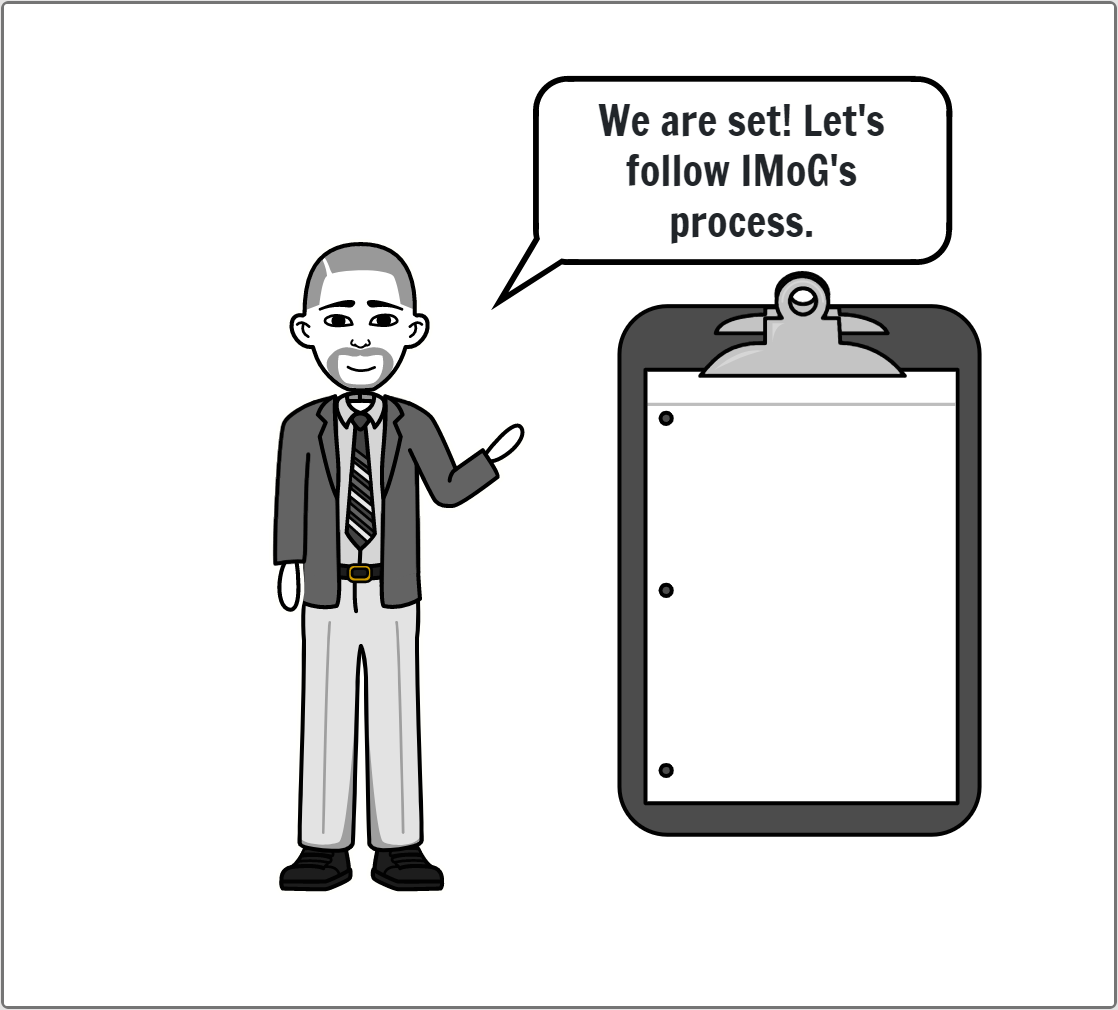}};
				\node[anchor=west, align=left] at (18.7,15.2) {\Large \textbf{Innovation Identification}};
				\node[anchor=west, align=left] at (18.7,14) {\Large \textbf{Feature and Function} \\\Large \textbf{Identification}};
				\node[anchor=west, align=left] at (18.7,12.59) {\Large \textbf{Requirements Elicitation} \\(Quality Requirements and Constraints)};
				\draw (18.4,11.9) -- (25,11.9);
				\node[anchor=west, align=left] at (18.7,11.3) {\Large \textbf{Solution Space Exploration}};
				\draw (18.4,10.8) -- (25,10.8);
				\node[anchor=west, align=left] at (18.7,9.8) {\Large \textbf{Extracting and saving} \\\Large \textbf{Insights for future} \\\Large \textbf{IMoG Innovations}};
				\node[anchor=west, align=left] at (18.7,8.4) {\Large \textbf{Roadmap Writing}};
				\node[anchor=west, align=left] at (18.7,7) {\Large \textbf{Maintaining and} \\\Large \textbf{Updating the Model} \\\Large \textbf{and Roadmap}};
			\end{tikzpicture}
		}
	\end{minipage}

	\vspace{0.45cm}
	\begin{minipage}{\textwidth}
		\centering
		\resizebox{0.7\textwidth}{!}{
			\begin{tikzpicture}[every node/.style={inner sep=0,outer sep=0}]
				\node[anchor=south west] at (0.12,0.12) {\includegraphics{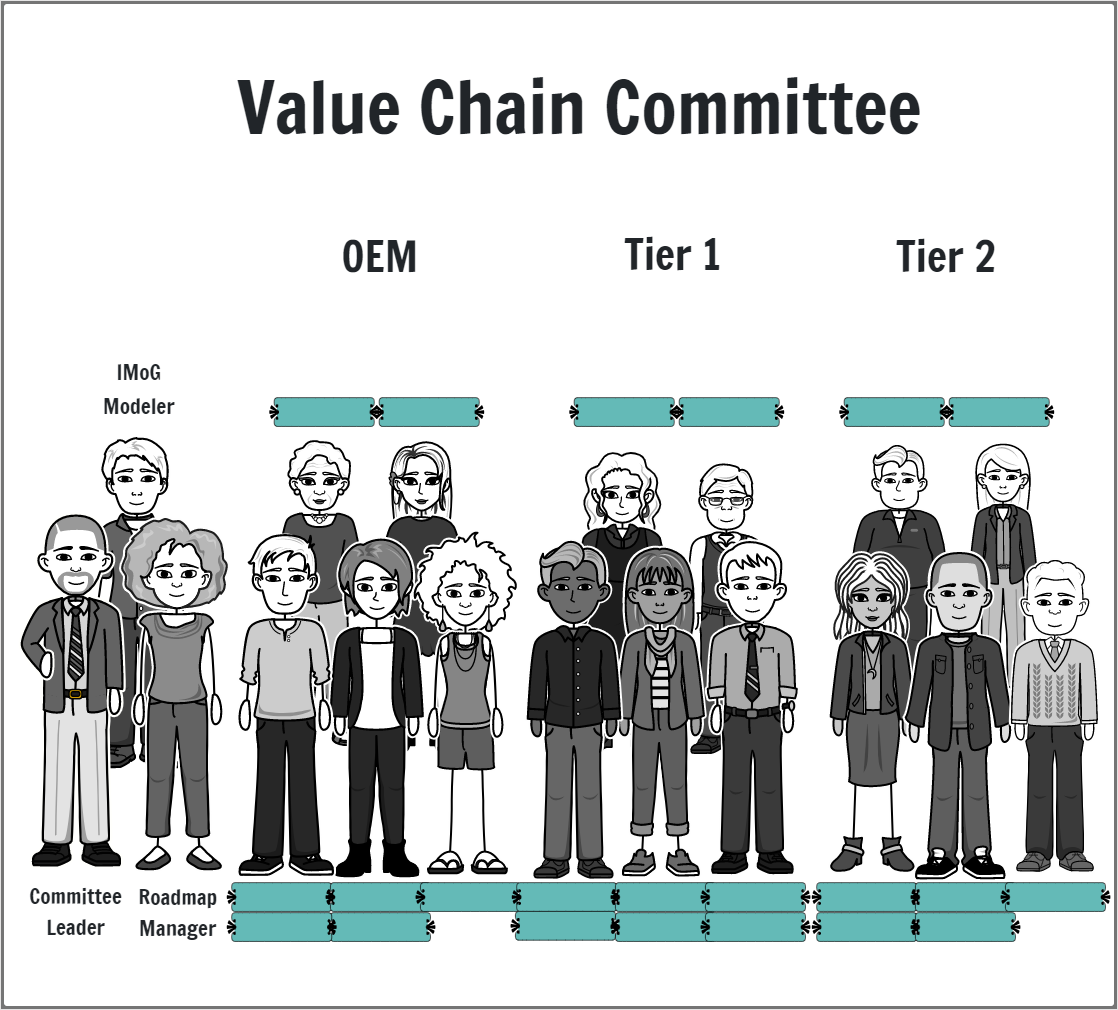}};
				\node at (7.55,3.25) {\small \textbf{Roadmap}};
				\node at (7.55,2.97) {\small \textbf{Manager}};
				\node at (8.7,16.1) {\small \textbf{Roadmap}};
				\node at (8.7,15.75) {\small \textbf{Manager}};
				\node at (12.57,3.25) {\small \textbf{Requirements}};
				\node at (12.57,2.95) {\small \textbf{Engineer}};
				\node at (10.1,3.25) {\small \textbf{System}};
				\node at (10.1,2.97) {\small \textbf{Architect}};
				\node at (11.5,16.1) {\small \textbf{Domain}};
				\node at (11.5,15.75) {\small \textbf{Expert}};
				\node at (10.2,2.45) {\small \textbf{Corporation}};
				\node at (10.2,2.1) {\small \textbf{Representative}};
				\node at (7.55,2.45) {\small \textbf{Corporation}};
				\node at (7.55,2.1) {\small \textbf{Representative}};
				\node at (17.7,3.25) {\small \textbf{Roadmap}};
				\node at (17.7,2.97) {\small \textbf{Manager}};
				\node at (20.15,3.25) {\small \textbf{Requirements}};
				\node at (20.15,2.97) {\small \textbf{Engineer}};
				\node at (15.1,3.25) {\small \textbf{System}};
				\node at (15.1,2.97) {\small \textbf{Architect}};
				\node at (16.65,16.1) {\small \textbf{System}};
				\node at (16.65,15.75) {\small \textbf{Architect}};
				\node at (19.45,16.1) {\small \textbf{Domain}};
				\node at (19.45,15.75) {\small \textbf{Expert}};
				\node at (15.1,2.45) {\small \textbf{Corporation}};
				\node at (15.1,2.1) {\small \textbf{Representative}};
				\node at (17.7,2.45) {\small \textbf{Corporation}};
				\node at (17.7,2.1) {\small \textbf{Representative}};
				\node at (20.15,2.45) {\small \textbf{Corporation}};
				\node at (20.15,2.1) {\small \textbf{Representative}};
				\node at (23.8,16.1) {\small \textbf{Roadmap}};
				\node at (23.8,15.75) {\small \textbf{Manager}};
				\node at (23.1,3.25) {\small \textbf{Requirements}};
				\node at (23.1,2.97) {\small \textbf{Engineer}};
				\node at (28.1,3.25) {\small \textbf{Requirements}};
				\node at (28.1,2.97) {\small \textbf{Engineer}};
				\node at (26.6,16.1) {\small \textbf{System}};
				\node at (26.6,15.75) {\small \textbf{Architect}};
				\node at (25.55,3.3) {\small \textbf{Domain}};
				\node at (25.55,2.97) {\small \textbf{Expert}};
				\node at (23.1,2.45) {\small \textbf{Corporation}};
				\node at (23.1,2.1) {\small \textbf{Representative}};
				\node at (25.75,2.45) {\small \textbf{Corporation}};
				\node at (25.75,2.1) {\small \textbf{Representative}};
			\end{tikzpicture}
		}
	\end{minipage}
	\caption{Roles storyboard (part 2).}
	\label{fig:storyboard:roles:2}
\end{figure}

\begin{minipage}{0.8\textwidth}
	Let's illustrate the process with an example storyboard (see Figures \ref{fig:storyboard:roles:1} and \ref{fig:storyboard:roles:2}).
	A manager (see the Figure on the right) from a known car manufacturer wants to dive into future mobility aspects and explore new areas for potential investments.
	He likes the aspect of e-scooters as part of future mobility services and decides to think through this innovation together with the automotive value chain.
	He starts a new committee with himself as the committee leader and publicly invites partners of the automotive value chain to join the committee.
	Several members of the automotive value chain join the innovation exploration.
	Some of them will also decline.
	He also creates a public invitation to increase the number of partners and thus the relevance of the committee.
	Some additional OEMs, Tier1 and Tier2 join the consortium.
	Before starting the committee he requests the committee corporations to assign the internal roles.
	Each OEM, Tier1 and Tier2 assign the roles of the Roadmap Manager, Requirement Engineer, System Architect and Domain Expert.
	Additionally, the committee leader invites suitable people to take over the role of the IMoG modeler and the role of the roadmap manager.
	Then the committee is ready to explore the innovation by following IMoG's process.
\end{minipage}
\hfill
\begin{minipage}{0.19\textwidth}
	\begin{tikzpicture}
		\node[anchor=south] (img) at (0,0) {\includegraphics[width=0.45\textwidth]{figures/role1.png}};
		\node at (0,-0.3) {\small OEM};
		\node at (0,-0.7) {\small Committee Leader};
	\end{tikzpicture}
\end{minipage}

\begin{figure}
	\begin{minipage}{0.475\textwidth}
		\resizebox{\textwidth}{!}{%
			\begin{tikzpicture}[every node/.style={inner sep=0,outer sep=0}]
				\node[anchor=south west]  at (0.12,0.12) {\includegraphics{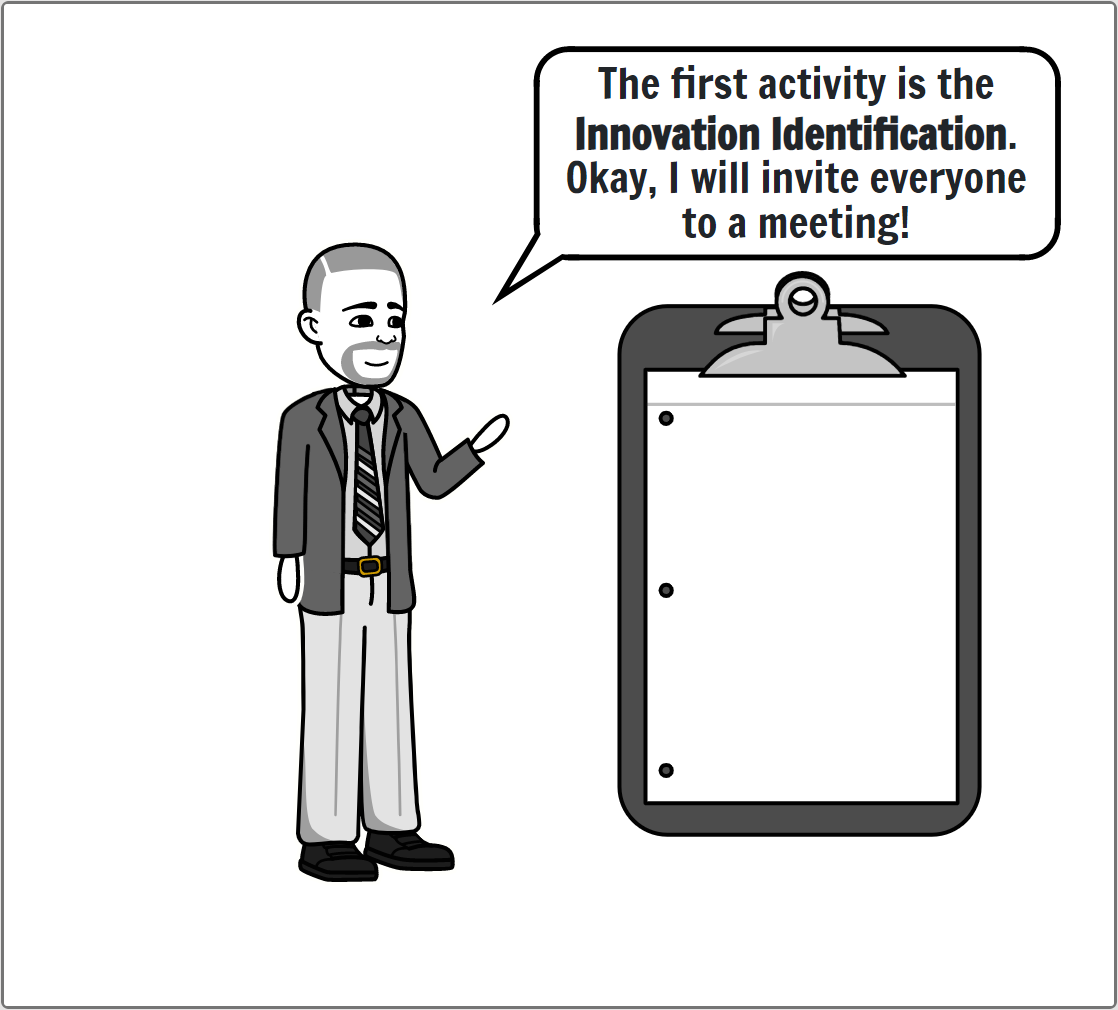}};
				\node[anchor=west, align=left] at (18.7,15.2) {\Large \textbf{Innovation Identification}};
				\node[anchor=west, align=left] at (18.7,14) {\Large \textbf{Feature and Function} \\\Large \textbf{Identification}};
				\node[anchor=west, align=left] at (18.7,12.59) {\Large \textbf{Requirements Elicitation} \\(Quality Requirements and Constraints)};
				\draw (18.4,11.9) -- (25,11.9);
				\node[anchor=west, align=left] at (18.7,11.3) {\Large \textbf{Solution Space Exploration}};
				\draw (18.4,10.8) -- (25,10.8);
				\node[anchor=west, align=left] at (18.7,9.8) {\Large \textbf{Extracting and saving} \\\Large \textbf{Insights for future} \\\Large \textbf{IMoG Innovations}};
				\node[anchor=west, align=left] at (18.7,8.4) {\Large \textbf{Roadmap Writing}};
				\node[anchor=west, align=left] at (18.7,7) {\Large \textbf{Maintaining and} \\\Large \textbf{Updating the Model} \\\Large \textbf{and Roadmap}};
			\end{tikzpicture}
		}
	\end{minipage}
	\hfill
	\begin{minipage}{0.475\textwidth}
		\includegraphics[width=\textwidth]{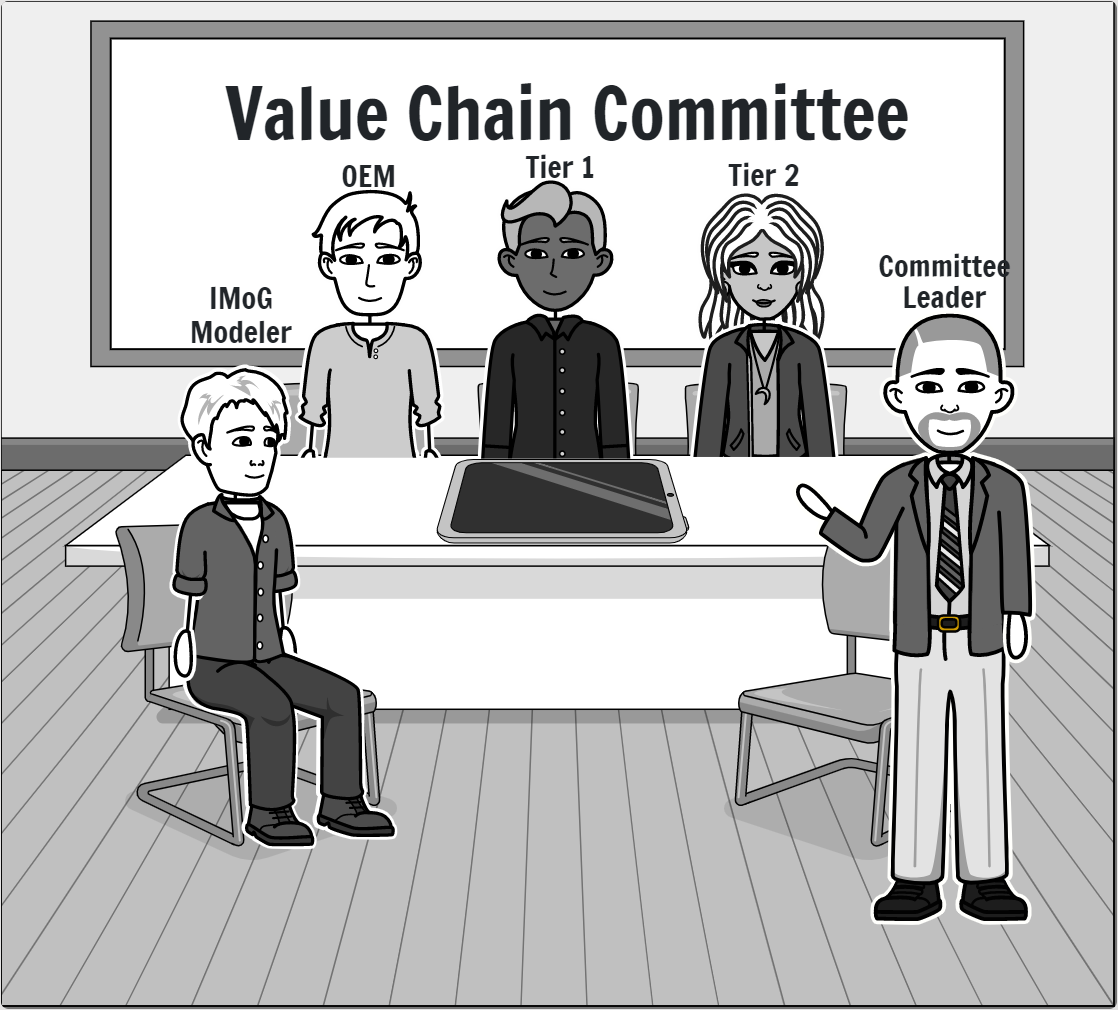}
	\end{minipage}

	\vspace{0.45cm}
	\begin{minipage}{0.475\textwidth}
		\includegraphics[width=\textwidth]{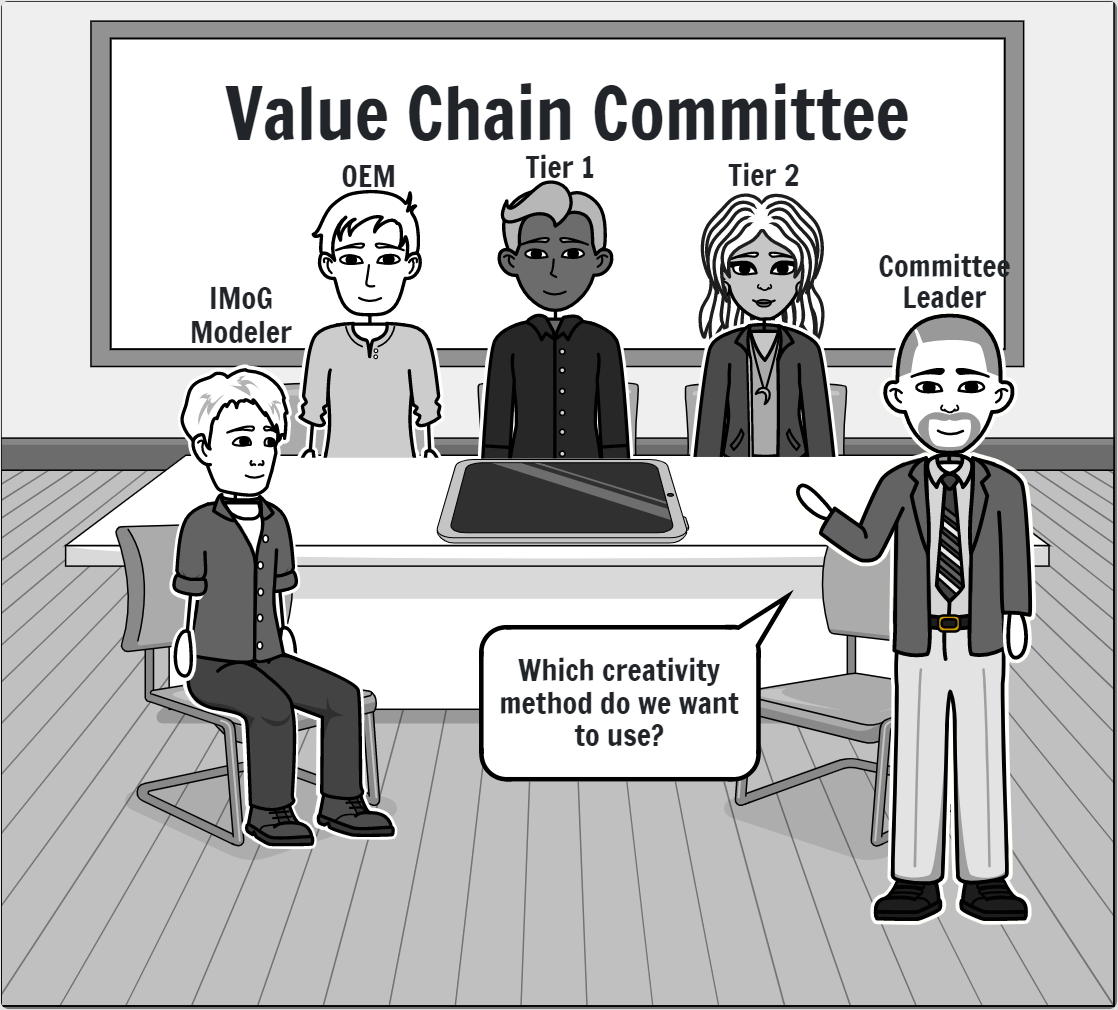}
	\end{minipage}
	\hfill
	\begin{minipage}{0.475\textwidth}
		\includegraphics[width=\textwidth]{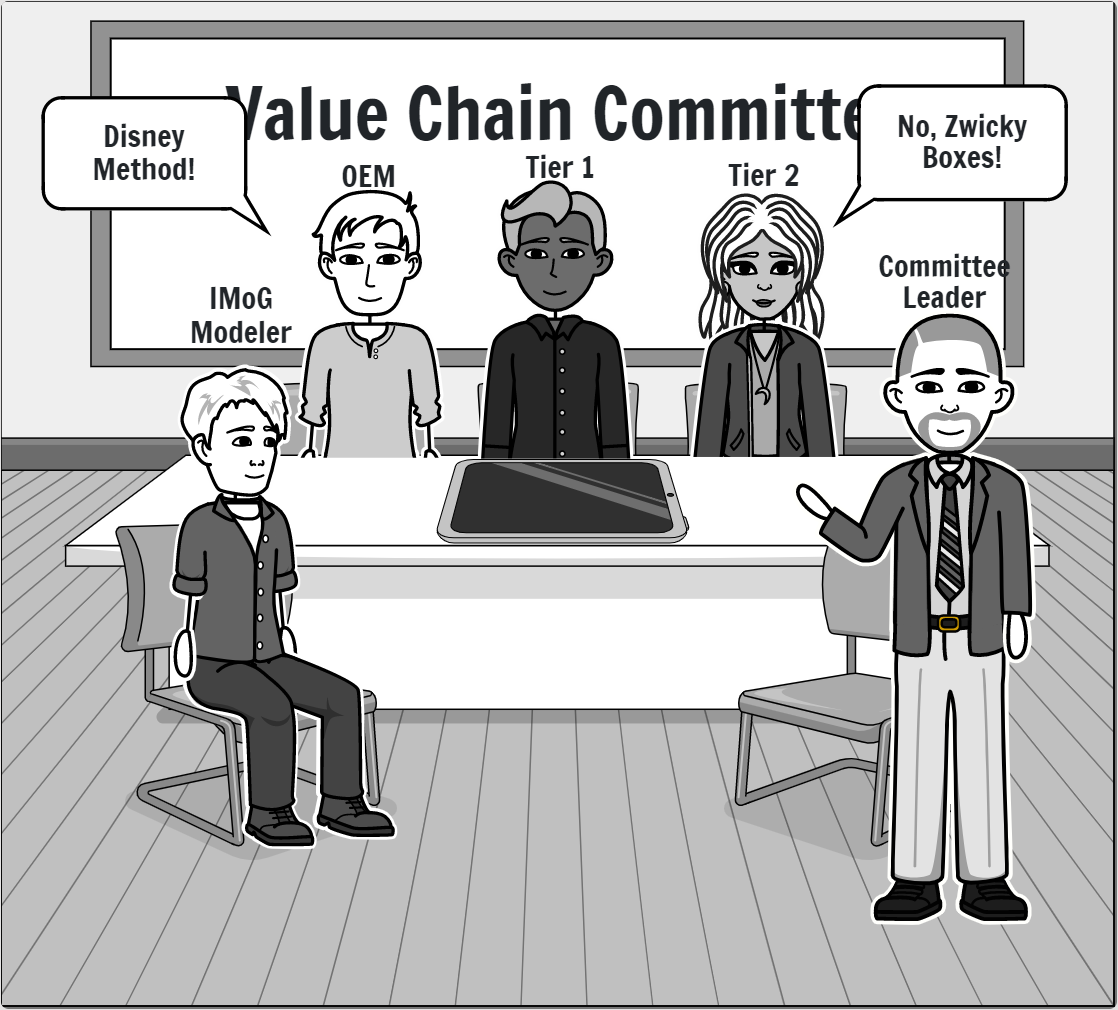}
	\end{minipage}

	\vspace{0.45cm}
	\begin{minipage}{0.475\textwidth}
		\includegraphics[width=\textwidth]{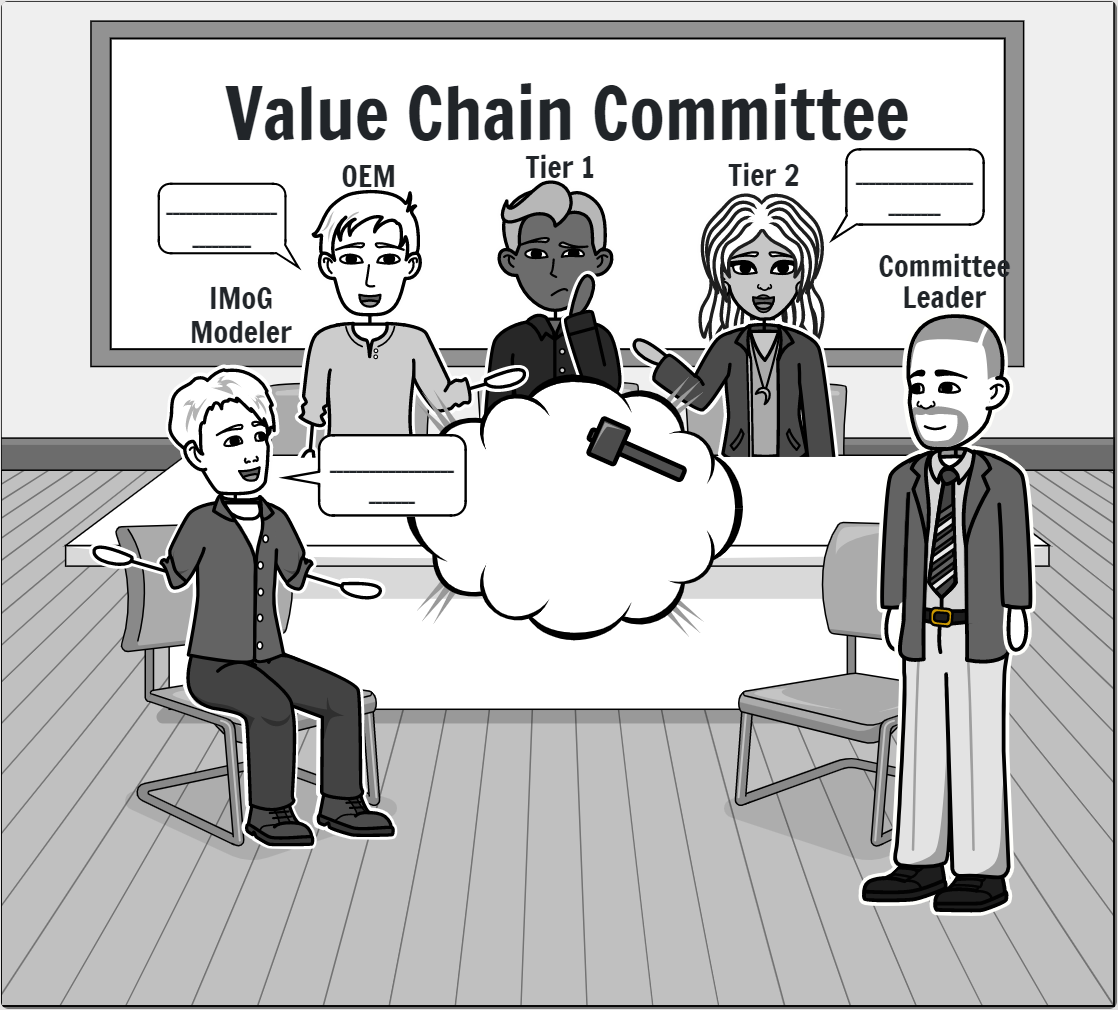}
	\end{minipage}
	\hfill
	\begin{minipage}{0.475\textwidth}
		\includegraphics[width=\textwidth]{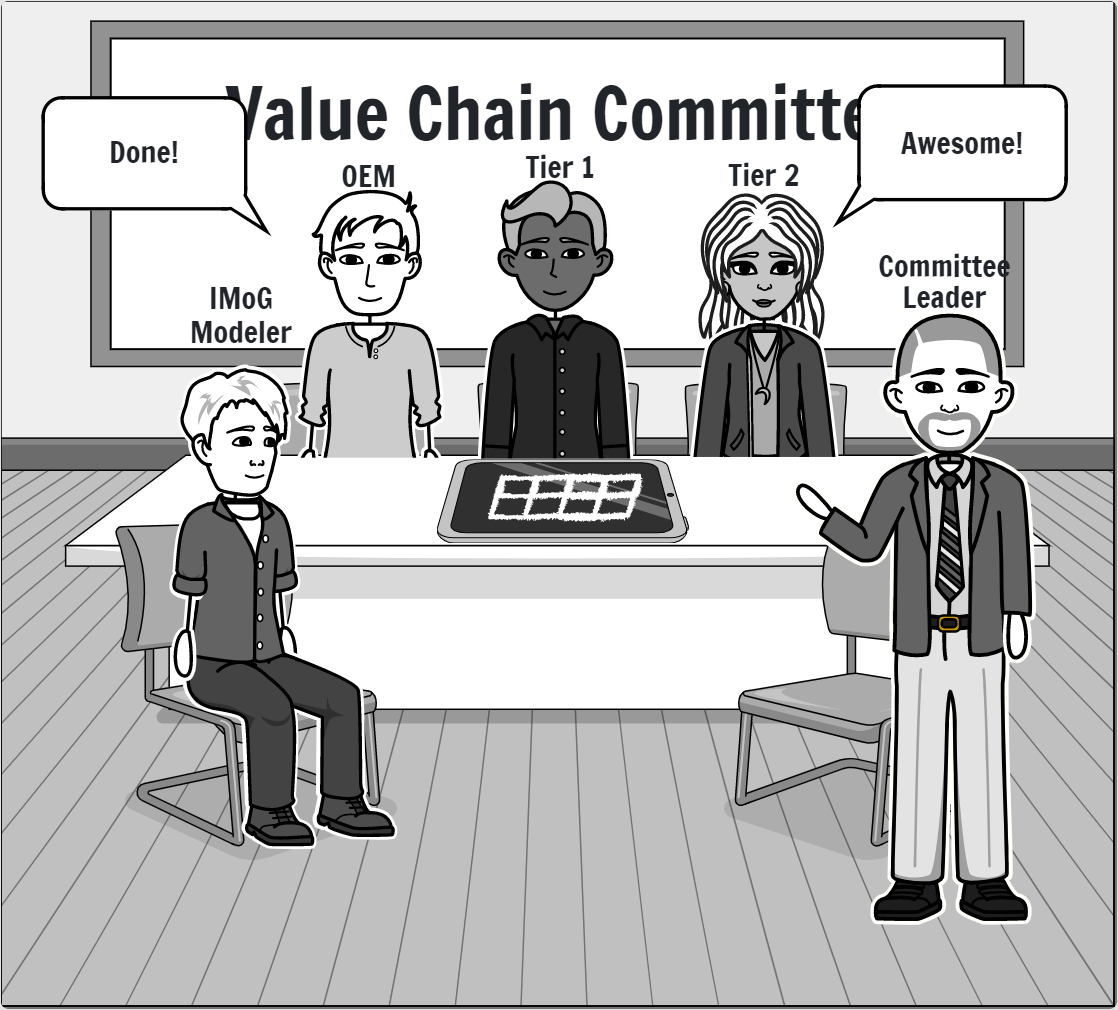}
	\end{minipage}
	\caption{The first activity: Innovation Identification}
	\label{fig:storyboard:strategy:1}
\end{figure}

\begin{figure}
	\begin{minipage}{0.475\textwidth}
		\includegraphics[width=\textwidth]{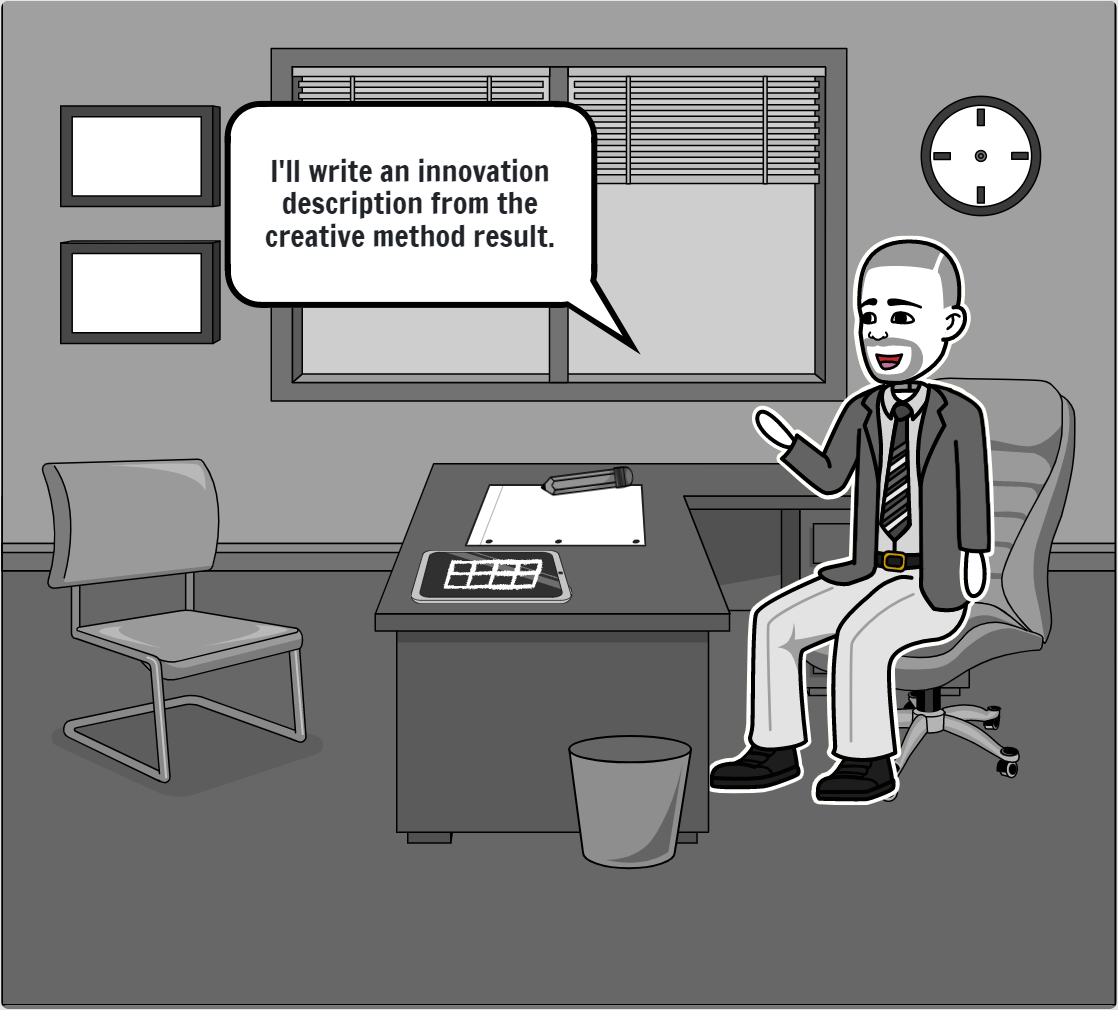}
	\end{minipage}
	\hfill
	\begin{minipage}{0.475\textwidth}
		\includegraphics[width=\textwidth]{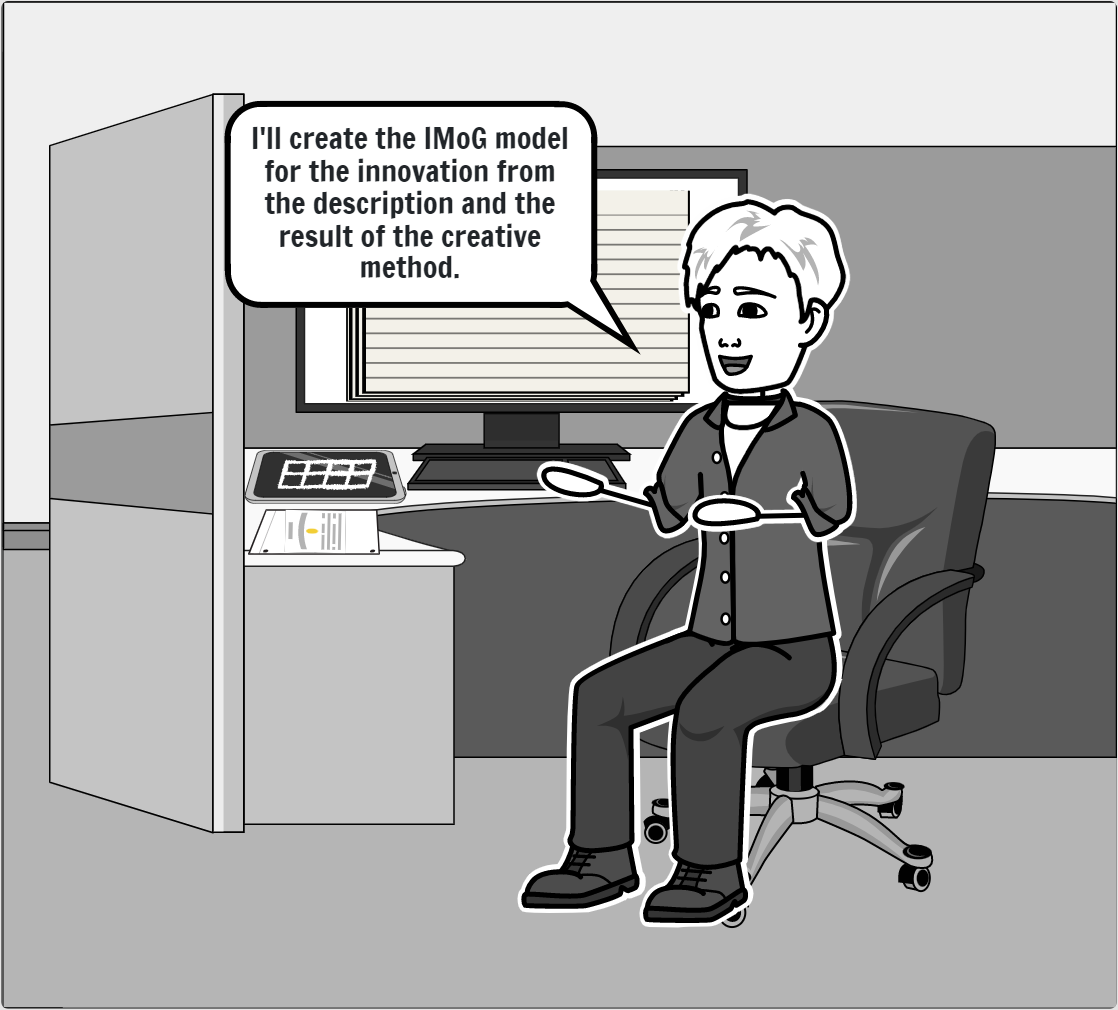}
	\end{minipage}

	\vspace{0.45cm}
	\begin{minipage}{0.475\textwidth}
		\includegraphics[width=\textwidth]{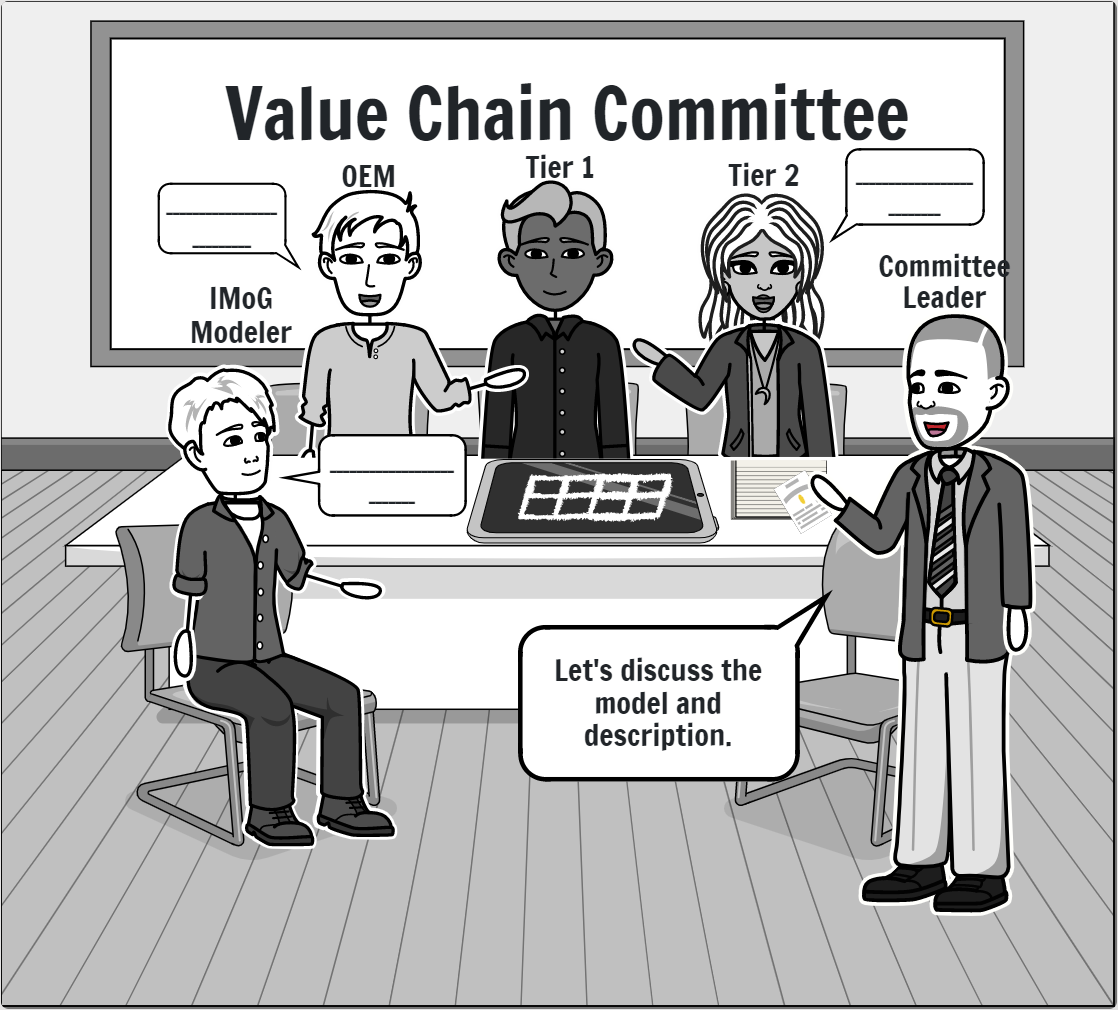}
	\end{minipage}
	\hfill
	\begin{minipage}{0.475\textwidth}
		\includegraphics[width=\textwidth]{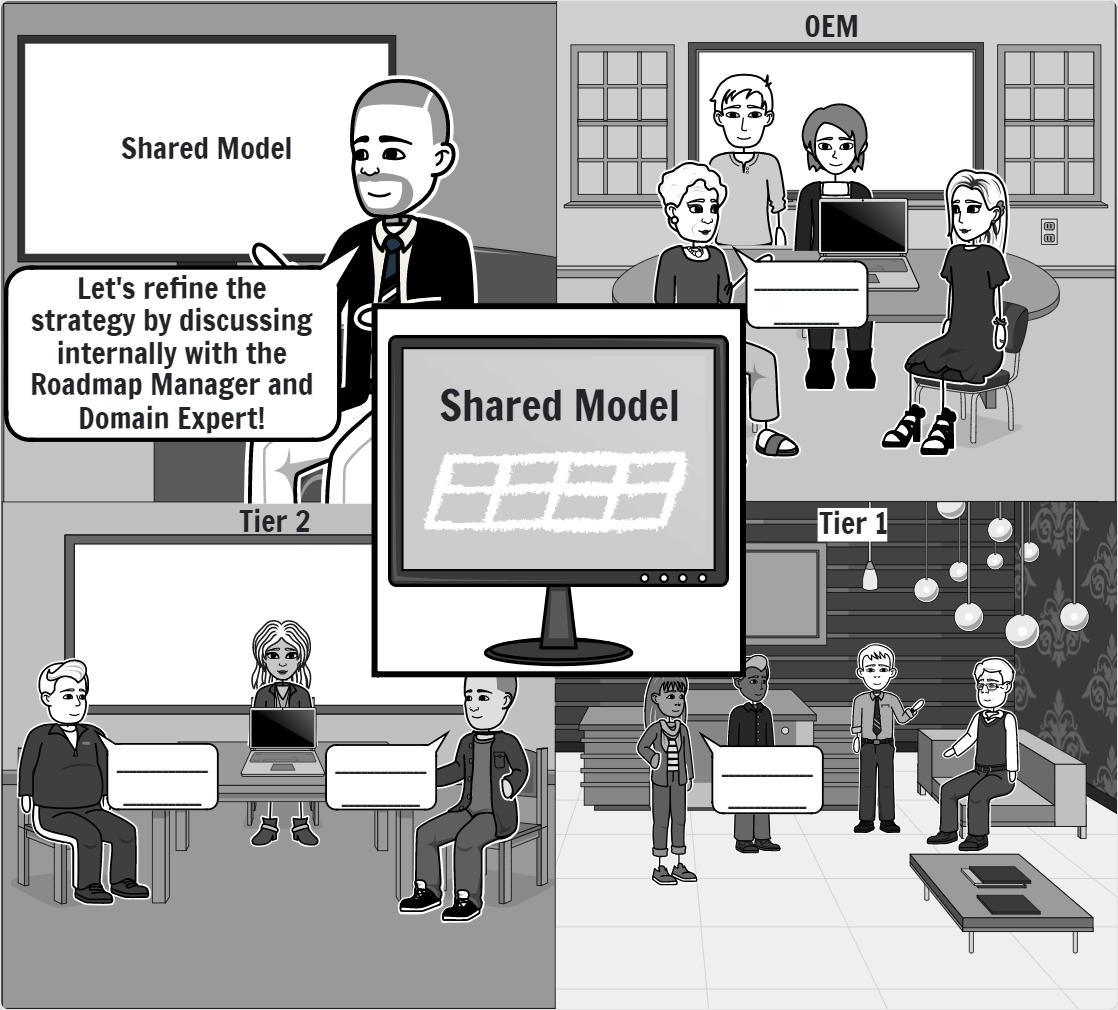}
	\end{minipage}

	\vspace{0.45cm}
	\begin{minipage}{0.475\textwidth}
		\includegraphics[width=\textwidth]{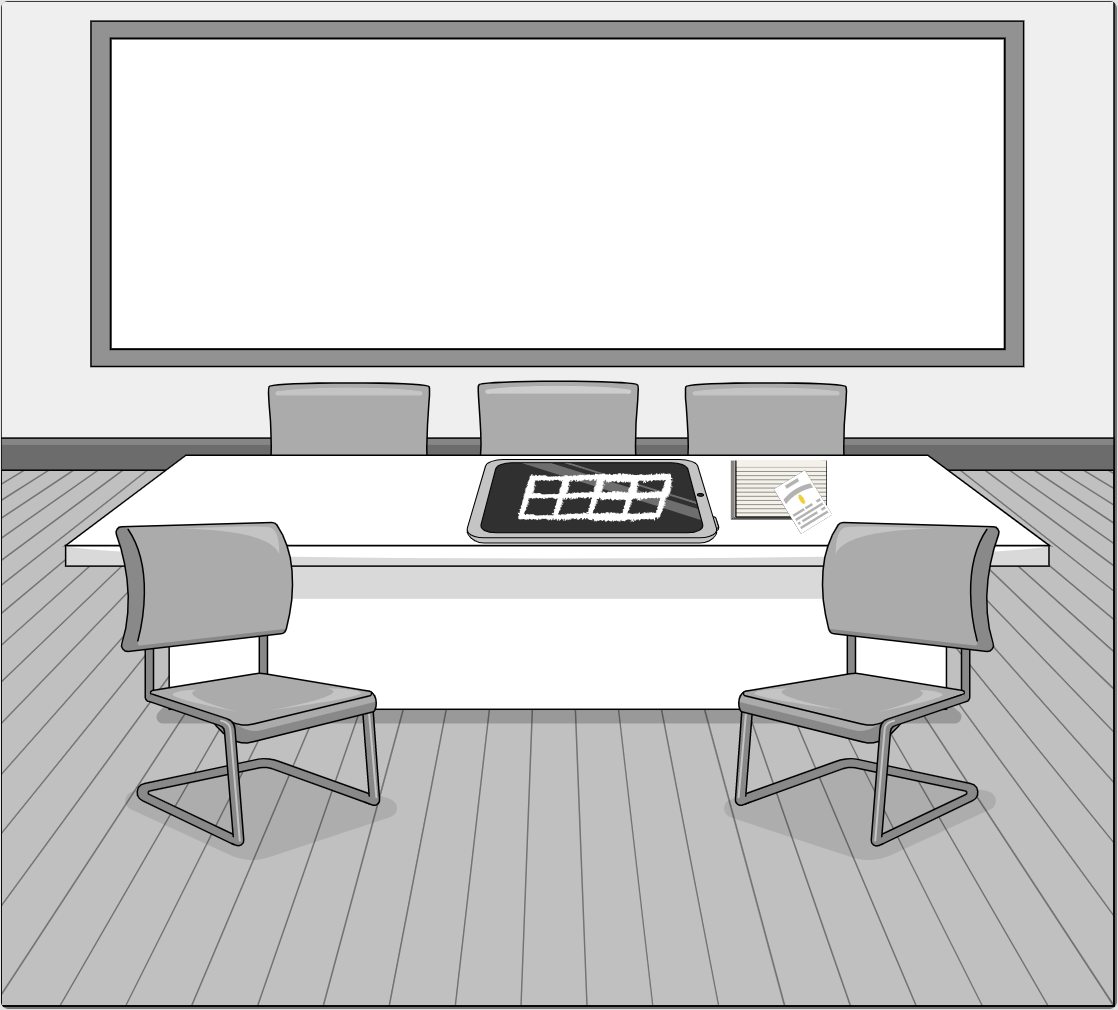}
	\end{minipage}
	\hfill
	\caption{The first activity: Innovation Identification (part 2)}
	\label{fig:storyboard:strategy:2}
\end{figure}

The committee starts with the first activity -- the Innovation Identification (see Figures \ref{fig:storyboard:strategy:1} and \ref{fig:storyboard:strategy:2}).
The committee leader invites the committee to a meeting to identify the innovation.
The committee meets and chooses a fitting creativity method to identify the innovation they want to explore.
In this case, they agree on using the creativity method \hyperref[Zwicky Boxes]{https://en.wikipedia.org/wiki/Morphological\_analysis\_(problem-solving)}.
The committee carries out the creativity method until they are satisfied with their result.
Based on the outcome, the committee leader writes an innovation description.
Afterwards, the IMoG modeler takes the creativity method outcome and the innovation description to create a draft of the Strategy Perspective.
Both results are discussed and refined in the committee and internally until they are satisfied.
Then the Strategy Perspective and the Innovation Identification activity is finished.

\begin{figure}
	\begin{minipage}{0.475\textwidth}
		\resizebox{\textwidth}{!}{%
			\begin{tikzpicture}[every node/.style={inner sep=0,outer sep=0}]
				\node[anchor=south west] at (0.12,0.12) {\includegraphics{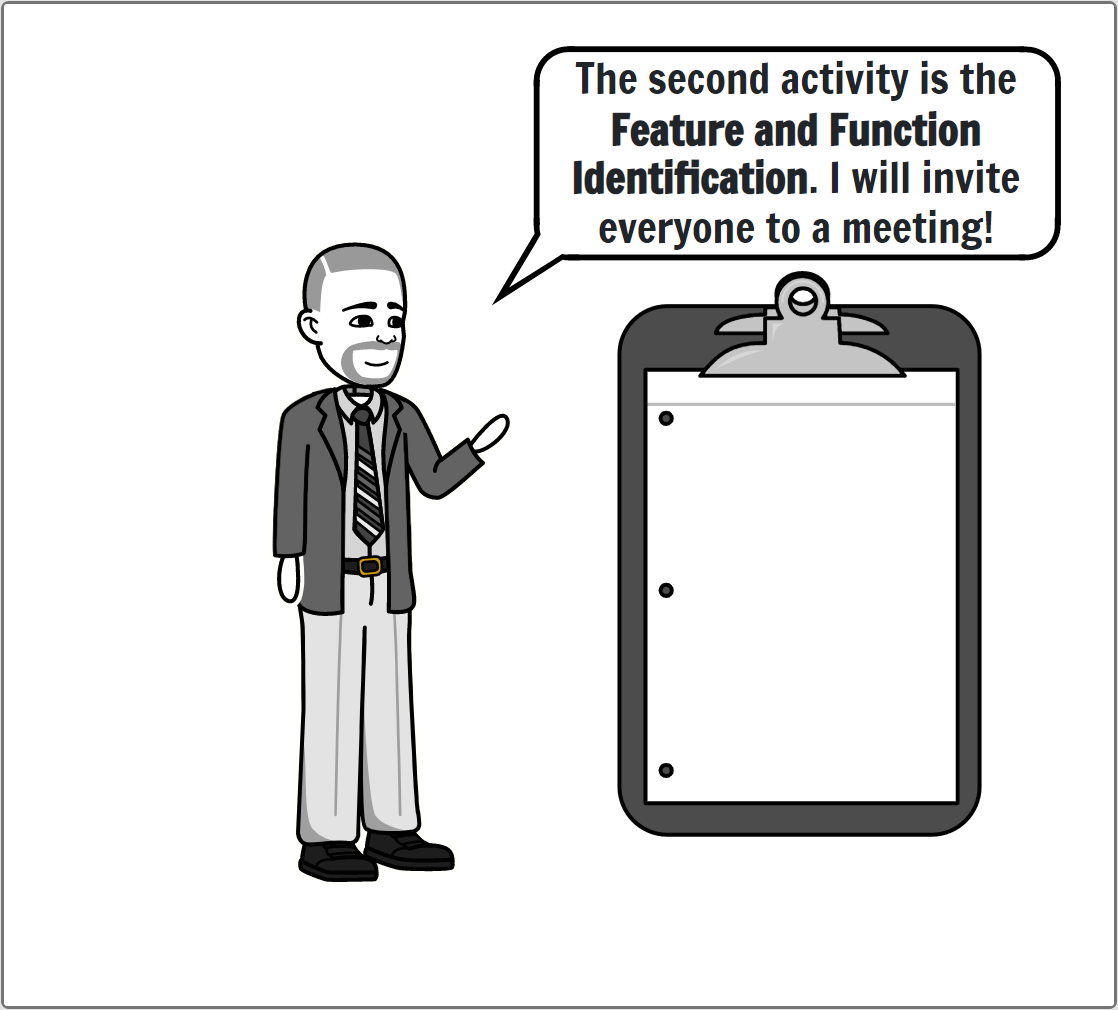}};
				\node[anchor=west, align=left] at (18.7,15.2) {\Large \textbf{Innovation Identification}};
				\node[anchor=west, align=left] at (18.7,14) {\Large \textbf{Feature and Function} \\\Large \textbf{Identification}};
				\node[anchor=west, align=left] at (18.7,12.59) {\Large \textbf{Requirements Elicitation} \\(Quality Requirements and Constraints)};
				\draw (18.4,11.9) -- (25,11.9);
				\node[anchor=west, align=left] at (18.7,11.3) {\Large \textbf{Solution Space Exploration}};
				\draw (18.4,10.8) -- (25,10.8);
				\node[anchor=west, align=left] at (18.7,9.8) {\Large \textbf{Extracting and saving} \\\Large \textbf{Insights for future} \\\Large \textbf{IMoG Innovations}};
				\node[anchor=west, align=left] at (18.7,8.4) {\Large \textbf{Roadmap Writing}};
				\node[anchor=west, align=left] at (18.7,7) {\Large \textbf{Maintaining and} \\\Large \textbf{Updating the Model} \\\Large \textbf{and Roadmap}};
				\node at (18.4,15.2) {\Large \color{green!70!black} \CheckmarkBold};
			\end{tikzpicture}
		}
	\end{minipage}
	\hfill
	\begin{minipage}{0.475\textwidth}
		\includegraphics[width=\textwidth]{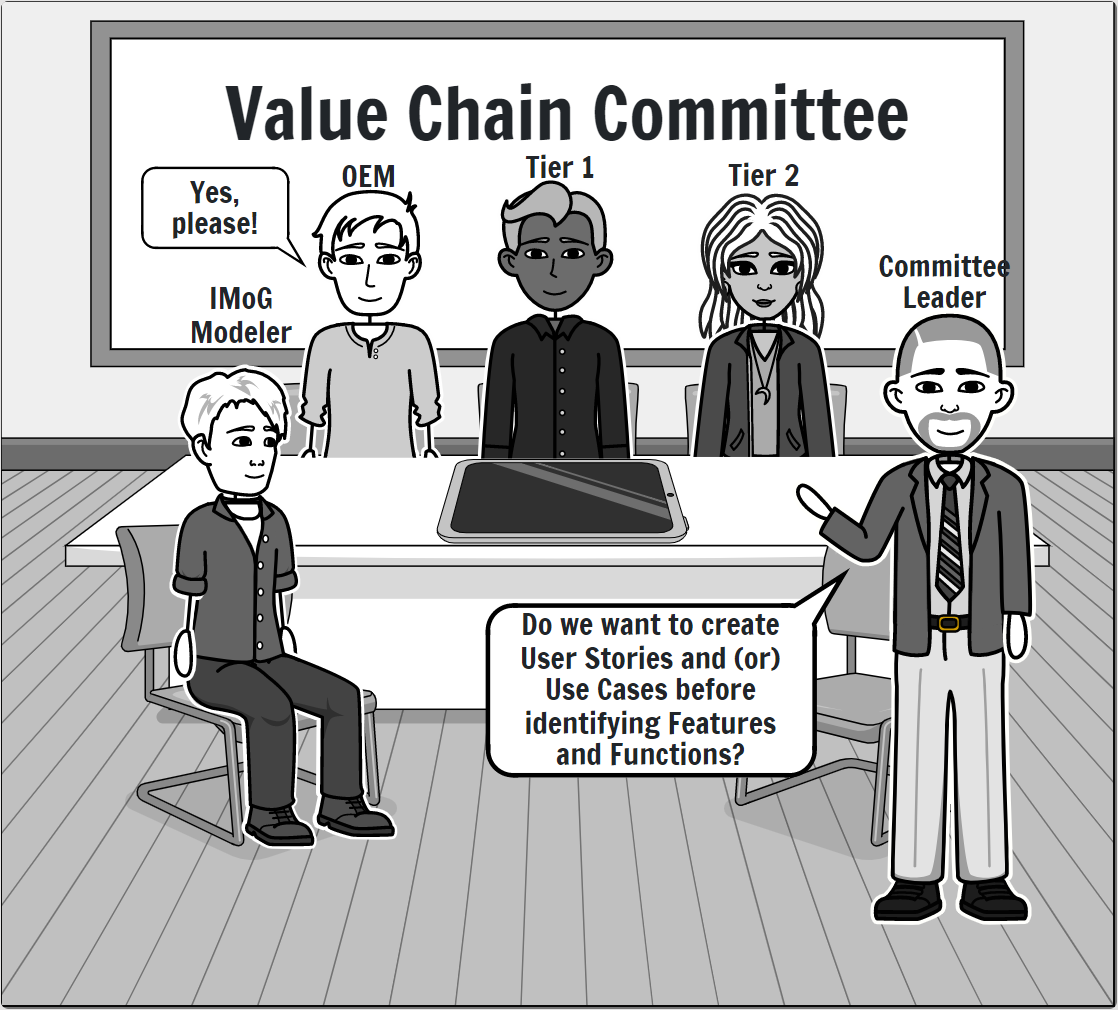}
	\end{minipage}

	\vspace{0.45cm}
	\begin{minipage}{0.475\textwidth}
		\includegraphics[width=\textwidth]{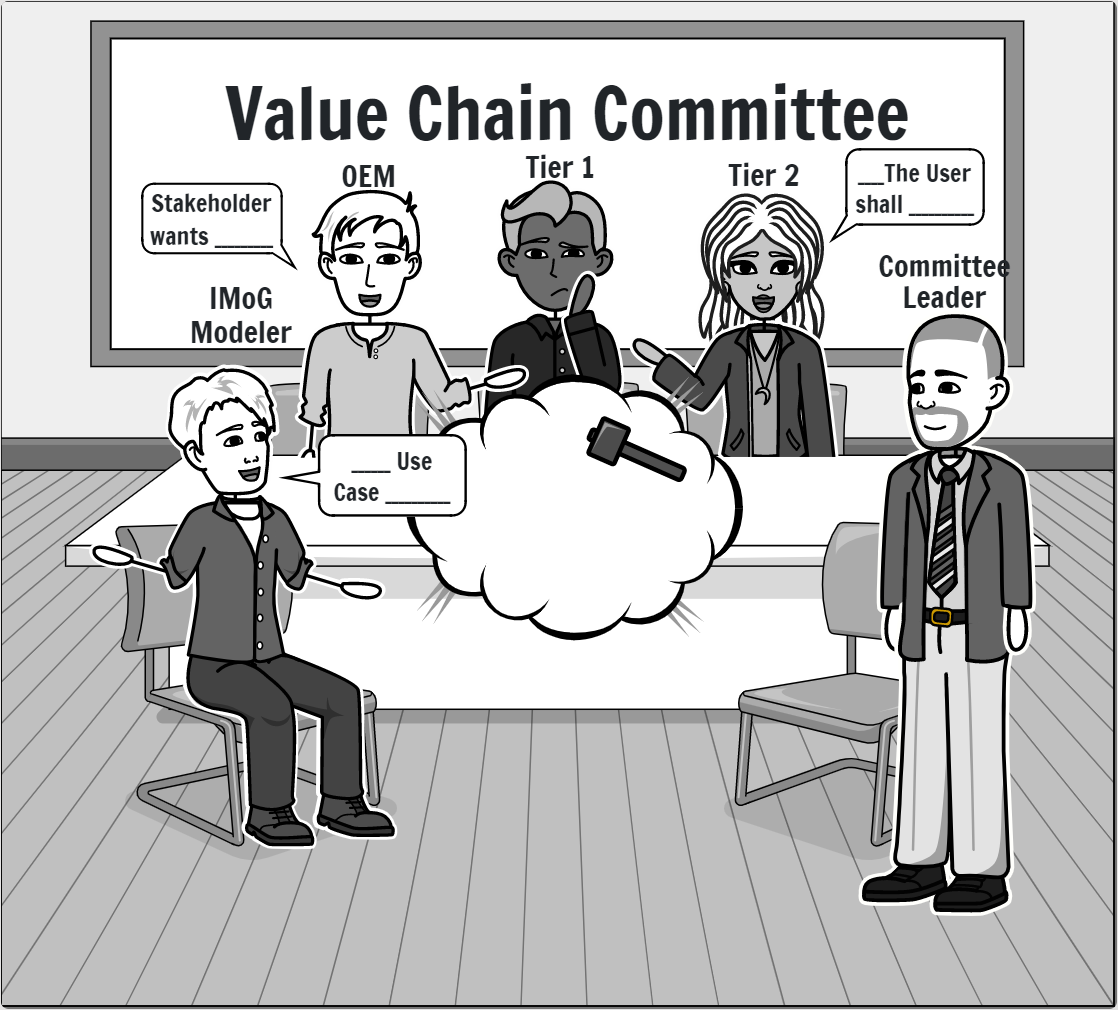}
	\end{minipage}
	\hfill
	\begin{minipage}{0.475\textwidth}
		\includegraphics[width=\textwidth]{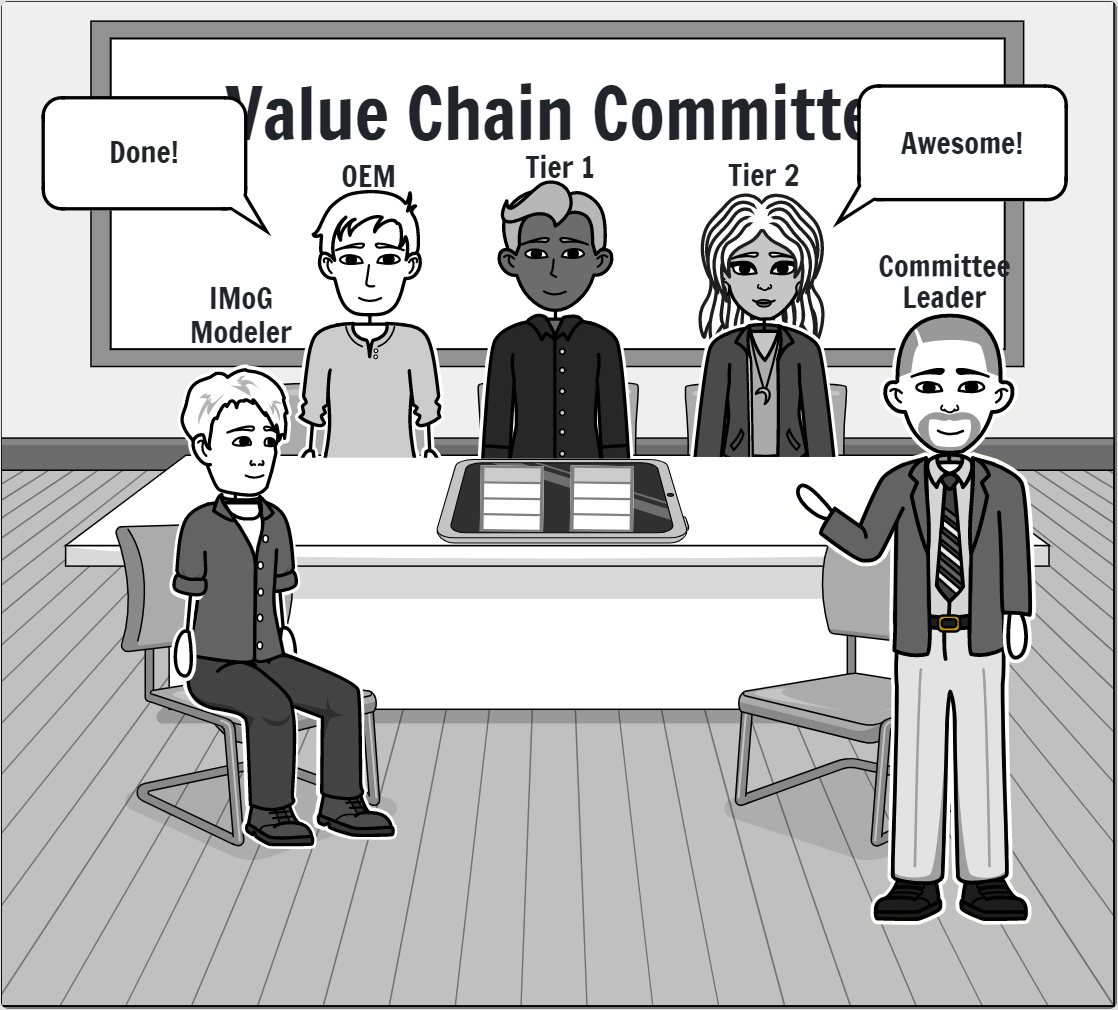}
	\end{minipage}

	\vspace{0.45cm}
	\begin{minipage}{0.475\textwidth}
		\includegraphics[width=\textwidth]{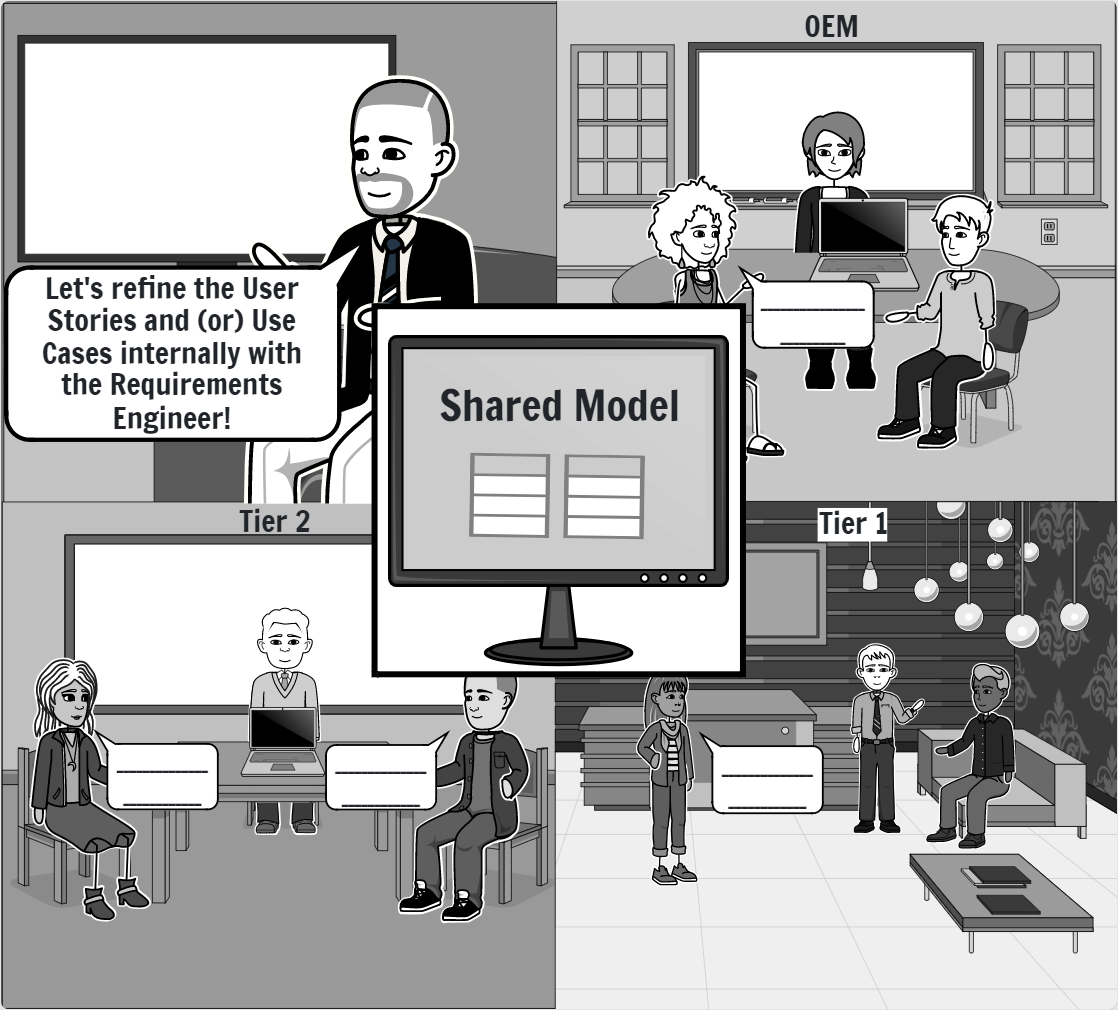}
	\end{minipage}
	\hfill
	\begin{minipage}{0.475\textwidth}
		\includegraphics[width=\textwidth]{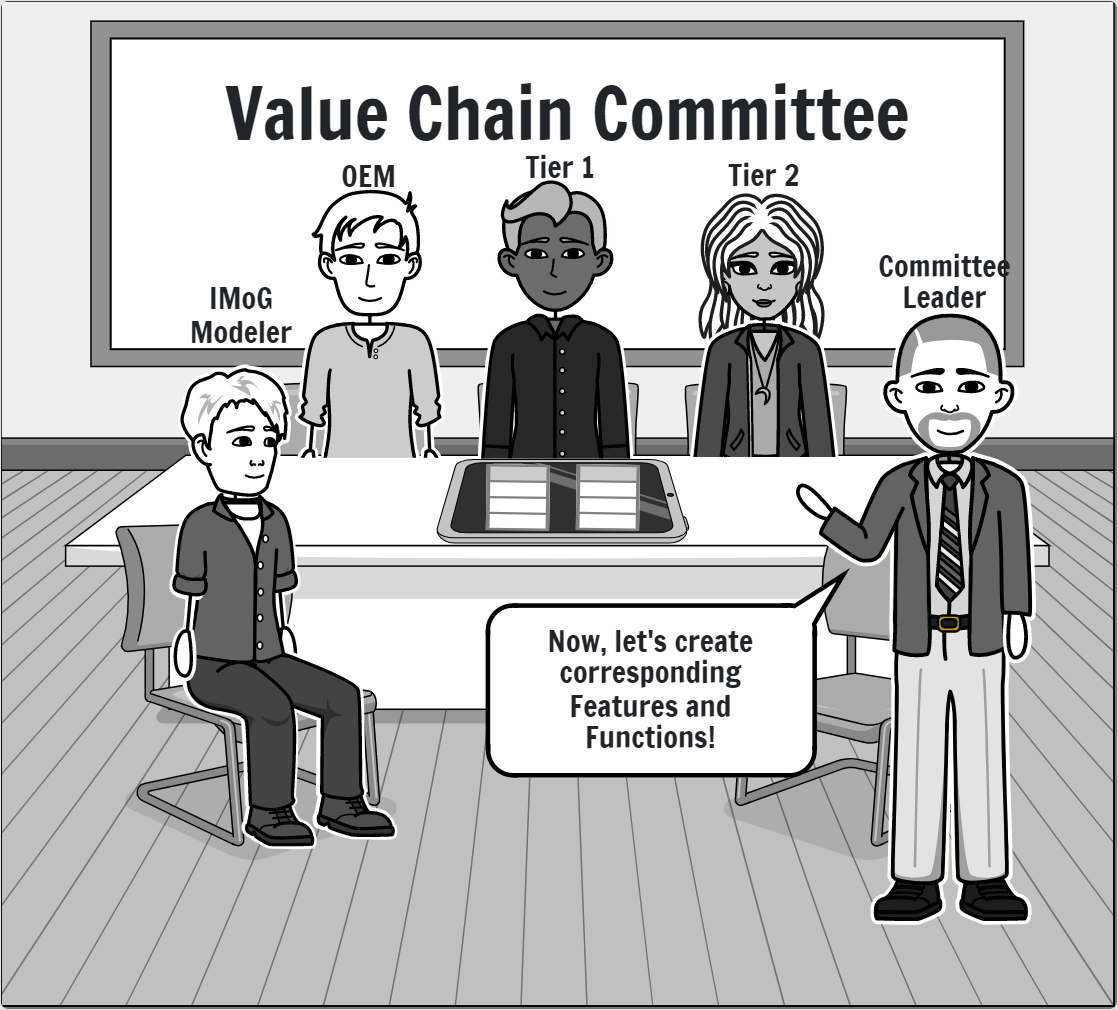}
	\end{minipage}
	\caption{The second activity: Feature and Function Identification}
	\label{fig:storyboard:fp:1}
\end{figure}

\begin{figure}
	\begin{minipage}{0.475\textwidth}
		\includegraphics[width=\textwidth]{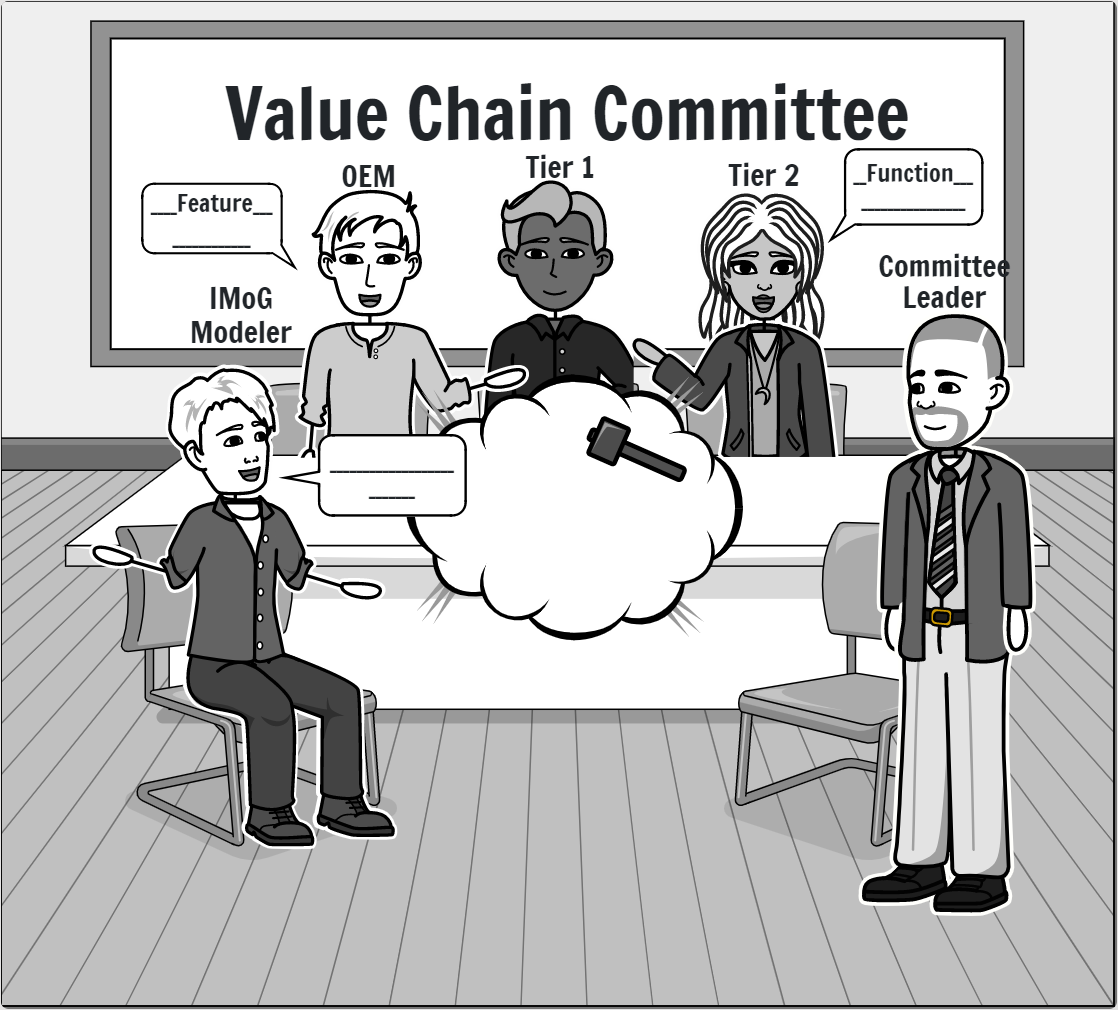}
	\end{minipage}
	\hfill
	\begin{minipage}{0.475\textwidth}
		\includegraphics[width=\textwidth]{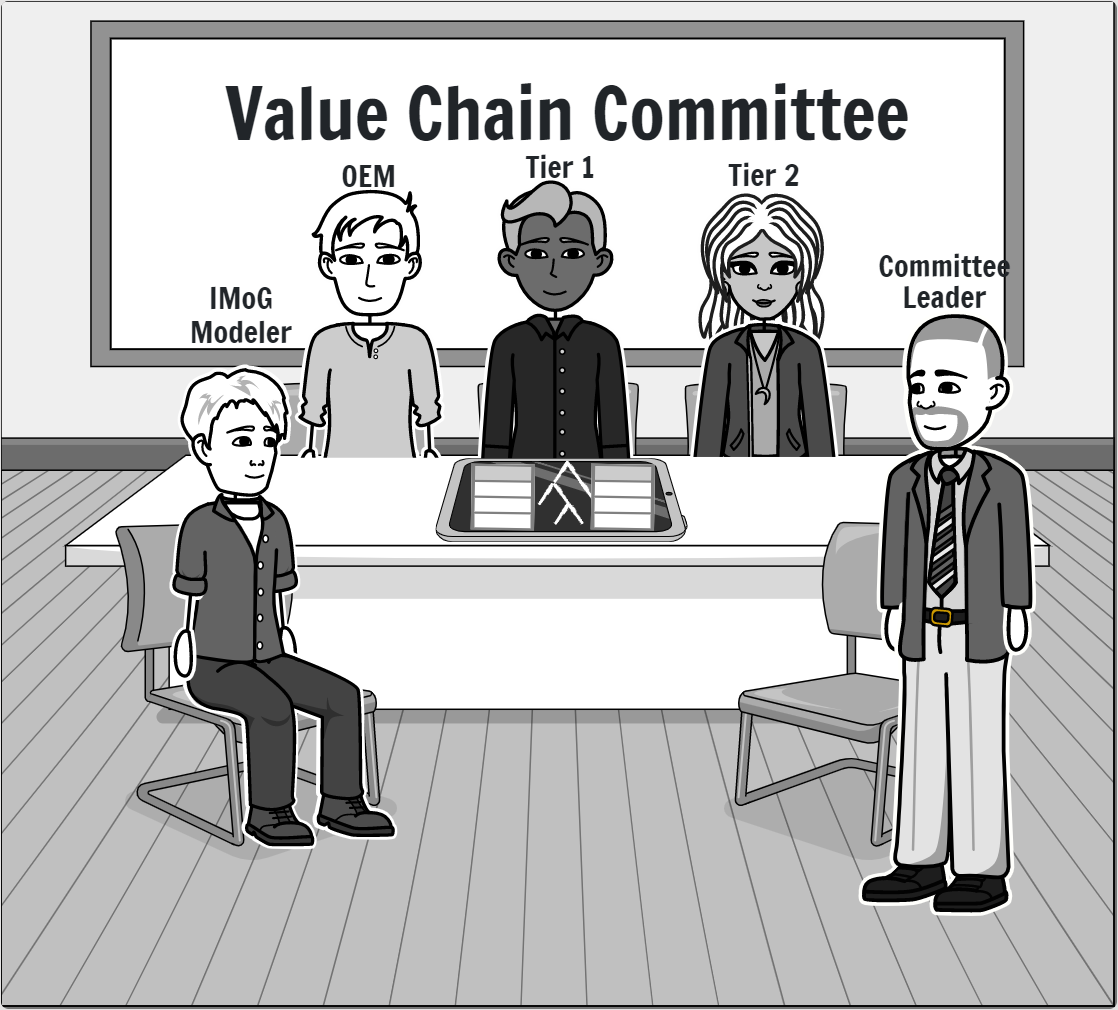}
	\end{minipage}

	\vspace{0.45cm}
	\begin{minipage}{0.475\textwidth}
		\includegraphics[width=\textwidth]{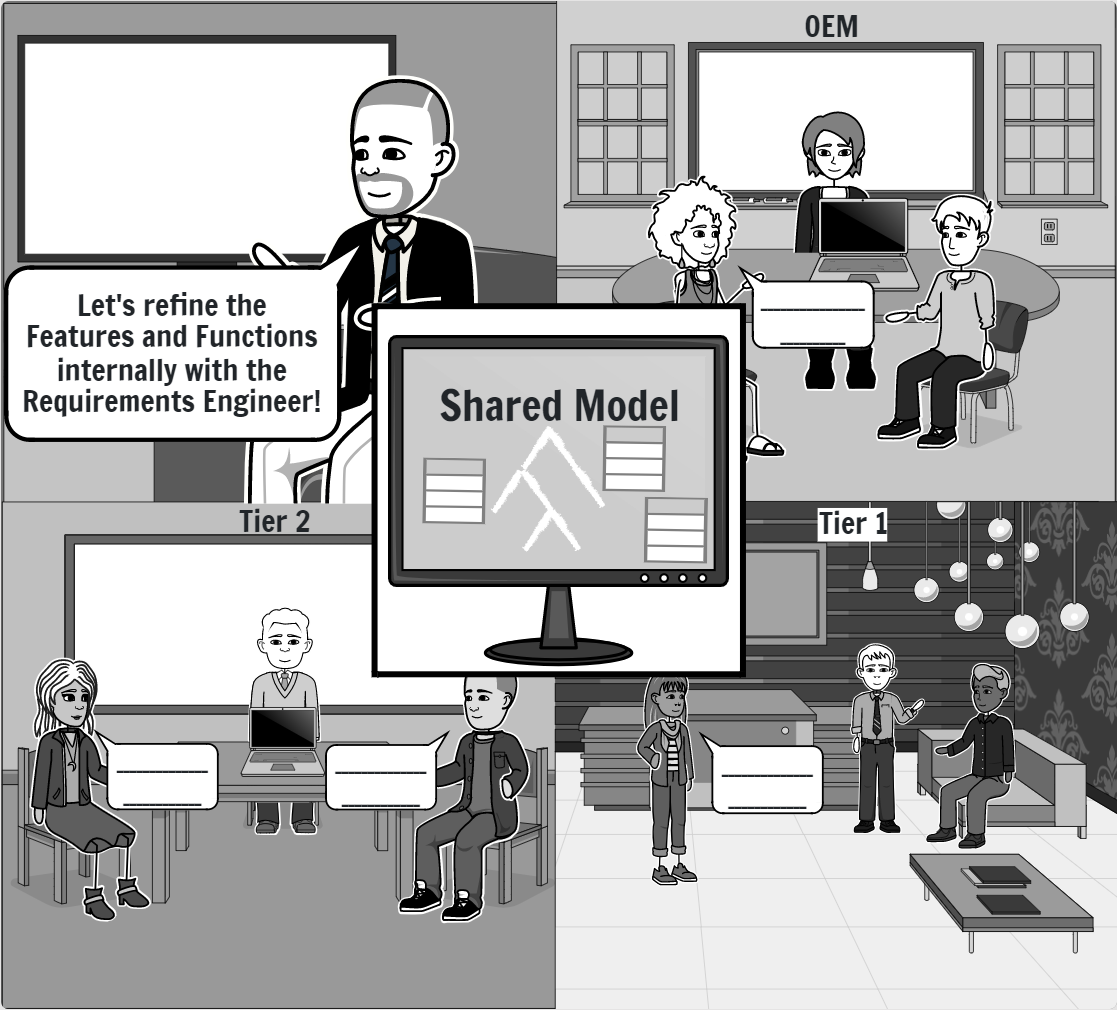}
	\end{minipage}
	\hfill
	\begin{minipage}{0.475\textwidth}
		\includegraphics[width=\textwidth]{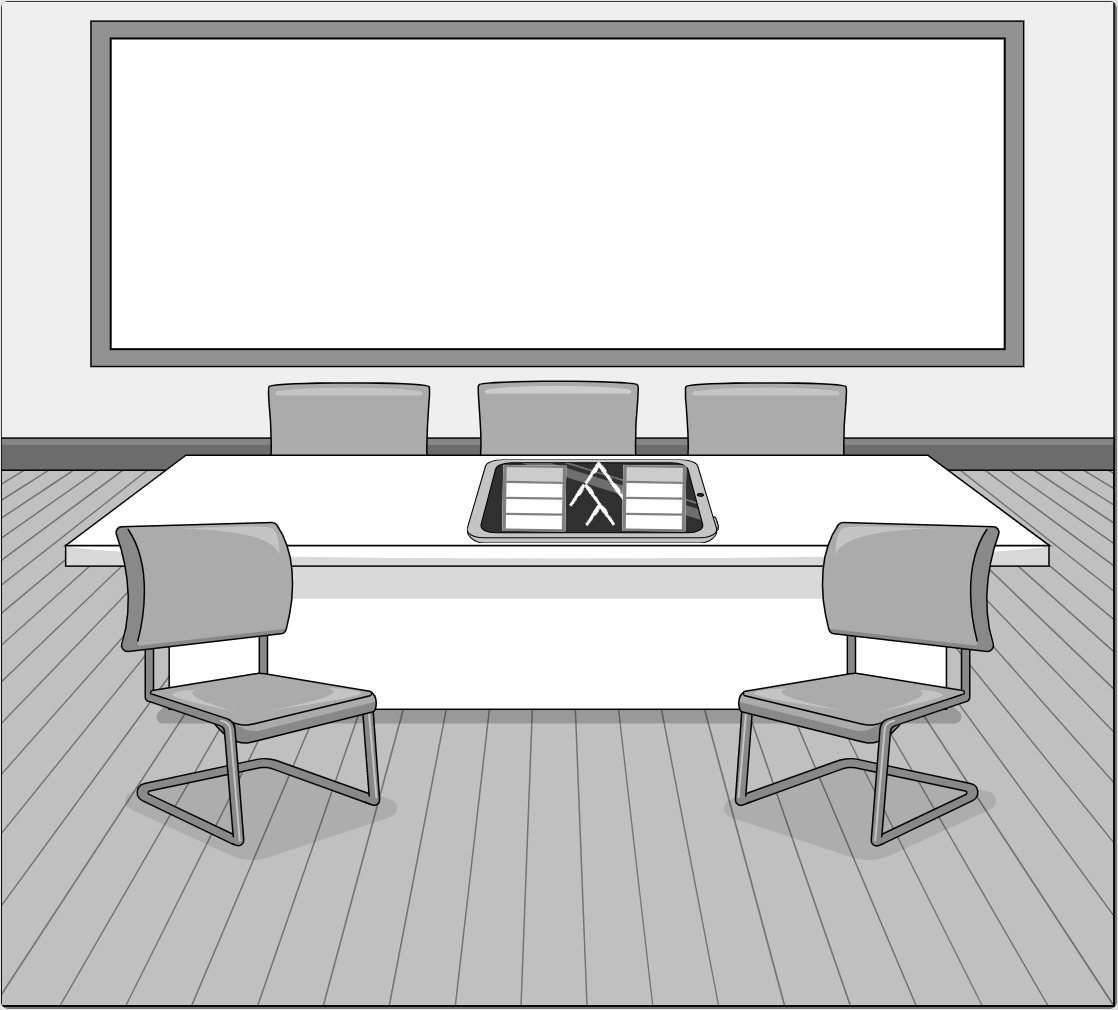}
	\end{minipage}
	\caption{The second activity: Feature and Function Identification (part 2)}
	\label{fig:storyboard:fp:2}
\end{figure}

The second activity is the Feature and Function Identification (see Figures \ref{fig:storyboard:fp:1} and \ref{fig:storyboard:fp:2}).
The committee leader invites the committee to a meeting to identify the features and functions.
The committee decides if they want to create user stories or use cases before identifying the features and functions.
In this case they agree to create them.
The committee starts to define user stories and use cases.
The user stories and use cases are refined internally in each company with their requirements engineer and consolidated in the committee.
Afterwards or before the elevation of the user stories and use cases, the features and function are defined and put into relation in a feature model.
Similarly to the user stories and use cases, the feature model is refined internally in each company with their requirements engineer and consolidated in the committee.
Then, the Feature and Function Identification activity is finished.

\begin{figure}
	\begin{minipage}{0.475\textwidth}
		\resizebox{\textwidth}{!}{%
			\begin{tikzpicture}[every node/.style={inner sep=0,outer sep=0}]
				\node[anchor=south west] at (0.12,0.12) {\includegraphics{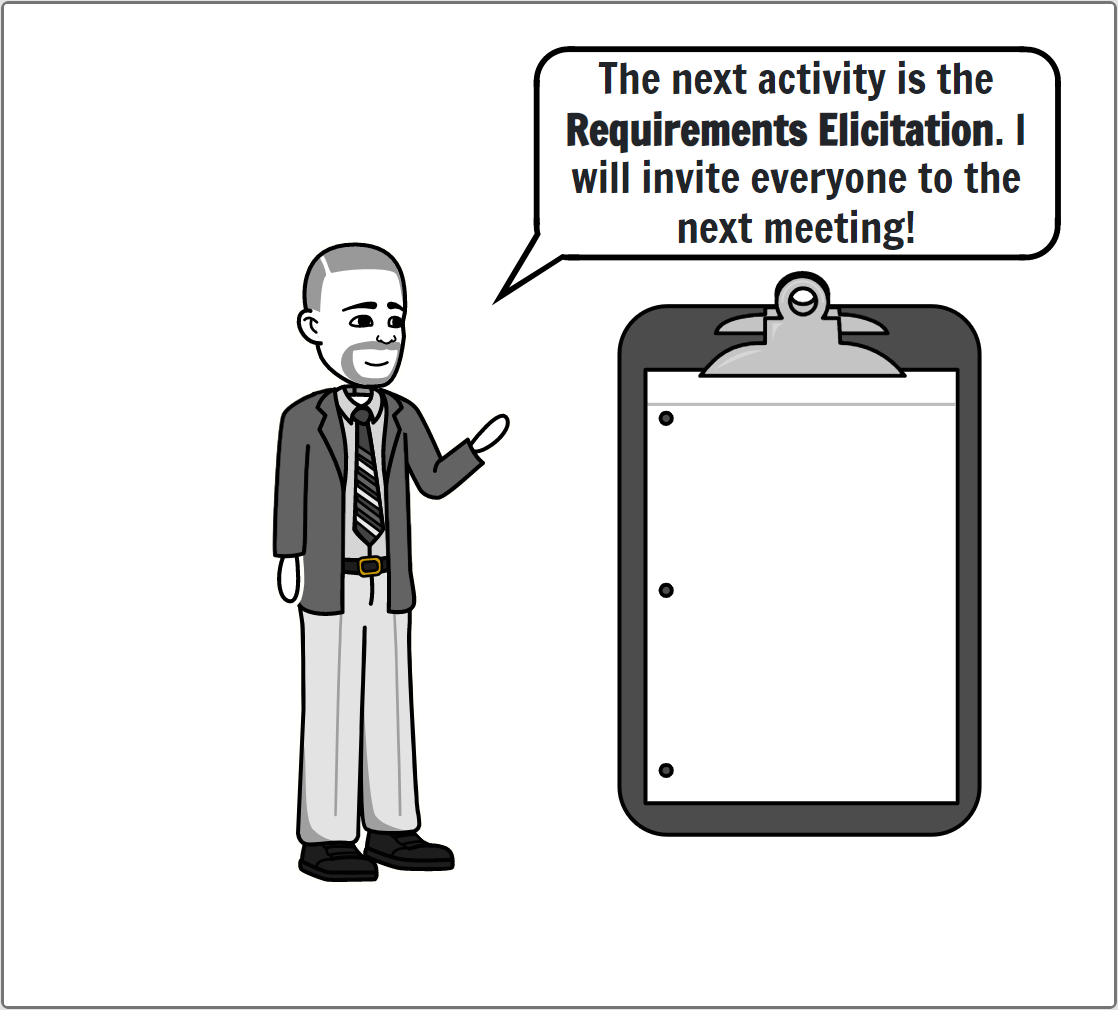}};
				\node[anchor=west, align=left] at (18.7,15.2) {\Large \textbf{Innovation Identification}};
				\node[anchor=west, align=left] at (18.7,14) {\Large \textbf{Feature and Function} \\\Large \textbf{Identification}};
				\node[anchor=west, align=left] at (18.7,12.59) {\Large \textbf{Requirements Elicitation} \\(Quality Requirements and Constraints)};
				\draw (18.4,11.9) -- (25,11.9);
				\node[anchor=west, align=left] at (18.7,11.3) {\Large \textbf{Solution Space Exploration}};
				\draw (18.4,10.8) -- (25,10.8);
				\node[anchor=west, align=left] at (18.7,9.8) {\Large \textbf{Extracting and saving} \\\Large \textbf{Insights for future} \\\Large \textbf{IMoG Innovations}};
				\node[anchor=west, align=left] at (18.7,8.4) {\Large \textbf{Roadmap Writing}};
				\node[anchor=west, align=left] at (18.7,7) {\Large \textbf{Maintaining and} \\\Large \textbf{Updating the Model} \\\Large \textbf{and Roadmap}};
				\node at (18.4,15.2) {\Large \color{green!70!black} \CheckmarkBold};
				\node at (18.4,14) {\Large \color{green!70!black} \CheckmarkBold};
			\end{tikzpicture}
		}
	\end{minipage}
	\hfill
	\begin{minipage}{0.475\textwidth}
		\includegraphics[width=\textwidth]{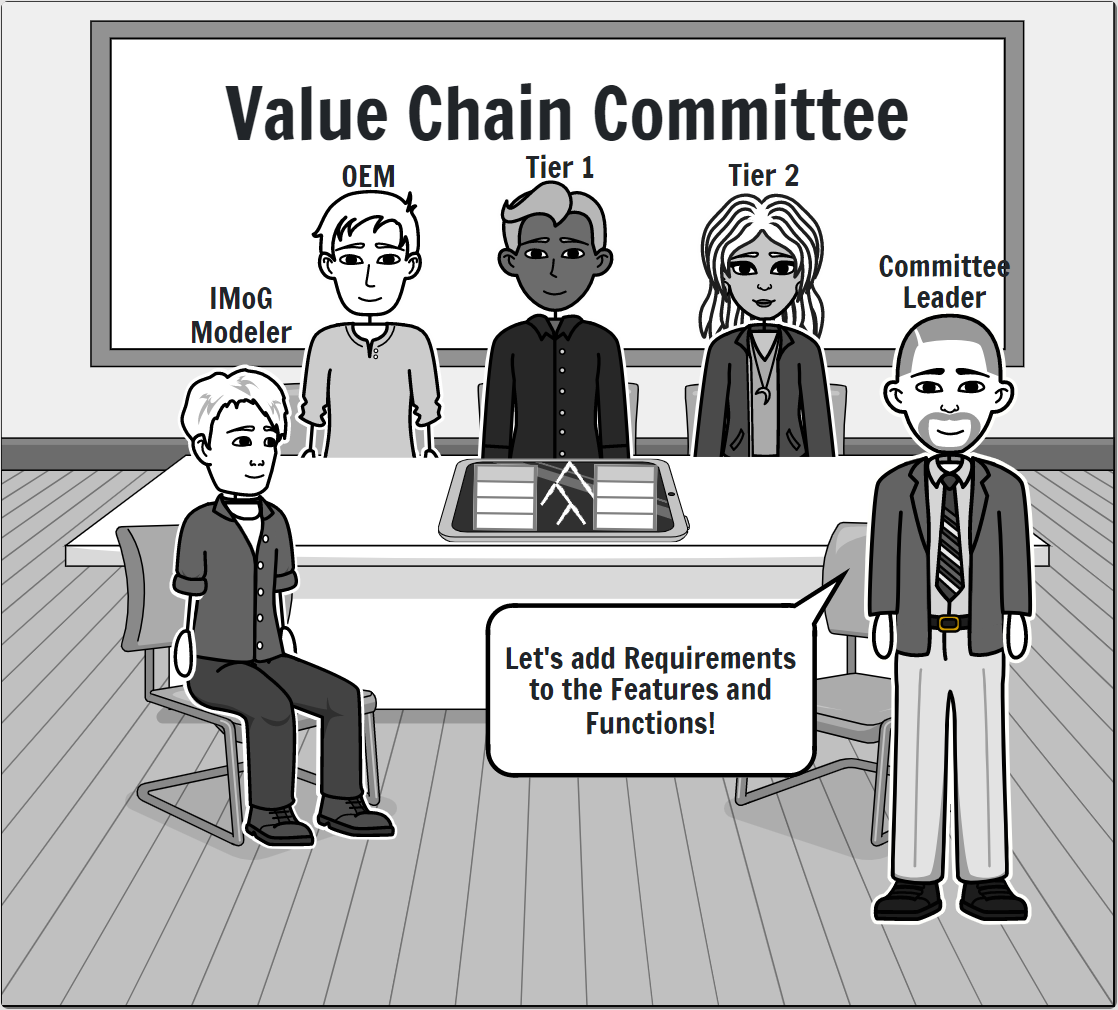}
	\end{minipage}

	\vspace{0.45cm}
	\begin{minipage}{0.475\textwidth}
		\includegraphics[width=\textwidth]{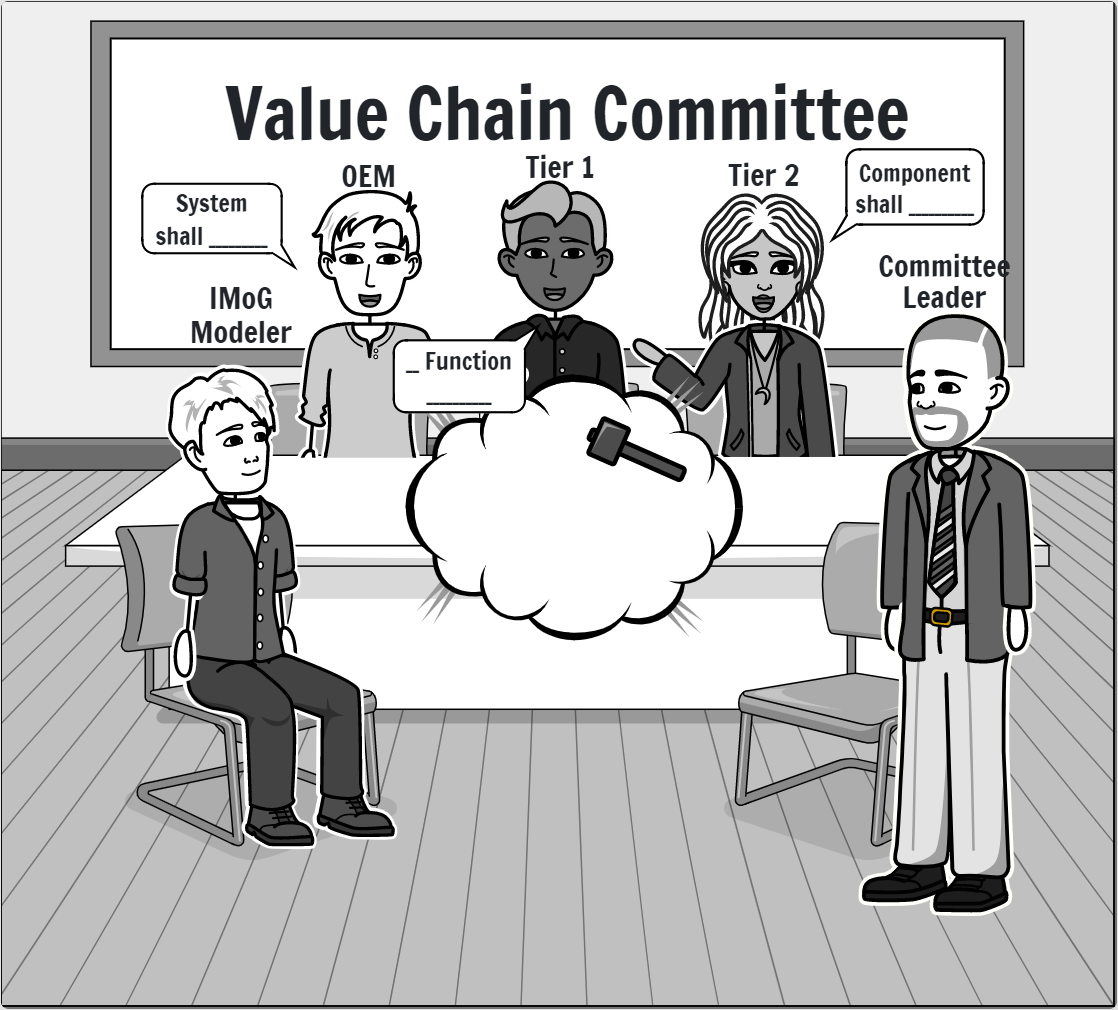}
	\end{minipage}
	\hfill
	\begin{minipage}{0.475\textwidth}
		\includegraphics[width=\textwidth]{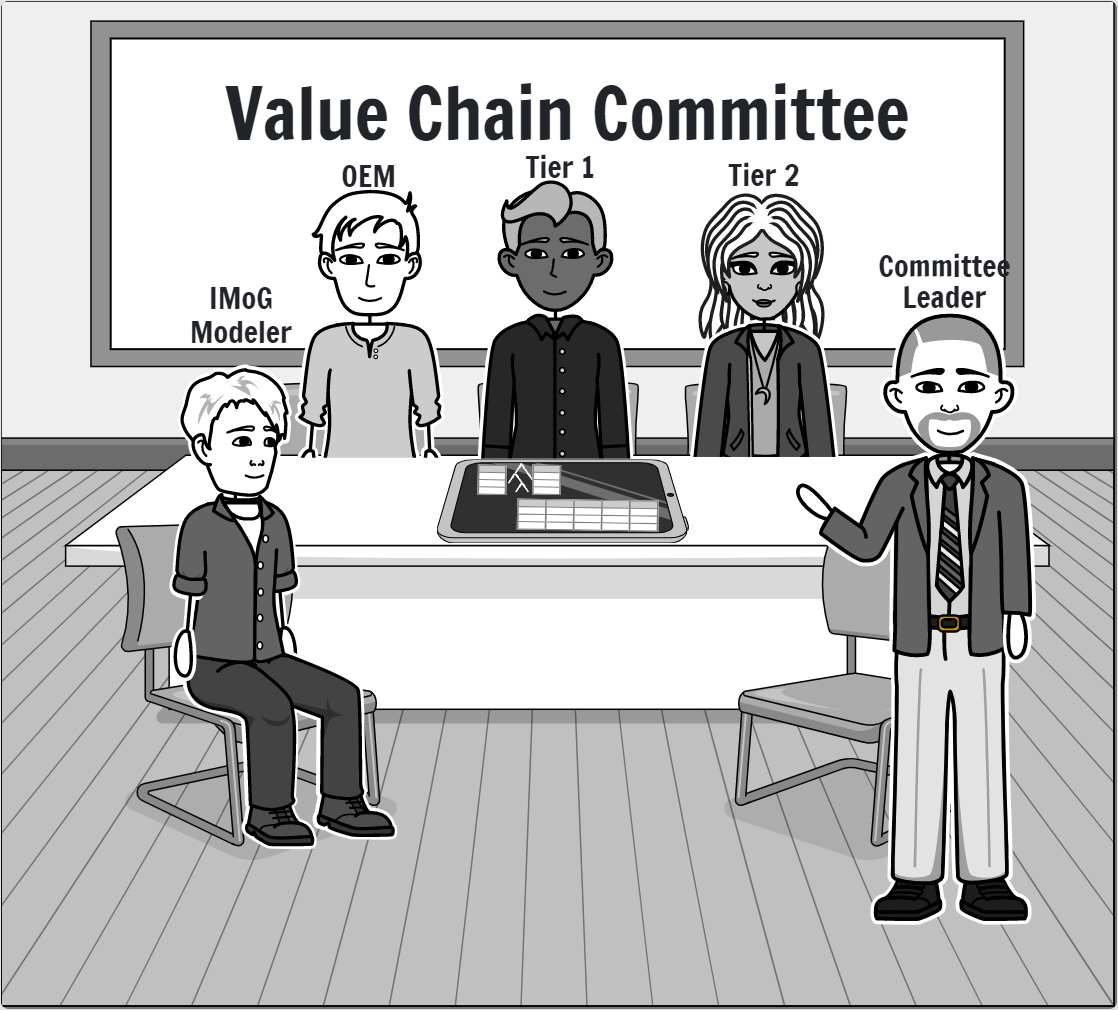}
	\end{minipage}

	\vspace{0.45cm}
	\begin{minipage}{0.475\textwidth}
		\includegraphics[width=\textwidth]{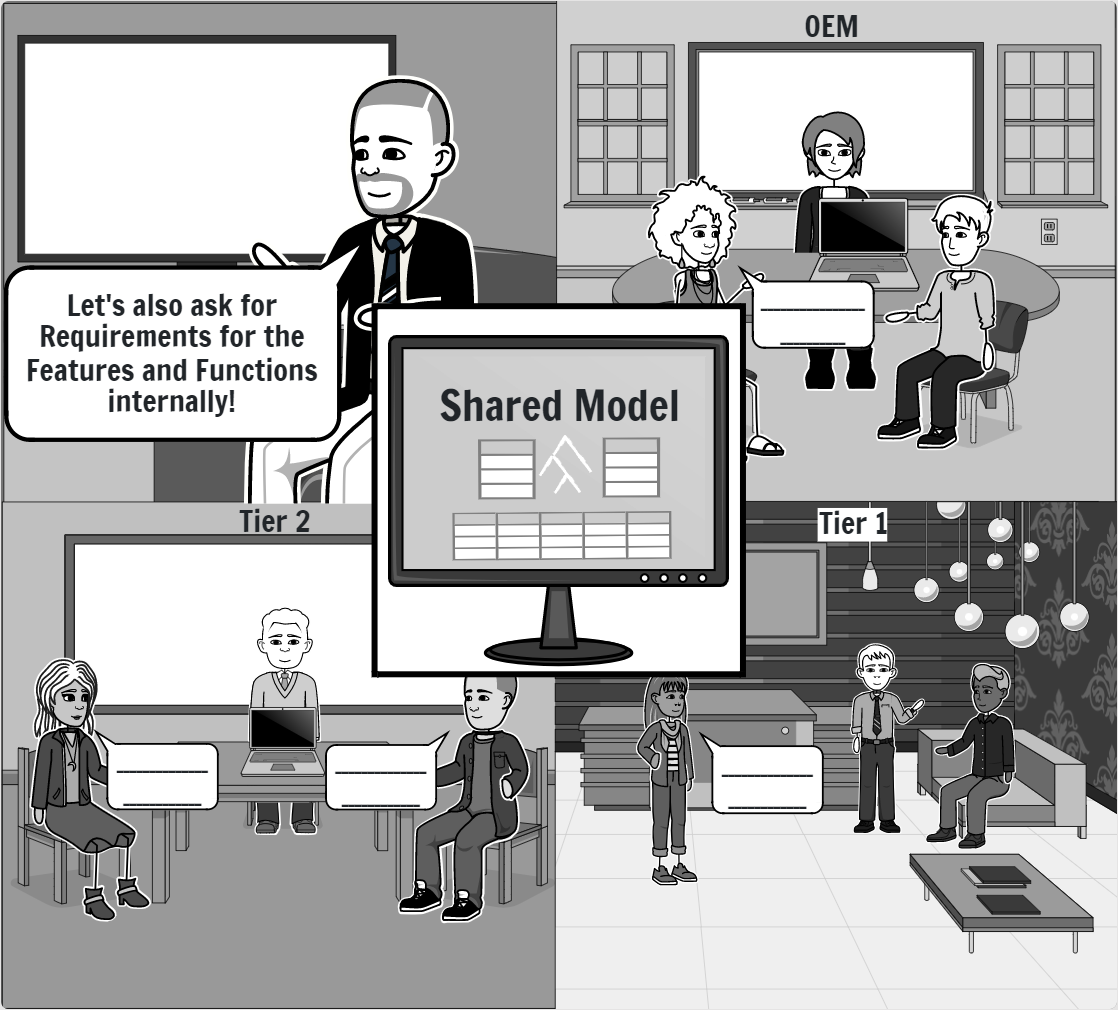}
	\end{minipage}
	\hfill
	\begin{minipage}{0.475\textwidth}
		\includegraphics[width=\textwidth]{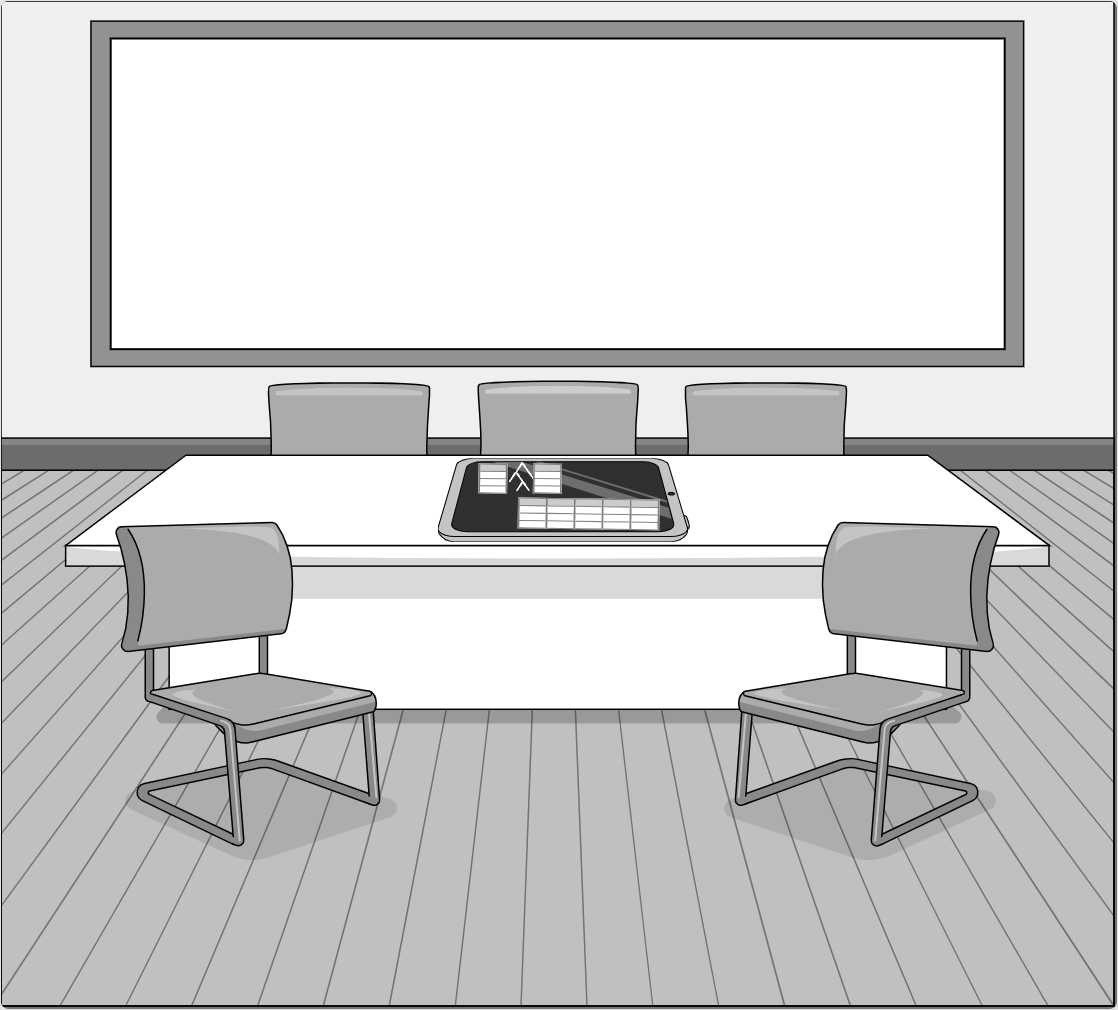}
	\end{minipage}
	\caption{The third activity: Requirements Elicitation}
	\label{fig:storyboard:qp:1}
\end{figure}

\begin{figure}
	\begin{minipage}{0.475\textwidth}
		\includegraphics[width=\textwidth]{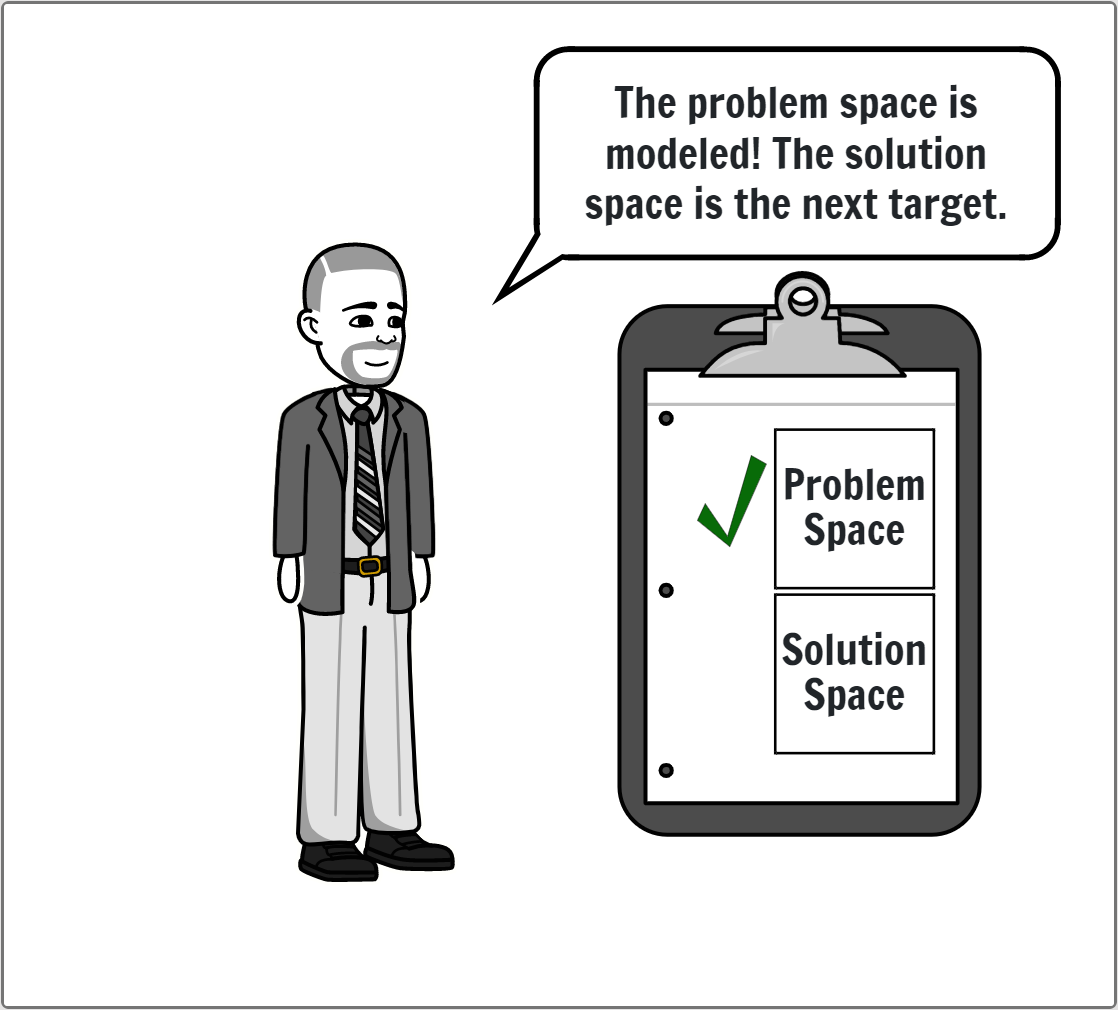}
	\end{minipage}
	\hfill
	\begin{minipage}{0.475\textwidth}
	\end{minipage}
	\caption{The third activity: Requirements Elicitation (part 2)}
	\label{fig:storyboard:qp:2}
\end{figure}

The next activity is the Requirements Elicitation.
The committee leader invites the committee to a meeting.
Then the committee meets to add the requirements to the features and functions (see Figures \ref{fig:storyboard:qp:1} and \ref{fig:storyboard:qp:2}).
They write down the requirements and constraints that were raised in the last two activities and now structurally elicit missing quality requirements and constraints for the features and functions.
Once they are done, the requirements and constraints are refined and completed internally with the requirements engineer.
Afterwards, the Requirements Elicitation activity and the modeling of the problem space is done.

\begin{figure}
	\begin{minipage}{0.475\textwidth}
		\resizebox{\textwidth}{!}{%
			\begin{tikzpicture}[every node/.style={inner sep=0,outer sep=0}]
				\node[anchor=south west] at (0.12,0.12) {\includegraphics{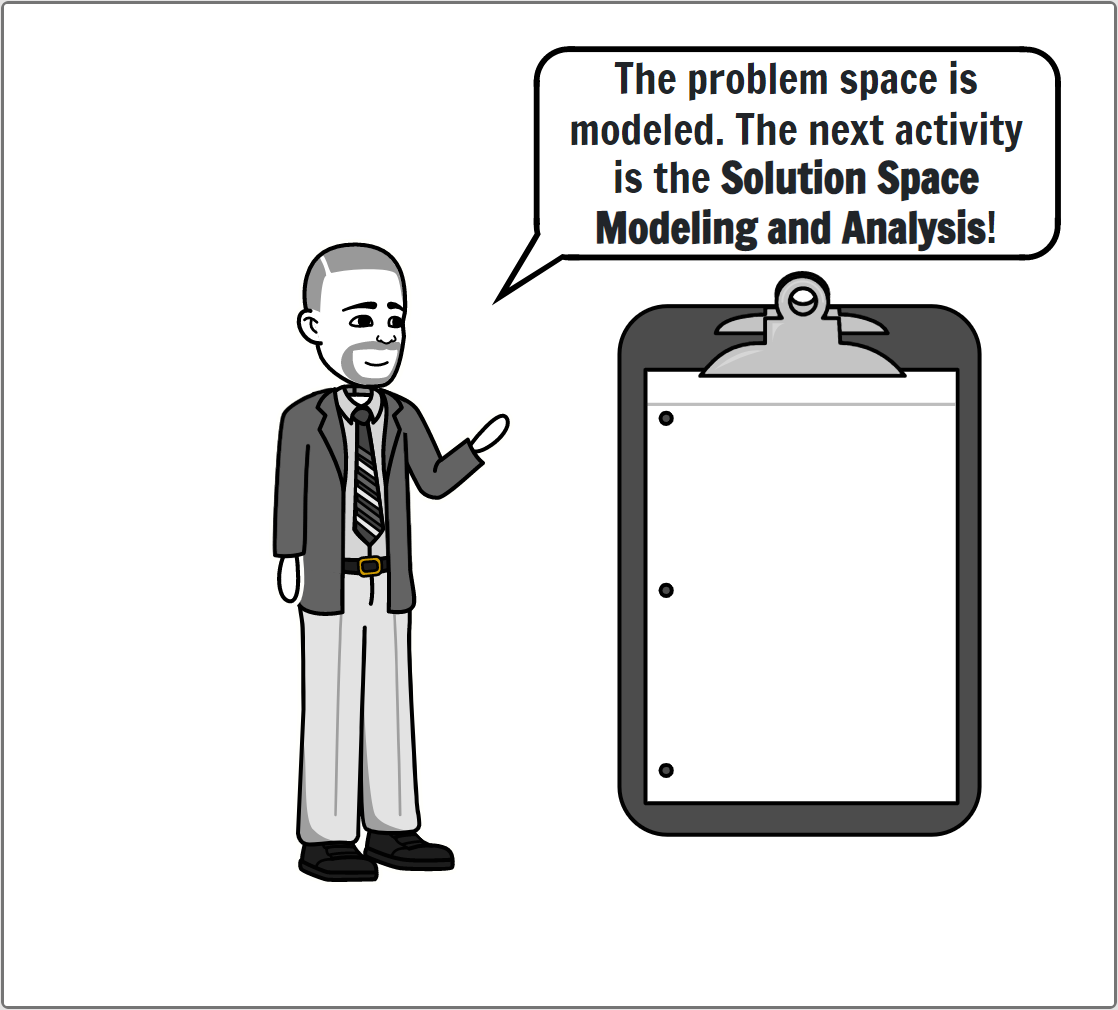}};
				\node[anchor=west, align=left] at (18.7,15.2) {\Large \textbf{Innovation Identification}};
				\node[anchor=west, align=left] at (18.7,14) {\Large \textbf{Feature and Function} \\\Large \textbf{Identification}};
				\node[anchor=west, align=left] at (18.7,12.59) {\Large \textbf{Requirements Elicitation} \\(Quality Requirements and Constraints)};
				\draw (18.4,11.9) -- (25,11.9);
				\node[anchor=west, align=left] at (18.7,11.3) {\Large \textbf{Solution Space Exploration}};
				\draw (18.4,10.8) -- (25,10.8);
				\node[anchor=west, align=left] at (18.7,9.8) {\Large \textbf{Extracting and saving} \\\Large \textbf{Insights for future} \\\Large \textbf{IMoG Innovations}};
				\node[anchor=west, align=left] at (18.7,8.4) {\Large \textbf{Roadmap Writing}};
				\node[anchor=west, align=left] at (18.7,7) {\Large \textbf{Maintaining and} \\\Large \textbf{Updating the Model} \\\Large \textbf{and Roadmap}};
				\node at (18.4,15.2) {\Large \color{green!70!black} \CheckmarkBold};
				\node at (18.4,14) {\Large \color{green!70!black} \CheckmarkBold};
				\node at (18.4,12.59) {\Large \color{green!70!black} \CheckmarkBold};
			\end{tikzpicture}
		}
	\end{minipage}
	\hfill
	\begin{minipage}{0.475\textwidth}
		\resizebox{\textwidth}{!}{%
			\begin{tikzpicture}[every node/.style={inner sep=0,outer sep=0}]
				\node[anchor=south west] at (0.12,0.12) {\includegraphics{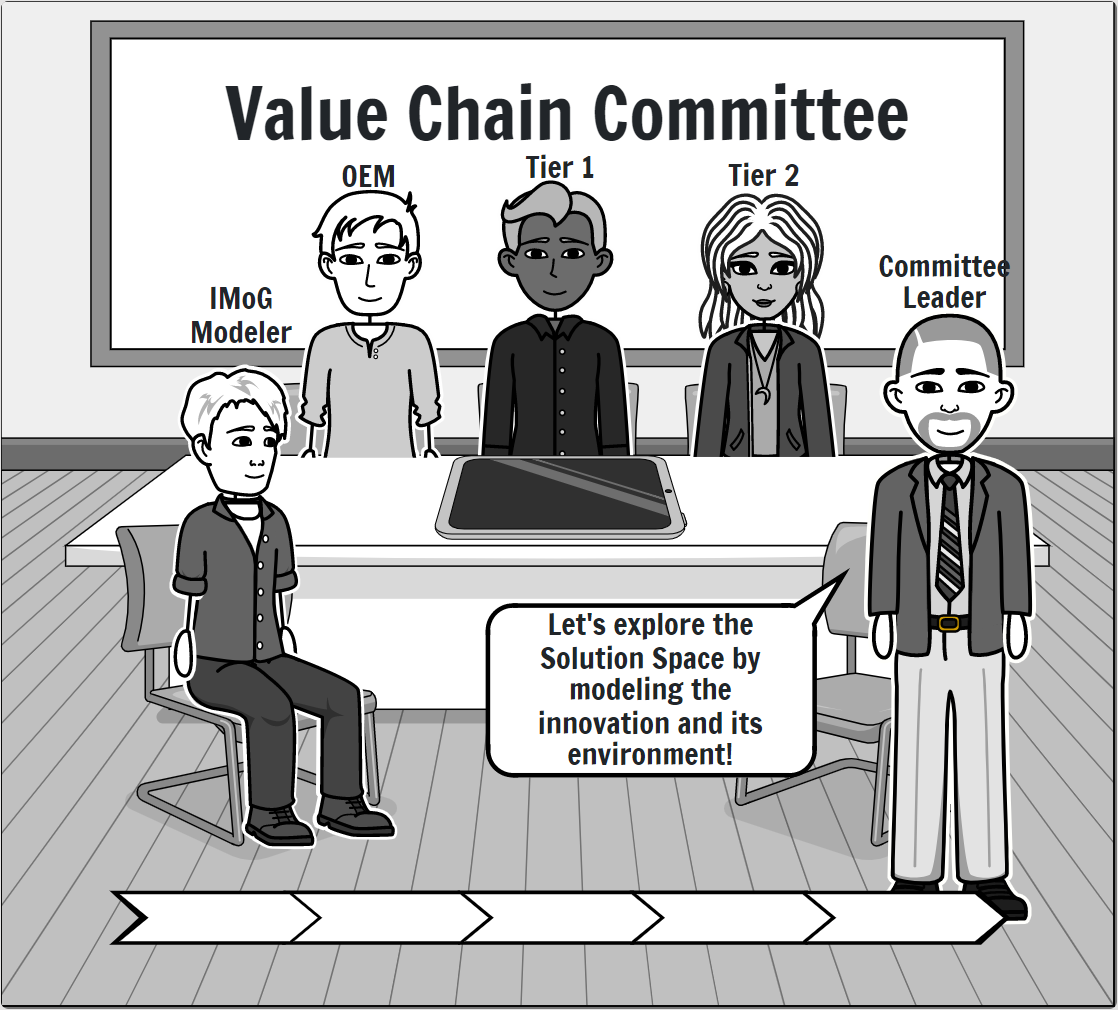}};
				\node[anchor=center, align=center] at (5.55,2.45) {\large \textbf{Context Level}\\\large \textbf{ Modeling}};
				\node[anchor=center, align=center] at (10.5,2.45) {\large \textbf{System}\\\large \textbf{Decomposition}\\\large \textbf{and FP Mapping}};
				\node[anchor=center, align=center] at (14.9,2.45) {\large \textbf{Effect Chain and}\\\large \textbf{Impact Analysis}};
				\node[anchor=center, align=center] at (19.7,2.45) {\large \textbf{Requirements}\\\large \textbf{Elicitation for}\\\large \textbf{Solutions}};
				\node[anchor=center, align=center] at (23.9,2.45) {\large \textbf{Alternative (and}\\\large \textbf{KPI) Exploration}};
			\end{tikzpicture}
		}
	\end{minipage}

	\vspace{0.45cm}
	\begin{minipage}{0.475\textwidth}
		\includegraphics[width=\textwidth]{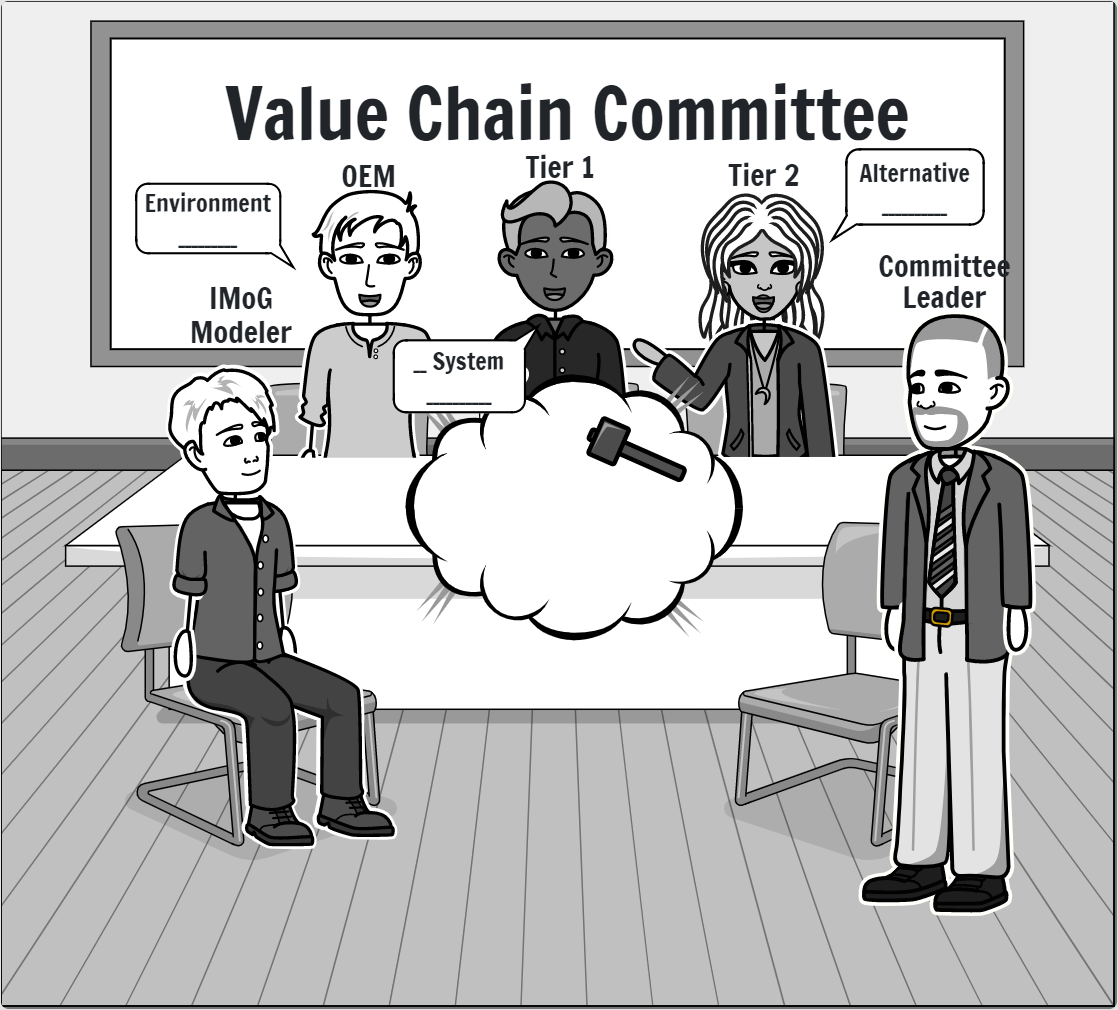}
	\end{minipage}
	\hfill
	\begin{minipage}{0.475\textwidth}
		\includegraphics[width=\textwidth]{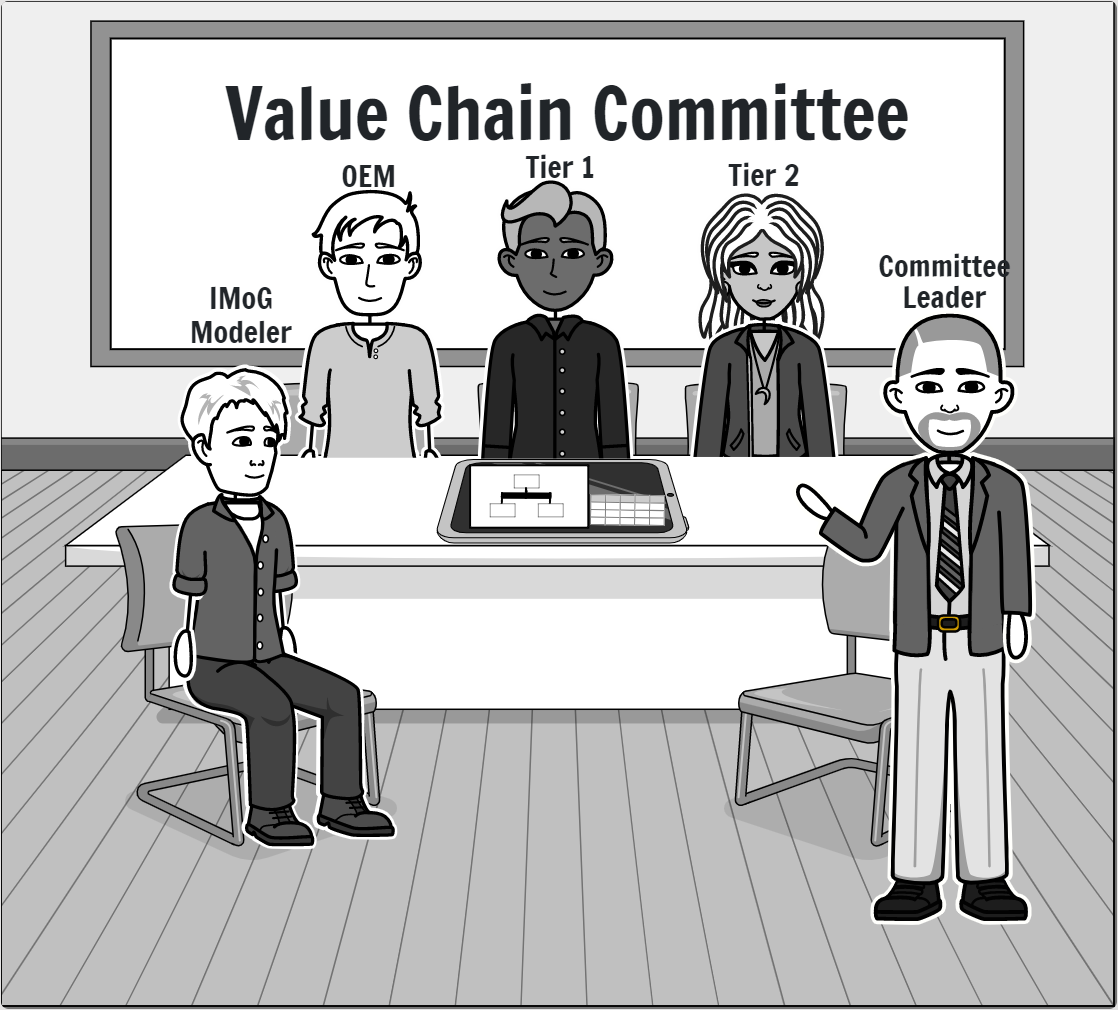}
	\end{minipage}

	\vspace{0.45cm}
	\begin{minipage}{0.475\textwidth}
		\resizebox{\textwidth}{!}{%
			\begin{tikzpicture}[every node/.style={inner sep=0,outer sep=0}]
				\node[anchor=south west] {\includegraphics{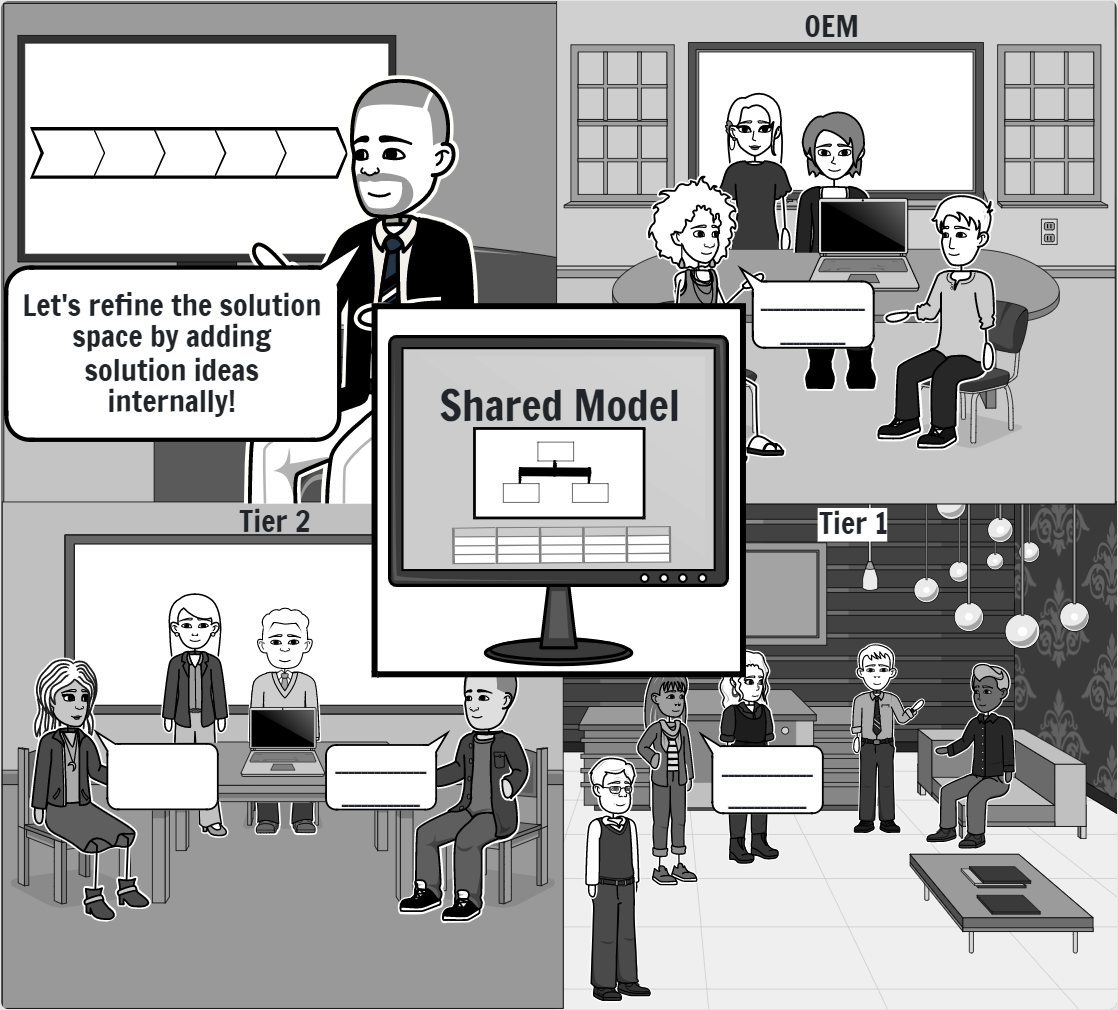}};
				\node at (1.8,23) {\scriptsize \textbf{Context}};
				\node at (1.8,22.7) {\scriptsize \textbf{Level}};
				\node at (1.8,22.4) {\scriptsize \textbf{Modeling}};
				\node at (3.3,23.15) {\scriptsize \textbf{System}};
				\node at (3.5,22.85) {\scriptsize \textbf{Decomp.}};
				\node at (3.4,22.55) {\scriptsize \textbf{and FP}};
				\node at (3.3,22.25) {\scriptsize \textbf{Mapping}};
				\node at (5,23.15) {\scriptsize \textbf{Effect}};
				\node at (5.1,22.85) {\scriptsize \textbf{Chain and}};
				\node at (5,22.55) {\scriptsize \textbf{Impact}};
				\node at (4.9,22.25) {\scriptsize \textbf{Analysis}};
				\node at (6.65,23.15) {\scriptsize \textbf{Requirem.}};
				\node at (6.7,22.85) {\scriptsize \textbf{Elicitation}};
				\node at (6.55,22.55) {\scriptsize \textbf{for}};
				\node at (6.55,22.25) {\scriptsize \textbf{Solutions}};
				\node at (8.25,23) {\scriptsize \textbf{Alternative}};
				\node at (8.3,22.7) {\scriptsize \textbf{(and KPI)}};
				\node at (8.25,22.4) {\scriptsize \textbf{Exploration}};
			\end{tikzpicture}
		}
	\end{minipage}
	\hfill
	\begin{minipage}{0.475\textwidth}
		\includegraphics[width=\textwidth]{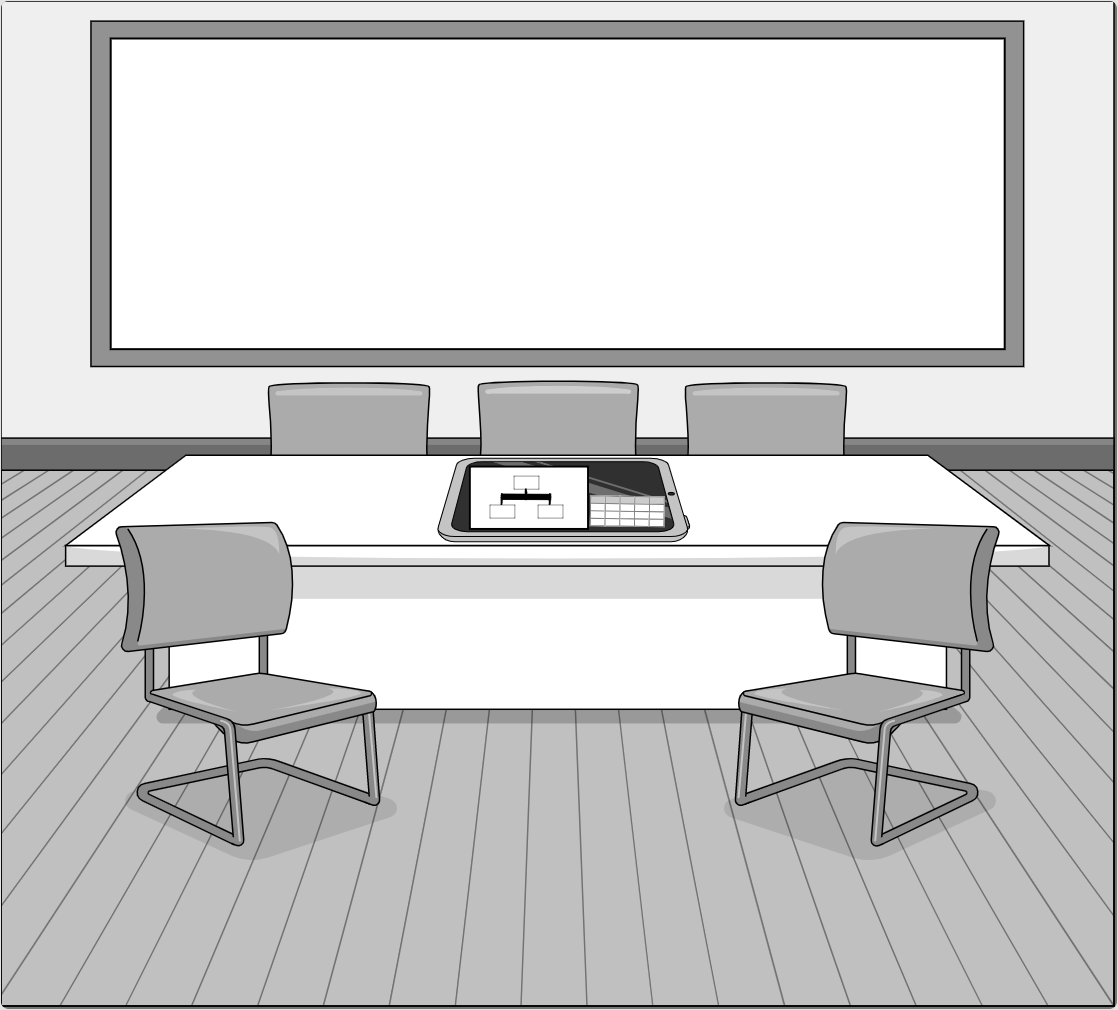}
	\end{minipage}
	\caption{The fourth activity: Solution Space Modeling and Analysis}
	\label{fig:storyboard:structural:1}
\end{figure}

\begin{figure}
	\begin{minipage}{0.475\textwidth}
		\includegraphics[width=\textwidth]{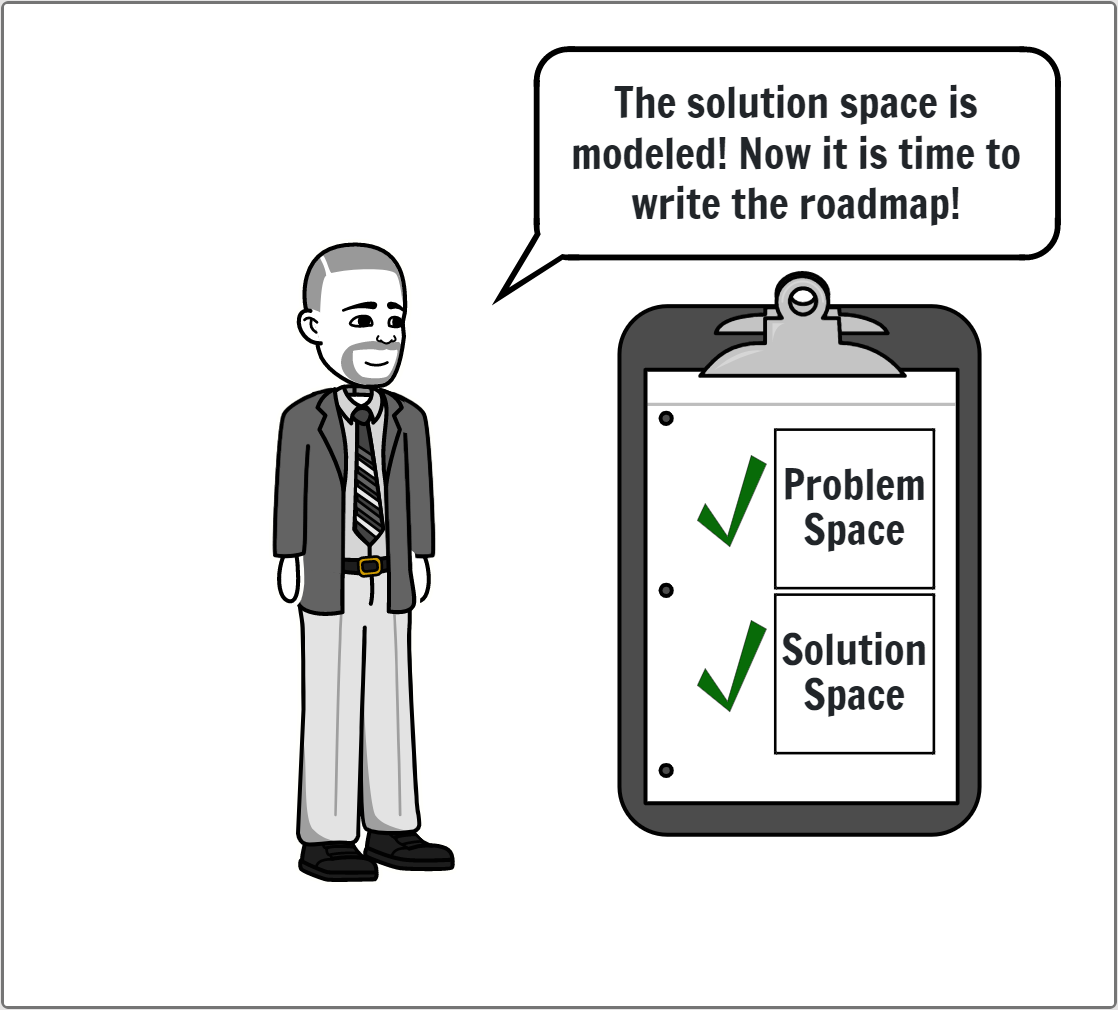}
	\end{minipage}
	\hfill
	\begin{minipage}{0.475\textwidth}
	\end{minipage}
	\caption{The fourth activity: Solution Space Modeling and Analysis (part 2)}
	\label{fig:storyboard:structural:2}
\end{figure}

The next activity is the Solution Space Modeling and Analysis (see Figures \ref{fig:storyboard:structural:1} and \ref{fig:storyboard:structural:2}).
The committee leader invites the committee to another meeting.
The committee explores the solution space of the innovation by modeling the innovation and its environment.
The exploration includes several steps like the context level modeling, the system decomposition and feature mapping, the effect chain and impact analysis, the requirements elicitation for solutions and the alternative and key performance indicator exploration.
Once the exploration is drafted, the refinement of the innovation model takes place internally within the companies.
In this activity, the system architect takes the main lead and gets support by the requirements engineer and the domain expert.
When this activity is finished, the solution space is also finished.

\begin{figure}
	\begin{minipage}{0.475\textwidth}
		\resizebox{\textwidth}{!}{%
			\begin{tikzpicture}[every node/.style={inner sep=0,outer sep=0}]
				\node[anchor=south west] at (0.12,0.12) {\includegraphics{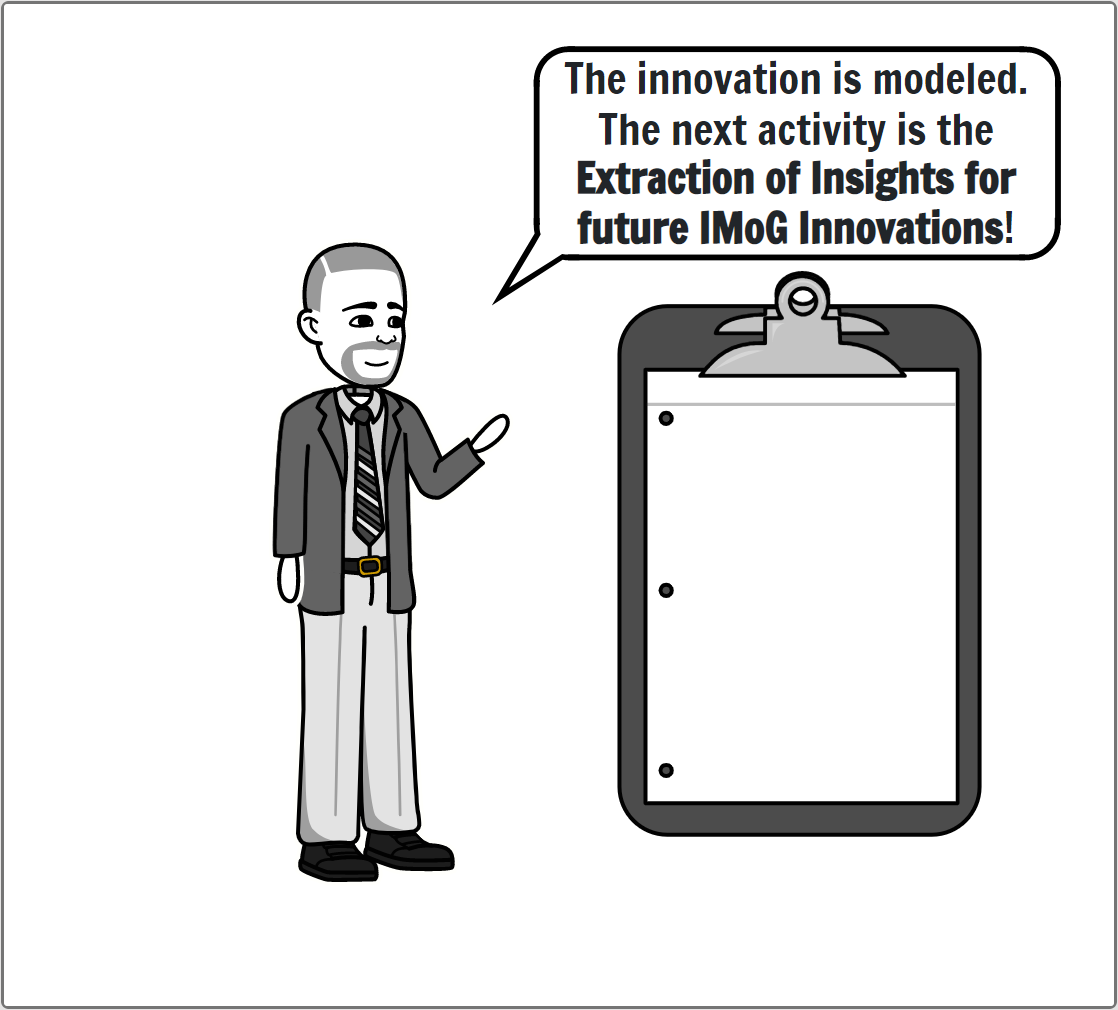}};
				\node[anchor=west, align=left] at (18.7,15.2) {\Large \textbf{Innovation Identification}};
				\node[anchor=west, align=left] at (18.7,14) {\Large \textbf{Feature and Function} \\\Large \textbf{Identification}};
				\node[anchor=west, align=left] at (18.7,12.59) {\Large \textbf{Requirements Elicitation} \\(Quality Requirements and Constraints)};
				\draw (18.4,11.9) -- (25,11.9);
				\node[anchor=west, align=left] at (18.7,11.3) {\Large \textbf{Solution Space Exploration}};
				\draw (18.4,10.8) -- (25,10.8);
				\node[anchor=west, align=left] at (18.7,9.8) {\Large \textbf{Extracting and saving} \\\Large \textbf{Insights for future} \\\Large \textbf{IMoG Innovations}};
				\node[anchor=west, align=left] at (18.7,8.4) {\Large \textbf{Roadmap Writing}};
				\node[anchor=west, align=left] at (18.7,7) {\Large \textbf{Maintaining and} \\\Large \textbf{Updating the Model} \\\Large \textbf{and Roadmap}};
				\node at (18.4,15.2) {\Large \color{green!70!black} \CheckmarkBold};
				\node at (18.4,14) {\Large \color{green!70!black} \CheckmarkBold};
				\node at (18.4,12.59) {\Large \color{green!70!black} \CheckmarkBold};
				\node at (18.4,11.3) {\Large \color{green!70!black} \CheckmarkBold};
			\end{tikzpicture}
		}
	\end{minipage}
	\hfill
	\begin{minipage}{0.475\textwidth}
		\includegraphics[width=\textwidth]{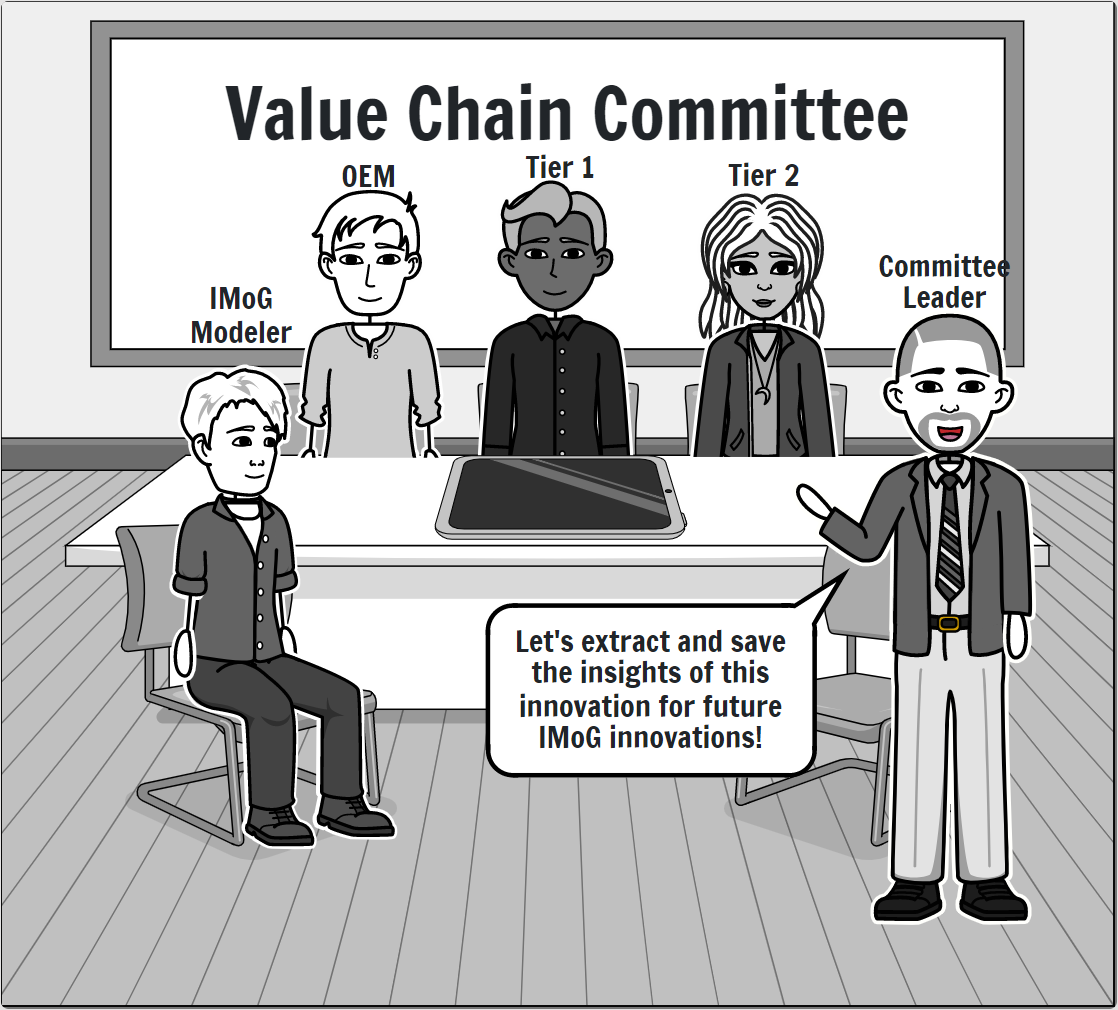}
	\end{minipage}

	\vspace{0.45cm}
	\begin{minipage}{0.475\textwidth}
		\includegraphics[width=\textwidth]{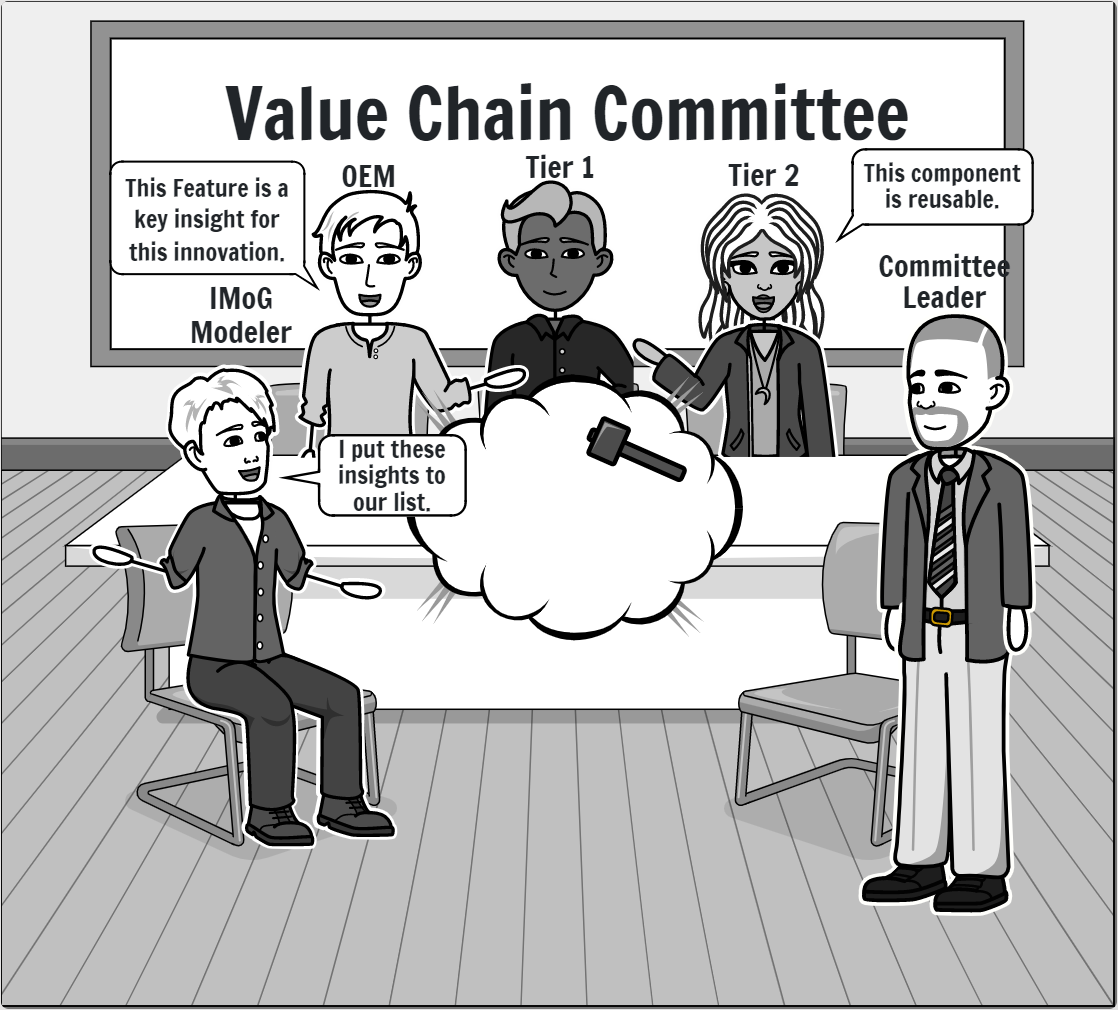}
	\end{minipage}
	\hfill
	\begin{minipage}{0.475\textwidth}
		\includegraphics[width=\textwidth]{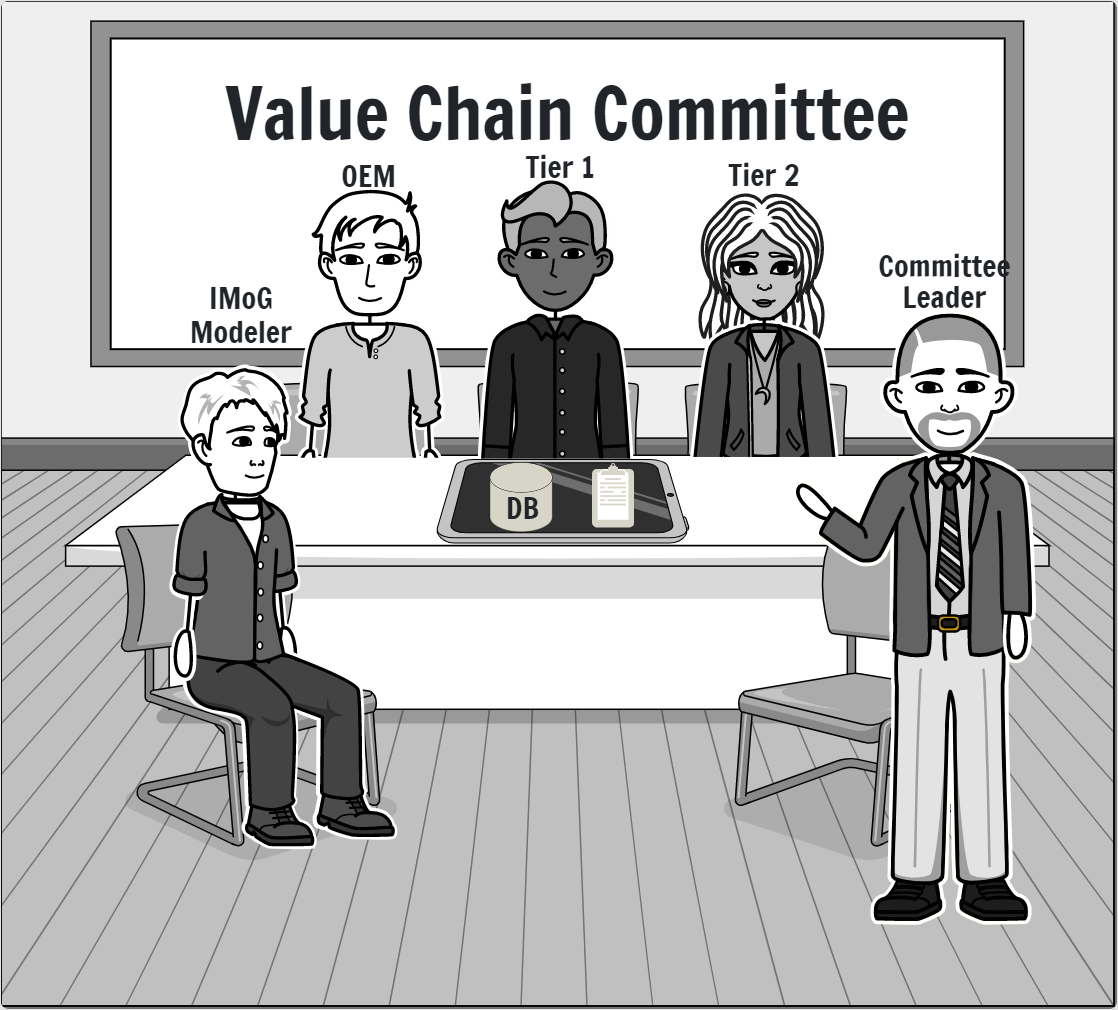}
	\end{minipage}

	\vspace{0.45cm}
	\begin{minipage}{0.475\textwidth}
		\includegraphics[width=\textwidth]{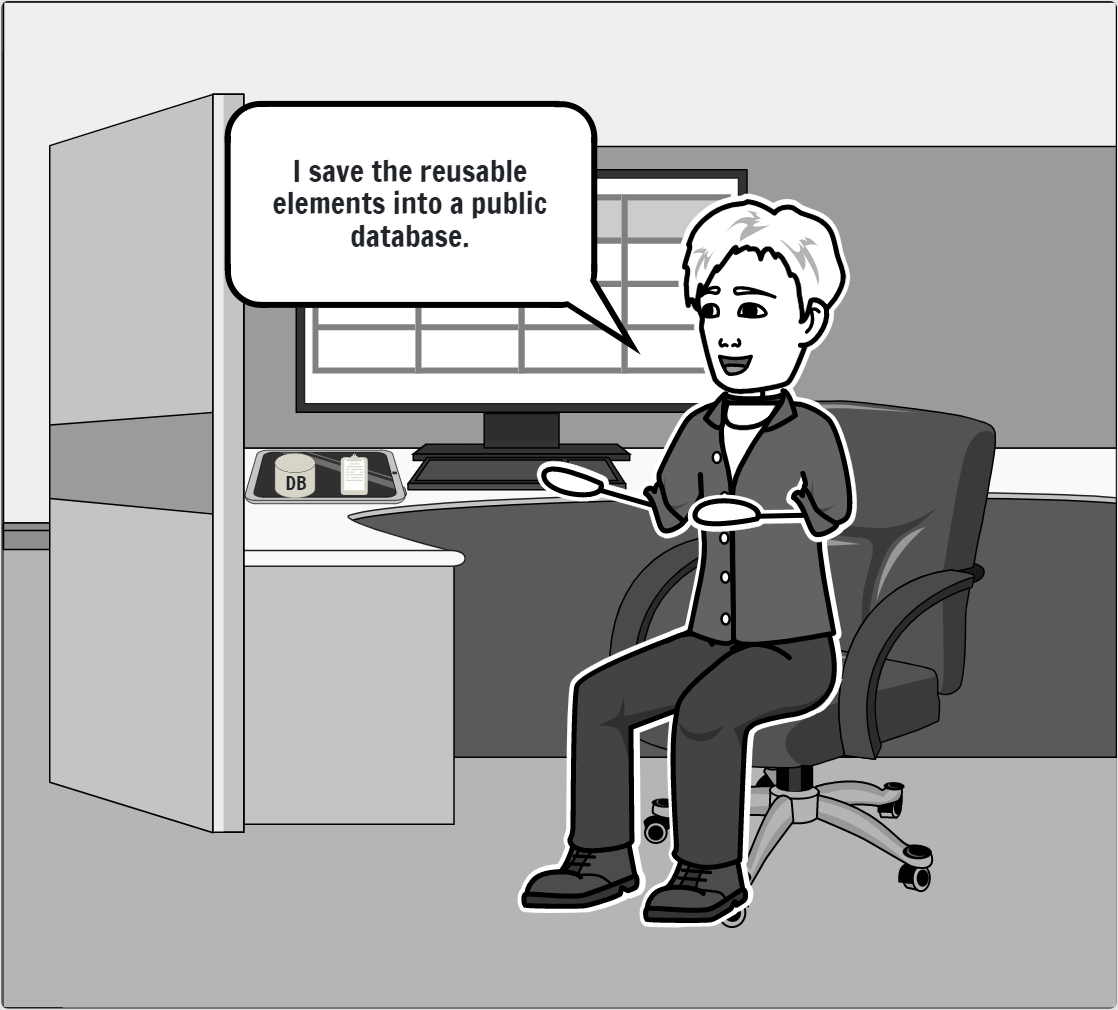}
	\end{minipage}
	\hfill
	\begin{minipage}{0.475\textwidth}
		\includegraphics[width=\textwidth]{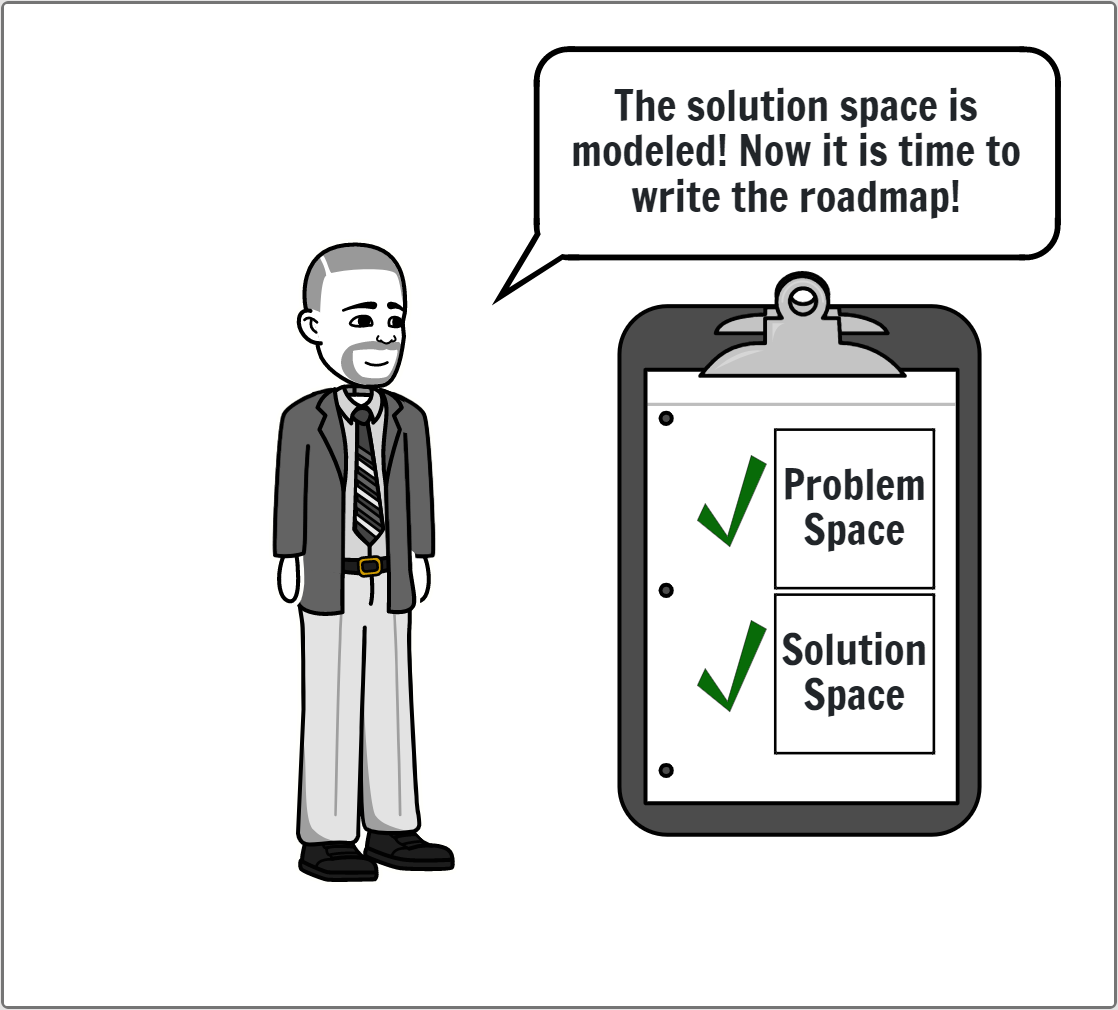}
	\end{minipage}
	\caption{The fifth activity: Extraction of Insights for future IMoG Innovations}
	\label{fig:storyboard:kp:1}
\end{figure}

\begin{figure}
	\begin{minipage}{0.475\textwidth}
		\includegraphics[width=\textwidth]{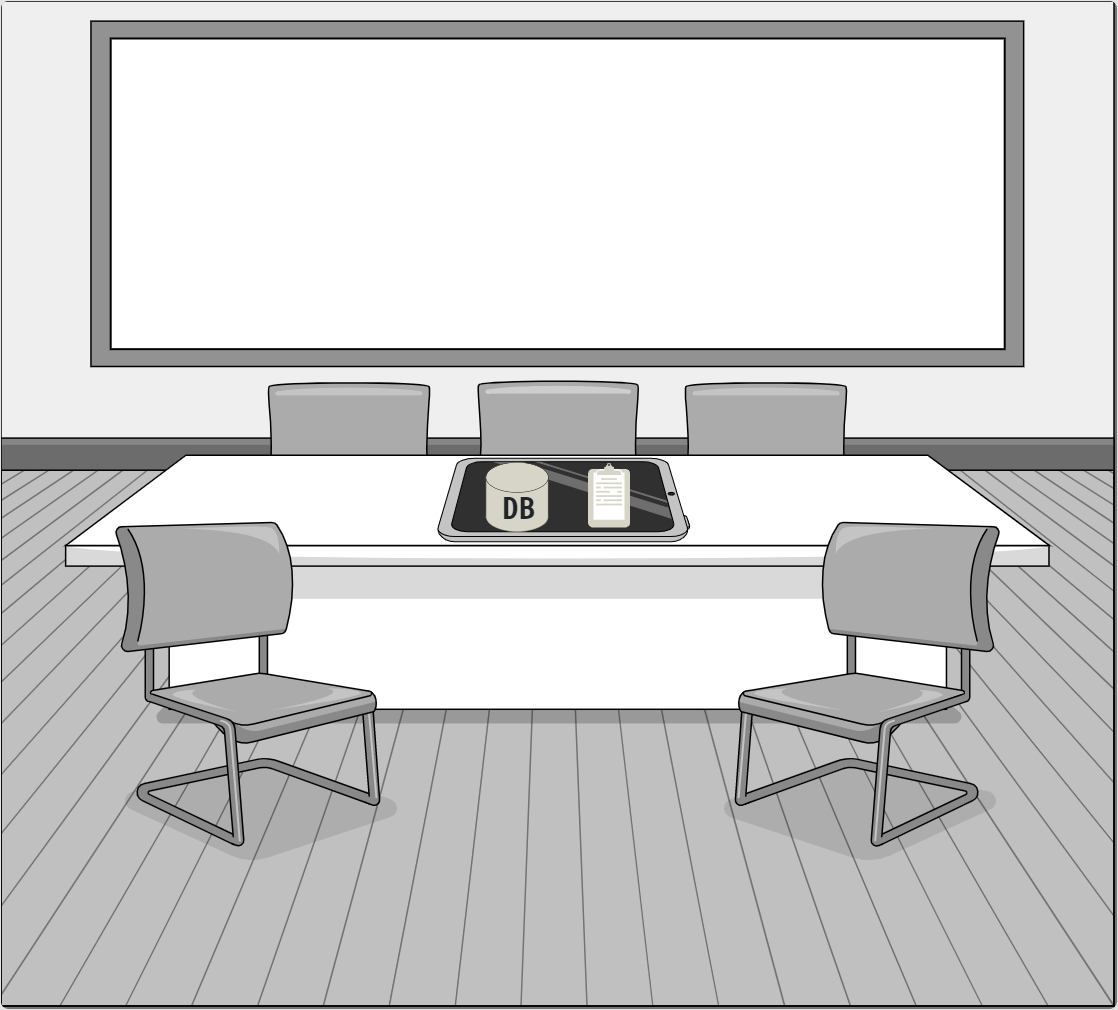}
	\end{minipage}
	\hfill
	\begin{minipage}{0.475\textwidth}
	\end{minipage}
	\caption{The fifth activity: Extraction of Insights for future IMoG Innovations (part 2)}
	\label{fig:storyboard:kp:2}
\end{figure}

The next activity is the extraction of insights for future IMoG innovations (see Figures \ref{fig:storyboard:kp:1} and \ref{fig:storyboard:kp:2}).
The committee meets again to discuss.
With the problem space and solution space modeled, the key elements of the innovation are identified and listed to be used in the innovation roadmap and in future IMoG models.
Once the elements are identified, the IMoG modeler puts them into the public database to be reused.
Now it is time to write the roadmap!

\begin{figure}
	\begin{minipage}{0.475\textwidth}
		\resizebox{\textwidth}{!}{%
			\begin{tikzpicture}[every node/.style={inner sep=0,outer sep=0}]
				\node[anchor=south west] at (0.12,0.12) {\includegraphics{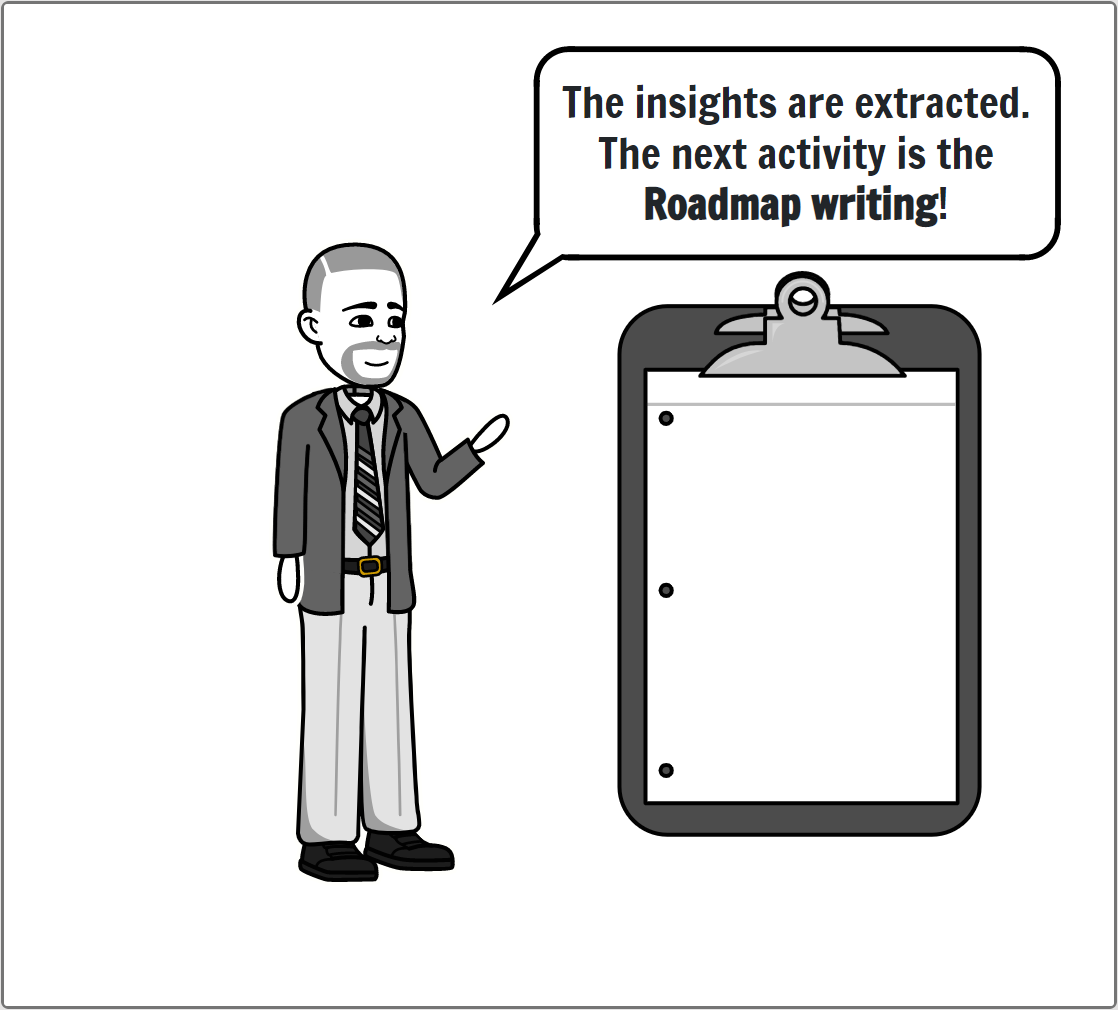}};
				\node[anchor=west, align=left] at (18.7,15.2) {\Large \textbf{Innovation Identification}};
				\node[anchor=west, align=left] at (18.7,14) {\Large \textbf{Feature and Function} \\\Large \textbf{Identification}};
				\node[anchor=west, align=left] at (18.7,12.59) {\Large \textbf{Requirements Elicitation} \\(Quality Requirements and Constraints)};
				\draw (18.4,11.9) -- (25,11.9);
				\node[anchor=west, align=left] at (18.7,11.3) {\Large \textbf{Solution Space Exploration}};
				\draw (18.4,10.8) -- (25,10.8);
				\node[anchor=west, align=left] at (18.7,9.8) {\Large \textbf{Extracting and saving} \\\Large \textbf{Insights for future} \\\Large \textbf{IMoG Innovations}};
				\node[anchor=west, align=left] at (18.7,8.4) {\Large \textbf{Roadmap Writing}};
				\node[anchor=west, align=left] at (18.7,7) {\Large \textbf{Maintaining and} \\\Large \textbf{Updating the Model} \\\Large \textbf{and Roadmap}};
				\node at (18.4,15.2) {\Large \color{green!70!black} \CheckmarkBold};
				\node at (18.4,14) {\Large \color{green!70!black} \CheckmarkBold};
				\node at (18.4,12.59) {\Large \color{green!70!black} \CheckmarkBold};
				\node at (18.4,11.3) {\Large \color{green!70!black} \CheckmarkBold};
				\node at (18.4,9.8) {\Large \color{green!70!black} \CheckmarkBold};
			\end{tikzpicture}
		}
	\end{minipage}
	\hfill
	\begin{minipage}{0.475\textwidth}
		\includegraphics[width=\textwidth]{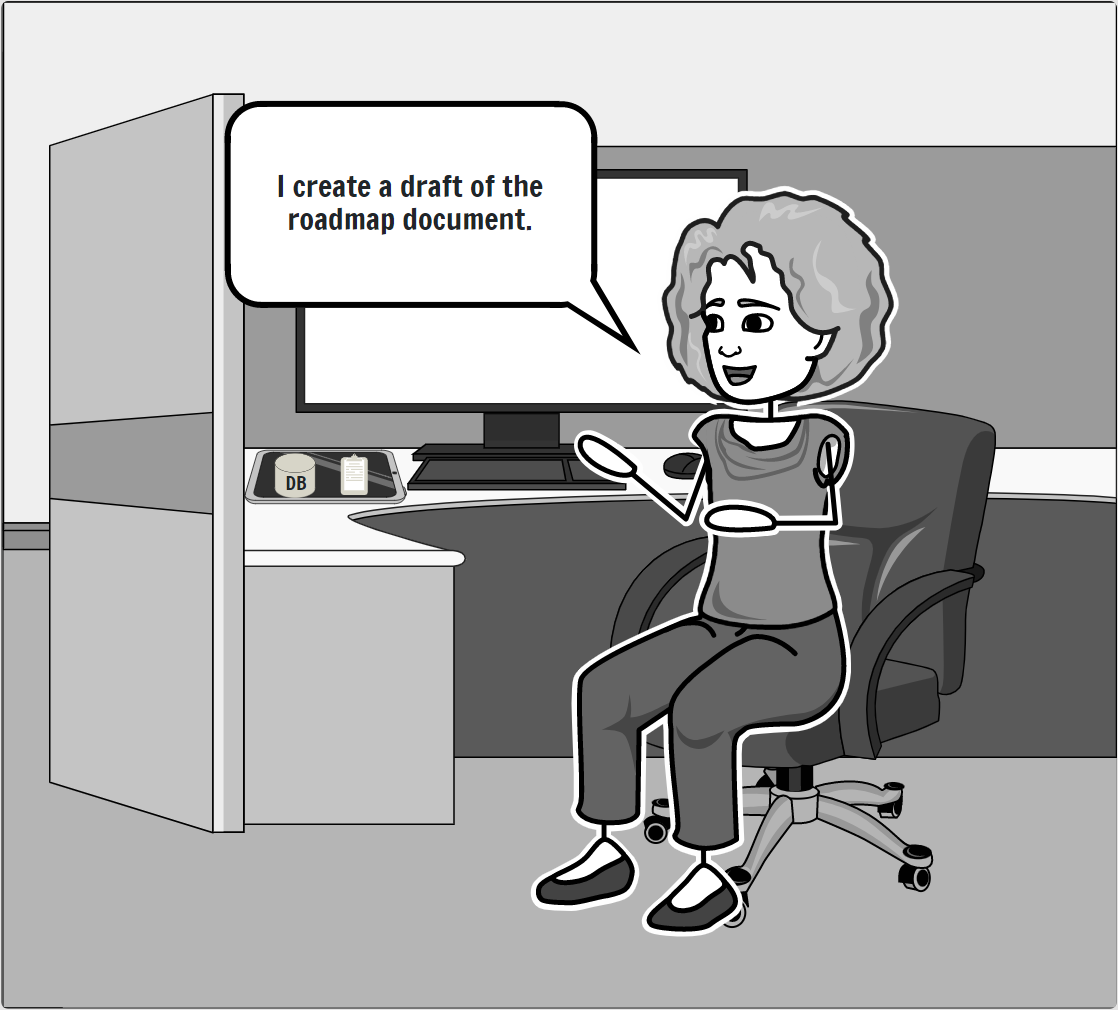}
	\end{minipage}

	\vspace{0.45cm}
	\begin{minipage}{0.475\textwidth}
		\includegraphics[width=\textwidth]{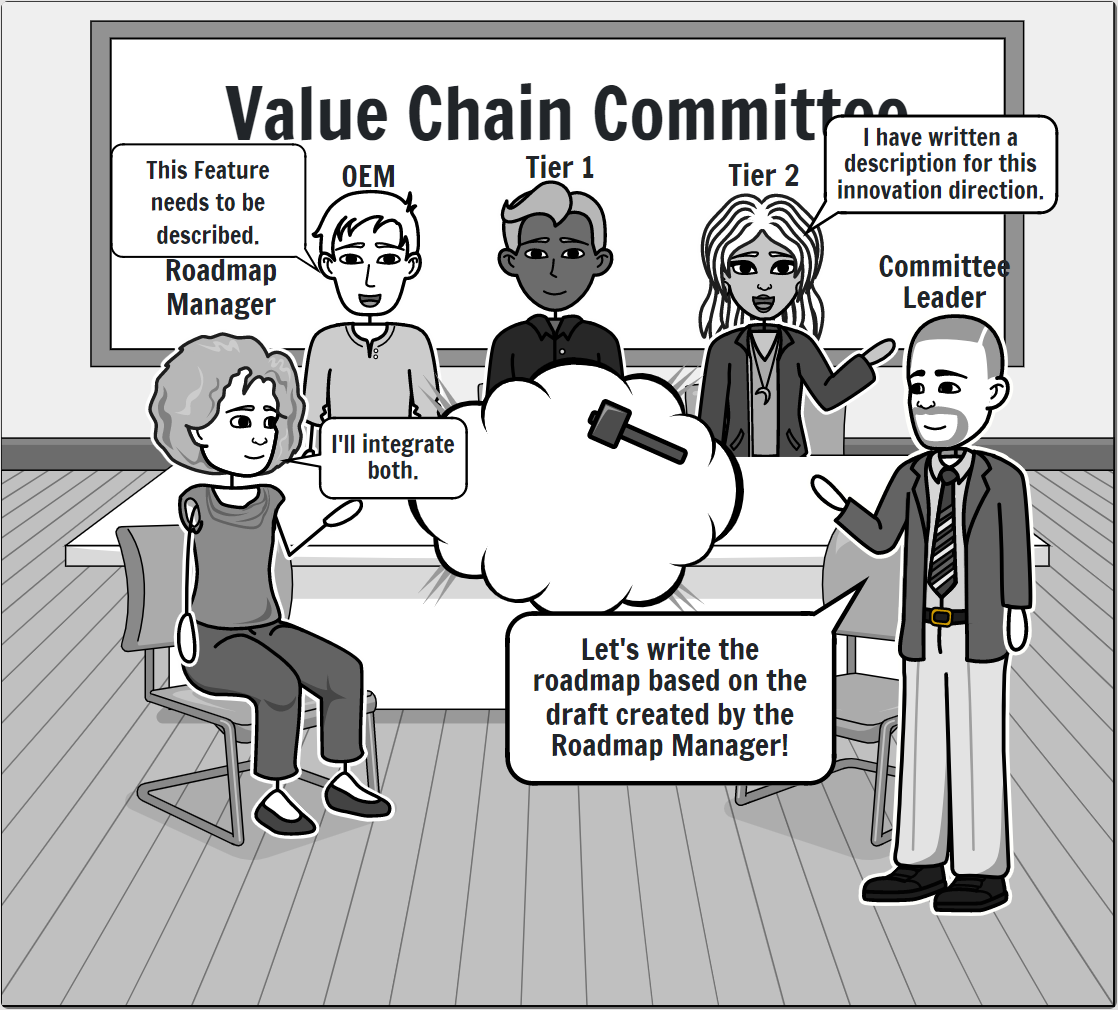}
	\end{minipage}
	\hfill
	\begin{minipage}{0.475\textwidth}
		\includegraphics[width=\textwidth]{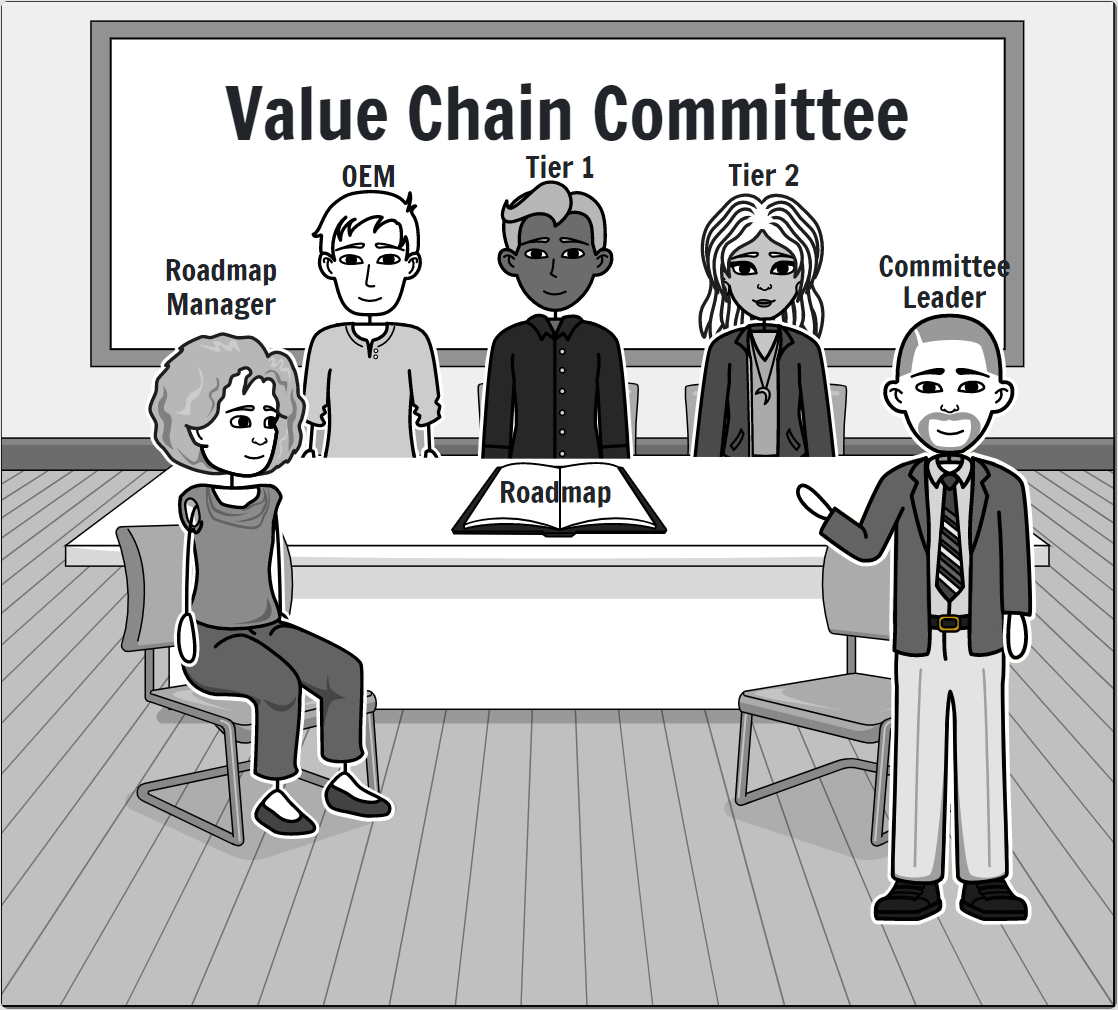}
	\end{minipage}

	\vspace{0.45cm}
	\begin{minipage}{0.475\textwidth}
		\includegraphics[width=\textwidth]{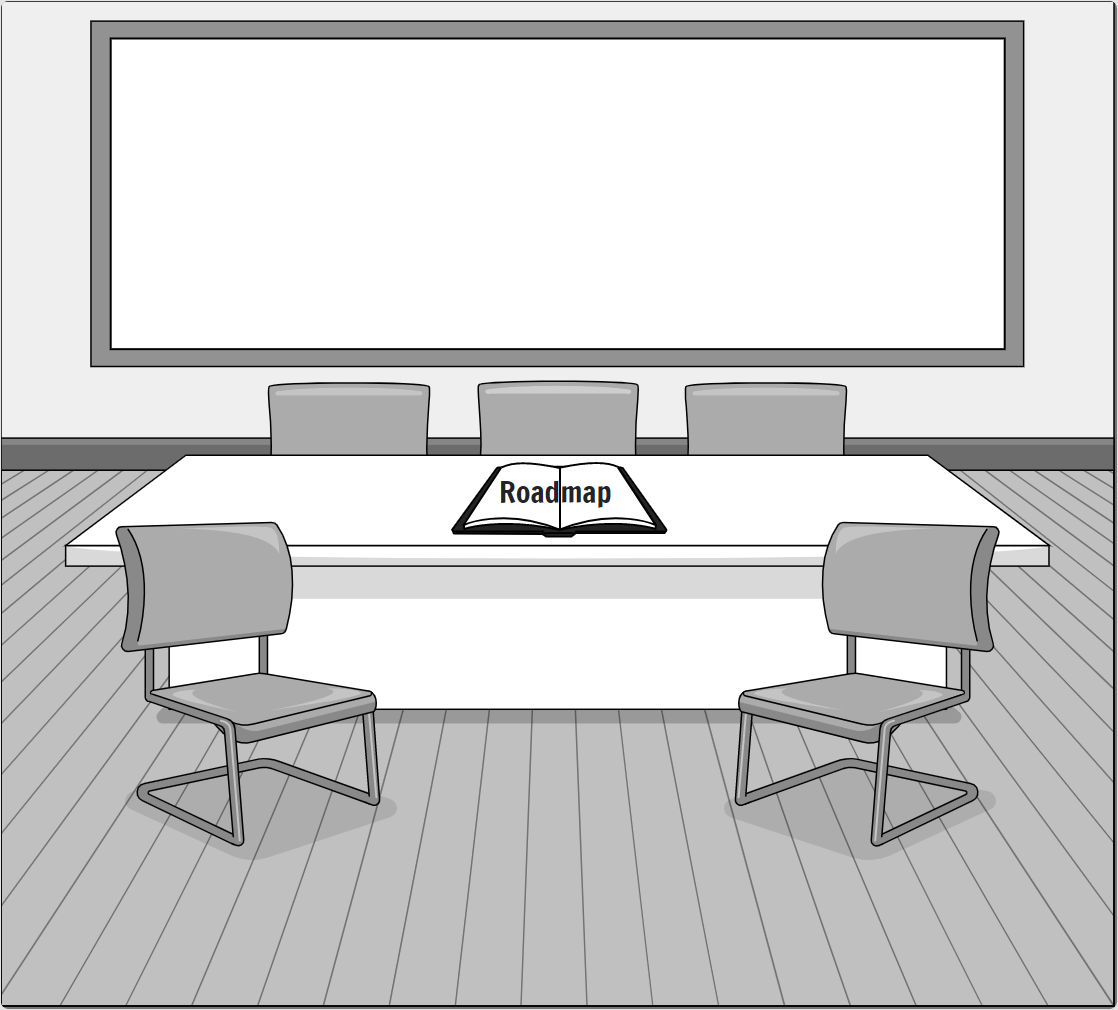}
	\end{minipage}
	\hfill
	\begin{minipage}{0.475\textwidth}
	\end{minipage}
	\caption{The sixth activity: Roadmap Writing}
	\label{fig:storyboard:rm:1}
\end{figure}

The next activity is the roadmap writing (see Figure \ref{fig:storyboard:rm:1}).
The roadmap manager of the committee creates a draft of the roadmap.
The committee discusses the draft and refines it until the roadmap is finished.
With this activity finished the general direction is known to the value chain, which can now start their own development processes outside the committee to make the innovation come true.
The main committee activities are also ended with this activity.

\begin{figure}
	\begin{minipage}{0.475\textwidth}
		\resizebox{\textwidth}{!}{%
			\begin{tikzpicture}[every node/.style={inner sep=0,outer sep=0}]
				\node[anchor=south west] at (0.12,0.12) {\includegraphics{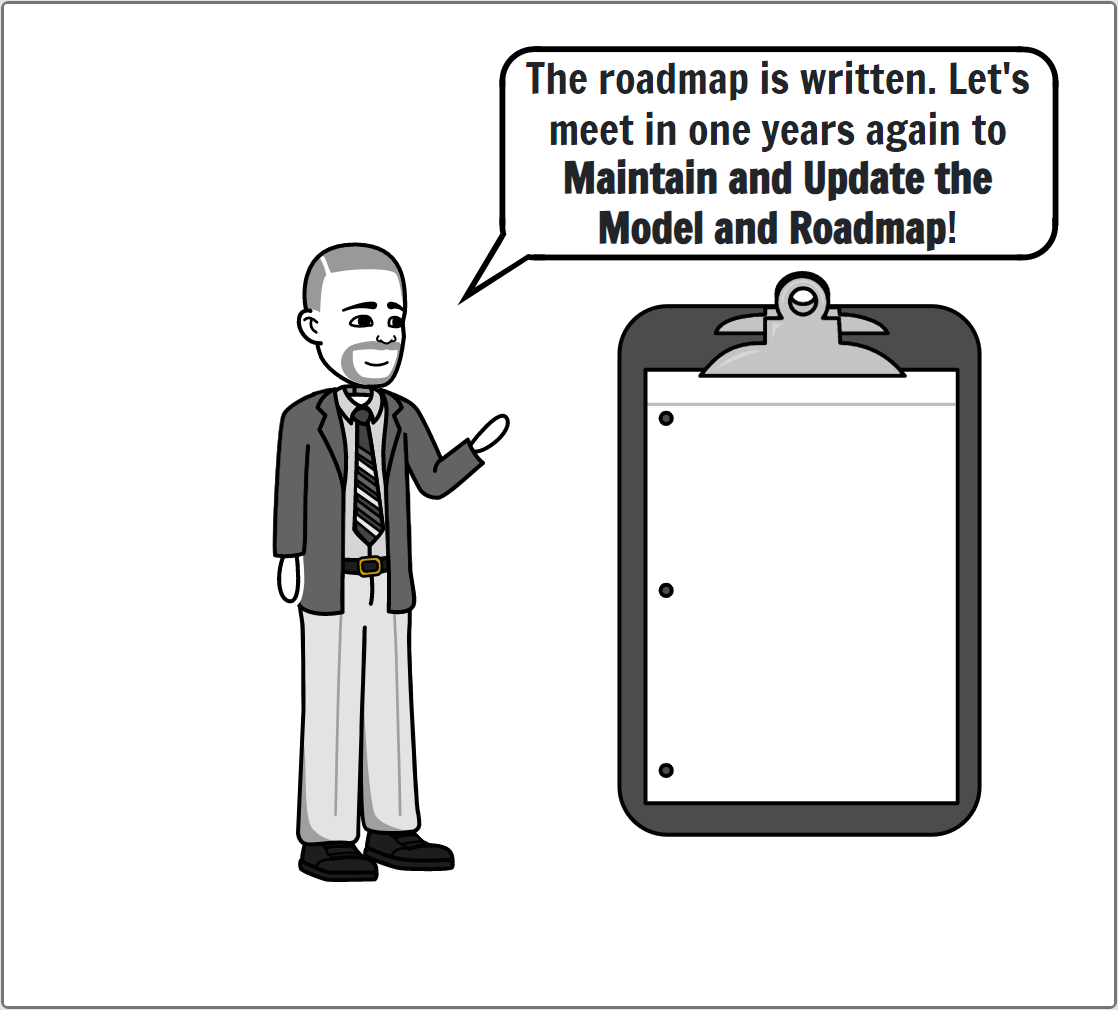}};
				\node[anchor=west, align=left] at (0.12,0.12) at (18.7,15.2) {\Large \textbf{Innovation Identification}};
				\node[anchor=west, align=left] at (18.7,14) {\Large \textbf{Feature and Function} \\\Large \textbf{Identification}};
				\node[anchor=west, align=left] at (18.7,12.59) {\Large \textbf{Requirements Elicitation} \\(Quality Requirements and Constraints)};
				\draw (18.4,11.9) -- (25,11.9);
				\node[anchor=west, align=left] at (18.7,11.3) {\Large \textbf{Solution Space Exploration}};
				\draw (18.4,10.8) -- (25,10.8);
				\node[anchor=west, align=left] at (18.7,9.8) {\Large \textbf{Extracting and saving} \\\Large \textbf{Insights for future} \\\Large \textbf{IMoG Innovations}};
				\node[anchor=west, align=left] at (18.7,8.4) {\Large \textbf{Roadmap Writing}};
				\node[anchor=west, align=left] at (18.7,7) {\Large \textbf{Maintaining and} \\\Large \textbf{Updating the Model} \\\Large \textbf{and Roadmap}};
				\node at (18.4,15.2) {\Large \color{green!70!black} \CheckmarkBold};
				\node at (18.4,14) {\Large \color{green!70!black} \CheckmarkBold};
				\node at (18.4,12.59) {\Large \color{green!70!black} \CheckmarkBold};
				\node at (18.4,11.3) {\Large \color{green!70!black} \CheckmarkBold};
				\node at (18.4,9.8) {\Large \color{green!70!black} \CheckmarkBold};
				\node at (18.4,8.4) {\Large \color{green!70!black} \CheckmarkBold};
			\end{tikzpicture}
		}
	\end{minipage}
	\hfill
	\begin{minipage}{0.475\textwidth}
		\includegraphics[width=\textwidth]{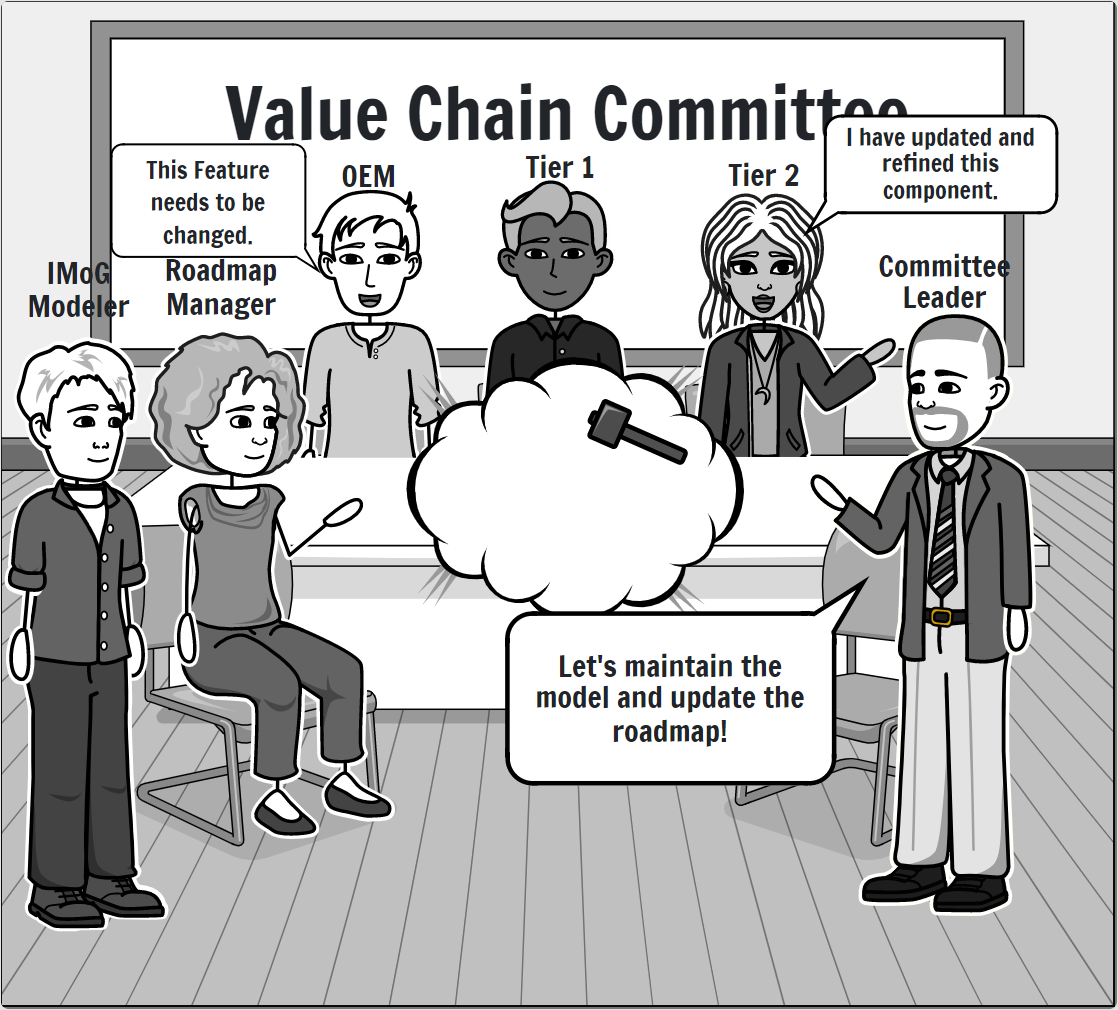}
	\end{minipage}

	\vspace{0.45cm}
	\begin{minipage}{0.475\textwidth}
		\includegraphics[width=\textwidth]{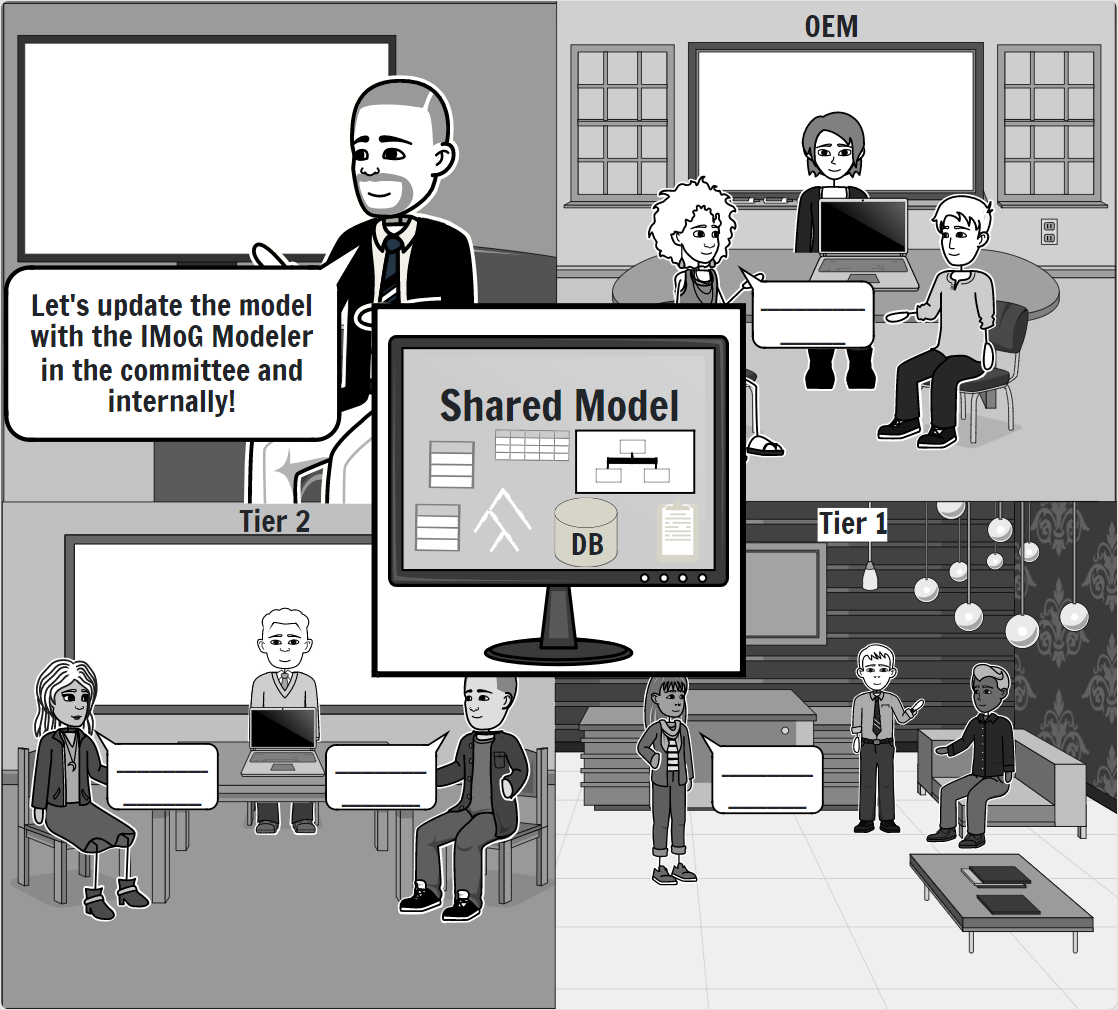}
	\end{minipage}
	\hfill
	\begin{minipage}{0.475\textwidth}
		\includegraphics[width=\textwidth]{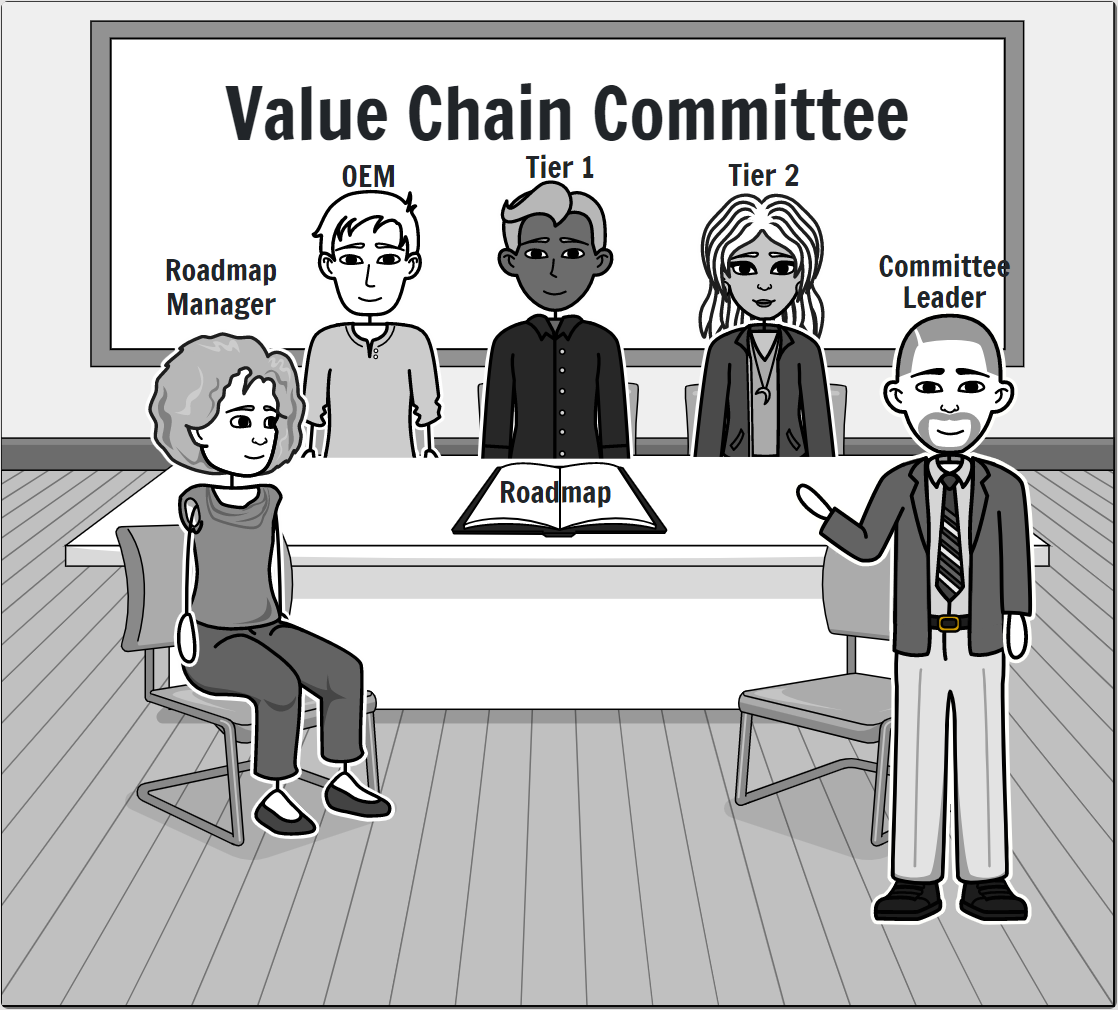}
	\end{minipage}

	\vspace{0.45cm}
	\begin{minipage}{0.475\textwidth}
		\includegraphics[width=\textwidth]{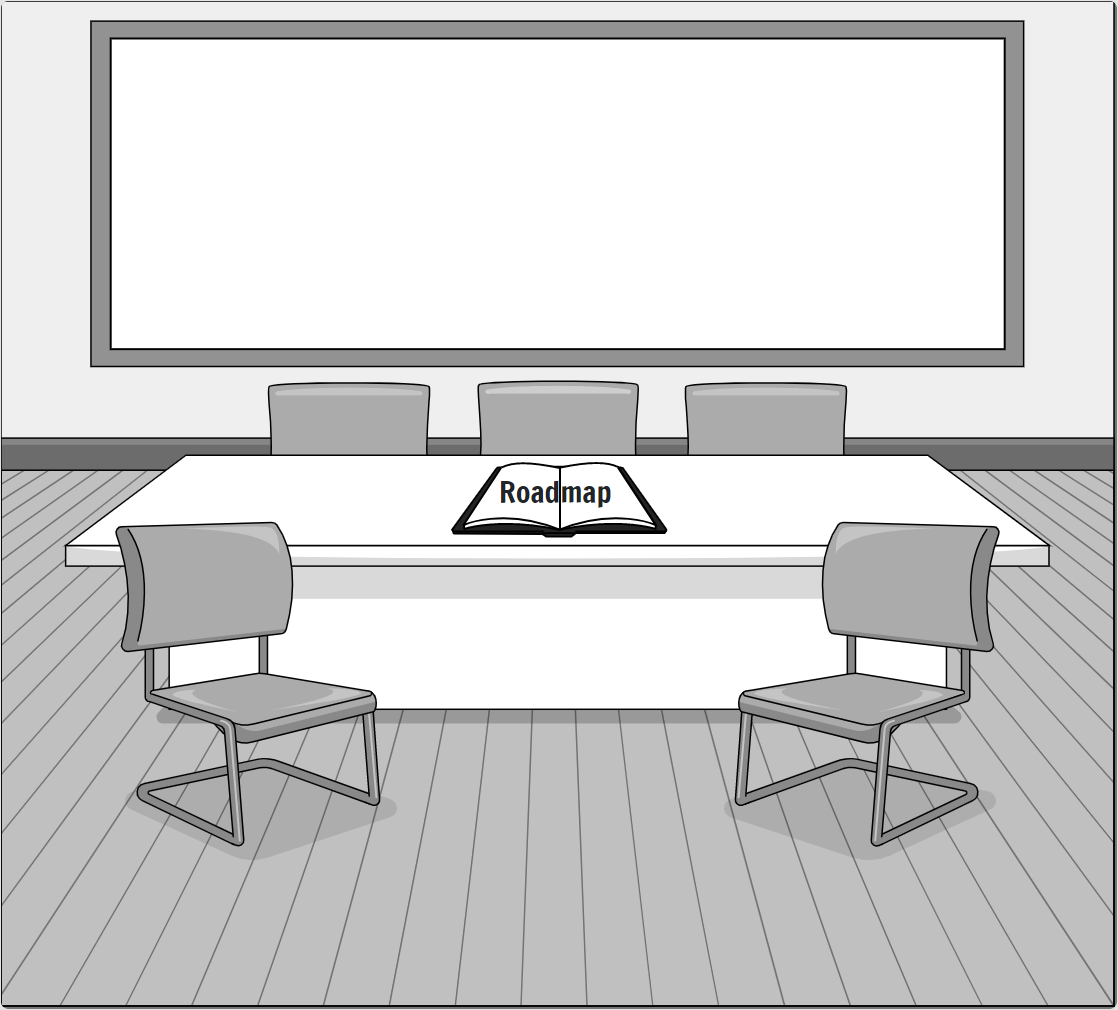}
	\end{minipage}
	\hfill
	\begin{minipage}{0.475\textwidth}
	\end{minipage}
	\caption{The seventh activity: Maintenance and Update of the Model and the Roadmap}
	\label{fig:storyboard:ru:1}
\end{figure}

The last activity left open is the maintenance and update of the model and the roadmap (see Figures \ref{fig:storyboard:ru:1}).
In predefined intervals (like every two years), the committee takes place and realign the model and the roadmap with their gained knowledge over the time frame.
With this activity they can also take action for important changes that the whole value chain needs to know.
The model is also refined internally with all important roles.
When this activity is done, the committee meets then again at the next predefined date for the maintenance and update.

%% file: content/methodology.tex
\chapter{Innovation Modeling Grid}
\label{chap:methodology}


We developed a methodology called \enquote{Innovation Modeling Grid} (IMoG) to accelerate the innovation development process along the automotive value chain.
The methodology IMoG provides a structure and defines elements to model the problem and the solution space of innovations.
IMoG defines a structure to reduce the time spend on the \enquote{What and how to model?} question and to help the modeler to focus on their innovation instead.
Furthermore, a process and a dedicated tooling supporting the methodology is required to handle the methodology this public roadmapping context.
A dedicated tooling for IMoG is currently in progress, but out of the scope of this document.

Section \ref{sec:imog:overview} presents the design principles of the methodology IMoG.
IMoG itself is described in Section \ref{sec:imog:methodology}.
An FAQ is given in Section \ref{sec:imog:faq}.
IMoG's process is described in Chapter \ref{chap:process}.

\section{Design Principles of IMoG}
\label{sec:imog:overview}

We developed IMoG under the context that an automotive value chain committee creates and maintains a public microelectronic roadmap.
This context shaped IMoG and we raised the following design principles:
\begin{itemize}
	\item \textbf{Abstract Innovations}:
	The focus of IMoG lies on describing abstract innovations that are shared in a public committee.
	These innovations are represented by a mix of informal and formal elements to remain beneficial to all participants.
	IMoG models are expected to include fewer details than development and engineering models.
	Therefore, complex modeling concepts are left out.
	This includes, for example, the concept of \enquote{Ports} to model communication interfaces of solutions and check their consistency.
	However, this does not mean that any kind of detail is too much.
	It is expected that IMoG contains sufficient details of the crucial parts of the innovation where the highest uncertainty and risk lies.
	Instead of ports, it is recommended to describe one communication channel with sufficient details.
	\item \textbf{Problem space vs solution space}:
	IMoG divides the modeling into a problem description and a solution description \cite{czarnecki2000generative, olsen2015lean}:
	The problem space should mainly contain information about the problem with as little information as necessary about the possible solutions.
	The solution space covers on the other side the possible solutions.
	Furthermore, a map between the problem space elements and the solution space elements is necessary for basic tracing.
	Natural language constraints, quality requirements and general conditions complete this tracing by giving the option to add further information.
	In the context of a road mapping committee, this problem-solution distinction is suitable, because it eliminates the frequently asked question whether a particular \enquote{Function}, \enquote{Block}, or \enquote{Requirement} describes in the IMoG model the target state or the actually designed state and thus helps reducing the cognitive load.
	\item \textbf{Support of Decomposition / Refinement / Variability}:
	IMoG distinguishes between three core concepts: Decomposition, Refinement and Variability.
	Decomposition describes a partitioning of an element into its parts.
	Refinement describes a more fine grained specification of an element.
	Variability describes possible alternatives of an element.
	Variability tends to create the wish for measurement and assessment, thus suitable comparison parameters should be defined.
	These concepts are less important for describing the problem space.
	However, they are invaluable to understand and apply for the solution space.
	\item \textbf{Other modeling dimensions}:
	IMoG applies other concepts to manage innovation modeling as well.
	The concepts of abstraction levels and perspectives help with the separation of concerns by focusing on certain aspects of the innovation.
	Furthermore, abstraction levels and types help with the support of filtering mechanisms to hide temporally unneeded details.
	The concept of availability describes when certain elements are available and ready-to-use, which plays an integral part in road mapping.
\end{itemize}

\section{Innovation Modeling Grid Methodology}
\label{sec:imog:methodology}

\begin{figure}
	\includegraphics[width=\linewidth]{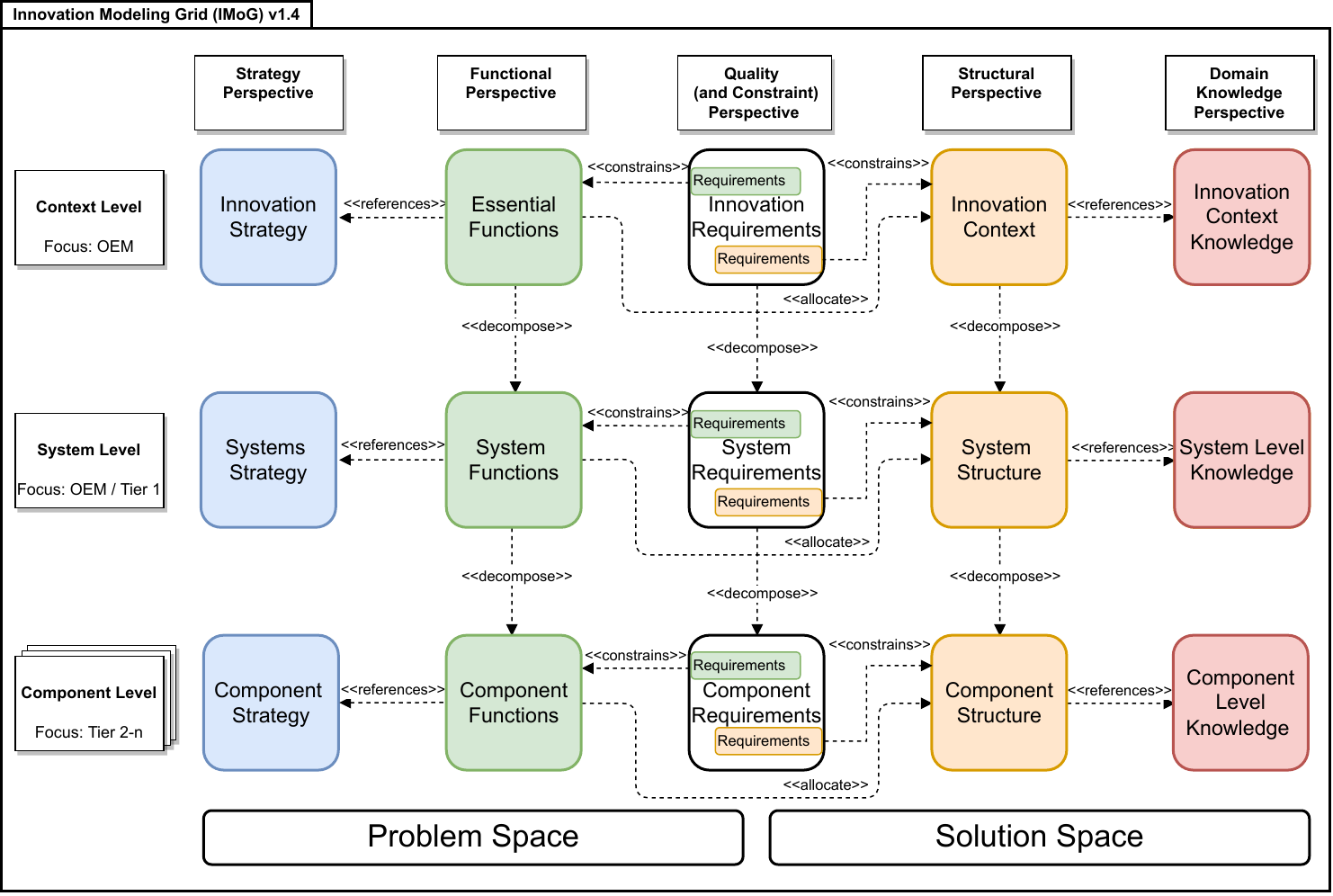}
	\caption{IMoG version 1.4.
		It contains three abstraction levels (rows) and five perspectives (columns).
		Each perspective and abstraction level is interconnected with its neighbor cell.}
	\label{fig:imog1.4}
\end{figure}

The Innovation Modeling Grid (IMoG) is depicted as a matrix in Figure \ref{fig:imog1.4}.
Each row represents an abstraction level, which can be understood as separating and designing the details of the innovation at different detail levels.
IMoG proposes three abstraction levels:
\begin{itemize}
	\item The \textbf{Context Level} describes the innovation as a whole system embedded into its environment.
	In the automotive domain, this level is particularly interesting for the OEM(s) in the committee.
	\item The \textbf{System Level} represents the innovation systems and its parts and is primarily relevant for Tier 1 suppliers in the classic automotive environment.
	\item The \textbf{Component Level} consists of the components of the system.
\end{itemize}
The modeler can also add more abstraction levels if needed.
However, the three abstraction levels should be a sufficient starting point for the most innovations.

IMoG follows a classical approach to distinguish between the problem space and the solution space, and proposes to analyze the spaces through so-called perspectives.
A perspective describes an aspect of the innovation, for example the strategy, the features or the structure of the innovation.
In IMoG, each perspective is represented as a column.
The Strategy Perspective, the Functional Perspective, and partly the Quality Perspective relate to the problem space and focus on describing aspects of the problem without many technical details.
On the other hand, the Structural Perspective, the Domain Knowledge Perspective, and the latter part of the Quality Perspective relate to the solution space and describe potential technical solutions corresponding to the problem in an abstract manner.
In the context of innovations this solution space are kept abstract as the knowledge about the future is only vague.

The five perspectives and the three levels of abstraction are arranged into a grid, where each cell in the grid represents a model of an aspect of the innovation on a specific abstraction level.
A grid cell is called a \textit{view}.
Note that not all grid cells need to be filled:
When a modeler does not see a purpose for one view, then there is no issue omitting this view.
This may happen if a breakdown of a model is not further required.
The IMoG meta model recommends for each perspective a set of model elements.
The corresponding details are out of the scope of this article.
Each perspective is presented in the following.
Afterwards, the interconnection (and thus the arrows in Figure \ref{fig:imog1.4}) between the perspectives are described.

\subsection{Strategy Perspective}
\label{sec:imog:methodology:strategy}
The identification of an innovation usually starts with creativity techniques, sketches and discussions.
These discussions are the starting point of the Strategy Perspective.
The Roadmap Manager of the committee (see Section \ref{sec:process:roles}) takes the result and writes an innovation description as well as the strategy description behind the innovation.
The description targets the innovation strategy, which may contain a vision, rationales, images, goals and diagrams.
Those descriptions can contain identifiable elements to enable referencing and tracing.
Additionally, the description contains companies’ intends and their stakes in the innovation.
The description and identifiable elements encompass enough information to start the modeling activities on the other Perspectives.
The filled Strategy Perspective constitutes a part of the artifacts of the \enquote{Innovation Identification} activity.

\begin{figure}
	\centering
	\includegraphics[width=0.85\linewidth]{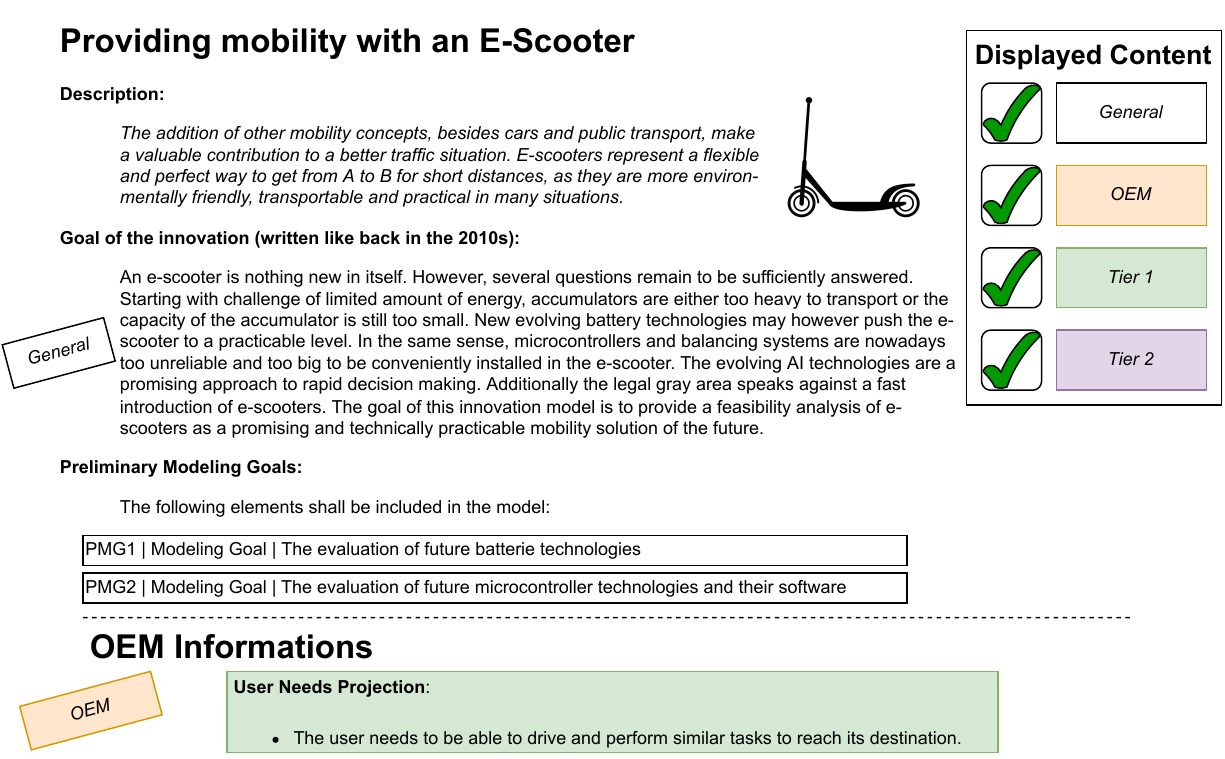}
	\caption{Strategy Perspective: part of the innovation description of the e-scooter.}
	\label{fig:imog:e-scooter:strategy}
\end{figure}

In the following, we illustrate the process steps with the innovation \enquote{Providing mobility with an e-scooter} (see Figure \ref{fig:imog:e-scooter:strategy}).
The Strategy Perspective of the e-scooter innovation includes a description with a vision and what the innovation is about, the goals written as text as well as goals listed as elements for cross referencing, information from the car manufacturer (OEM) regarding their estimated customer needs, their concern and possibly some additional bubble diagram for a better explanation of their interest and information from the other suppliers (Tier 1 and Tier 2) including their interest, diagrams, etc.

\subsection{Functional Perspective}
\label{sec:imog:methodology:fp}
The Functional Perspective describes the required features (end-user visible characteristics) and functions (traceable tasks or actions that a system shall perform) of the innovation.
The features and functions of the Functional Perspective represent a derivative of the well-known feature models from \cite{kang1990feature}.
The Functional Perspective’s input is the strategy description.
Optionally, User Stories or Use Cases can be created if the committee determines the need for more information on each feature and function.

\begin{figure}
	\includegraphics[width=\linewidth]{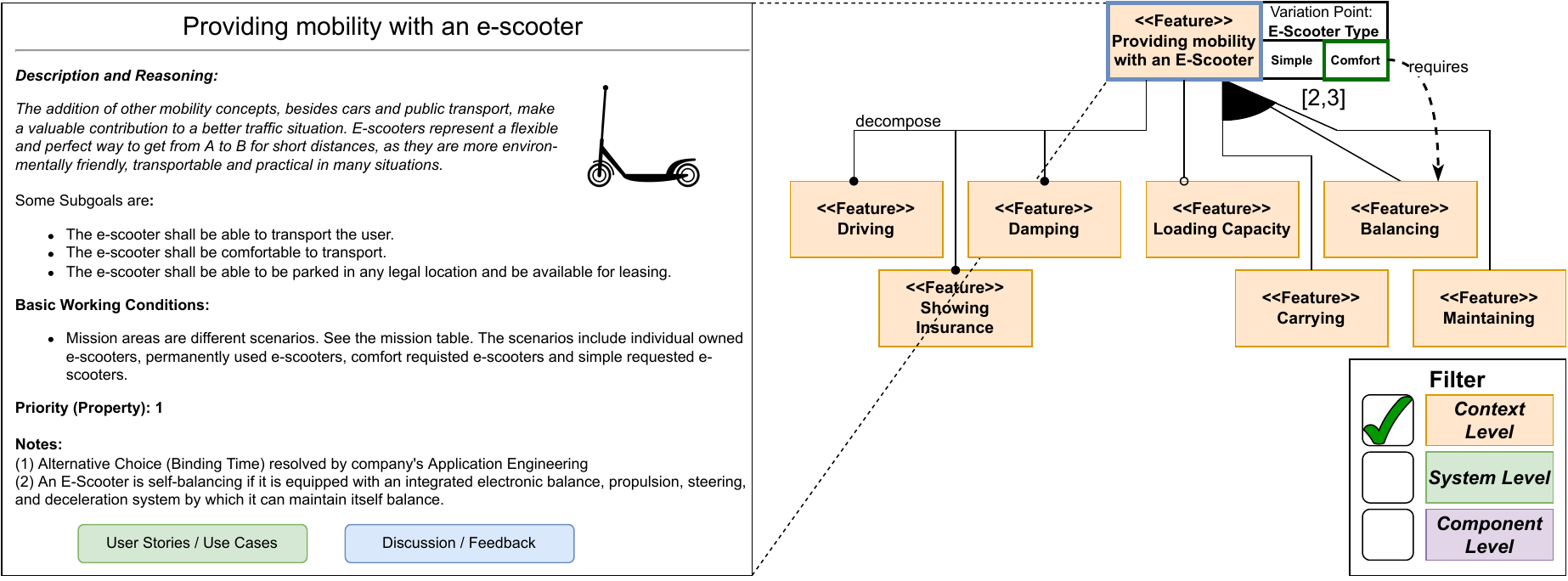}
	\caption{Functional Perspective: a part of the feature model of the e-scooter.}
	\label{fig:imog:e-scooter:fp}
\end{figure}

Considering the e-scooter example, the Functional Perspective model may look as it is depicted in Figure \ref{fig:imog:e-scooter:fp}.
It starts with \enquote{Providing mobility with an e-scooter} as its root feature, decomposed into several other features using well-known relations.
The mandatory relations are depicted using an arrow with a black circle at its end, and optional relations as an arrow with a white circle at its end. Finally, an or-relation with cardinality is depicted as a black arc with several arrows going out of it with its cardinality interval) and a constraint relation (\enquote{requires}).
For more details about these described relations, see \cite{kang1990feature,czarnecki2000generative}.
The Variation Point representation represents a labeled alternative relation which graphical depiction is IMoG specific.
The e-scooter feature description can be viewed on the left.
It includes a detailed textual description, aligned goals, basic working conditions and other properties, like notes, priorities or links to user stories and use cases.
The functions of the model are left out of the image.
One specialty of the Functional Perspective is the detailed description of each feature and function, which helps to understand what they actually represent.

\subsection{Quality Perspective}
\label{sec:imog:methodology:qp}
Based on the strategy description and the features and functions, the Quality Perspective captures further quality requirements and constraints of each feature and function.
Requirement diagrams and requirement tables are suitable representations of the Quality Perspective.
The strategy description, the features and functions and the requirements and constraints build together the problem space.
It shall noted, that the Quality Perspective also contains the quality requirements and constraints of the solution space, which are referenced on the solutions on the Structural Perspective.

\begin{figure}
	\includegraphics[width=\linewidth]{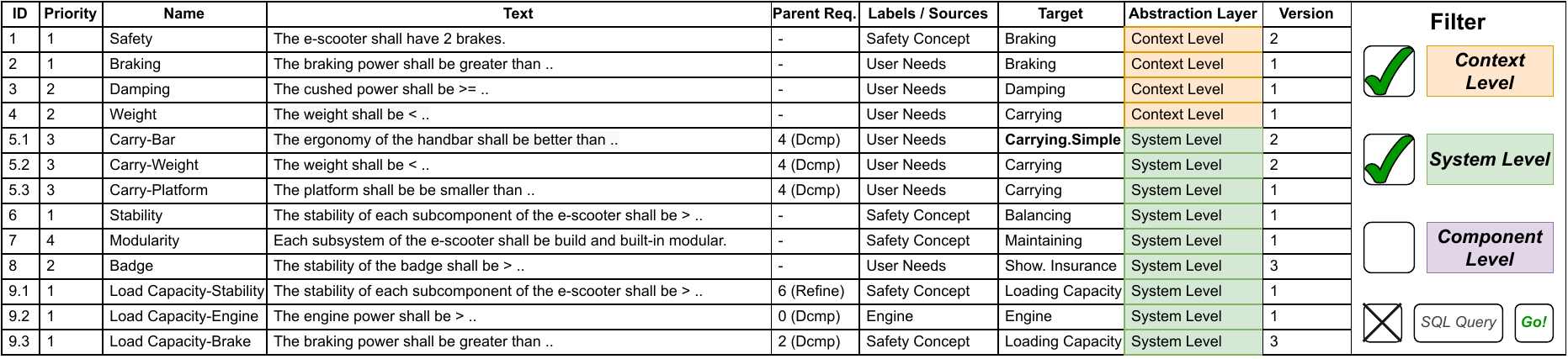}
	\caption{Quality Perspective: a typical table of requirements with many attributes, which reference features or functions of the Functional Perspective or solution blocks of the Structural Perspective.
		The details -- like the meaning of the attributes -- can be chosen depending on each innovation and are not further elaborated here.
		As depicted on the right side of the image, filter functionality is of special importance for the Quality Perspective.}
	\label{fig:imog:e-scooter:qp}
\end{figure}

The e-scooter innovation's Quality Perspective is depicted in Figure \ref{fig:imog:e-scooter:qp}.
It contains the quality requirements for the problem space and for the solution space.

\subsection{Structural Perspective}
\label{sec:imog:methodology:structural}
\begin{figure}
	\centering
	\includegraphics[width=0.75\linewidth]{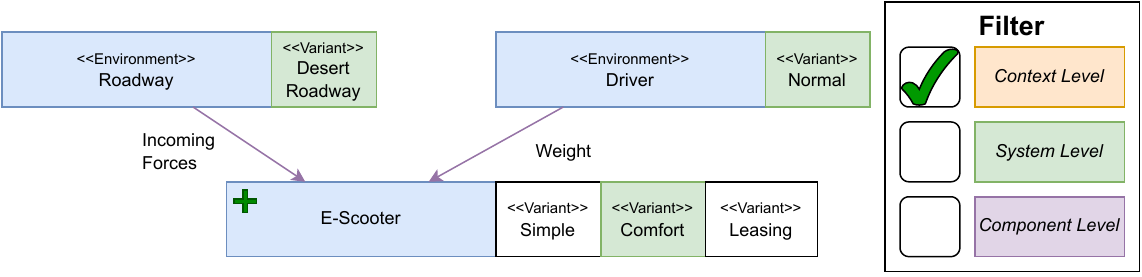}
	\caption{Structural Perspective - Context Level: A simple context model for the e-scooter.}
	\label{fig:imog:e-scooter:structural1}
\end{figure}

The Structural Perspective targets the modeling of the solution space.
It is worth mentioning, that the word \enquote{Structural} does not mean here the relations of solution blocks to each other alone, but also includes properties and values of these solution blocks.
The context level of the Structural Perspective contains the environment, the relation and the effects between the environment and the innovation.
A simple environment description may for example contain the street, the driver and the e-scooter (see Figure \ref{fig:imog:e-scooter:structural1}).
It contains the innovation (e-scooter) with the driver and roadway blocks (blue rectangles with a name and optionally a stereotype over the name).
Each block has variants attached, that specify similar forms of solutions.
The variants are depicted (as green and white boxes) next to the solution blocks with the stereotype <<Variant>>.
Each of these blocks own properties, which further refine the block.
The properties are crucial information for analyzing and evaluating the solution space.
Furthermore, relations like `Incoming Forces' and `weight' are modeled as unidirectional (purple) arrows, where purple represents the color for relations stereotyped as <<effect>>.
The solution blocks of the different abstraction levels are left out of the model.

\begin{figure}
	\includegraphics[width=\linewidth]{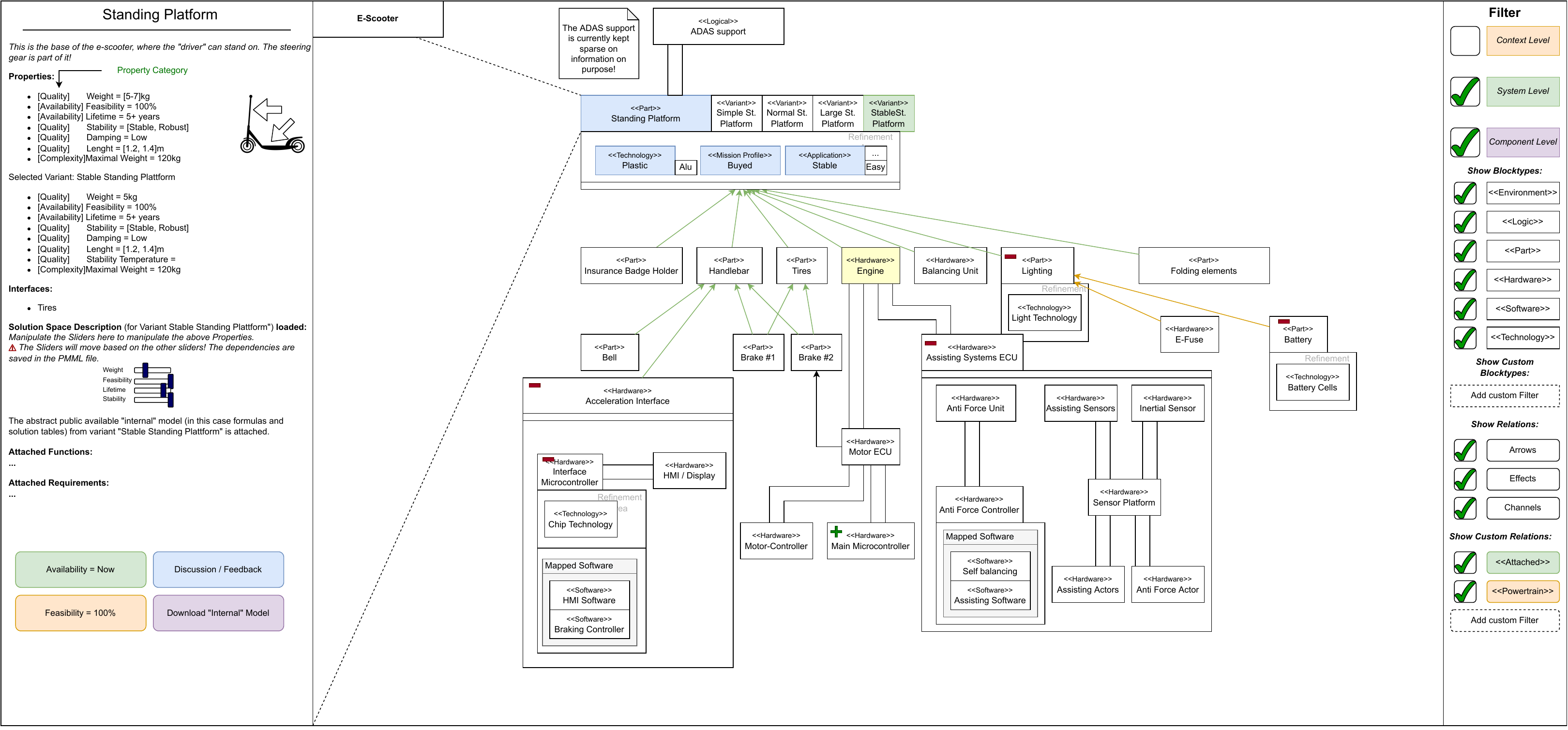}
	\caption{Structural Perspective - System Level:  The decomposition of the e-scooter into its system parts.
		It contains many blocks, variants, relations and channels (for modeling communication).
		This figure shall only give a glance at what may be included in the Structural Perspective.}
	\label{fig:imog:e-scooter:structural2}
\end{figure}

The system level contains the decomposition of the innovation into components,  while also including the software and the hardware elements.
Software and hardware elements as well as architectures and mappings between them are included in the system level.
The component level encompasses the system atoms which are decomposed from the system blocks.
These may include sensor descriptions with parameters, functions, properties or abstract technologies.
The atoms may include sensor descriptions with parameters, functions, properties or abstract technologies.
When creating a solution space any form of constraints and parameters of chosen technologies are particularly of interest.
Furthermore, requirements can be added to any solution block on any abstraction level.
These requirements are placed on the Quality Perspective and referenced on the corresponding solution blocks on the Structural Perspective.
An example of a system model can be viewed in Figure \ref{fig:imog:e-scooter:structural2}:
it decomposes the e-scooter block known from Figure \ref{fig:imog:e-scooter:structural1} into several parts of the e-scooter.
The model elements are designed specifically for the microelectronic context.

\subsection{Domain Knowledge Perspective}
\label{sec:imog:methodology:kp}
\begin{figure}
	\includegraphics[width=\linewidth]{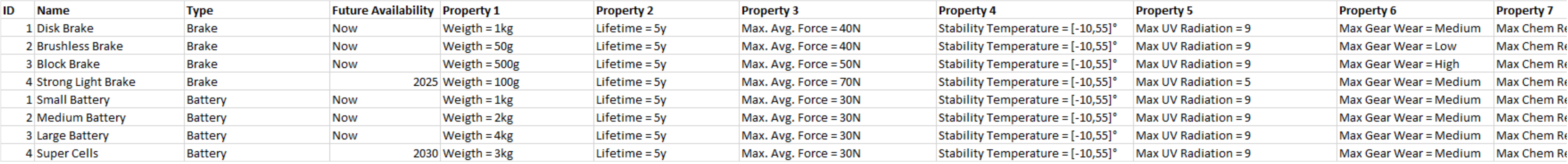}
	\caption{Domain Knowledge Perspective: A glance at the database view of the Domain Knowledge Perspective.}
	\label{fig:imog:e-scooter:kp}
\end{figure}

The insights from the innovation and the reusable element are collected and stored on the Domain Knowledge Perspective.
The elements of the Domain Knowledge Perspective enable references to the finished innovation model in future innovation models (see Figure \ref{fig:imog:e-scooter:kp}).
Furthermore, the Domain Knowledge Perspective may contain a component database in a knowledge representation.
The database may, for example, contain sensor characteristics and constraints from road traffic regulations, with each element owning an id, a name, a type, an estimated year of availability and several properties depending on the context of innovation.

In essence, this perspective is used to refine the model with existing knowledge and constraints.
Afterwards, the gained insights can be used to write the roadmap!

\subsection{Connecting Perspectives}
\label{sec:imog:methodology:connecting}
All perspectives were presented in detail.
However, their interconnection needs to be mentioned.
These interconnections are already visible in Figure \ref{fig:imog1.4} and shall be described here briefly.
The elements of the Strategy Perspective can be referred by the features and functions, building the interconnection between the Strategy Perspective and Functional Perspective (represented by the <<references>> relation in Figure \ref{fig:imog1.4}).
The constraints are part of the Quality Perspective and own a target reference to the corresponding features and functions.
The same holds for the requirements mapped on the Structural Perspective's solution blocks.
Thus the Quality Perspective has traces to both Functional Perspective and Structural Perspective (represented by the <<constrains>> relation in Figure \ref{fig:imog1.4}).
Each feature and function should be mapped on one or several solution blocks (represented by the <<allocate>> relation in Figure \ref{fig:imog1.4}).
This allocation is crucial, because it represents the interconnection of the problem space with the solution space.
Finally, there is the reference between the solution blocks of the Structural Perspective and the Domain Knowledge Perspective  (represented by the <<references>> relation in Figure \ref{fig:imog1.4}).
Thus all perspectives are interconnected to each other.
Worth to note is, that the IMoG modeler has to take care of not introducing inconsistencies (e.g., a requirement that is mapped on a feature or function, which is then allocated on a solution block that owns a contradicting requirement).

\subsection{Reviewing IMoG: Pros and Cons}
\label{sec:imog:methodology:proscons}

IMoG’s definition comes with strengths, some limitations and some recommendations from the authors.
These strengths, limitations are presented in the following including some recommendations and the experience from the authors.

One strength of IMoG is that it is well defined and owns a concise meta model for innovations.
This is illustrated by the following points.
First, the distinction between problem and solution space reduces the thinking overhead when exploring innovations.
One can first focus on the problem and its needs before diving into solutions.
This distinction was evaluated in IMoG’s definition phase as very helpful.
Second, the recommended elements of IMoG are on an appropriate level of abstraction for modeling innovations.
Elements that are required in engineering phases are left out.
This is especially true for innovations discussed in committees.
Third, IMoG’s perspectives and abstraction levels represent a good choice.
These perspectives and abstraction levels are not too many or too detailed, however, they do capture the important aspects of innovations, like strategies, features and functions, requirements and constraints as well as solutions and properties.
And fourth, the handling of availability and variability is supported as well, which is crucial for modeling future innovation while coping with the design space.

Another strength is IMoG’s flexibility.
It is possible to start an innovation considering market pulls as well as with considering technology pushs.
A market pull is understood as an innovation that is driven by a demand of the market.
A market pull in IMoG starts with the identification of the innovation from the Strategy Perspective and then slowly moves over the functions and requirements to the solutions.
A technology push is understood as an innovation that is driven by the development of a new technology.
A technology push in IMoG starts with modeling the technology on the solution level and then slowly explores the possible demands on the Quality Perspective, Functional Perspective and Strategy Perspective.
Another sign of IMoG’s flexibility lies in its domain agnostic-ness.
IMoG targets microelectronic innovations discussed in a committee, however many elements of IMoG are abstract enough to be used in any context.
Elements like software and hardware are more system related, however, still very abstract.
Thus, IMoG can be applied to similar (enough) problems.

Furthermore, IMoG is easy to apply with an IMoG expert.
By guiding through the exploration process substantial time can be saved as the people creating the idea do not have to bother with the modeling elements and modeling decisions.
This was also validated in the evaluation studies conducted by the authors.
IMoG’s validity, usefulness and adequacy were all positively evaluated.

IMoG also has some limitations.
Its high abstraction is the cost of flexibility.
Without having an IMoG expert, it is challenging to find a suitable path through the grid for a specific innovation, because multiple paths may seem valuable.
Furthermore, detailed behavioral models are out of scope of IMoG.
This can be considered a strength as state machines and alike are often too much detail for innovations.
And if the detailed behavioral models are really required, they may be added as an attachment.
On the other side, the high level of abstraction is definitely a challenge when transforming the IMoG model into a product level model.
A transformation approach is required here that takes the innovation’s context into account (e.g. see Broy et al. in \cite{broy2009toward}).
Overall, IMoG is difficult to apply without guidance from an IMoG expert.

Another limitation of IMoG is the scalability known from other modeling languages.
Its graphical nature does not scale very well in large diagrams, however, innovation modeling tend to have a small amount of elements.
Therefore, scalability was not yet identified as a big problem.

Intellectual Property protection is of high importance in committees.
This limitation is not tackled by IMoG, however, it does not restrict the use of further approaches tackling this issue while using IMoG.

From the experience of the authors, the following three recommendations support the application of IMoG:
First, it is recommended to interpret the abstraction levels as filtering mechanisms.
This thinking helps to apply abstraction levels only when they provide a clear advantage and not just \enquote{$\ldots$ because IMoG says so}.
Furthermore, it is recommended to search for an IMoG expert before starting the innovation modeling in a committee.
Without one, the whole modeling phase may become quite challenging and inefficient.
This may include improvised, ineffective meetings with inconsistent diagram exchanges.
Finally, it is recommended to make use of the Glossary and FAQ, that was created for IMoG as well as for every perspective.

\section{FAQ}
\label{sec:imog:faq}

\includegraphics[width=\linewidth]{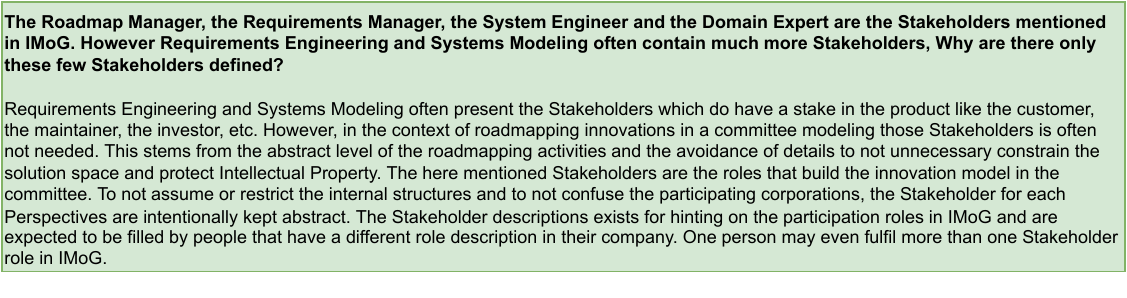}
\includegraphics[width=\linewidth]{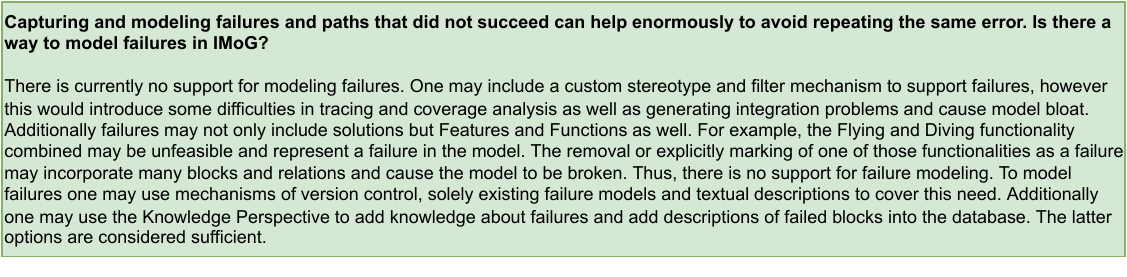}
\includegraphics[width=\linewidth]{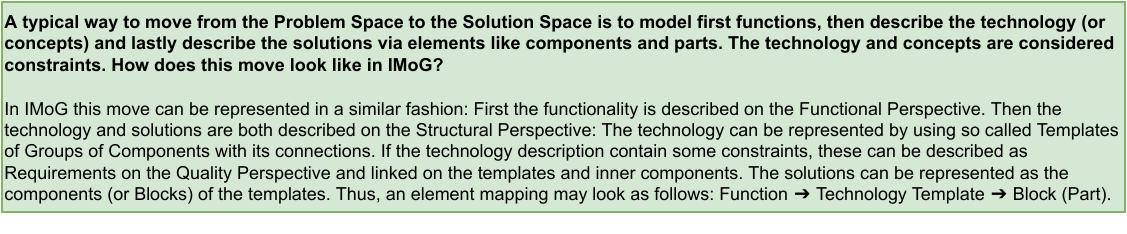}
\includegraphics[width=\linewidth]{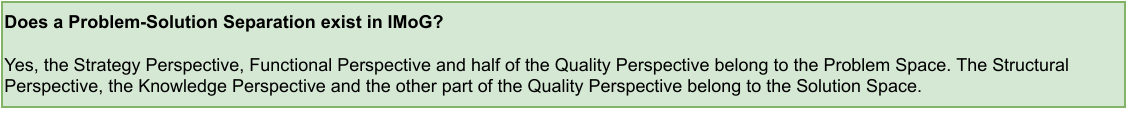}
\includegraphics[width=\linewidth]{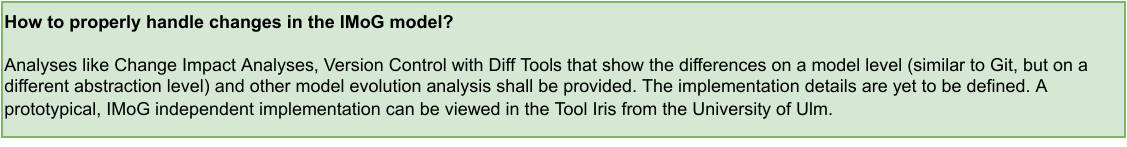}
\includegraphics[width=\linewidth]{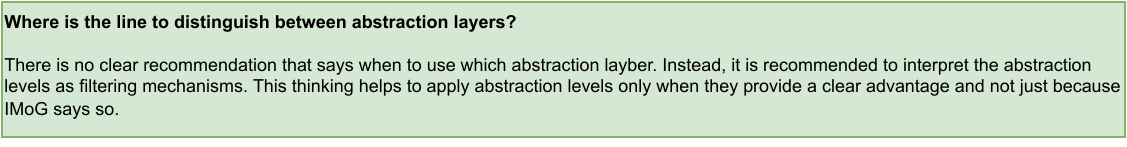}
\includegraphics[width=\linewidth]{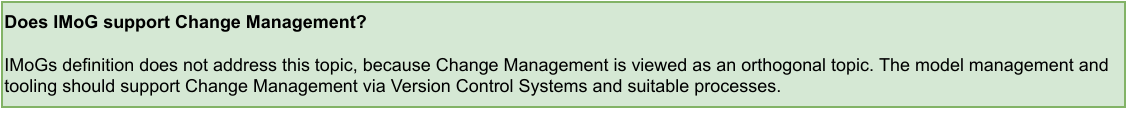}

%% file: content/strategy_perspective.tex
\chapter{Strategy Perspective}
\label{chap:strat}

\begin{figure}[b!]
	\centering
	\begin{tikzpicture}
		\newcommand\scf{0.9} 
		\node[anchor=south west] {\includegraphics[width=\scf\linewidth]{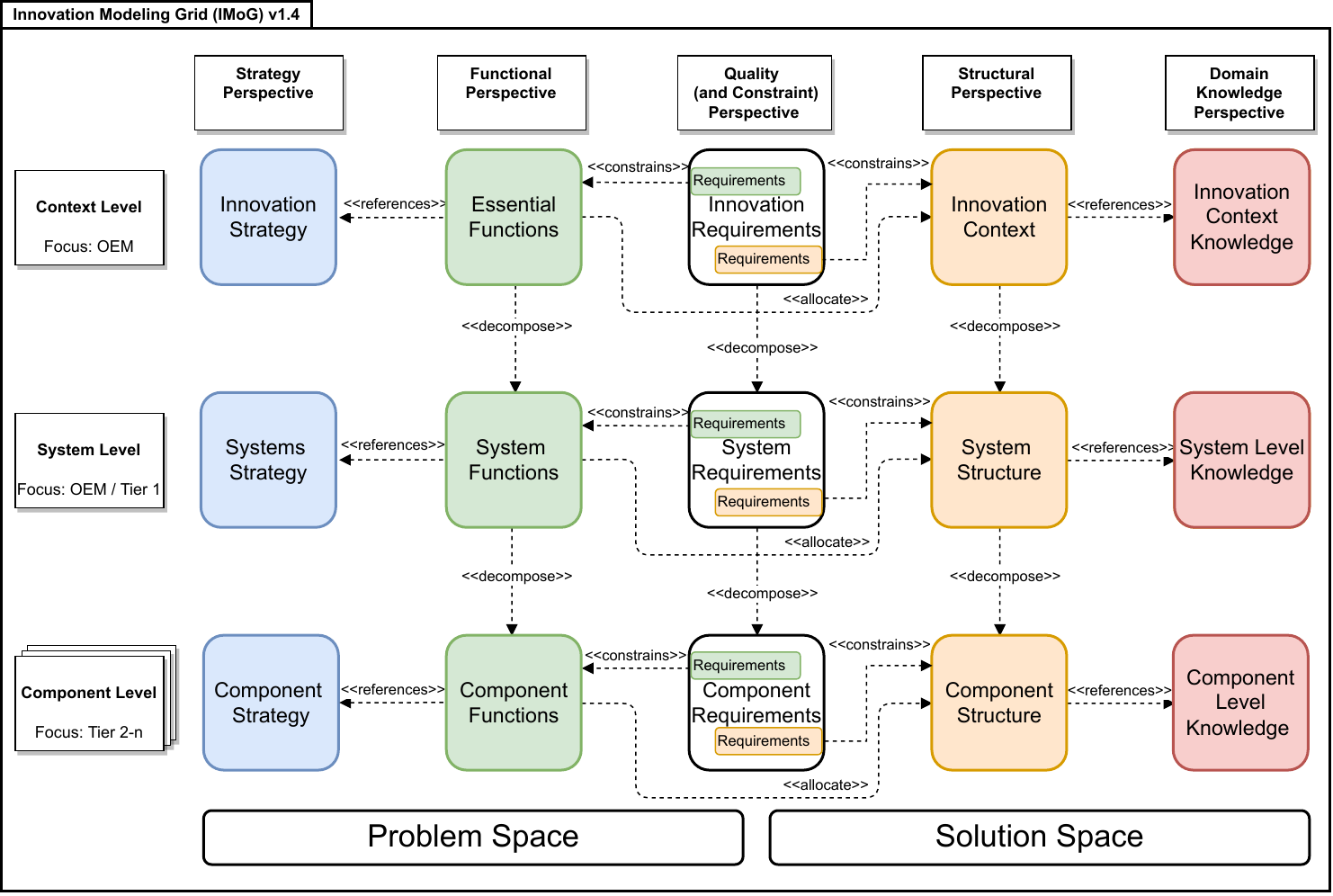}};
		\draw[ultra thick, blue, rounded corners] (\scf*2.2,\scf*1.2) rectangle (\scf*4,\scf*9.5);
		\path[fill=white,opacity=0.5] (\scf*4.9,\scf*1.2) rectangle (\scf*6.7,\scf*9.5);
		\path[fill=white,opacity=0.5] (\scf*7.55,\scf*1.2) rectangle (\scf*9.4,\scf*9.5);
		\path[fill=white,opacity=0.5] (\scf*10.25,\scf*1.2) rectangle (\scf*12.05,\scf*9.5);
		\path[fill=white,opacity=0.5] (\scf*12.9,\scf*1.2) rectangle (\scf*14.7,\scf*9.5);
	\end{tikzpicture}
	\caption{Location of the Strategy Perspective in IMoG}
	\label{fig:strat:imog}
\end{figure}

The Strategy Perspective is the first perspective in IMoG applications and targets the capturing of the chosen innovation (see Figure \ref{fig:strat:imog}).
The purpose of the Strategy Perspective is the capturing of the innovation idea as well as the capturing of the interests and strategies of the stakeholders.
Regarding, the IMoG process, the Strategy Perspective is the generated artifact of the first activity: \enquote{Innovation Identification}.
The identification itself is not part of the Strategy Perspective.
It is recommended to use creativity techniques to identify the innovation and use afterwards the Strategy Perspective to describe the innovation extensively.
The choice of creativity technique should be chosen based on the preferences and experience of the involved stakeholders in the committee.
One key principle of the Strategy Perspective is to not bother the committee with modeling restriction and give as much freedom as possible to capture the early innovation.
Thus, the only guidelines of the Strategy Perspective care of labeling and referencing, to allow a tracing of information.
Overall, the Strategy Perspective can be seen like the presentation of the innovation to externals.

\begin{figure}[h]
	\centering
	\includegraphics[width=0.7\linewidth]{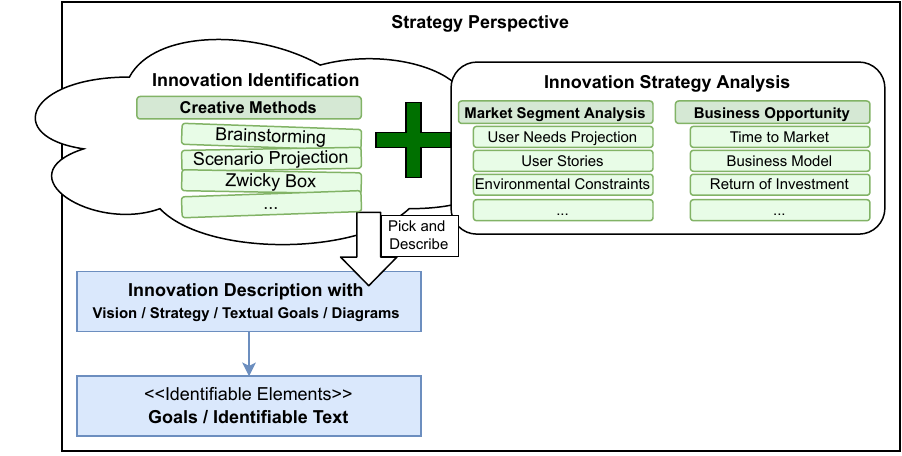}
	\caption{Activities considered for the Strategy Perspective}
	\label{fig:strat:process}
\end{figure}

The innovation identification activity and the capturing of the innovation can be imagined as followed (see Figure \ref{fig:strat:process}):
The identification of an innovation starts with many discussions, sketches and non formal descriptions.
For such activities, creativity techniques like brainstorming, scenario projection and zwicky boxes are suited.
Based on these ideas, marketing analysts take the idea and perform market segment analysis and analyze business opportunities.
This may include looking on user needs, environmental constraints, business models, time to market predictions and more.
The outcome of these creativity results and analyses is the starting point of the Strategy Perspective.
The committee leader takes the outcome and writes a draft of the innovation description.
The description is then refined with the committee.
The description may contain a vision, an explanation about the overall strategy, goals and diagrams.
Identifiable elements can also be added to the content of the Strategy Perspective to allow referencing and tracing of goals, text phrases etc.
The description and identifiable elements encompasses enough information to start the real modeling activities on the other perspectives.

The chapter is structured as followed:
In Section \ref{sec:strat:me} the meta model and its model elements are presented.
In Section \ref{sec:strat:e-scooter} an example of the Strategy Perspective is given.
The strengths and limitations of the Strategy Perspective are discussed in Section \ref{sec:strat:eval}.
A FAQ finalizes the description in Section \ref{sec:strat:faq}.

\section{Model elements}
\label{sec:strat:me}

The meta model of the Strategy Perspective is kept simple.
When discussing strategy it is uncommon to just start modeling activities.
Instead the committee is mostly interested on the description of their interests.
This meta model tries to encompass this view by only introducing descriptions (in form of HTML divs) and traceable (identifiable) elements in the Strategy Perspective model.
The descriptions can be labeled to allow filtering them out.
There are no relations defined for connecting different information on the Strategy Perspective.
However, the identifiable elements are defined for the purpose of perspective cross-referencing (including references to functions, requirements and structure).

\begin{figure}
	\centering
	\includegraphics[width=0.7\linewidth]{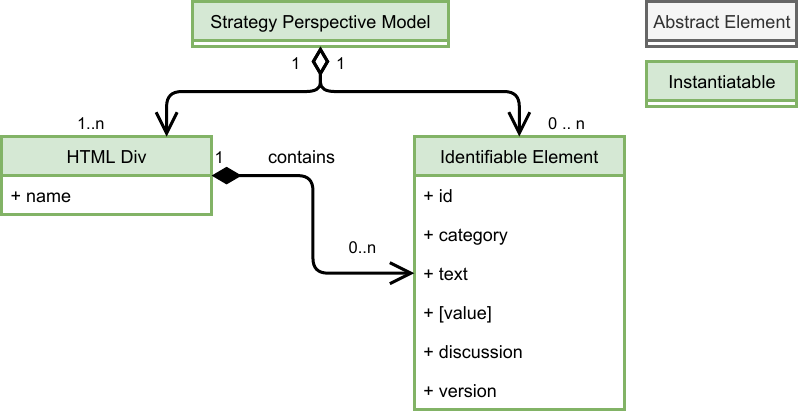}
	\caption{The model elements of the Strategy Perspective.}
	\label{fig:strat:me}
\end{figure}

\rule{\textwidth}{1pt}
Meta Model Element:
\begin{center}
	\includegraphics[width=0.3\linewidth]{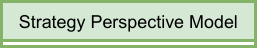}
\end{center}

Description:

\fcolorbox{gray!30!black}{gray!20!white}{
	\begin{minipage}{0.955 \textwidth}
		\large \textbf{Strategy Perspective Model}\\
		\normalsize The Strategy Perspective Model is the underlying content of the Strategy Perspective of an innovation. It contains all HTML Divs and identifiable elements.
	\end{minipage}
}

Example: A full Strategy Perspective Model example is shown in Section \ref{sec:strat:e-scooter}.

\rule{\textwidth}{1pt}
Meta Model Element:
\begin{center}
	\includegraphics[width=0.5\linewidth]{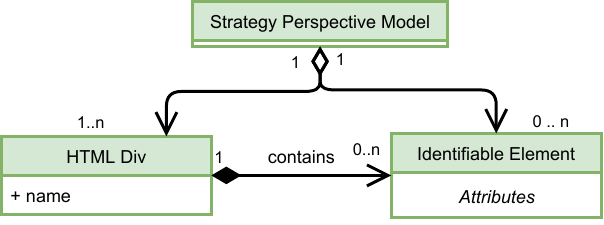}
\end{center}

Description:

\fcolorbox{gray!30!black}{gray!20!white}{
	\begin{minipage}{0.955 \textwidth}
		\large \textbf{HTML div} \\
		\normalsize The HTML Div is the container for all descriptions, textual goals, diagrams, etc. Additionally, they contain the Identifiable Elements embedded in their descriptions. The HTML Divs can be named to allow filtering them out by the tooling.
	\end{minipage}
}

Example: The round rectangle represents an example HTML div with a name, descriptions, an image and identifiable elements.
\begin{center}
	\includegraphics[width=0.7\linewidth]{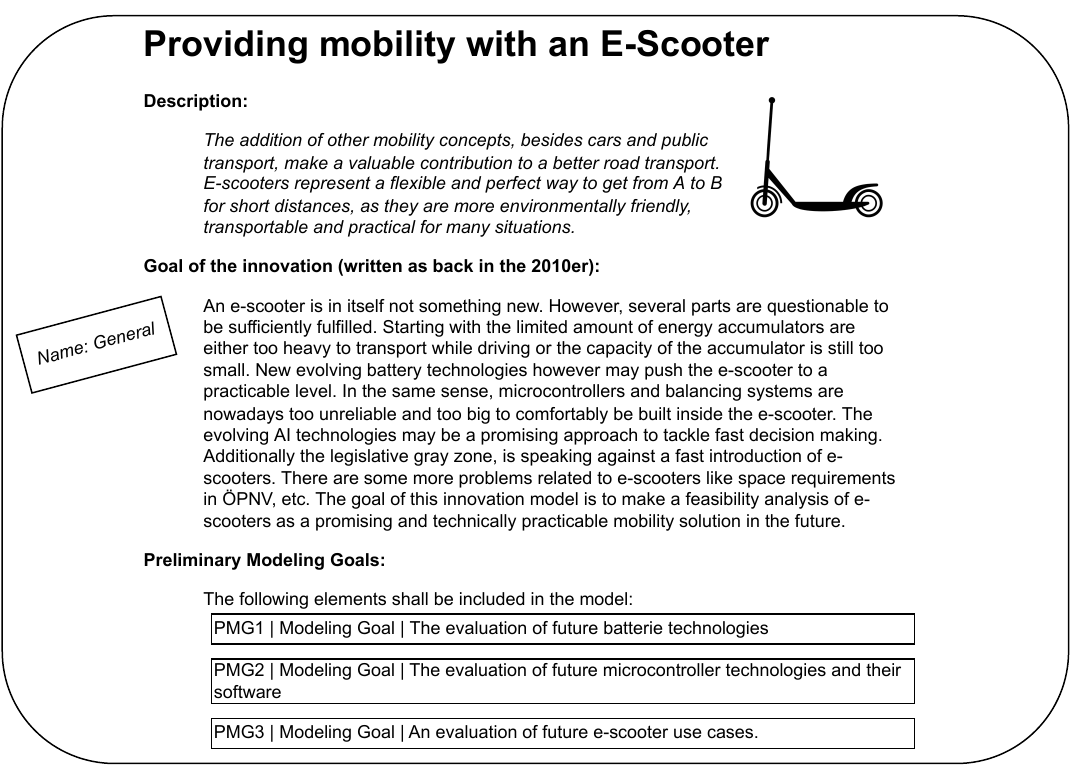}
\end{center}

\rule{\textwidth}{1pt}
Meta Model Element:
\begin{center}
	\includegraphics[width=0.5\linewidth]{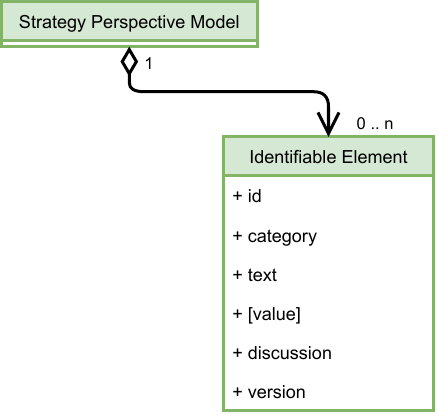}
\end{center}

Description:

\fcolorbox{gray!30!black}{gray!20!white}{
	\begin{minipage}{0.955 \textwidth}
		\large \textbf{Identifiable element} \\
		\normalsize The identifiable element is designed for tracing over perspectives. It defines the following attributes:
		\begin{itemize}
			\item The \textit{id} represents the identifier. It can be a number, a string or any other value the tooling allows.
			\item The \textit{category} attribute allows to customly group identifiable elements by strings. Example categories (maybe proposed by the tooling) are:
			\begin{itemize}
				\item Modeling Goals
				\item Sub Goals
				\item Marketing Strategies (e.g. \enquote{Production + Sales OEM})
				\item Parameters + Characteristics (2 Brakes)
				\item Chosen E-Scooter values (Speed > 100km/h)
				\item Tier 1 specialized part (Microcontroller Z)
				\item Mindmap element
				\item Technology demand
			\end{itemize}
			\item The \textit{text} represents the content.
			\item The optional \textit{value} attribute allows to enhance the element with a value. The value can be used for checking consistency between the identifiable element of the Strategy Perspective and any other element that includes a value or property. Note, that on early strategy considerations values are seldom available.
			\item The \textit{discussion} and \textit{version} fields enhance the Block description by allowing discussions and version control.
		\end{itemize}
	\end{minipage}
}

Example: The following three examples represent Identifiable Elements with an id, a category and a text.
The optional value is not set and there exists no discussion and version number.
(The empty non optional attributes are not shown here).
\begin{center}
	\includegraphics[width=0.7\linewidth]{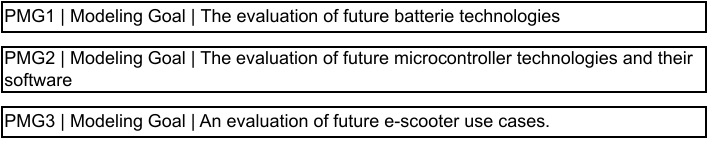}
\end{center}

\section{E-Scooter example}
\label{sec:strat:e-scooter}

The example of the Strategy Perspective describes the innovation of \enquote{Providing mobility with an e-scooter}.

The example is divided into two views:
The Strategy Description View describes the innovation textually and graphically from the OEM view, Tier 1 view, Tier 2 view and from a general view.
It contains descriptions, goals, business models, aspects, important notes and diagrams from creativity methods.
Additionally, some text phrases and goals are made identifiable to allow a mapping and tracing on the other perspectives.
The Strategy List View shows the identifiable elements from the Strategy Perspective, that were used in the Strategy Description View.
The descriptions are left out.
This view is used to focus and remark the identifiable elements.

\textbf{Strategy Description View}
\begin{center}
	\includegraphics[width=1\linewidth]{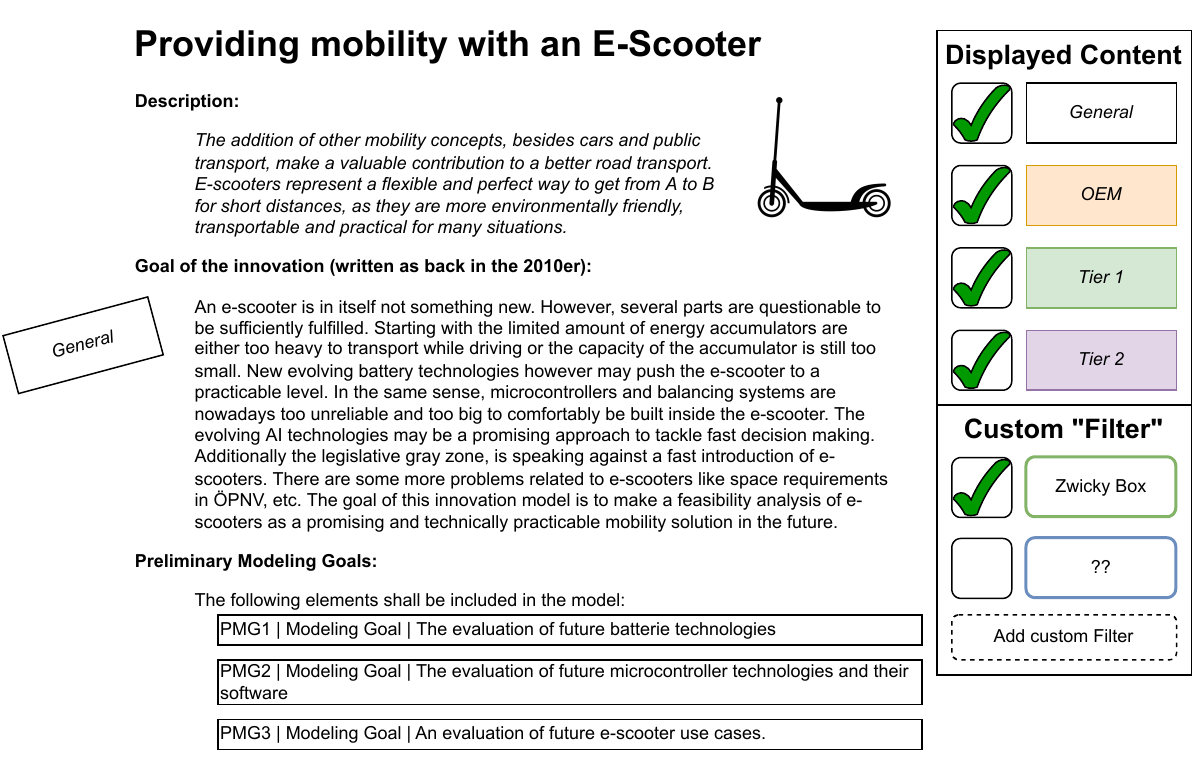}
	\includegraphics[width=0.8\linewidth]{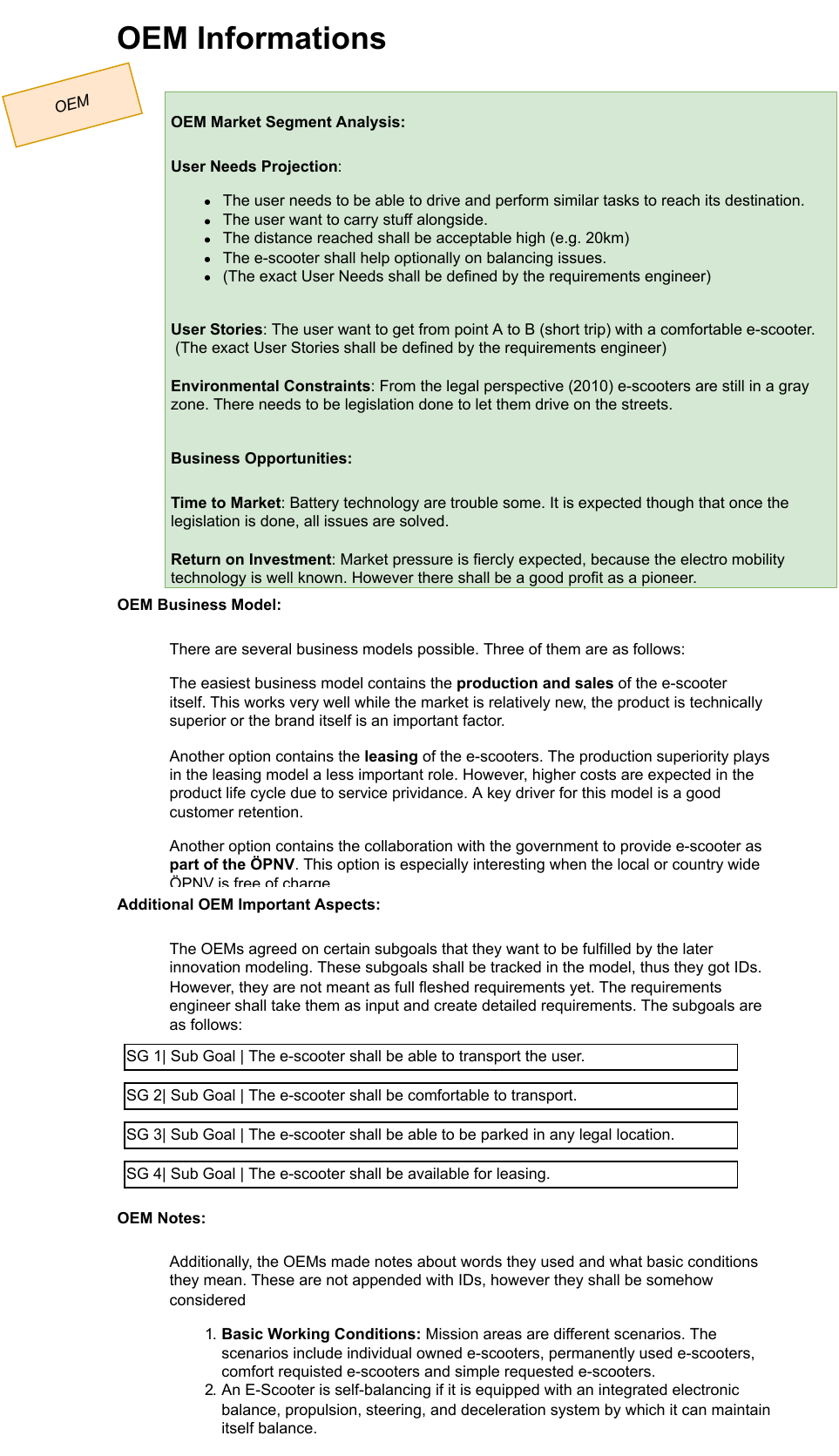}
	\includegraphics[width=1\linewidth]{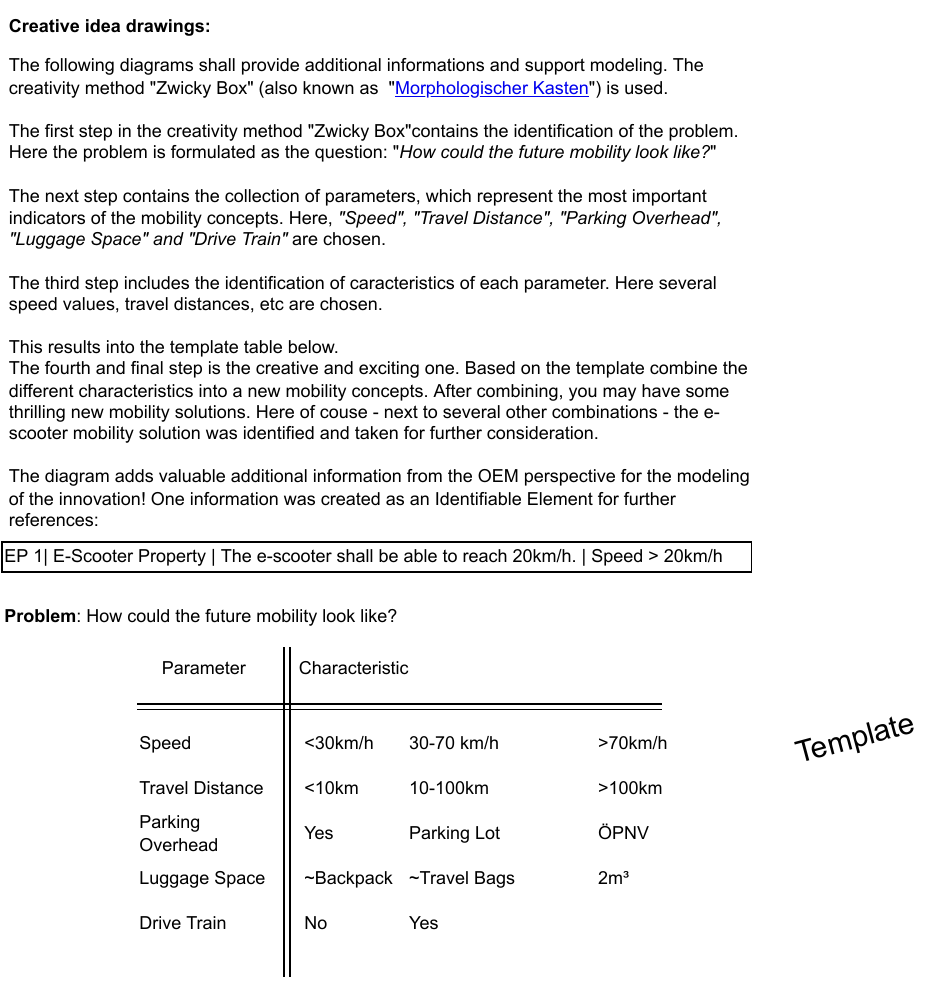}
	\includegraphics[width=1\linewidth]{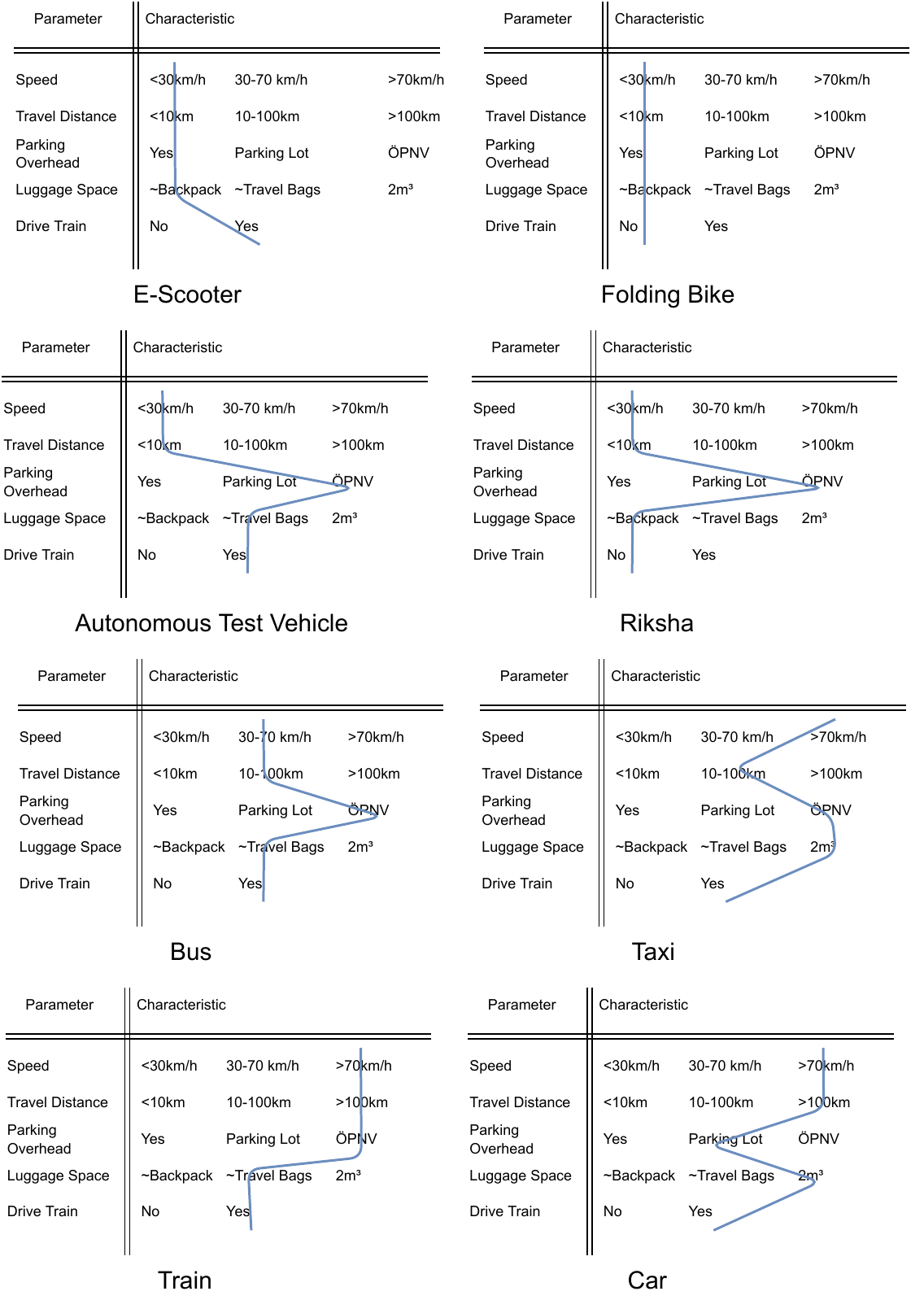}
	\includegraphics[width=1\linewidth]{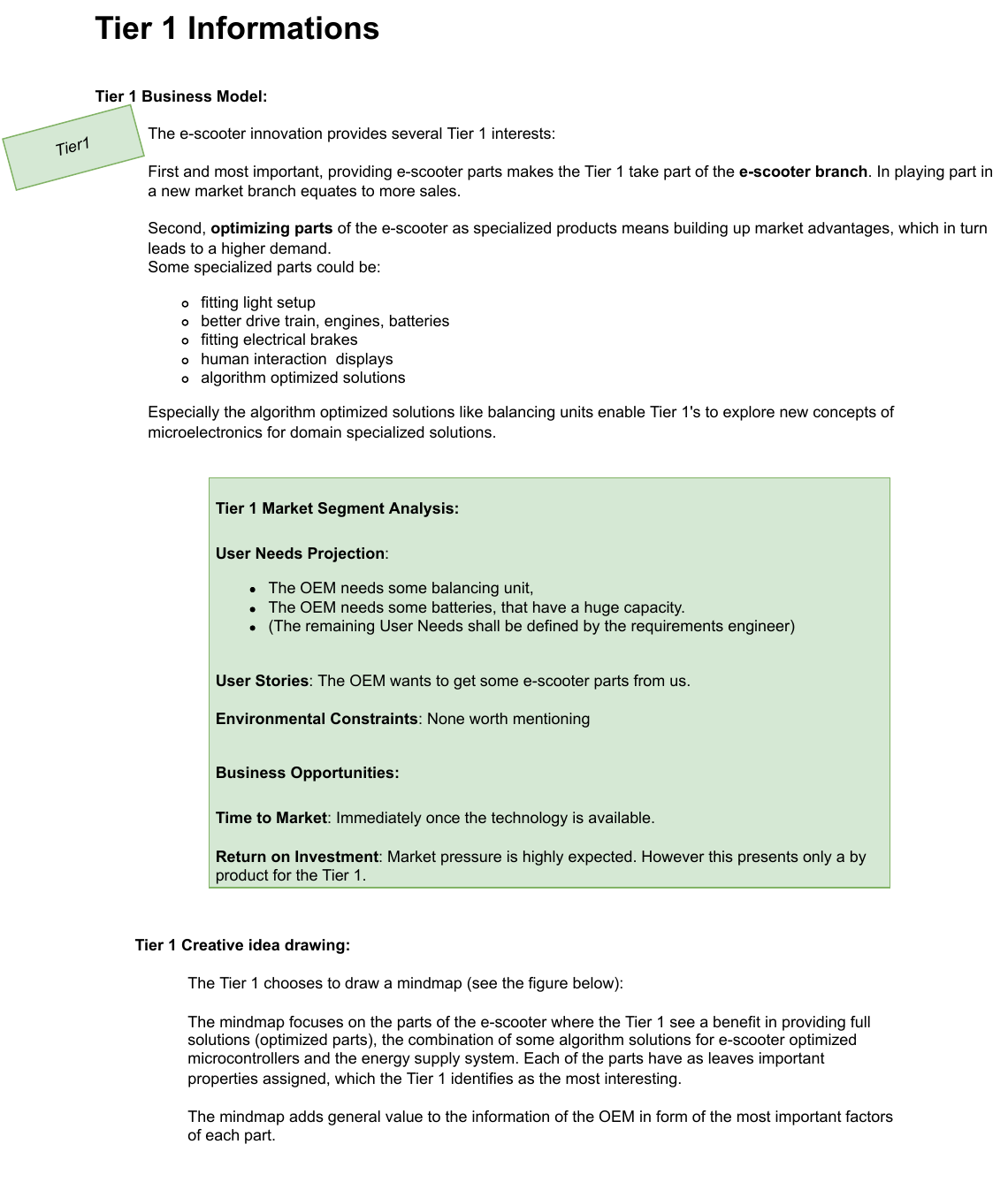}
	\includegraphics[width=1\linewidth]{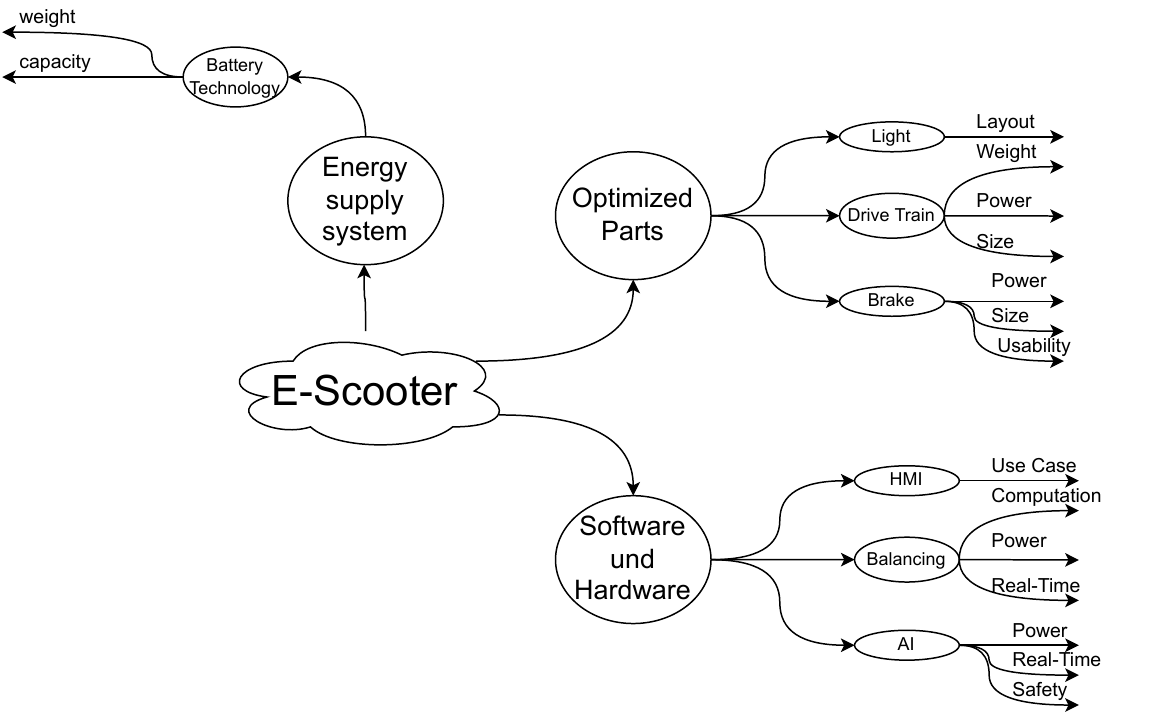}
	\includegraphics[width=0.95\linewidth]{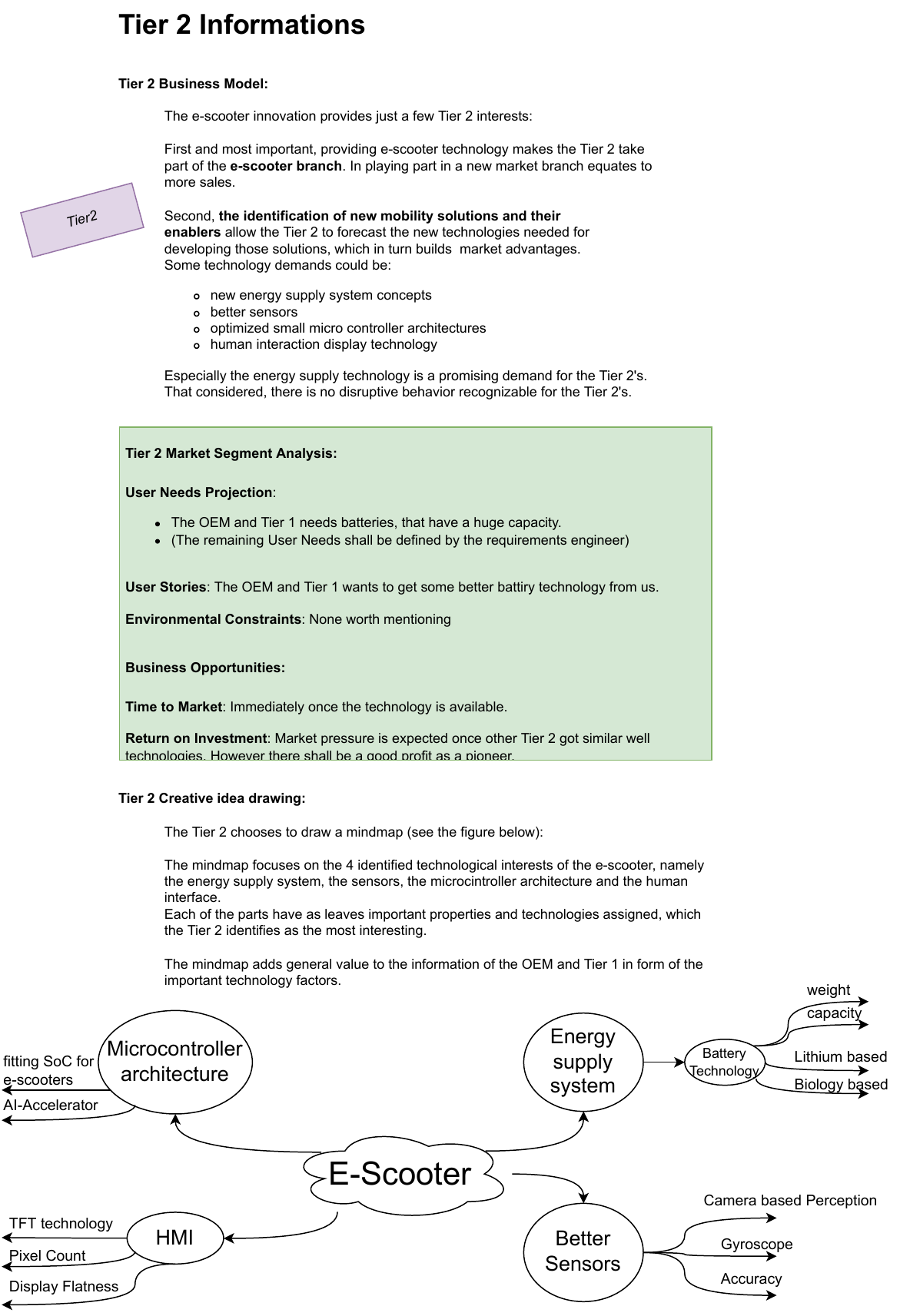}
	\includegraphics[width=1\linewidth]{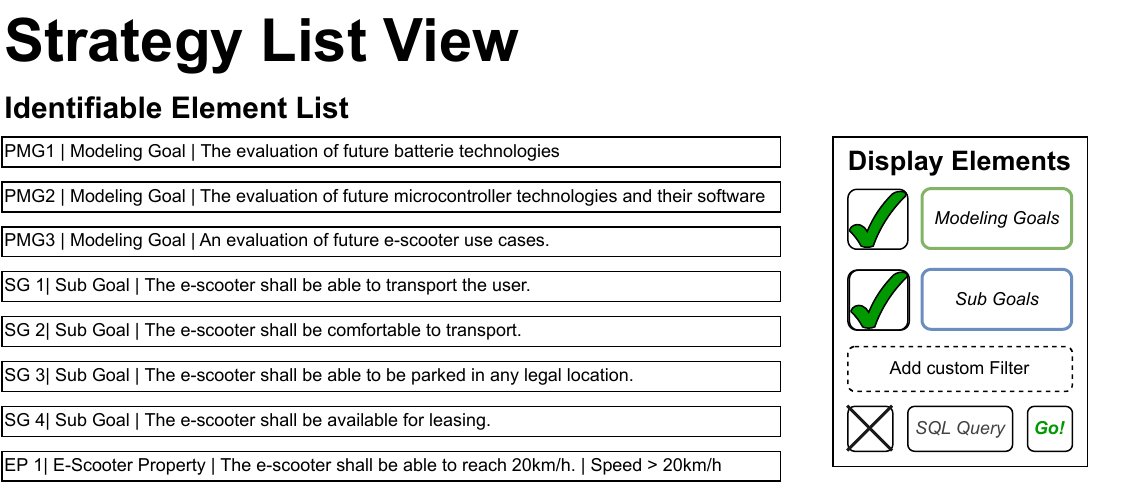}
\end{center}

\section{Strategy Perspective: Strengths and Limitations}
\label{sec:strat:eval}

The Strategy Perspective is kept abstract on purpose.
It contains (mostly) non formal descriptions making it easy to kick start the Strategy Perspective by directly starting with the creative methods results.
Additionally, the Strategy Perspective does not restrict IMoG to use any specific creativity methodology.
While the burden to choose a creativity methodology for an innovation is shifted to the user, the interchangeability of taking a proven methodology for the stakeholders instead of a predefined methodology is considered an advantage.
These decisions to focus on abstract and interchangeable modeling make the Strategy Perspective simple to model and visualize.
On the other hand, the abstract and informal model of the Strategy Perspective builds the basis for basic analysis other than tracing.
This is not per se a con but rather a limitation that was traded in for flexibility through informality.

\section{Strategy Perspective FAQ}
\label{sec:strat:faq}

The FAQ splits up into the categories:
\begin{enumerate}
	\item Questions and answers about the general Strategy Perspective model elements
	\item Questions and answers about the relation of the Strategy Perspective to other perspectives
\end{enumerate}

\textbf{Strategy model elements:}

\includegraphics[width=1\linewidth]{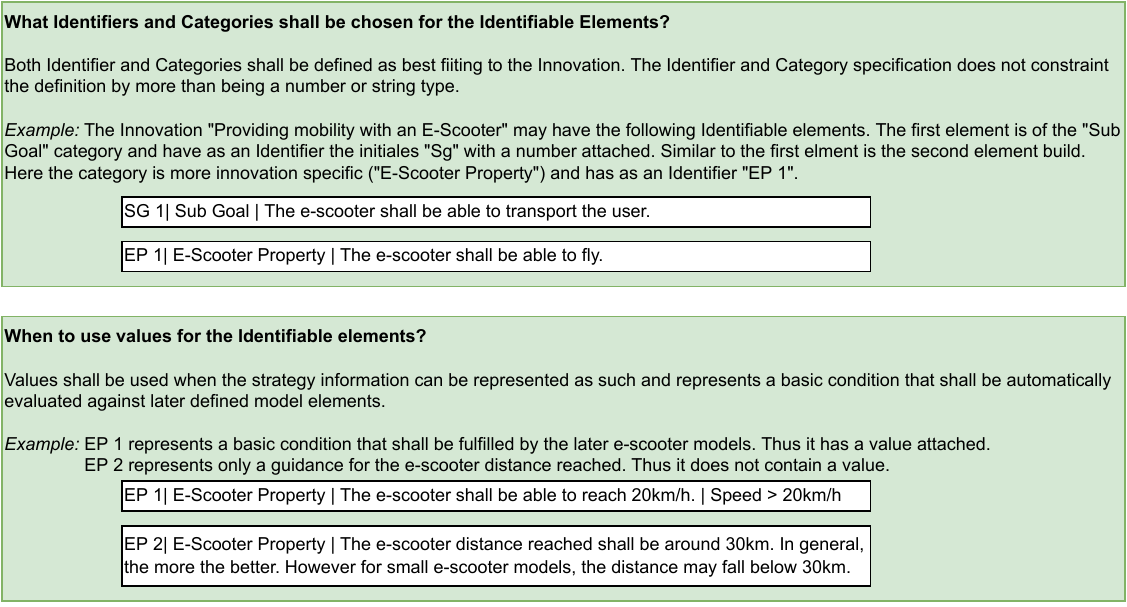}

\textbf{Relations to the other perspectives:}

\includegraphics[width=1\linewidth]{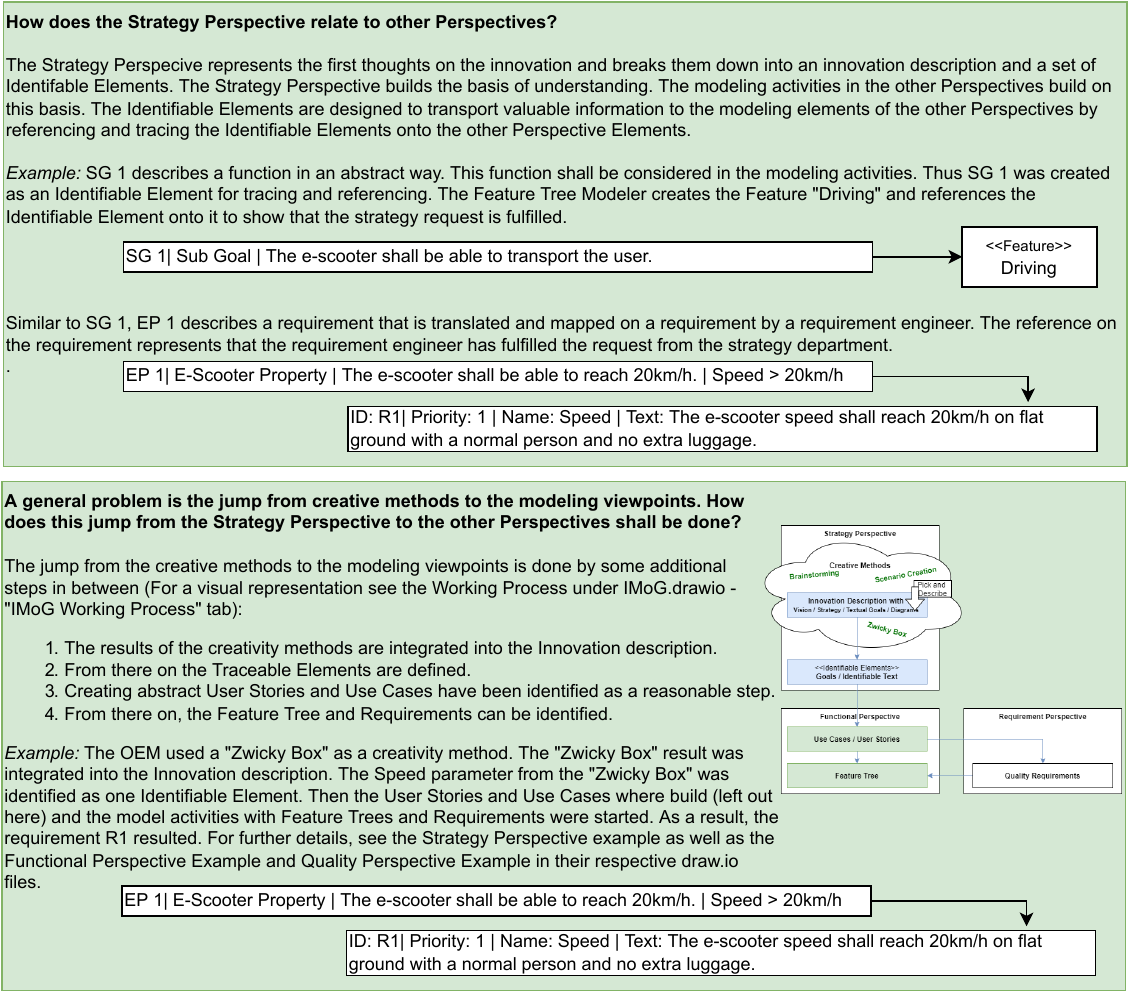}

%% file: content/functional_perspective.tex
\chapter{Functional Perspective}
\label{chap:fp}

\begin{figure}[b!]
	\centering
	\begin{tikzpicture}
		\newcommand\scf{0.9} 
		\node[anchor=south west] {\includegraphics[width=\scf\linewidth]{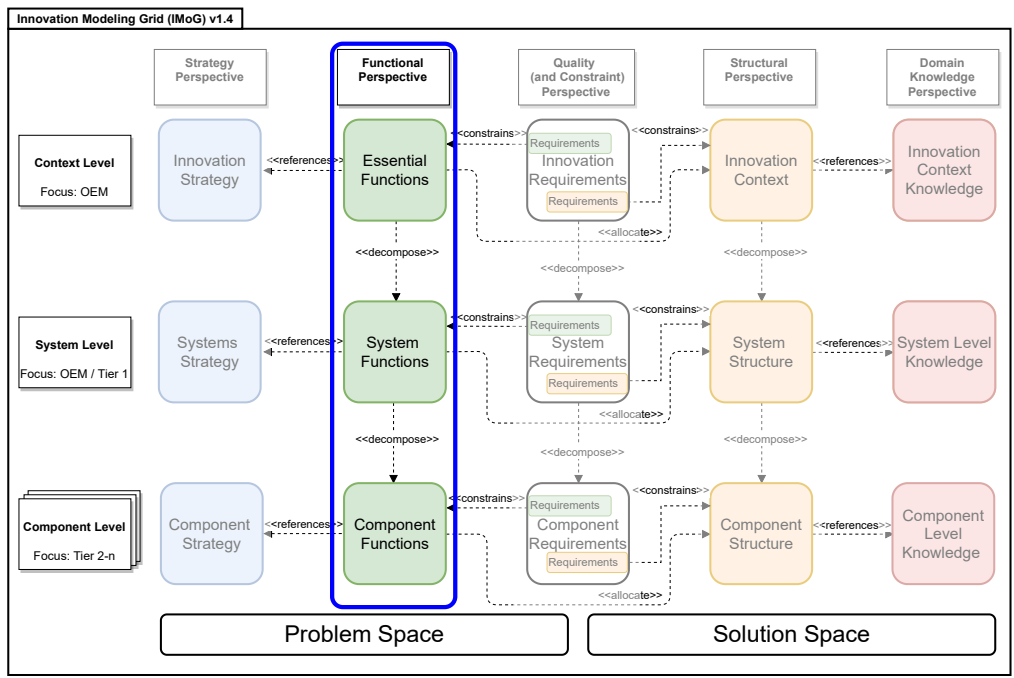}};
		\path[fill=white,opacity=0.5] (\scf*2.2,\scf*1.2) rectangle (\scf*4,\scf*9.5);
		\draw[ultra thick, blue, rounded corners] (\scf*4.9,\scf*1.2) rectangle (\scf*6.7,\scf*9.5);
		\path[fill=white,opacity=0.5] (\scf*7.55,\scf*1.2) rectangle (\scf*9.4,\scf*9.5);
		\path[fill=white,opacity=0.5] (\scf*10.25,\scf*1.2) rectangle (\scf*12.05,\scf*9.5);
		\path[fill=white,opacity=0.5] (\scf*12.9,\scf*1.2) rectangle (\scf*14.7,\scf*9.5);
	\end{tikzpicture}
	\caption{Location of the Functional Perspective in IMoG}
	\label{fig:fp:imog}
\end{figure}

The Functional Perspective is the second perspective in IMoG and describes the features (end-user visible characteristics) and functions (traceable tasks or actions that a system shall perform) of the innovation (see Figure \ref{fig:fp:imog}).
The purpose of the Functional Perspective is to capture the features and functions required for the innovation before diving deep into solutions.
The Functional Perspective is thus used in the early phases of innovation modeling and is part of capturing the problem space.

The Functional Perspective model bases on the well-known feature trees \cite{kang1990feature}.
Feature Trees are a subclass of Feature Models, which restrict themselves to tree structures.
This restriction reduces the complexity and increases the ability to maintain an overview.
Feature Trees put a high focus on variability and are often used in product line engineering.
The high-level innovation modeling in IMoG on the other side, does not focus too much on variability and the modeling of details are expected to be after the innovation modeling, being part of the subsequent design and engineering phases.
That noted, having the ability to model a bit of variability may support in expressing innovation dependencies.

IMoG also makes some extensions to feature trees, however these are mostly of cosmetic nature and can be directly translated to the default feature trees.
This translation allows the use of available Feature Tree analysis tools.


The chapter is structured as followed:
In Section \ref{sec:fp:me} the meta model and its model elements are presented.
In Section \ref{sec:fp:e-scooter} an example of the Functional Perspective is given.
The strengths and limitations of the Functional Perspective are discussed in Section \ref{sec:fp:eval}.
A FAQ finalizes the description in Section \ref{sec:fp:faq}.

\section{Model elements}
\label{sec:fp:me}

\begin{figure}
	\centering
	\includegraphics[width=1\linewidth]{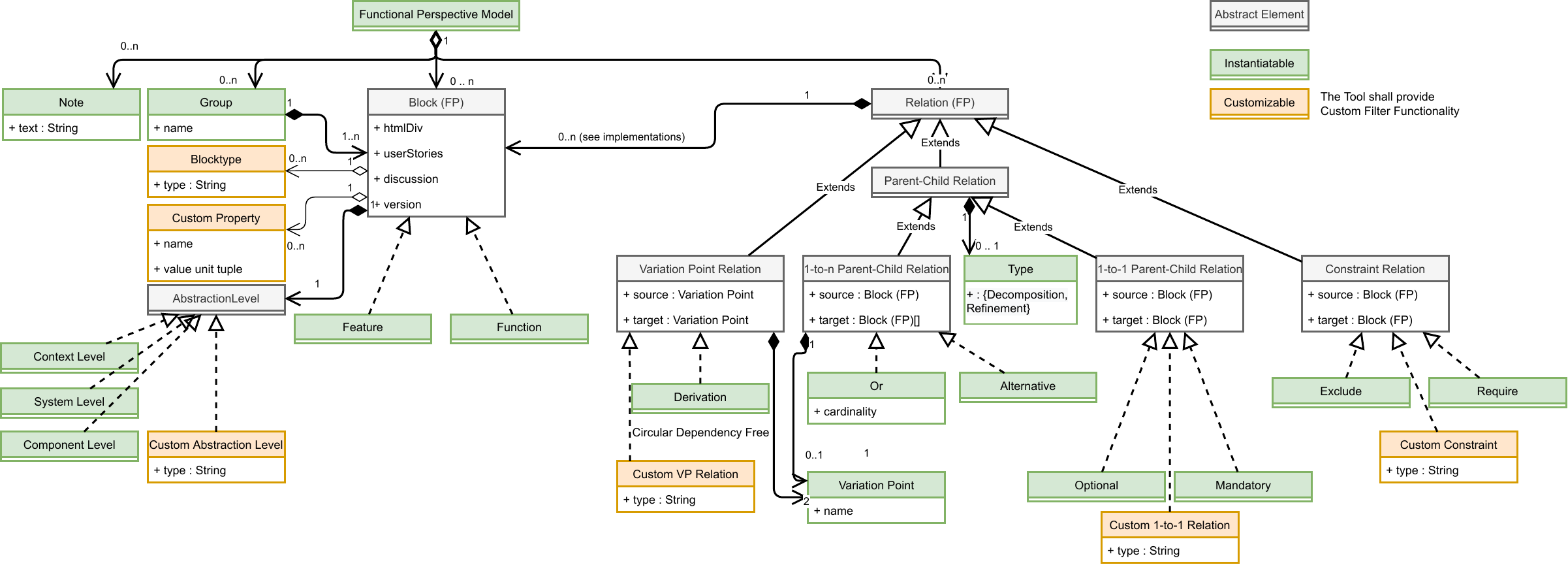}
	\caption{The model elements of the Functional Perspective.}
	\label{fig:fp:me}
\end{figure}

The meta model (see Figure \ref{fig:fp:me}) builds on the FODA (Feature Tree model and Feature Diagram model, \cite{kang1990feature}) and includes all relevant concepts of FODA.
The meta model has - as the top level unit of the Functional Perspective - the Functional Perspective Model.
It contains a set of Blocks (FP) with Relations (FP) between them.
Additionally Groups of Blocks (FP) are contained.
Blocks (FP) represent the basic units of functionality known from Feature Trees and are extended with several attributes, Blocktypes and an Abstraction Level (which can be either Context Level, System Level, Component Level or of Type custom Abstraction Level).
A detailed description of the attributes can be found in the Block (FP) description.
Unlike the original FODA model definition \cite{kang1990feature}, where functions are explicitely not part of the Feature Tree, Blocks (FP) are here further categorized into features and functions for specifying what explicitly a 'Block' means.
A feature represents a logical unit of behavior that is too abstract to be mapped on structural solutions, while a function represents a mappable unit onto the structural solutions.
For a flexible mapping, each feature shall have a set of functions.
The ability to model functions allows the seamless tracing from features onto functions and later onto solutions on the Structural Perspective.

Several types of Relations between Blocks (FP) can be made, including and extending the typical relations from Feature Trees.
The Relations (FP) split up into Parent-Child Relations, Constraint Relations and Variation Point Relations.
The Parent-Child Relations include the Alternative-Relation, the Or-Relation, the Mandatory-Relation and the Optional-Relation known from Feature Trees.
Additionally the Parent-Child Relations can be optionally labeled as 'Refinement' or 'Decomposition' or a custom Parent-Child Relation can be used.
The Constraint Relations include the known extensions to Feature Trees to express restrictions on configurations: The Require and Exclude relations.
If not enough, custom Constraint Relations can be added.
The last extension made to Relations (FP) are the Variation Point Derivation Relation to represent similar alternative choices.
The following model elements Section will dive into more details.

\rule{\textwidth}{1pt}
Meta Model Element:
\begin{center}
	\includegraphics[width=0.3\linewidth]{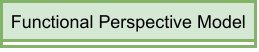}
\end{center}

Description:

\fcolorbox{gray!30!black}{gray!20!white}{
	\begin{minipage}{0.955 \textwidth}
		\large \textbf{Functional Perspective Model}\\
		\normalsize The \textit{Functional Perspective Model} is the diagram of the Functional Perspective of an innovation. It contains all model elements of the Functional Perspective.
	\end{minipage}
}

Example: A full Functional Perspective Model example is shown in Section \ref{sec:fp:e-scooter}.

\rule{\textwidth}{1pt}
{\Large Meta Model Base}

\rule{\textwidth}{1pt}
Meta Model Element:
\begin{center}
	\includegraphics[width=1\linewidth]{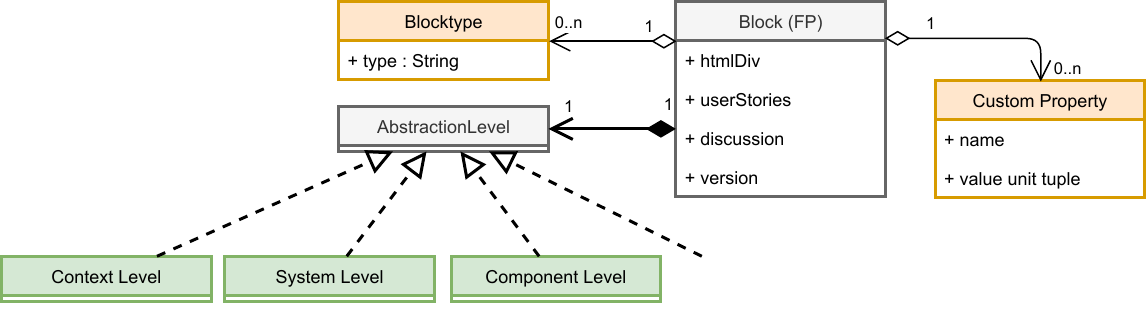}
\end{center}

Description:

\fcolorbox{gray!30!black}{gray!20!white}{
	\begin{minipage}{0.955 \textwidth}
		\large \textbf{Block (FP)} \small \textbf{(Read: Block on the Functional Perspective)} \\
		\normalsize The \textit{Block (FP)} is the abstract Block element of the Functional Perspective which is implemented by \textit{Features} and \textit{Functions}.
		It defines the attributes of the Blocks:
		\begin{itemize}
			\item The Abstraction Level of the Block defines the level of abstraction the Block represents.
			It can be either \textit{Context Level, System Level, Component Level} or from the type \textit{Custom Abstraction Level}.
			\item An optional \textit{Custom Block Type} can refine the category of the Block further.
			\item The \textit{HTMLDiv} represents the description of the Block to solve the problem of lack of clarity by adding information next to Feature Trees.
			The description shall answer shortly \enquote{What the Block shall provide?}, the Reasoning behind the Block and its basic conditions to work and if the Block has alternative choices, then additionally the binding time of the choice.
			The binding time being part of the \textit{HTMLDiv} is considered enough here. It does not have to be a Block property like proposed in the original Feature Tree publication \cite{kang1990feature}.
			There is no template needed.
			Images or drafts provide valuable information.
			\item Optional Custom Block Properties can be defined for additional tooling analysis including filtering and consistency checks.
			\item The \textit{User Stories, discussion} and \textit{version} enhance the Block description.
		\end{itemize}
	\end{minipage}
}

Example: The following block shows an example of a Block (FP) with its attributes.
\begin{center}
	\includegraphics[width=\linewidth]{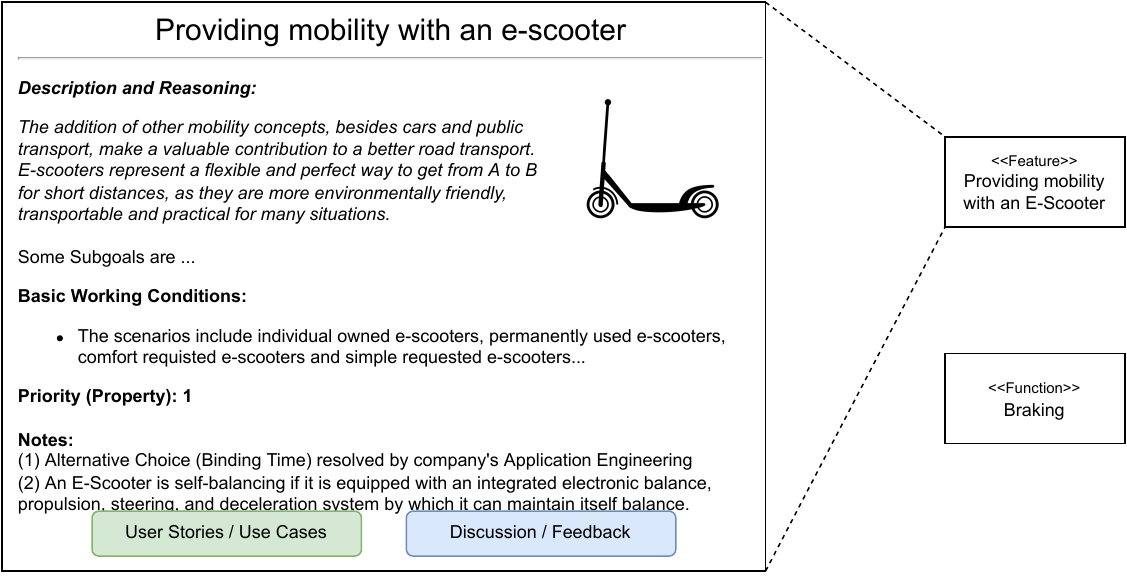}
\end{center}

\rule{\textwidth}{1pt}
Meta Model Element:
\begin{center}
	\includegraphics[width=\linewidth]{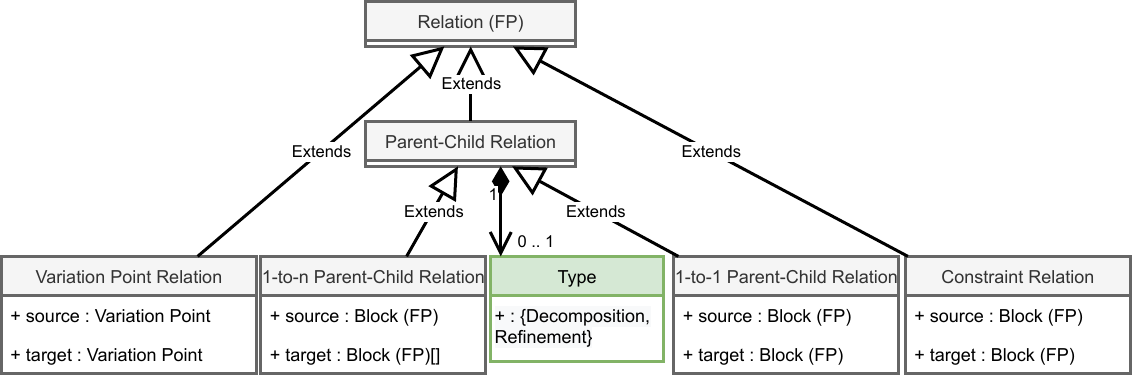}
\end{center}

Description:

\fcolorbox{gray!30!black}{gray!20!white}{
	\begin{minipage}{0.955 \textwidth}
		\large \textbf{Relation (FP)} \small \textbf{(Read: Relation on the Functional Perspective)} \\
		\normalsize The abstract \textit{Relation (FP)} describe relations between \textit{Blocks (FP)} or respectively \textit{Variation Points} on the Functional Perspective. \textit{Relations} are further categorized by
		\begin{itemize}
			\item \textit{1-to-1 Variation Point Relations}
			\item \textit{Parent-Child Relations} between Blocks (FP) of either category \textit{1-to-n Parent-Child} relation or category \textit{1-to-1 Parent-Child} relation.
			The \textit{Parent-Child Relations} can be specified by an optional type, which can be either of value \textit{Decomposition} or \textit{Refinement}.
			Note that the additional stereotypes are similar to the relation differentiation \textit{\{Specialization (Refinement), Decomposition, Parametrization\}} outside the model definition in the original Feature Tree publication \cite{kang1990feature}.
			\item 1-to1 Constraint Relations between Blocks (FP)
		\end{itemize}
	Each of the Relations are described in more detail on its own.
	\end{minipage}
}

Example: An example is shown for each relation under their respective description.

\rule{\textwidth}{1pt}
Meta Model Element:
\begin{center}
	\includegraphics[width=1\linewidth]{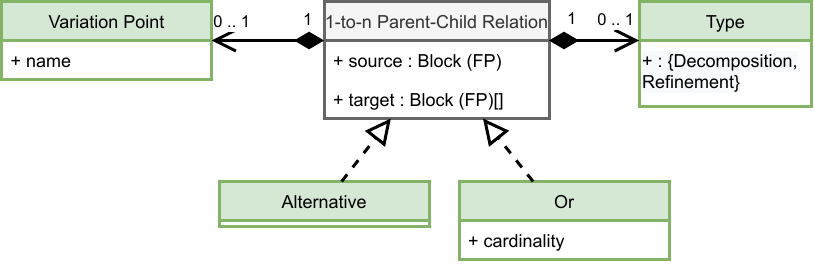}
\end{center}

Description:

\fcolorbox{gray!30!black}{gray!20!white}{
	\begin{minipage}{0.955 \textwidth}
		\large \textbf{1-to-n Parent-Child Relations} \\
		\normalsize The \textit{1-to-n Parent-Child Relations} build the Foundation for the well known \textit{Alternative} and \textit{Or} Relation from Feature Trees.
		\textit{Alternative} and \textit{Or} Relations can connect one parent Block (FP) with multiple child Blocks (FP).
		They are used to describe Decomposition or Refinement Choices of the Parent Block.
		Additionally, 1-to-n Parent-Child Relations can own Variation Points.
		\textit{Variation Points} represent the description of the choice or variability and are written optionally.
		For Refinement Relations, the special Refinement Representation using \textit{Variant Lists} can be used.
		However, the normal Representation and the Variant List Refinement Representation are both valid.
	\end{minipage}
}

\begin{center}
	\includegraphics[width=\linewidth]{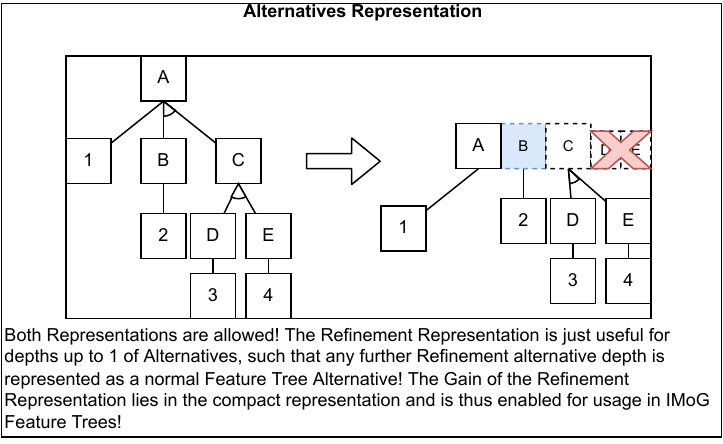}
\end{center}

Example:
\begin{center}
	\includegraphics[width=\linewidth]{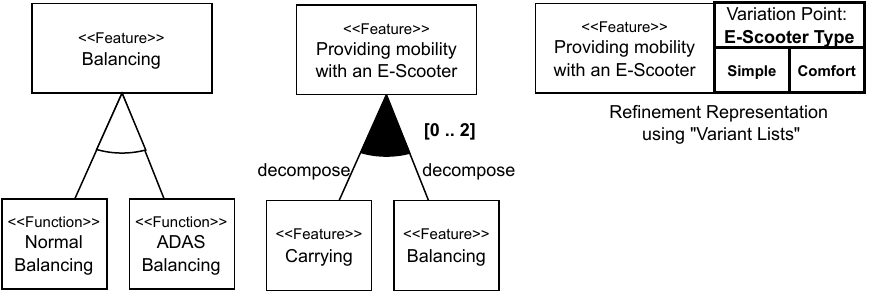}
\end{center}

\rule{\textwidth}{1pt}
Meta Model Element:
\begin{center}
	\includegraphics[width=\linewidth]{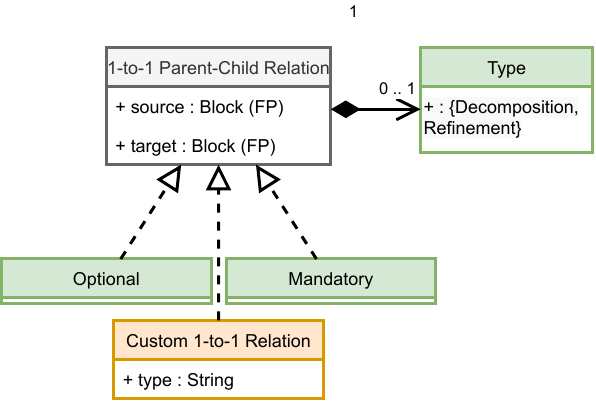}
\end{center}

Description:

\fcolorbox{gray!30!black}{gray!20!white}{
	\begin{minipage}{0.955 \textwidth}
		\large \textbf{1-to-1 Parent-Child Relation} \\
		\normalsize The \textit{1-to-1 Parent-Child} relation connects two Blocks (FP) with each other.
		Up to now there are three relations defined.
		The \textit{Optional} relation, the \textit{Mandatory} relation and the \textit{Custom 1-to-1} relation.
		The \textit{Custom 1-to-1} relation allows to describe additional relations between Blocks (FP).
		Custom 1-to-1 relations are not analyzed.
		The other relations are described under their respective description.
		As before mentioned, the\textit{ 1-to-1 Parent-Child} relations can be specified by an optional type, which can be either of value \textit{Decomposition} or \textit{Refinement}.
	\end{minipage}
}

Example: An example is shown for each relation under their respective description.

\rule{\textwidth}{1pt}
{\Large Model Elements}

\rule{\textwidth}{1pt}
\fbox{
	\begin{minipage}{0.955 \textwidth}
		\large \textbf{Model Elements:} \\
		\normalsize \setstretch{0.8}
		Block Types:
		\begin{enumerate}
			\item Features
			\item Functions
		\end{enumerate}
		Relation Types:
		\begin{enumerate}
			\item And Relation (Mandatory Sub-Features) with
			\begin{itemize}
				\item Refinement / Decomposition Relations
			\end{itemize}
			\item Optional Features (+ Optional Relations) with
			\begin{itemize}
				\item Refinement / Decomposition Relations
			\end{itemize}
			\item Xor Relation (Alternative or Variant) with
			\begin{itemize}
				\item Refinement / Decomposition Relations
				\item Variation Point
				\item Variant List (Refinement Representation)
				\item Cyclefree Variation Point Selection Derivation
			\end{itemize}
			\item Or Relation with Cardinalities with
			\begin{itemize}
				\item Refinement / Decomposition Relations
			\end{itemize}
			\item Constraint Relations (Require / Exclude)
			\begin{itemize}
				\item Constraint Grouping
			\end{itemize}
		\end{enumerate}
		Miscellaneous:
		\begin{enumerate}
			\item Notes
		\end{enumerate}
	\end{minipage}
}

In the following the model elements are introduced.
First the two Block Types are introduced and then the relations between the Blocks are presented.
Lastly the \enquote{Notes} element is introduced.

\rule{\textwidth}{1pt}
{\Large Block Types}

\rule{\textwidth}{1pt}
Meta Model Element:
\begin{center}
	\includegraphics{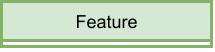}
\end{center}

Description:

\fcolorbox{gray!30!black}{gray!20!white}{
	\begin{minipage}{0.955 \textwidth}
		\large \textbf{Feature} \\
		\normalsize The \textit{Feature} is a Block (FP) of the Functional Perspective and is next to \textit{Functions} one of the two existing Block (FP) elements.
		A Feature defines a logical unit of behavior.
		It semantics originates from Feature Models \cite{kang1990feature}.
		However, Features are additionally understood here as actionable, uncountable items and shall be described like an activity.
		A Feature is considered as too abstract to be mapped on structural solutions.
		The Stereotype <<Feature>> can be omitted.
	\end{minipage}
}

Example:
\begin{center}
	\includegraphics{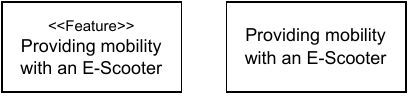}
\end{center}

\rule{\textwidth}{1pt}
Meta Model Element:
\begin{center}
	\includegraphics{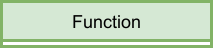}
\end{center}

Description:

\fcolorbox{gray!30!black}{gray!20!white}{
	\begin{minipage}{0.955 \textwidth}
		\large \textbf{Feature} \\
		\normalsize The \textit{Function} is a Block (FP) of the Functional Perspective and is next to \textit{Features} one of the two existing Block (FP) elements.
		A Function defines a logical unit of behavior that shall be implemented by structural components.
		Functions are understood here as actionable, uncountable items. and shall be described like an activity.
	\end{minipage}
}

Example:
\begin{center}
	\includegraphics{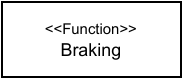}
\end{center}

\rule{\textwidth}{1pt}
{\Large Relations}

\rule{\textwidth}{1pt}
Meta Model Element:
\begin{center}
	\includegraphics{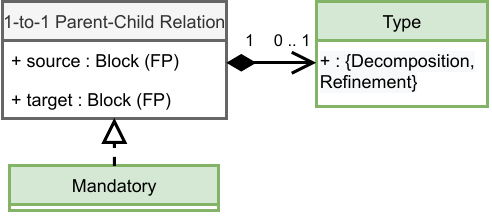}
\end{center}

Description:

\fcolorbox{gray!30!black}{gray!20!white}{
	\begin{minipage}{0.955 \textwidth}
		\large \textbf{Mandatory Relation} \\
		\normalsize The \textit{Mandatory} relation connects one parent (always the top one) Block (FP) with a child (always the bottom one) Block (FP).
		The \textit{Mandatory} relation describes that the child Block must be provided once the parent Block is part of the configuration.
		The relation has two additional stereotyped forms: The \textit{Mandatory-Decomposition} relation and the \textit{Mandatory-Refinement} relation.
		The \textit{Mandatory-Decomposition} relation describes, that the child Block is a decomposed element of the parent Block.
		The \textit{Mandatory-Refinement} relation on the other hand describes that the child Block is a refinement of the parent Block.
		If the Stereotype is omitted, then the \textit{Mandatory} relation is interpreted as a \textit{Mandatory-Decomposition} relation.
	\end{minipage}
}

Example:
\begin{center}
	\includegraphics{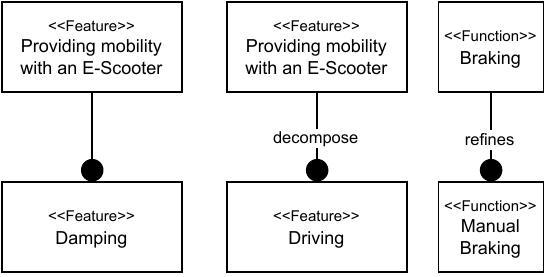}
\end{center}

\rule{\textwidth}{1pt}
Meta Model Element:
\begin{center}
	\includegraphics{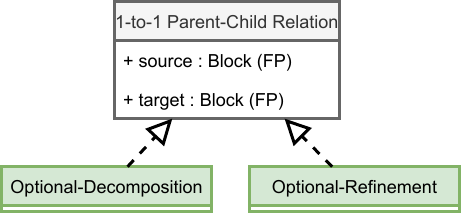}
\end{center}

Description:

\fcolorbox{gray!30!black}{gray!20!white}{
	\begin{minipage}{0.955 \textwidth}
		\large \textbf{Optional Relation} \\
		\normalsize The \textit{Optional} relation connects one parent (always the top one) Block (FP) with a child (always the bottom one) Block (FP).
		The \textit{Optional} relation describes that the child Block may be optionally provided once the parent Block is part of the configuration.
		The relation has two additional stereotyped forms: The \textit{Optional-Decomposition} relation and the \textit{Optional-Refinement} relation.
		The \textit{Optional-Decomposition} relation describes, that the child Block is a decomposed element of the parent Block.
		The \textit{Optional-Refinement} relation on the other hand describes that the child Block is a refinement of the parent Block.
		If the Stereotype is omitted, then the \textit{Optional} relation has no additional semantics assigned.
	\end{minipage}
}

Example:
\begin{center}
	\includegraphics{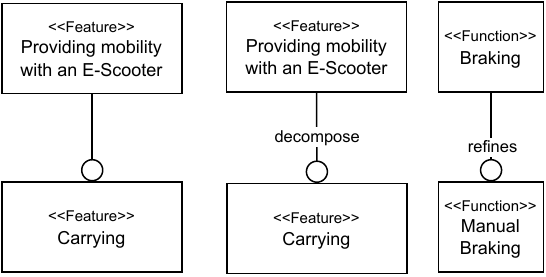}
\end{center}

\rule{\textwidth}{1pt}
Meta Model Element:
\begin{center}
	\includegraphics{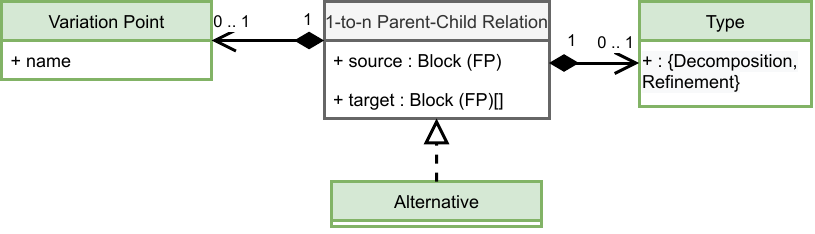}
\end{center}

Description:

\fcolorbox{gray!30!black}{gray!20!white}{
	\begin{minipage}{0.955 \textwidth}
		\large \textbf{Alternatives and Variation Points} \\
		\normalsize The \textit{Alternative} relation is a well known element from Feature Trees.
		It represents a choice of Blocks (FP) from which exactly one option can be taken.
		\textit{Variation Points} represent the description of the choice or variability and are written optionally with the \textit{Alternatives}.
		\textit{Alternatives} are used to describe Decomposition or Refinement relations of the parent Block.
		For Refinement relations, the special Refinement Representation using 'Variant Lists' can be used.
		However, the normal representation and the Variant List Refinement representation are both valid.

		As mentioned before, the \textit{Alternatives Relations} can be specified by an optional type, which can be either of value \textit{Decomposition} or \textit{Refinement}.
	\end{minipage}
}

Example:
\begin{center}
	\includegraphics{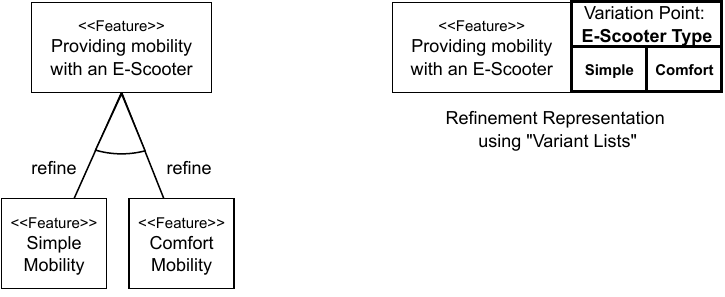}
\end{center}

\rule{\textwidth}{1pt}
Meta Model Element:
\begin{center}
	\includegraphics{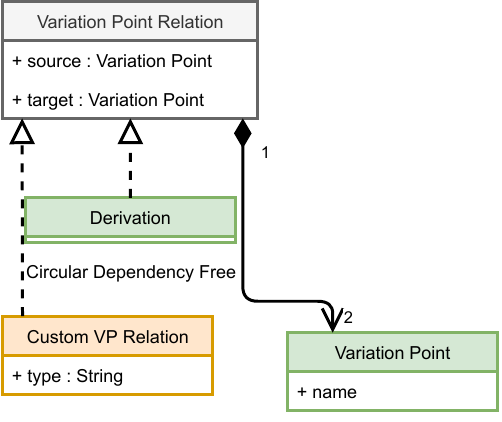}
\end{center}

Description:

\fcolorbox{gray!30!black}{gray!20!white}{
	\begin{minipage}{0.955 \textwidth}
		\large \textbf{Variation Point Relation} \\
		\normalsize The \textit{Variation Point} relation connects two \textit{Variation Points} with each other.
		Up to now there are only the \textit{Derivation} relation and the \textit{Custom VP} relation defined.
		The \textit{Derivation} relation represents the derivation of the choice of the target \textit{Variation Point} given that the choices are the same.
		The \textit{Derivation} relation can only derive from Variation Points with at least a level higher than the trees depth.
		This avoids cycles.
		With this defined, some global configuration can be defined and locally used on the fitting places.
		The \textit{Derivation} relation can replace several \textit{Require} relations.
		The \textit{Custom VP} relation allows to describe additional relations between Variation Points.
		Custom VP Relations are not analyzed.
	\end{minipage}
}

Example:
\begin{center}
	\includegraphics{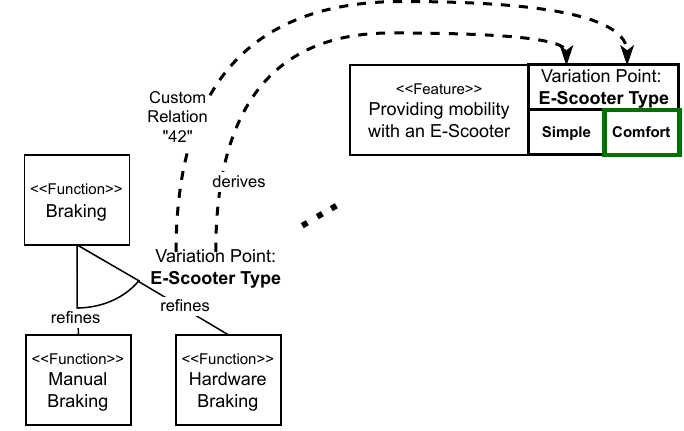}
\end{center}

\rule{\textwidth}{1pt}
Meta Model Element:
\begin{center}
	\includegraphics{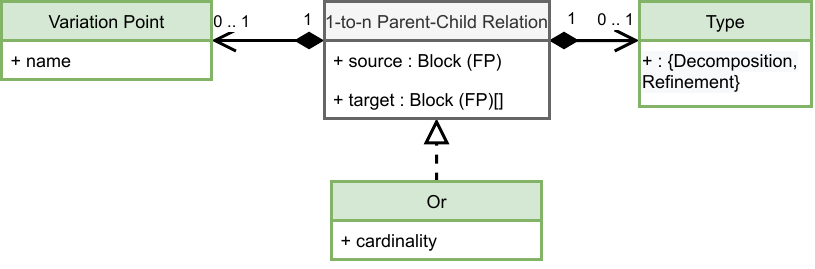}
\end{center}

Description:

\fcolorbox{gray!30!black}{gray!20!white}{
	\begin{minipage}{0.955 \textwidth}
		\large \textbf{Or Relation with Cardinality} \\
		\normalsize The \textit{Or} relation is a well known element from Feature Trees.
		The \textit{Or} relation represents a choice of Blocks (FP) from which one or more options can be taken.
		The \textit{Or} relation is a generalization of \textit{Alternatives}.
		The \textit{cardinality} describes how many Blocks can be chosen to create a valid configuration.
		\textit{Variation Points} represent the description of the choice or variability and are written optionally with the Or relations.
		As mentioned before, the Or relations can be specified by an optional type, which can be either of value \textit{Decomposition} or \textit{Refinement}.
		Or relations are mostly used to describe \textit{Decomposition} relations of the parent Block.
		\textit{Refinement} relations need to have a \textit{cardinality} of [1,1] to be valid thus the \textit{Alternative} relation is used for them instead.
	\end{minipage}
}

Example:
\begin{center}
	\includegraphics{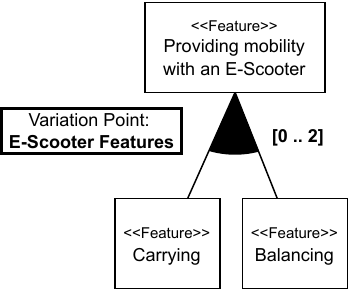}
\end{center}

\rule{\textwidth}{1pt}
Meta Model Element:
\begin{center}
	\includegraphics{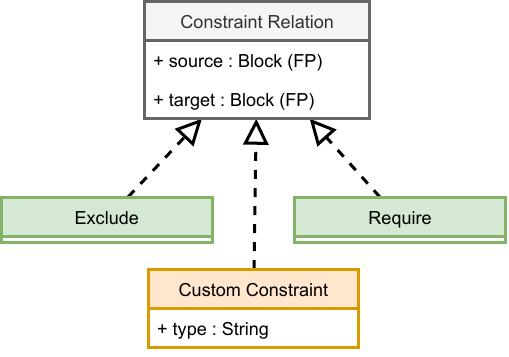}
\end{center}

Description:

\fcolorbox{gray!30!black}{gray!20!white}{
	\begin{minipage}{0.955 \textwidth}
		\large \textbf{Constraint Relation} \\
		\normalsize The \textit{Constraint} relation connects one Block (FP) with any other (even non child) Block (FP).
		The \textit{Constraint} relation describes a restriction to the space of configurations and thus how the constraint Block relates to the other Block if the other Block is part of the configuration.
		There are currently three Constraint Relation Types:
		\begin{itemize}
			\item The \textit{Require} relation A$\rightarrow$B specifies that if A is part of the configuration, then B must be part of the configuration too.
			\item The \textit{Exclude} relation A$\rightarrow$B specifies that if A is part of the configuration, then B is not allowed to be part of the configuration.
			\item The \textit{Custom Constraint} allows to describe additional constraints between Blocks (FP) and own a custom type.
			\textit{Custom Constraints} are not analysed.
		\end{itemize}
	\end{minipage}
}

Example:
\begin{center}
	\includegraphics{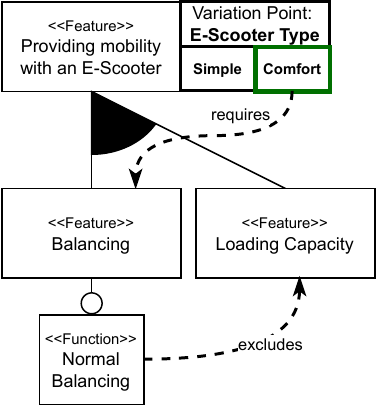}
\end{center}

\rule{\textwidth}{1pt}
Meta Model Element:
\begin{center}
	\includegraphics{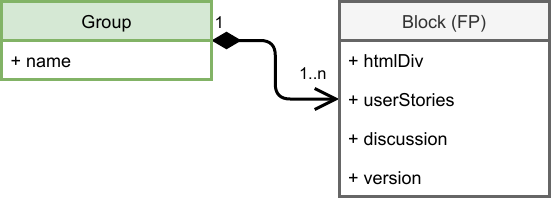}
\end{center}

Description:

\fcolorbox{gray!30!black}{gray!20!white}{
	\begin{minipage}{0.955 \textwidth}
		\large \textbf{Grouping} \\
		\normalsize The \textit{Grouping} is an additional usability feature.
		A \textit{Group} represents a set of together belonging Blocks (FP).
		If one Block of a Group is part of the configuration, then every other Block in the Group must be part of the configuration too.
		A Group can be rewritten as \textit{Require} relations between every two members of the Group.
		With one major difference: The Groups shall be toggable in the tool before the analysis is started, as such constraint-set is typically only wanted in experiments of configurations with the Feature Tree.
		The toggability in the tool represents a preprocess before the analysis and does not increase the expressiveness from Feature Trees nor does it increase the size of the problem of the analysis.
	\end{minipage}
}

Example:
\begin{center}
	\includegraphics{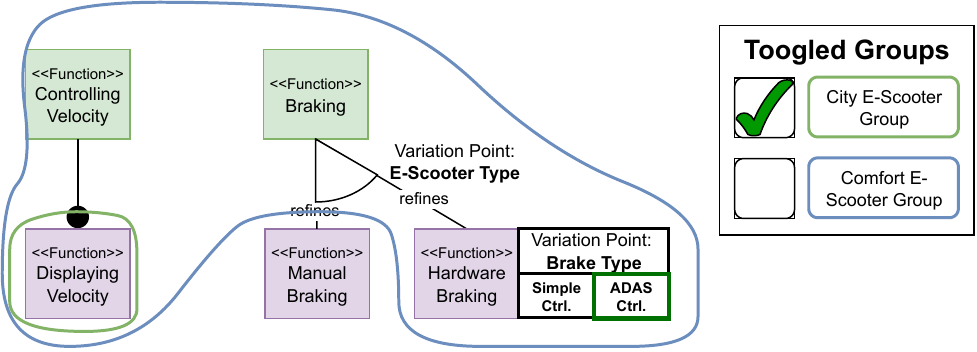}
\end{center}

\rule{\textwidth}{1pt}
{\Large Miscellaneous}

\rule{\textwidth}{1pt}
Meta Model Element:
\begin{center}
	\includegraphics{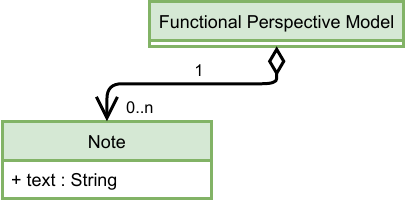}
\end{center}

Description:

\fcolorbox{gray!30!black}{gray!20!white}{
	\begin{minipage}{0.955 \textwidth}
		\large \textbf{Note} \\
		\normalsize The \textit{Note} can be used to add information to the model that can not or should not be modeled.
		Notes should be used sparsely!
	\end{minipage}
}

Example:
\begin{center}
	\includegraphics{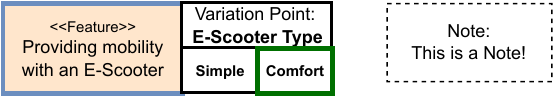}
\end{center}

\section{E-Scooter example}
\label{sec:fp:e-scooter}
\FloatBarrier

The example of the Functional Perspective describes the features and functions identified for the innovation of \enquote{Providing mobility with an e-scooter}.

\begin{figure}[h]
	\centering
	\includegraphics[width=\linewidth]{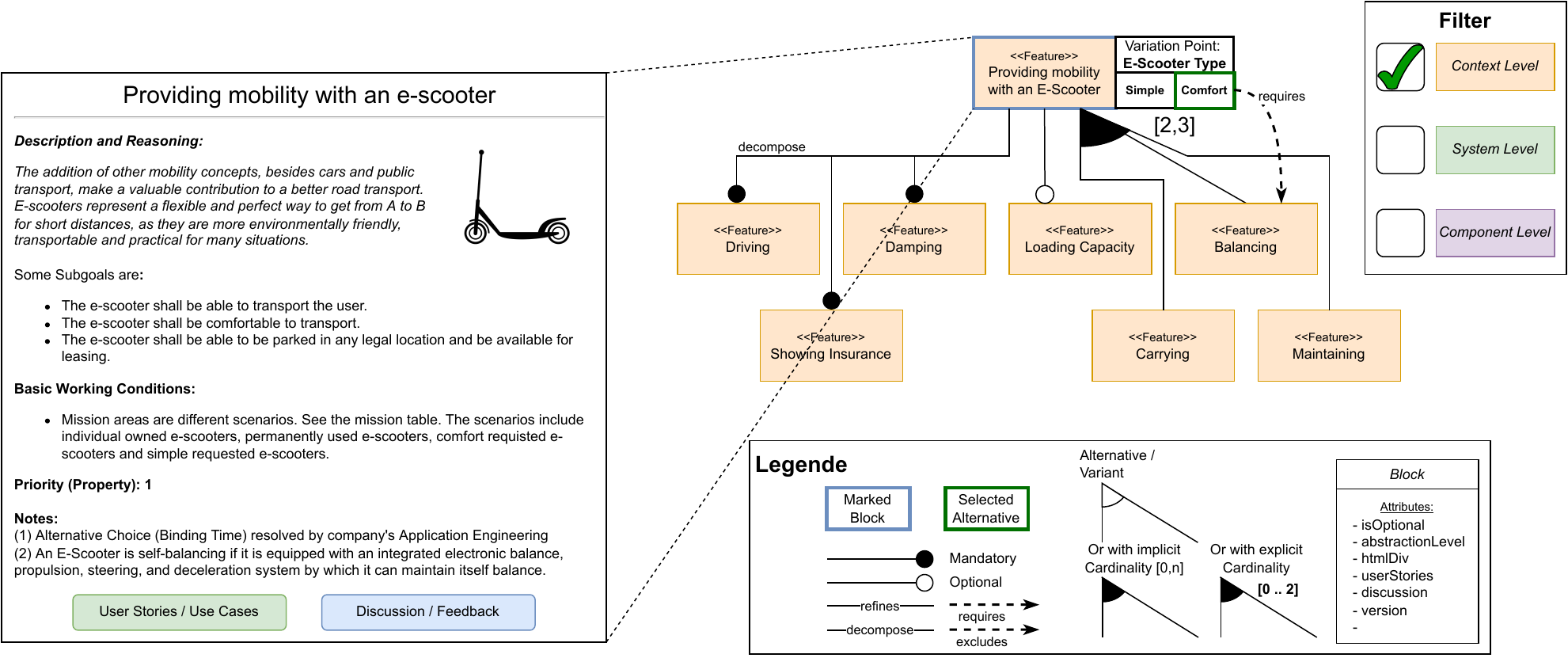}
	\caption{The identified features and functions for the innovation of \enquote{Providing mobility with an e-scooter}.}
	\label{fig:fp:example}
\end{figure}

Figure \ref{fig:fp:example} shows the feature model including the context level only.
The root feature \enquote{Providing mobility with an e-scooter} is marked here, showing the description and properties of the root block.
It includes the reasoning behind this innovation, some goals from the strategy perspective, some basic working conditions, notes as well as references to the use cases and user stories.
The root feature has three mandatory subfeatures, the \enquote{Driving} feature, the \enquote{Damping} feature and the \enquote{Showing Insurance} feature, and one optional \enquote{Loading Capacity} feature.
The root feature has a variation point focusing on the type of the e-scooter: either a simple and cheap version of the e-scooter or a comfort version.
Note, that the variation point can be represented like an Alternative relation with a name.
Finally, the e-scooter has three features as a choice, from which two must be at least be taken.
This choice includes the \enquote{Carrying} feature, the \enquote{Balancing} feature and the \enquote{Maintaining} feature.
The choice is more for showing a complete example than having a decent reasoning behind the choice.

\begin{sidewaysfigure}[!h]
	\centering
	\includegraphics[width=\linewidth]{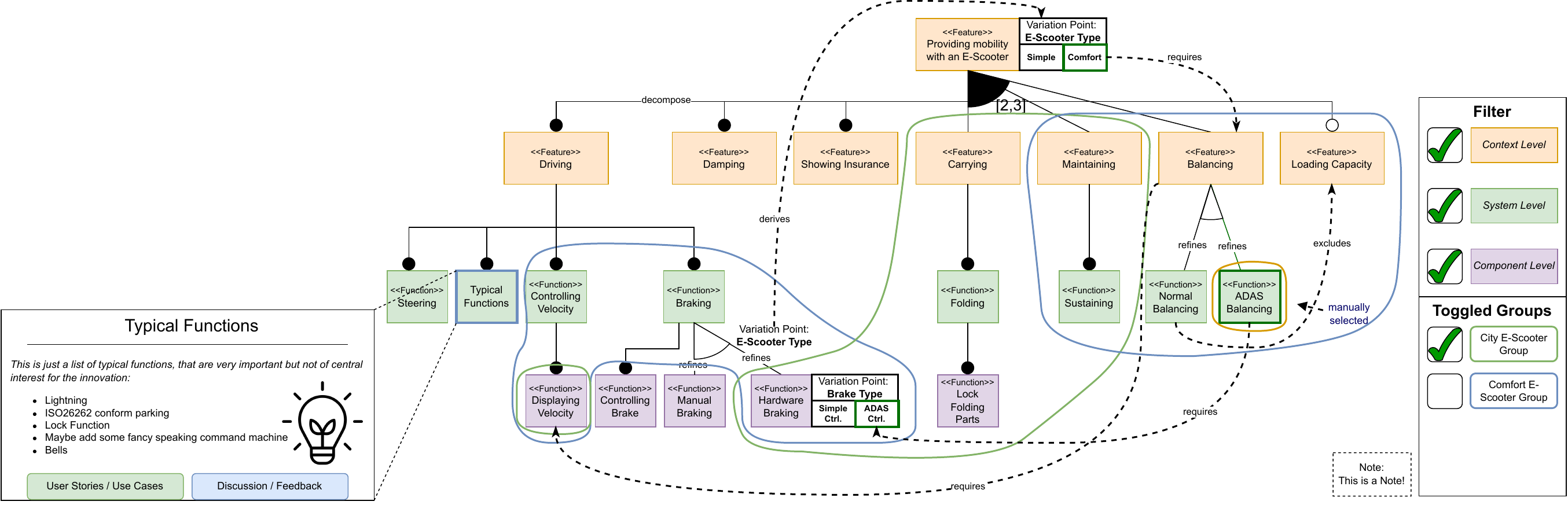}
	\caption{The identified features and functions for the innovation of \enquote{Providing mobility with an e-scooter} including the system level and component level.}
	\label{fig:fp:example2}
\end{sidewaysfigure}

Figure \ref{fig:fp:example2} now also includes the system level and component level (shown in green blocks and purple blocks respectively).
This figure shows also constraint relations, grouping and variation point relations.

\section{Functional Perspective: Strengths and Limitations}
\label{sec:fp:eval}

The Functional Perspective targets the modeling of the problem space basing on the well known feature trees.
This tailoring on features and functions from the problem space alone makes it easy to apply to any innovation.
The well known basis on feature trees makes the Functional Perspective straightforward to use for any experience modeler.
The basis on feature trees also enables to use the tooling and analyses from feature trees for the Functional Perspective.
The extensions provided in the Functional Perspective are for the purpose of adding further information, for filtering and for a better usability and are of cosmetic nature only.
The distinction between \textit{Features} and \textit{Functions} is well used in the automotive domain and thus supports well the domain of this project.
While these extensions do in fact represent a small learning overhead, the overhead is reasonable.
These extensions can be translated directly into basic feature trees.
That noted, a feature model expert can model the Functional Perspective without knowing about the extensions.
Feature trees are also great at modeling a mix of \textit{Decomposition} and \textit{Refinement} in one model.
Additionally, the Functional Perspective can be  constrained via Requirements from the Quality Perspective, leading to a very sophisticated model.

As limitation, the Functional Perspective is not designed to capture early structure and interfaces between features.
However, this is also good, because it stops the modeler from thinking too much about solutions and their interfaces.
It is also well known that feature models can grow very large due to the vast amount of variants, however in the context of abstract innovation modeling, these models should be manageable.
Also worth discussing is the purpose of feature models as a basis for the Functional Perspective.
Feature models are made for modeling variability and often used in product lifecycle management.
Innovation modeling on the other side is abstract and has not too much variability to model.
While this clash of purposes may be debatable, the IMoG applications showed that the Functional Perspective is good to use.

\FloatBarrier

\section{Functional Perspective FAQ}
\label{sec:fp:faq}

The FAQ splits up into the following categories:
\begin{itemize}
	\item Questions and Answers about the general Feature Trees
	\item Questions and Answers the Tooling
	\item Questions and Answers about the concepts an dependencies
	\item General Questions and Answers about the Functional Perspective
\end{itemize}

\subsection{Feature Tree Base}
\label{sec:fp:faq:ftreebase}

\includegraphics[width=\linewidth]{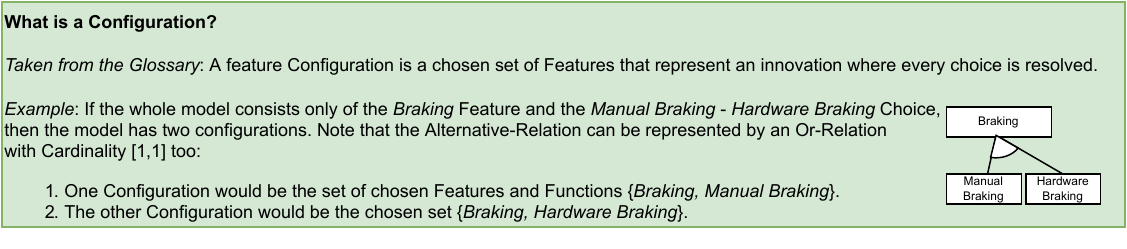}
\includegraphics[width=\linewidth]{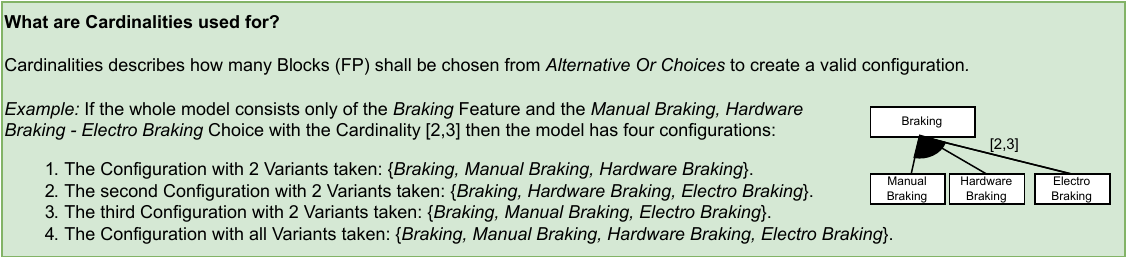}
\includegraphics[width=\linewidth]{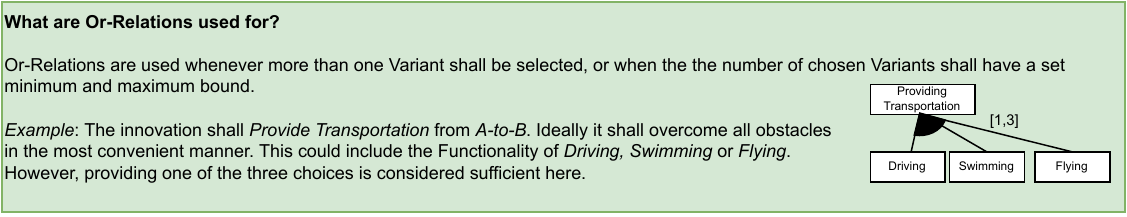}
\includegraphics[width=\linewidth]{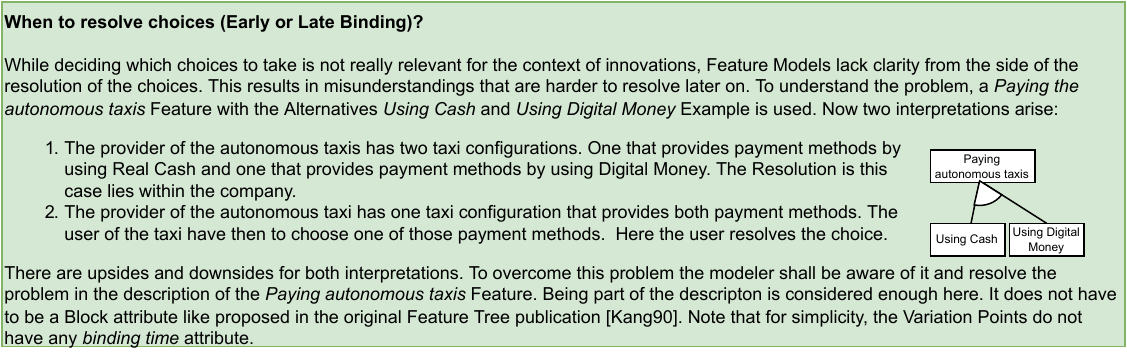}

\subsection{Tooling}
\label{sec:fp:faq:tooling}

\includegraphics[width=\linewidth]{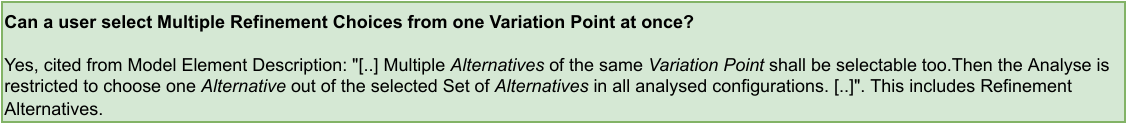}

\subsection{Concepts and Dependencies}
\label{sec:fp:faq:candd}

\fbox{
	\begin{minipage}{0.955 \textwidth}
		\small General FAQ
	\end{minipage}
}

\noindent
\includegraphics[width=\linewidth]{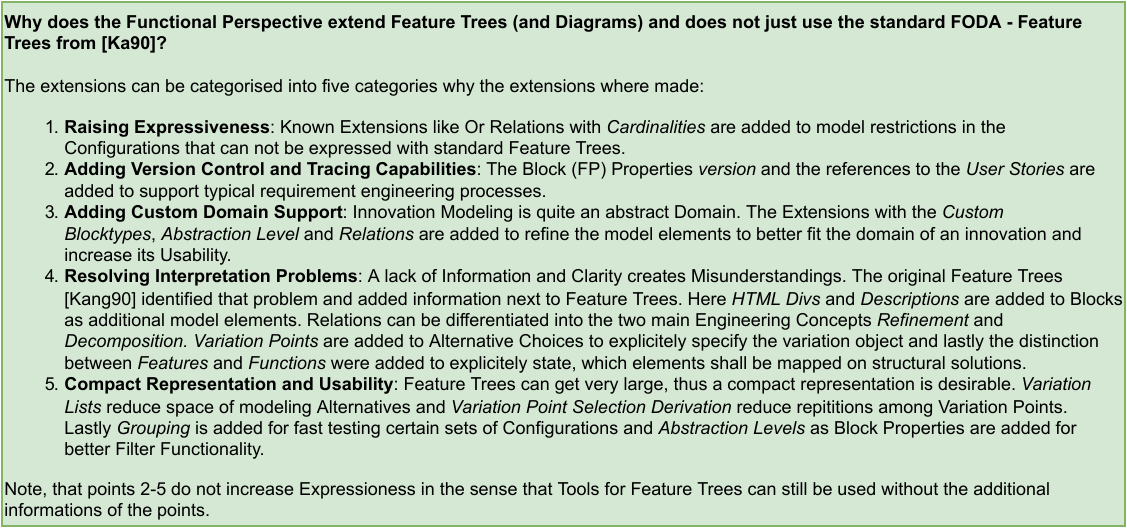}
\includegraphics[width=\linewidth]{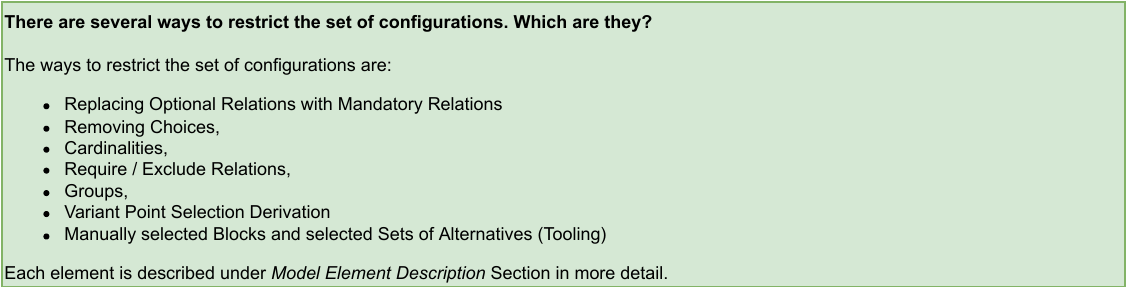}
\includegraphics[width=\linewidth]{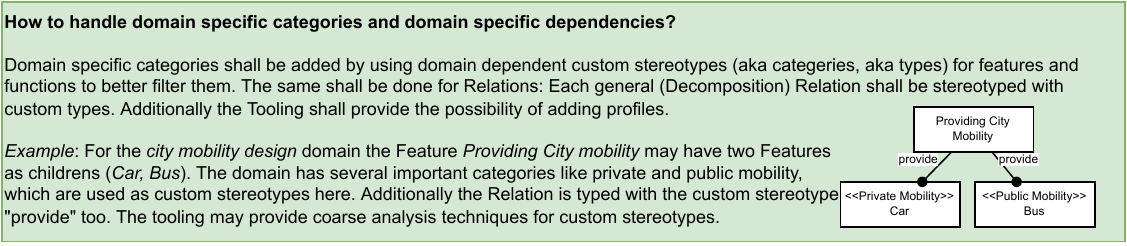}
\fbox{
	\begin{minipage}{0.955 \textwidth}
		\small Features and Functions
	\end{minipage}
}

\noindent
\includegraphics[width=\linewidth]{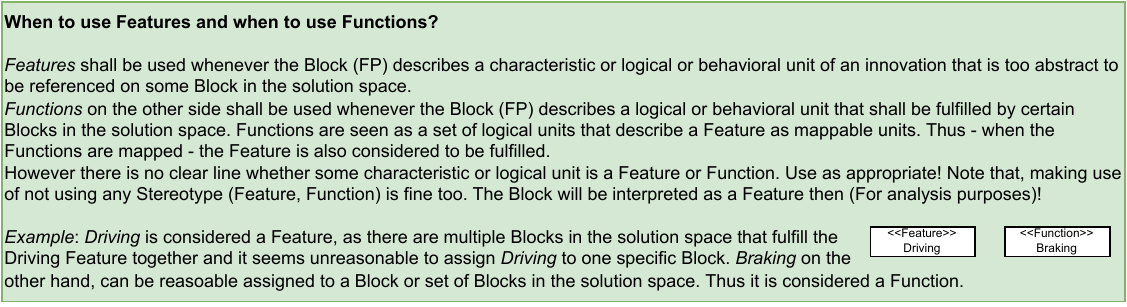}
\includegraphics[width=\linewidth]{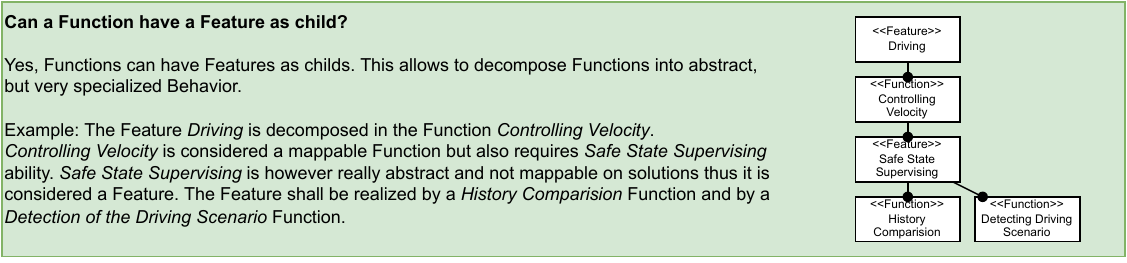}
\includegraphics[width=\linewidth]{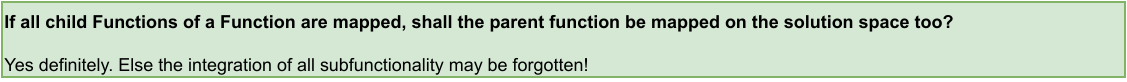}
\includegraphics[width=\linewidth]{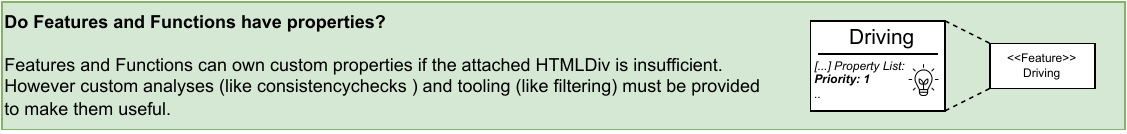}
\includegraphics[width=\linewidth]{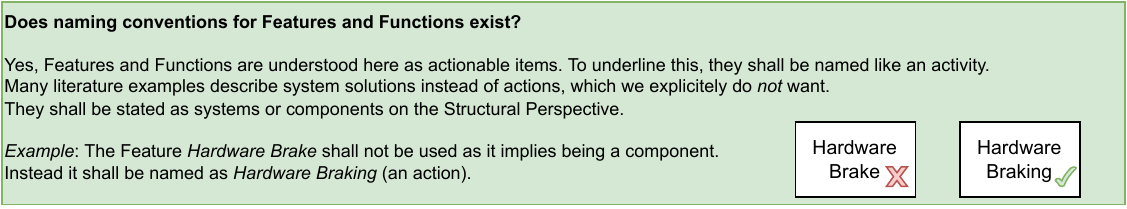}
\fbox{
	\begin{minipage}{0.955 \textwidth}
		\small Variation Points - Variant List Representation - Variation Point Selection Derivation - OVM
	\end{minipage}
}

\noindent
\includegraphics[width=\linewidth]{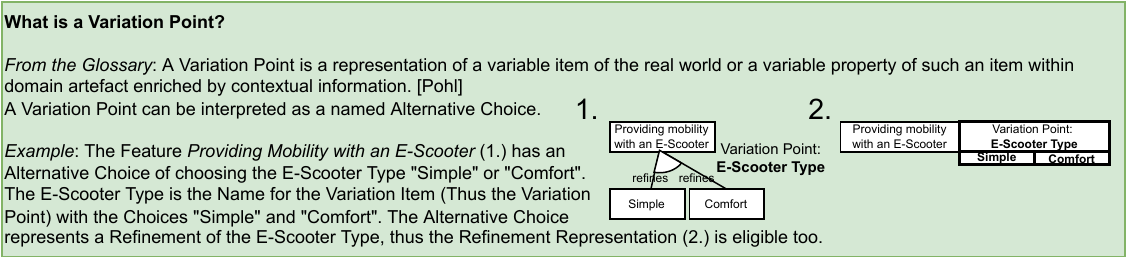}
\includegraphics[width=\linewidth]{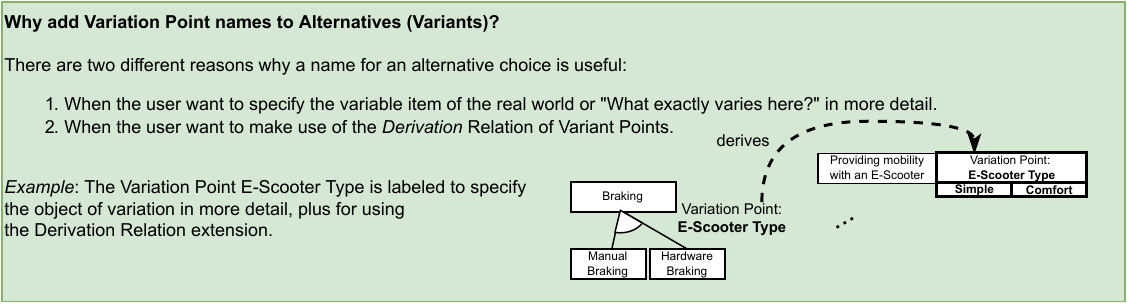}
\includegraphics[width=\linewidth]{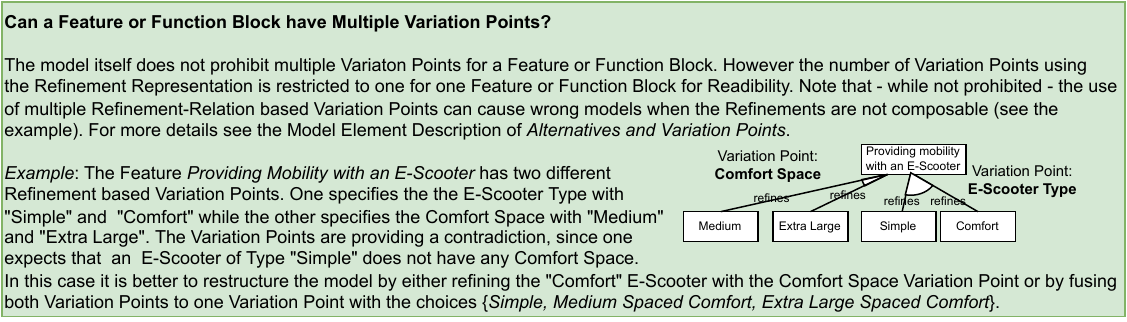}
\includegraphics[width=\linewidth]{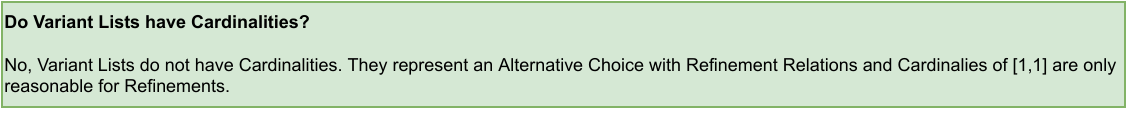}
\includegraphics[width=\linewidth]{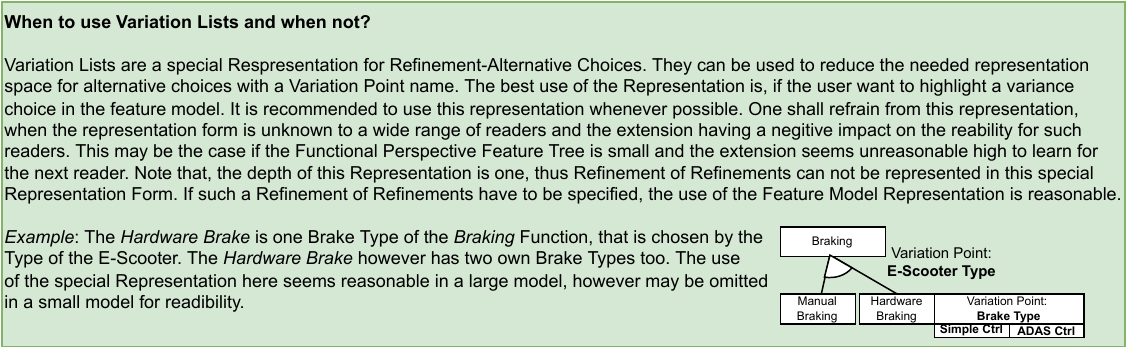}
\includegraphics[width=\linewidth]{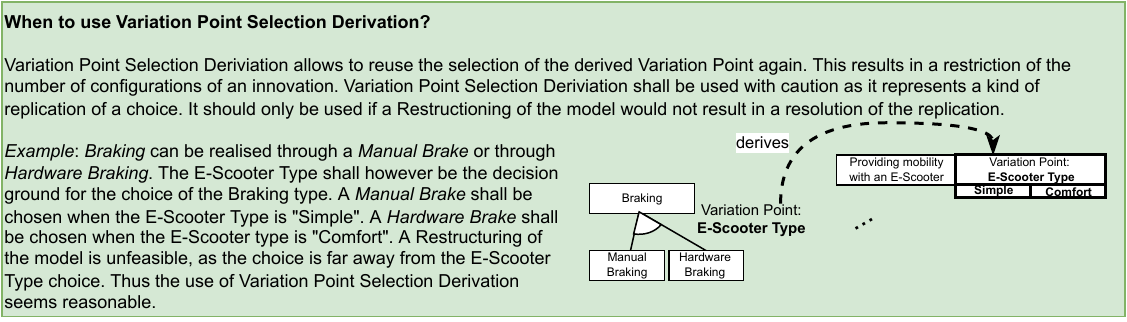}
\includegraphics[width=\linewidth]{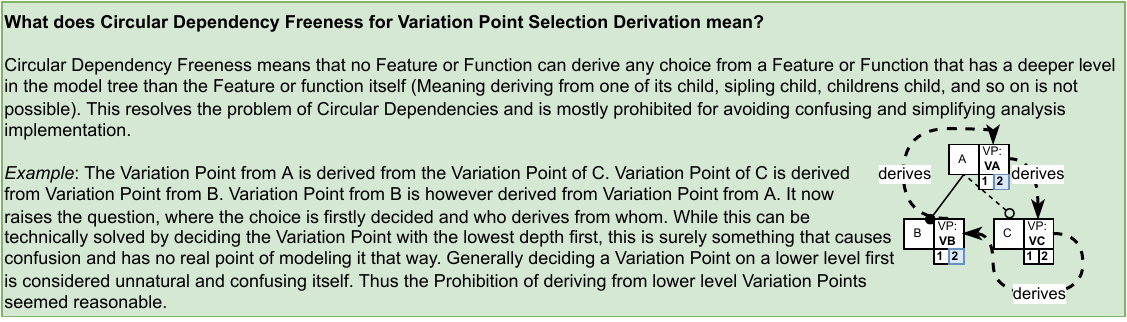}
\includegraphics[width=\linewidth]{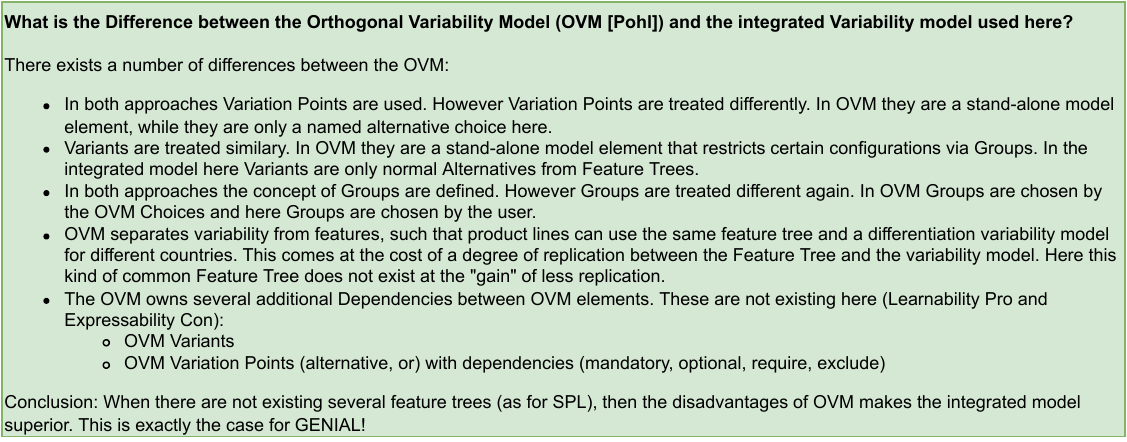}
\fbox{
	\begin{minipage}{0.955 \textwidth}
		\small Relations
	\end{minipage}
}

\noindent
\includegraphics[width=\linewidth]{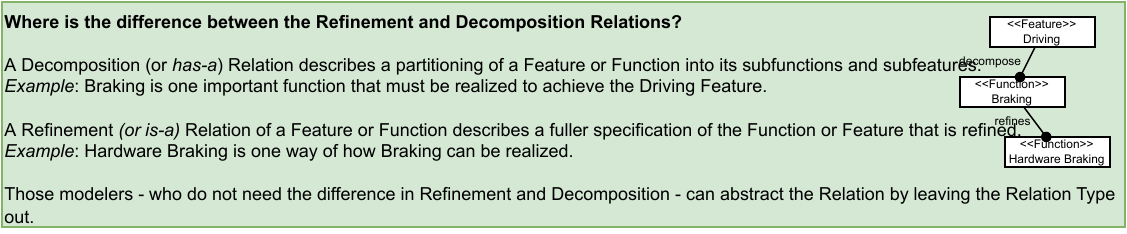}
\includegraphics[width=\linewidth]{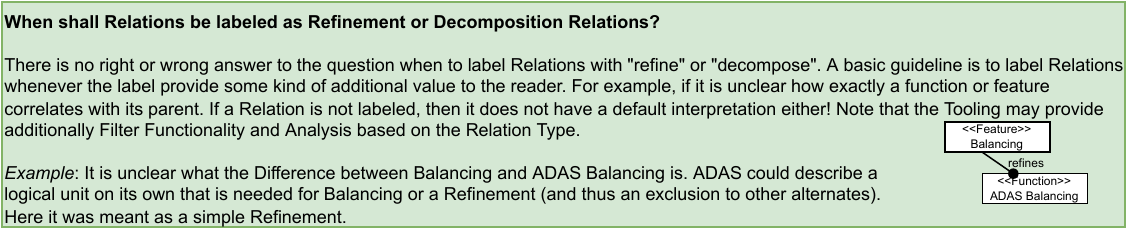}
\includegraphics[width=\linewidth]{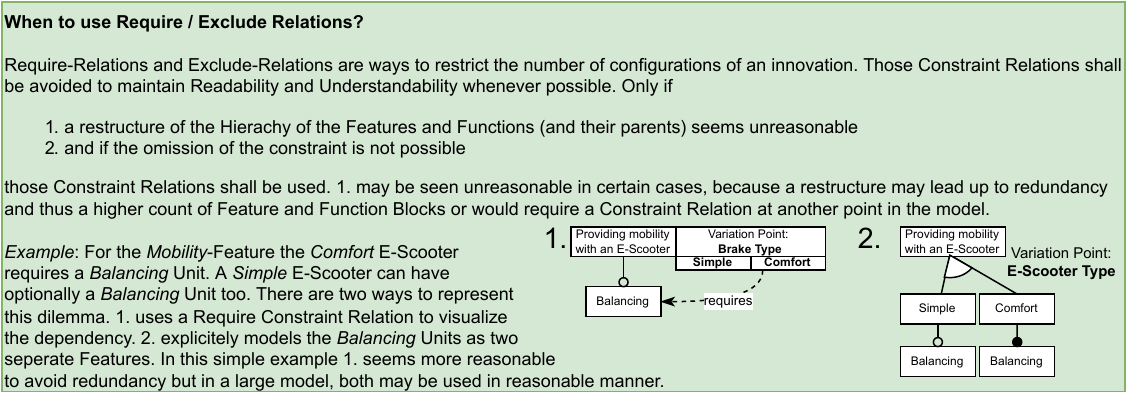}
\fbox{
	\begin{minipage}{0.955 \textwidth}
		\small Miscellaneous
	\end{minipage}
}

\noindent
\includegraphics[width=\linewidth]{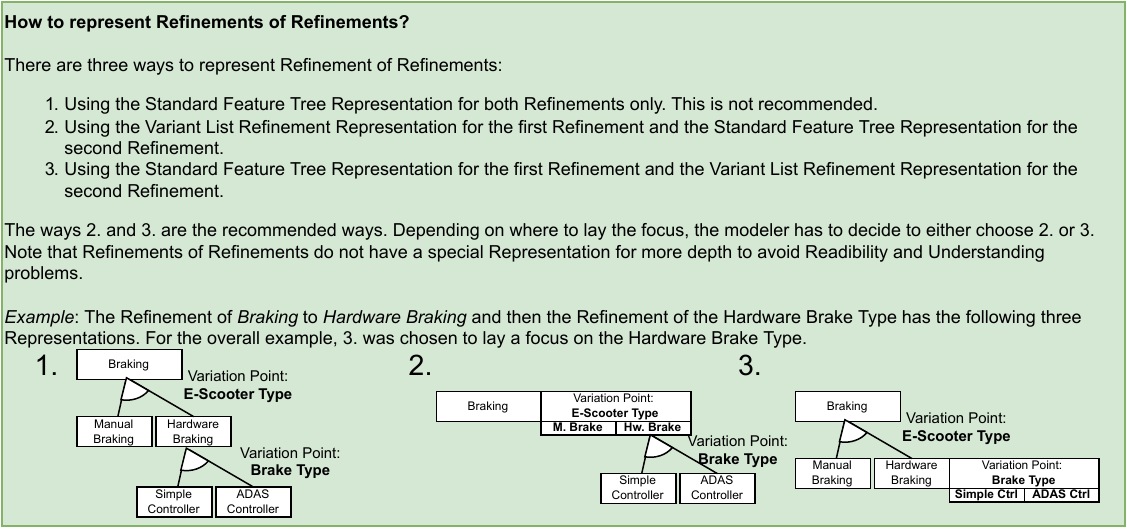}
\includegraphics[width=\linewidth]{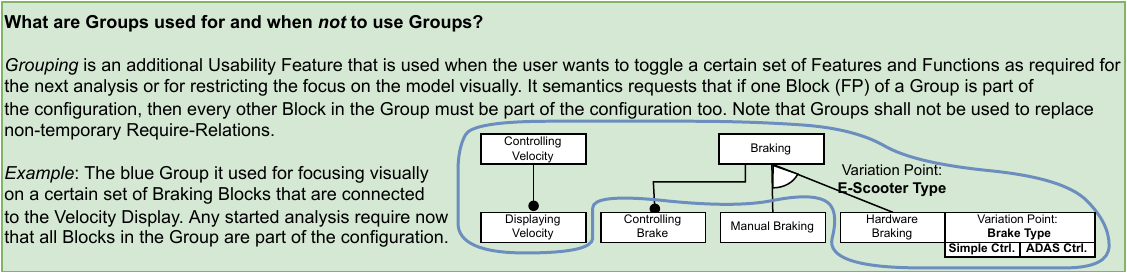}
\includegraphics[width=\linewidth]{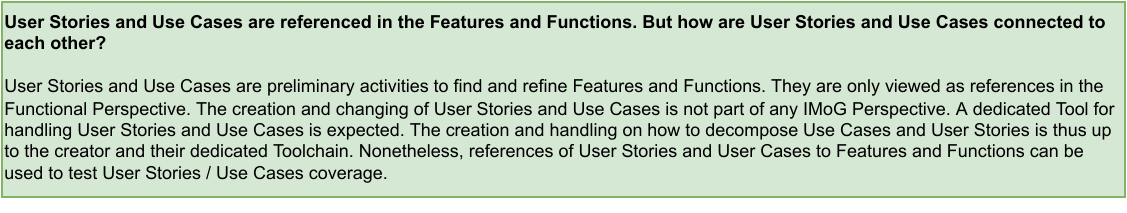}

\subsection{General Stuff}
\label{sec:fp:faq:stuff}

\includegraphics[width=\linewidth]{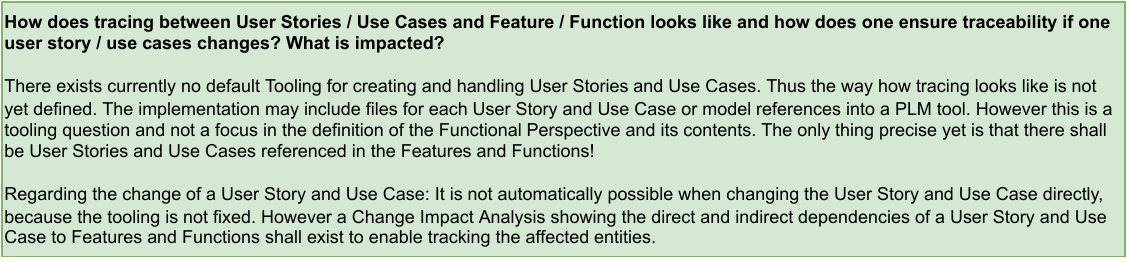}
\includegraphics[width=\linewidth]{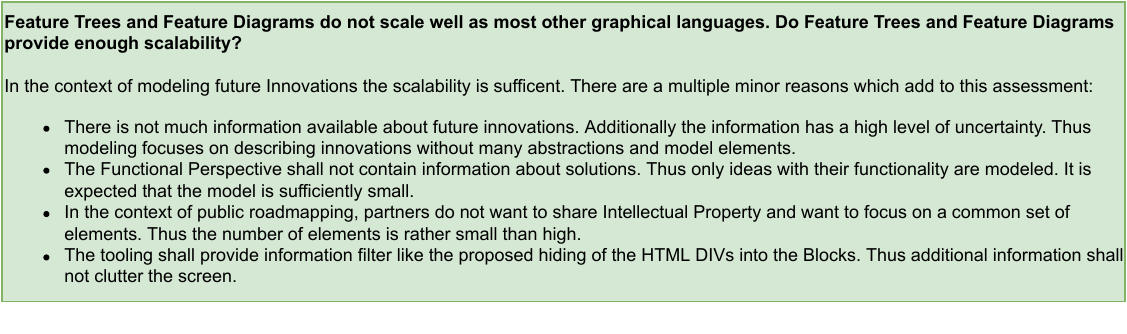}

%% file: content/quality_perspective.tex
\chapter{Quality Perspective}
\label{chap:qp}

\begin{figure}[b!]
	\centering
	\begin{tikzpicture}
		\newcommand\scf{0.9} 
		\node[anchor=south west] {\includegraphics[width=\scf\linewidth]{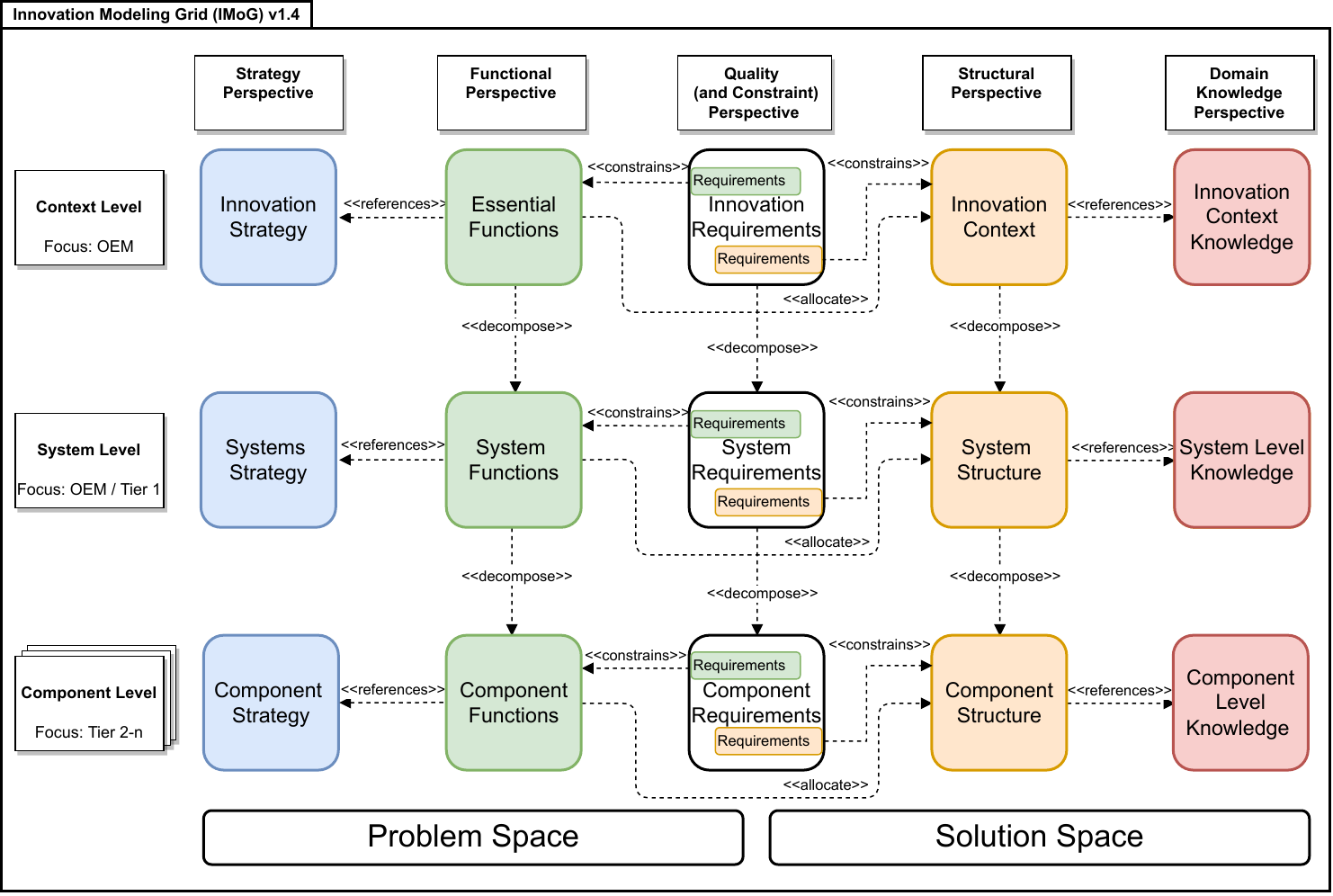}};
		\path[fill=white,opacity=0.5] (\scf*2.2,\scf*1.2) rectangle (\scf*4,\scf*9.5);
		\path[fill=white,opacity=0.5] (\scf*4.9,\scf*1.2) rectangle (\scf*6.7,\scf*9.5);
		\draw[ultra thick, blue, rounded corners] (\scf*7.55,\scf*1.2) rectangle (\scf*9.4,\scf*9.5);
		\path[fill=white,opacity=0.5] (\scf*10.25,\scf*1.2) rectangle (\scf*12.05,\scf*9.5);
		\path[fill=white,opacity=0.5] (\scf*12.9,\scf*1.2) rectangle (\scf*14.7,\scf*9.5);
	\end{tikzpicture}
	\caption{Location of the Quality Perspective in IMoG}
	\label{fig:qp:imog}
\end{figure}

The Quality Perspective is the third perspective in IMoG applications and targets the coverage of mostly non functional requirements for both the problem space and the solution space (see Figure \ref{fig:qp:imog}).
The Quality Perspective contains for example constraints from the legislation, robustness requirements and performance requirements.
Functional requirements should not exist in the Quality Perspective (hence its name) because the Functional Perspective already covers the features and functions of the innovation.
Exceptions that cannot be handled on the Functional Perspective are fine.
The Quality Perspective takes requirements from both sides: The problem space and solution space and represent an interface between the spaces.

The Quality Perspective contains two representations:
The table format for representing the requirements data.
The dependency view for representing the inter Perspectives relations as well as the decomposition of the requirements (parent-child relation)
With the Quality Perspective based only on generic tables, some recommendations for the data fields of each requirements were made.
Special extensions do not exist.


The chapter is structured as followed:
In Section \ref{sec:qp:me} the meta model and its model elements are presented.
In Section \ref{sec:qp:e-scooter} an example of the Quality Perspective is given.
The strengths and limitations of the Quality Perspective are discussed in Section \ref{sec:qp:eval}.
A FAQ finalizes the description in Section \ref{sec:qp:faq}.

\section{Model elements}
\label{sec:qp:me}
\FloatBarrier

\begin{sidewaysfigure}[h]
	\centering
	\includegraphics[width=\linewidth]{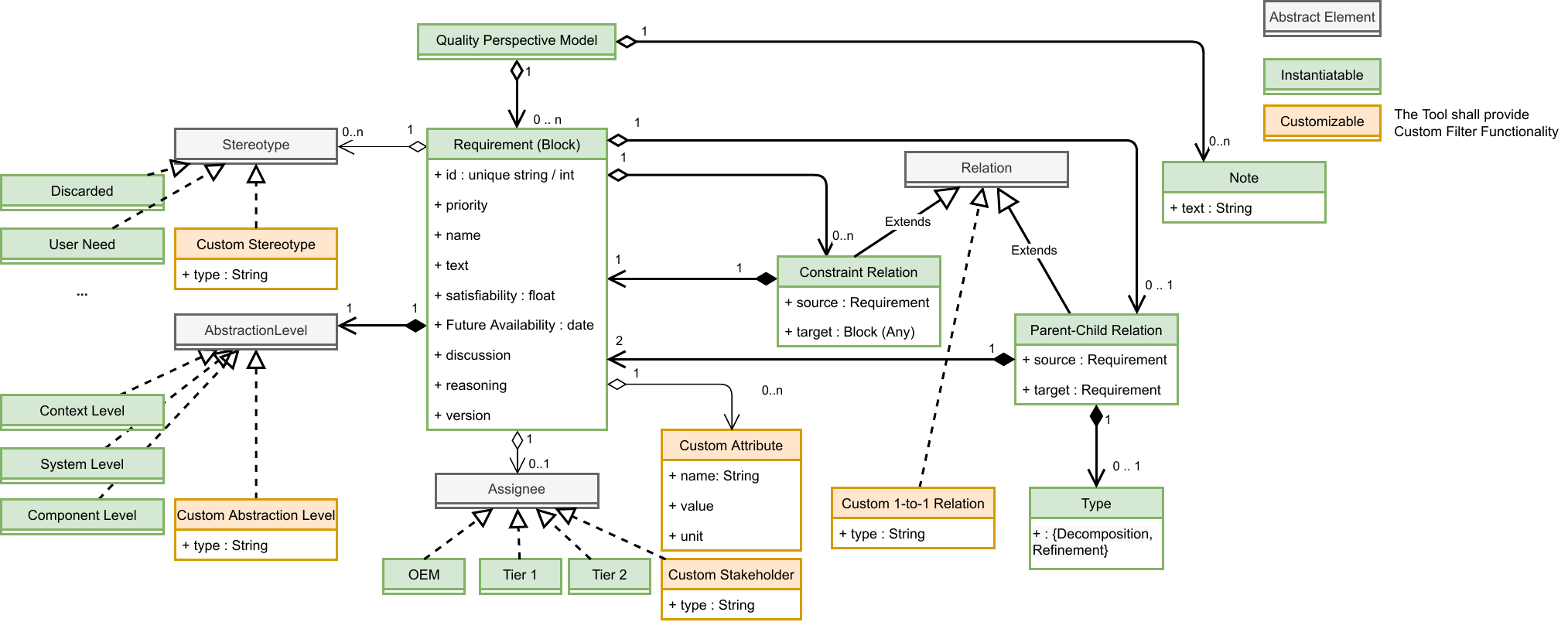}
	\caption{The model elements of the Quality Perspective.}
	\label{fig:qp:me}
\end{sidewaysfigure}

The meta model of the Quality Perspective (see Figure \ref{fig:qp:me}) consists only of two important model element: the \textit{Requirement} and the \textit{Requirement Relation}.
The \textit{Requirement} element has quite a few attributes that are recommended to fill.
This section will emphasize on presenting the attributes.

The \textit{Quality Perspective Model} contains as its main members the \textit{Requirement} element.
The \textit{Requirement} has a bunch of attributes.
Some of those are modeled as an own Block to allow the restriction of the data types in form of Enums (Variable types with defined possible values).
The Stereotype, the abstraction level and the assignee are part of them.
To keep them extensible, each of them has a customization interface.
Additionally, custom attributes can be defined.
Each requirement can have relations.
There are currently three types of relations:
\begin{itemize}
	\item the \textit{Parent-Child} relation to represent a requirement \textit{Decomposition} or requirement \textit{Refinement}.
	\item the \textit{Constraint} relation for describing which (problem space / solution space) Block shall satisfy the requirement and
	\item the \textit{Custom} requirement for extensibility.
\end{itemize}
Each attribute and relation is described in more detail in the following.

\FloatBarrier

\fbox{
	\begin{minipage}{0.955 \textwidth}
		\large \textbf{Model Elements:}
		\normalsize \setstretch{0.8}
		\begin{itemize}
			\item Quality Perspective Model
			\item Requirement
			\item Relations
			\item Defined Stereotypes
			\item Notes
		\end{itemize}
	\end{minipage}
}

\fbox{
	\begin{minipage}{0.955 \textwidth}
		\Large Model Elements
	\end{minipage}
}

In the following the model elements are introduced in four parts.
First the \textit{Quality Perspective Model} is introduced, then the \textit{Requirement} with all its attributes and associated Enums is described.
Afterwards, the \textit{Relations} are described.
Lastly, the defined \textit{Stereotypes} and \textit{Notes} are presented.

\rule{\textwidth}{1pt}
Meta Model Element:
\begin{center}
	\includegraphics[width=0.3\linewidth]{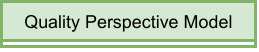}
\end{center}

Description:

\fcolorbox{gray!30!black}{gray!20!white}{
	\begin{minipage}{0.955 \textwidth}
		\large \textbf{Quality Perspective Model}\\
		\normalsize The \textit{Quality Perspective Model} is the diagram of the Quality Perspective of an innovation.
		It contains all model elements of the Quality Perspective.
	\end{minipage}
}

Example: A full Quality Perspective Model example is shown in Section \ref{sec:qp:e-scooter}.

\rule{\textwidth}{1pt}
Meta Model Element:
\begin{center}
	\includegraphics[width=\linewidth]{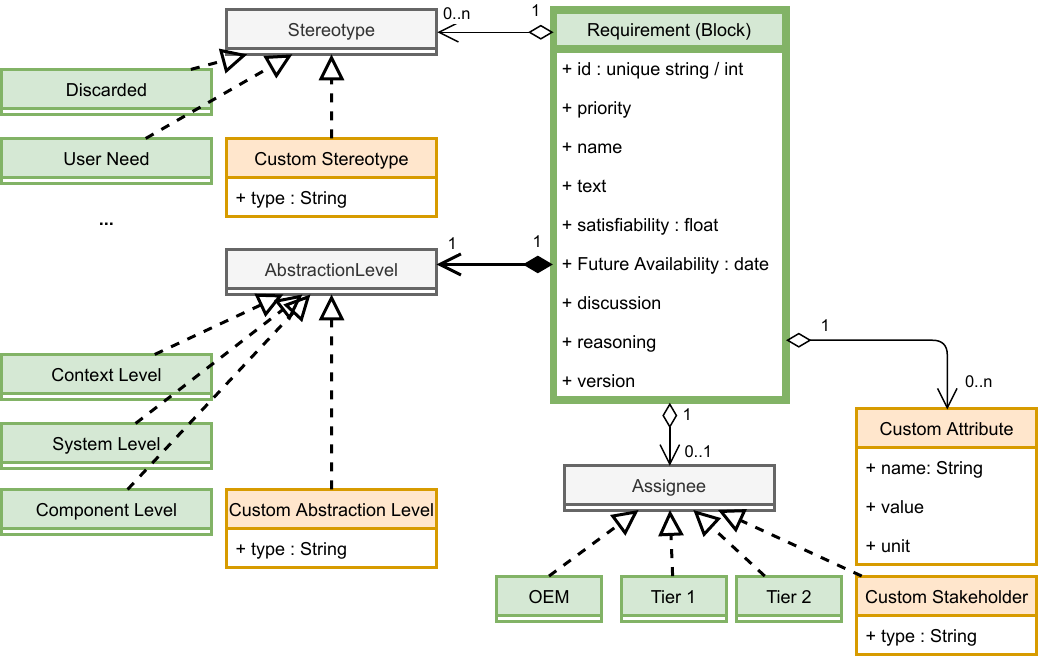}
\end{center}

Description:

\fcolorbox{gray!30!black}{gray!20!white}{
	\begin{minipage}{0.955 \textwidth}
		\large \textbf{Requirement (Block)} \\
		\normalsize The \textit{Requirement (Block)} is the main element of the Quality Perspective and represents any Quality Requirement, user need, constraint and so on.
		It is a quite complex element due to its importance.
		It defines many attributes.
		Each attribute can be imagined as a column in a requirements table.
		Except the outsourced description of the relations, each attribute is described in the following:
		\begin{itemize}
			\item The \textit{id} for a requirement to identify and trace it.
			It typically takes a unique number of a unique string as an scheme.
			Both are possible here.
			\item The optional \textit{priority} for giving the requirement an importance.
			The value interpretation has to be defined by the stakeholder.
			Default is an empty field.
			\item The \textit{name} of the requirement to better identify it (for humans).
			\item The \textit{text} of the requirement.
			It can be described as natural language or as formal sentences.
			If it is formulated as formal sentences then the requirement shall have a proper \textit{Stereotype} to allow analysis to identify and parse it.
		\end{itemize}
\end{minipage}
}

\fcolorbox{gray!30!black}{gray!20!white}{
	\begin{minipage}{0.955 \textwidth}
		\large \textbf{Requirement (Block) continued} \\
		\begin{itemize}
			\item The \textit{satisfiability} of the requirement.
			A numerical number between 0 and 1 which estimates the chance the requirement is fulfilled according to the year and other parameters.
			The satisfiability may be given by a formula instead of a static value.
			\item The \textit{Future Availability} describes the year date when the requirement will be relevant.
			In the days / years before the given date this Requirement shall be 'ignored'.
			\item The \textit{discussion} field provides a platform for discussing the requirements within the value chain.
			\item The \textit{reasoning} field allows to give a rationale for the requirement definition.
			\item The \textit{version} field is for proper version management and to identify updated requirements.
		\end{itemize}
		The requirement has additionally attributes with predefined value ranges (so called Enums).
		These are represented by an own entity in the meta model:
		\begin{itemize}
			\item The \textit{Abstraction Level} of the requirement defines the level of abstraction the requirement represents.
			It can be either \textit{Context Level, System Level, Component Level} or from the type \textit{Custom Abstraction Level}.
			\item Optional \textit{Stereotypes} can refine the category of the requirement further.
			There are some Stereotypes predefined, like \textit{Discarded} (Requirement), \textit{User Need} and so on.
			However one can define their own \textit{Custom} Stereotypes.
			A list of predefined Stereotype is presented in the FAQ.
			\item The optional \textit{Assignee} of the requirement represent the responsibility owner of the requirement.
			It can be either of the predefined type \textit{OEM}, \textit{Tier 1} or \textit{Tier 2} or set by a string to a \textit{Custom} Stakeholder.
			\item If all the above attributes are not enough, or there is an attribute missing for the specified domain, then one can define a \textit{Custom} Attribute with a \textit{name}, a \textit{value} and an \textit{unit}.
		\end{itemize}
	\end{minipage}
}

Example: An example with 4 requirements is shown below.
The requirements attributes are presented shortly:
\begin{itemize}
	\item The \textit{ids} are represented as numbers for this example.
	\item The \textit{priority} for three requirements were determined.
	The value 1 describes here the most important priority.
	\item The first safety requirement is expected to be in all cases fulfilled (thus \textit{satisfiability} is 1=100\%) to be conform to the German traffic rules.
	The other requirements should be fulfilled up to a certain degree according to the expectations.
	\item Each requirement has a \textit{name} and a \textit{text}.
	All of them are described as natural text with some being unfinished.
	\item The \textit{Future Availability} is set to Now. Meaning the requirements shall be already considered.
	\item The \textit{parent} requirement and the \textit{targets} represent relations, they are not further described here.
	\item Each requirement has \textit{Stereotypes}. The first requirement is considered to be part of the safety concept while the three other requirements are considered user needs.
	\item All four requirements are considered to be part of the \textit{Context Level} abstraction level.
	\item The \textit{assignee} of these requirements is either the \textit{OEM} or the \textit{Tier 1}.
	\item The \textit{reasoning} and the \textit{discussion} are not further detailed here.
	\item The \textit{version} number of each requirement.
\end{itemize}

\begin{center}
	\includegraphics[width=\linewidth]{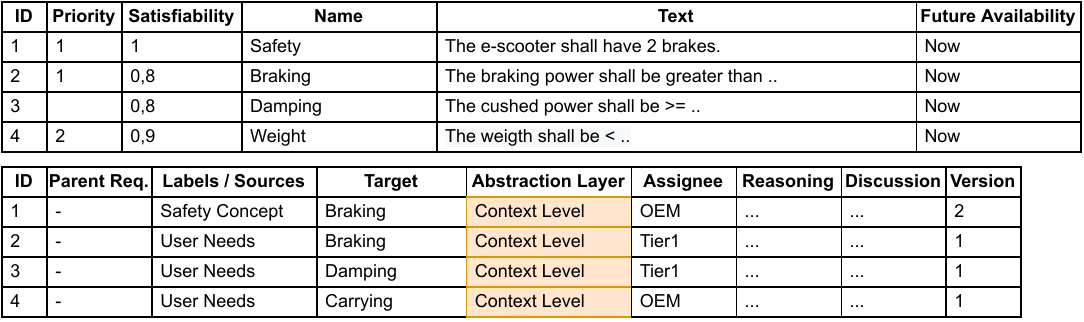}
\end{center}

\rule{\textwidth}{1pt}
Meta Model Element:
\begin{center}
	\includegraphics{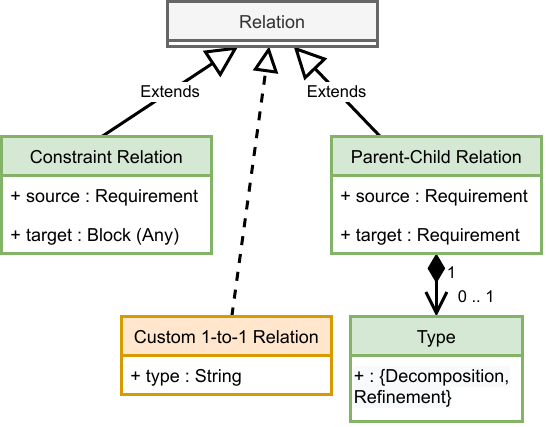}
\end{center}

Description:

\fcolorbox{gray!30!black}{gray!20!white}{
	\begin{minipage}{0.955 \textwidth}
		\large \textbf{Relation (QP)} \small \textbf{(Read: Relation on the Quality Perspective)} \\
		\normalsize The abstract \textit{Relation (QP)} describes Relations between requirements or between a requirement and a target Block on the Functional Perspective or on the Structural Perspective.
		Relations are further categorized into
		\begin{itemize}
			\item \textit{Constraint} relations to describe which Block on the Functional Perspective or the Structural Perspective shall fulfill the specified requirement.
			This constraint relation is often called <<satisfy>> relation when reversed.
			However, to avoid confusion, only <<constraint>> relations shall be used.
			\item \textit{Parent-Child} relations between requirements.
			The \textit{Parent-Child} relations can be specified by an optional type, which can be either of value \textit{Decomposition} or \textit{Refinement} to specify the relation type.
			\item \textit{Custom 1-to-1} relations between requirements can be described by using \textit{Custom} attributes of the requirement.
		\end{itemize}
	\end{minipage}
}

Example: An example for the \textit{Constraint} Relation and the \textit{Parent-Child} relation is shown below.
The \textit{Custom 1-to-1} relations are skipped for now.
\begin{center}
	\includegraphics[width=\linewidth]{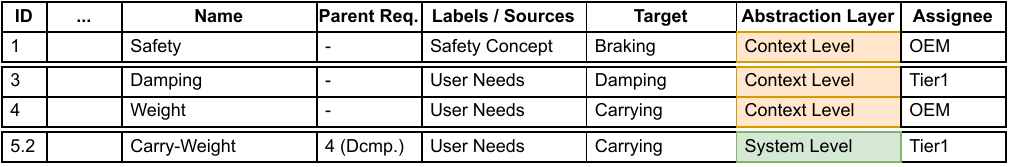}
\end{center}
The four requirements above have two important attributes that represent relations:
\begin{itemize}
	\item The \textit{Parent} requirement represent a \textit{Decomposition} or \textit{Refinement} relation.
	Only the \enquote{Carry-Weight} requirement has a parent: The weight requirement.
	The exact type of the relation is in this example not described.
	\item Each requirement has a \textit{target}.
	The targets describe in all four cases a function of the Functional Perspective (which can be found in the associated file).
\end{itemize}

The same requirements represented in the Relations View would be depicted as followed:
\begin{center}
	\includegraphics{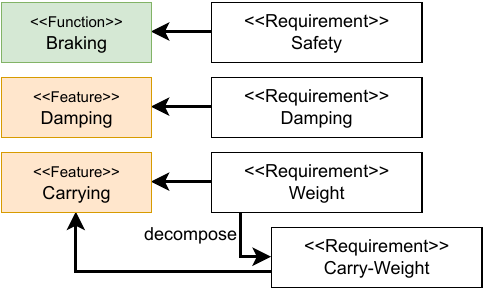}
\end{center}

\rule{\textwidth}{1pt}
Meta Model Element:
\begin{center}
	\includegraphics{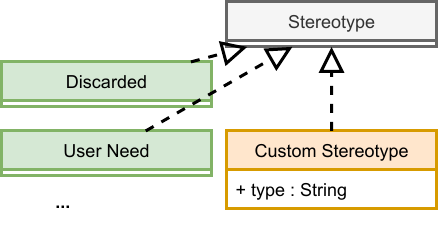}
\end{center}

Description:

\fcolorbox{gray!30!black}{gray!20!white}{
	\begin{minipage}{0.955 \textwidth}
		\large \textbf{Stereotypes} \\
		\normalsize The \textit{Stereotype} can refine the category of the requirement further.
		A requirement can have multiple stereotypes.
		However, it is recommended to not apply two that contradict each other or are of the same category.
		The following Stereotypes are predefined:
		\underline{Requirement Categorization Stereotypes:}
		\begin{itemize}
			\item \textbf{Quality Requirement}
			\begin{itemize}
				\item \textbf{Performance Requirement}
				\item Technical Professional Guess
				\item User Need (non functional)
			\end{itemize}
			\item \textbf{Constraint}
			\begin{itemize}
				\item Safety Requirement
				\item Security Requirement
				\item Legal Constraint
				\item Technology Requirement
			\end{itemize}
		\end{itemize}
		\underline{Requirement Status:}
		\begin{itemize}
			\item Discarded
			\item Proposed
			\item Confirmed (The default interpretation if no requirement status is given)
		\end{itemize}
		There is some overlap in the definitions of the categories, for example between Quality Requirement and User Need.
		If one can not decide which category to choose, then take the one that feels as best fit.
		The categories are only used for Filtering Purposes, thus a miscategorization is not that harmful.
		The three bold categories are of special interest of some suppliers.
		Maybe these shall be given some special treatment?

		It is possible to define \textit{Custom} Stereotypes.
	\end{minipage}
}

Example: ToDo!

\rule{\textwidth}{1pt}
Meta Model Element:
\begin{center}
	\includegraphics{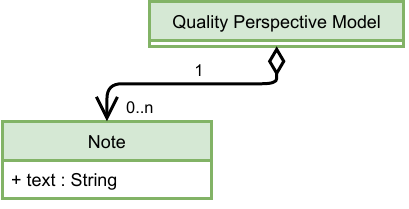}
\end{center}

Description:

\fcolorbox{gray!30!black}{gray!20!white}{
	\begin{minipage}{0.955 \textwidth}
		\large \textbf{Note} \\
		\normalsize The \textit{Note} can be used to add information to the model that can not or should not be modeled.
		Notes should be used sparsely!
	\end{minipage}
}

Example:
\begin{center}
	\includegraphics{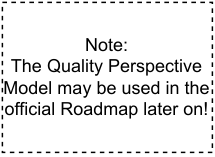}
\end{center}

\section{E-Scooter example}
\label{sec:qp:e-scooter}

The example of the Quality Perspective comprises the innovation requirements of \enquote{Providing mobility with an e-scooter} in two views called Requirements Table View and Relations View.

\begin{sidewaysfigure}[!h]
	\centering
	\includegraphics[width=\linewidth]{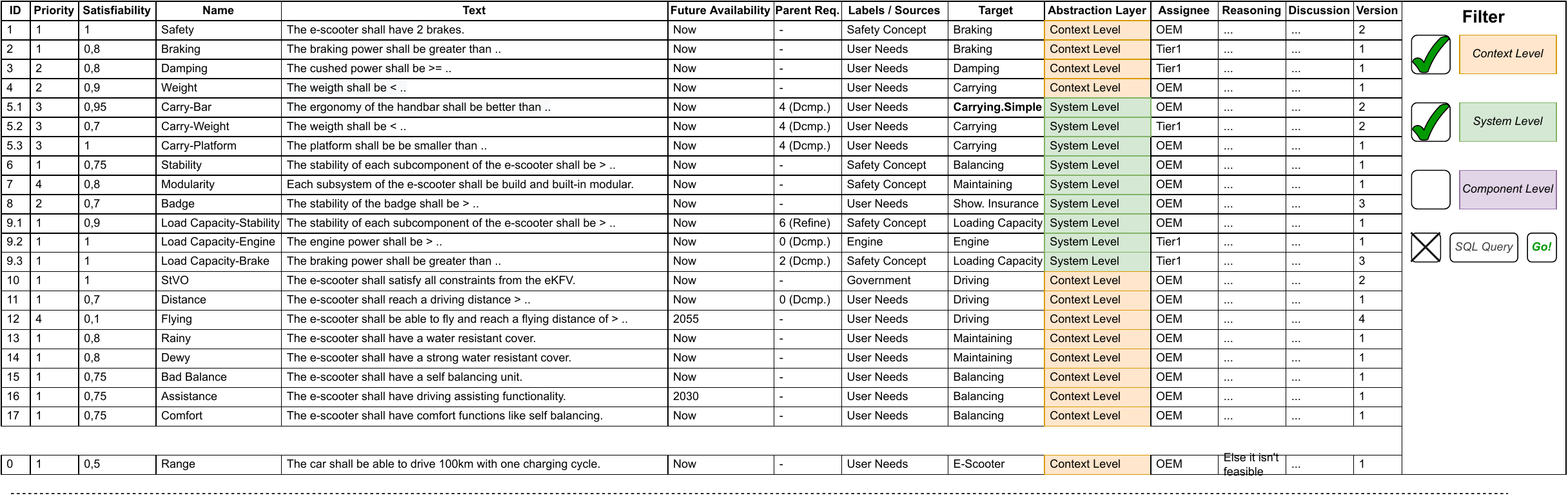}
	\caption{Requirements Table View: The requirements for the innovation of \enquote{Providing mobility with an e-scooter}.}
	\label{fig:qp:example1}
\end{sidewaysfigure}

The Requirements Table View (see Figure \ref{fig:qp:example1}) contains several identified quality requirements, user needs, constraints and so on for each abstraction level.
The Requirements Table View focuses on showing all details of each requirement.
The requirements of the component level are filtered out here (to keep the example small).
The requirements are listed as relational data bases thus SQL queries shall be executable.
The exact details of those requirements are not further explained.

\begin{figure}[!h]
	\centering
	\includegraphics[width=\linewidth]{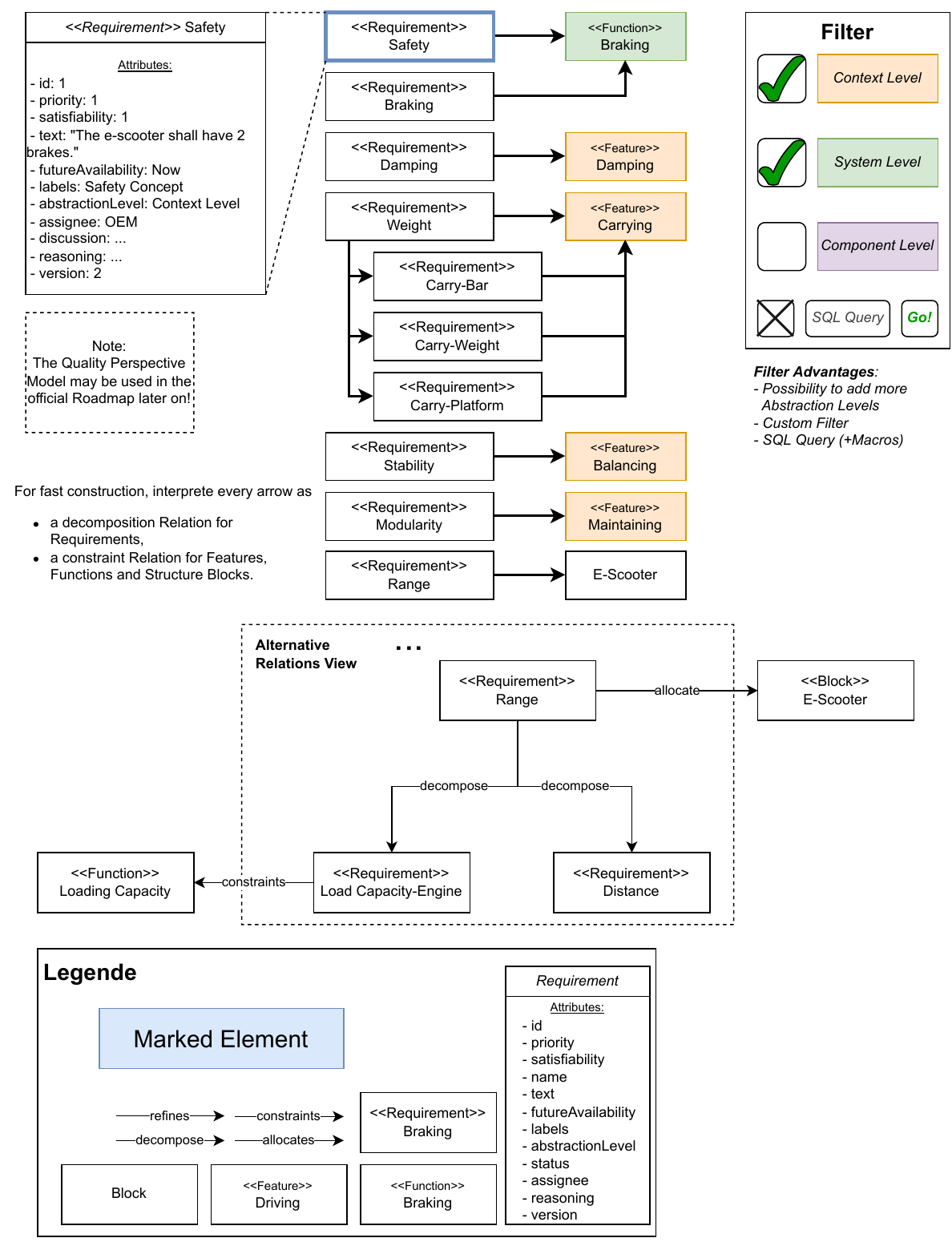}
	\caption{Relations View.}
	\label{fig:qp:example2}
\end{figure}

The Relations View (see Figure \ref{fig:qp:example2}) hides the attributes and shows the requirements only by their names.
The attributes shall be visible when marked with the mouse in an extra window.
This view focuses on presenting the Parent-Child relations between two requirements and the Constraint relations of the requirements to the Blocks of the other perspectives.

\section{Quality Perspective: Strengths and Limitations}
\label{sec:qp:eval}

The Quality Perspective contains simply requirements tables and relation views.
The Quality Perspective does not capture (many) functional requirements, because these requirements should be handled in the Functional Perspective.
Otherwise, there is nothing special about the Quality Perspective.
One noteworthy strength is the tracing of requirements to features, functions and solutions, because these links are already weaved into IMoG.

\FloatBarrier

\section{Quality Perspective FAQ}
\label{sec:qp:faq}

The FAQ splits up into one part:
\begin{itemize}
	\item Questions and answers about the general requirements on the Quality Perspective
\end{itemize}

\subsection{Requirements}
\label{sec:qp:faq:requirements}

\fbox{
	\begin{minipage}{0.955 \textwidth}
		\small General Requirements FAQ
	\end{minipage}
}

\noindent
\includegraphics[width=\linewidth]{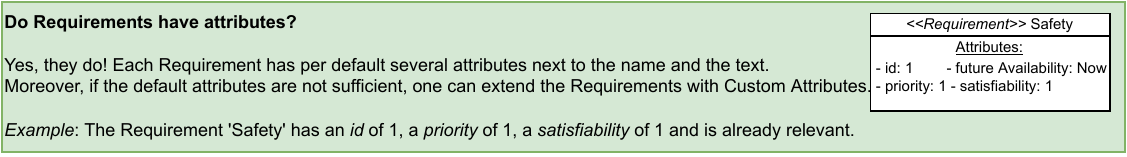}
\includegraphics[width=\linewidth]{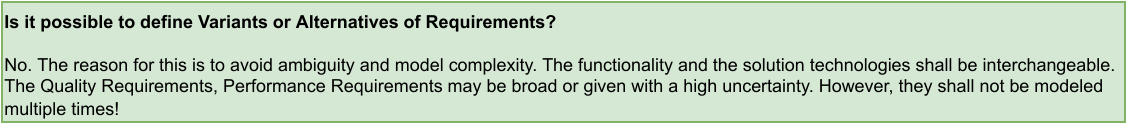}
\includegraphics[width=\linewidth]{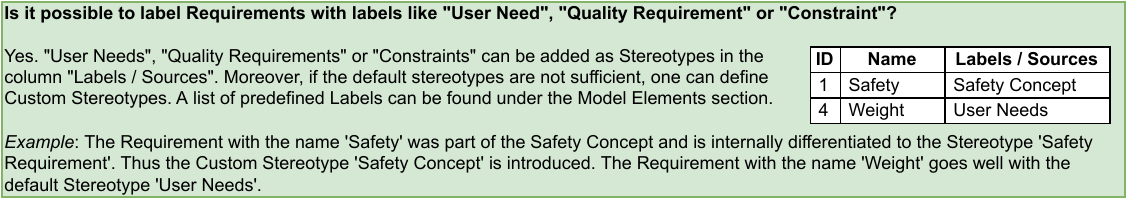}
\includegraphics[width=\linewidth]{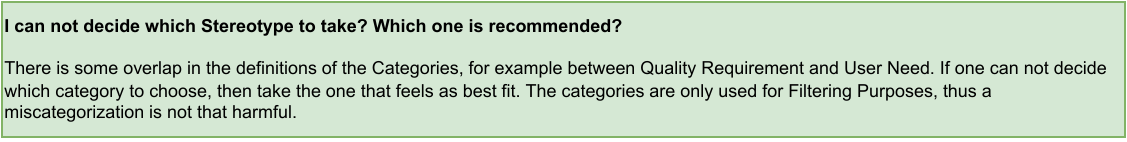}
\includegraphics[width=\linewidth]{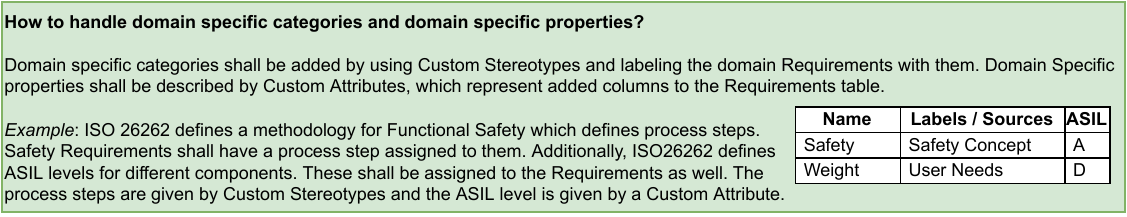}
\fbox{
	\begin{minipage}{0.955 \textwidth}
		\small Requirements Attributes
	\end{minipage}
}

\noindent
\includegraphics[width=\linewidth]{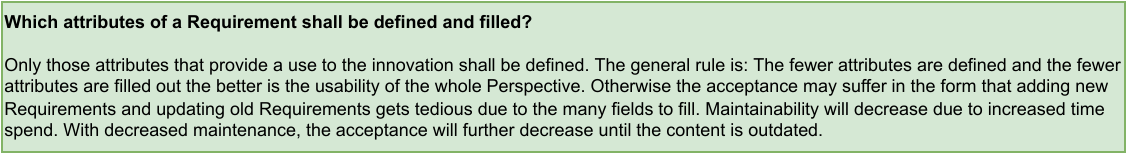}
\includegraphics[width=\linewidth]{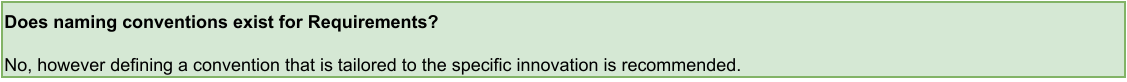}
\includegraphics[width=\linewidth]{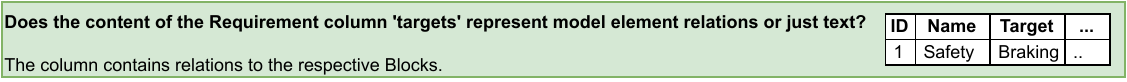}
\includegraphics[width=\linewidth]{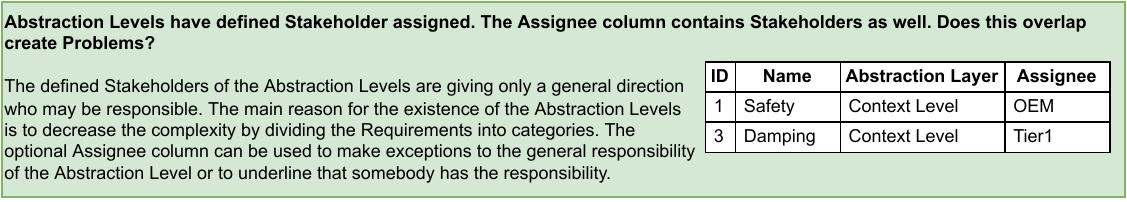}

%% file: content/structural_perspective.tex
\chapter{Structural Perspective}
\label{chap:strp}

\begin{figure}[b!]
	\centering
	\begin{tikzpicture}
		\newcommand\scf{0.9} 
		\node[anchor=south west] {\includegraphics[width=\scf\linewidth]{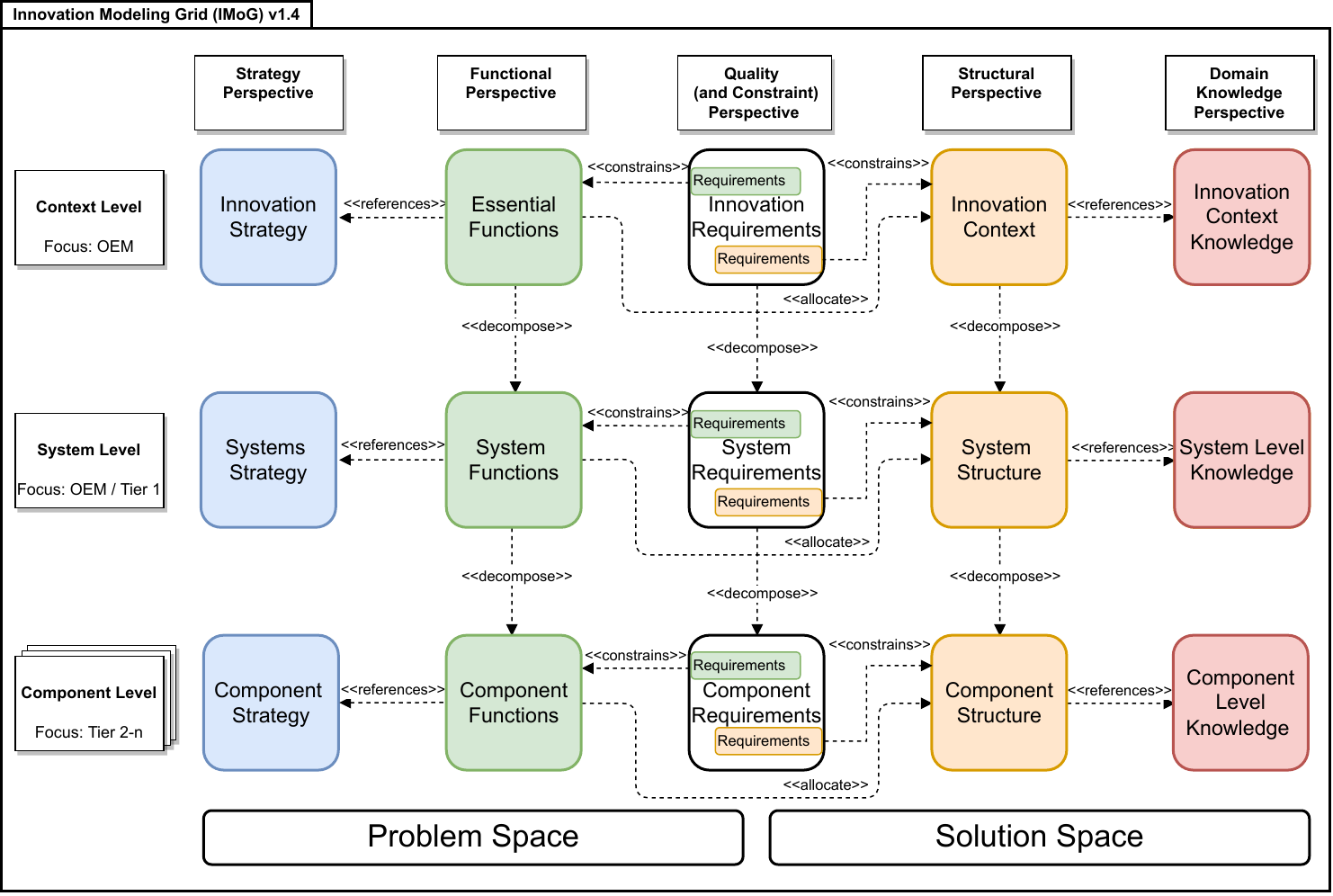}};
		\path[fill=white,opacity=0.5] (\scf*2.2,\scf*1.2) rectangle (\scf*4,\scf*9.5);
		\path[fill=white,opacity=0.5] (\scf*4.9,\scf*1.2) rectangle (\scf*6.7,\scf*9.5);
		\path[fill=white,opacity=0.5] (\scf*7.55,\scf*1.2) rectangle (\scf*9.4,\scf*9.5);
		\draw[ultra thick, blue, rounded corners] (\scf*10.25,\scf*1.2) rectangle (\scf*12.05,\scf*9.5);
		\path[fill=white,opacity=0.5] (\scf*12.9,\scf*1.2) rectangle (\scf*14.7,\scf*9.5);
	\end{tikzpicture}
	\caption{Location of the Structural Perspective in IMoG}
	\label{fig:strp:imog}
\end{figure}

The Structural Perspective is the fourth perspective in IMoG and focuses on the modeling of the solution space of the innovation in an abstract manner. (see Figure \ref{fig:strp:imog}).
Based on the functions, features and requirements from the problem space, the Structural Perspective draws the corresponding possible solutions.
The Structural Perspective is located at the later phases of innovation modeling shortly before the roadmap is written.

The Structural Perspective bases on the well-known concepts and representations of systems engineering.
The concepts of \textit{Decomposition}, \textit{Refinement} and \textit{Variation} that are used on the perspectives are here of high importance too.
Next to the three concepts, the concept of properties and their relations provide the possibility to describe and compare solutions in detail.
The description of solution alternatives is provided by variants for Blocks (SP) and Refinement Blocks.
The representation of the Structural Perspective also follows the principles of systems engineering by using hierarchical Blocks and arrows / channels as relations.


The chapter is structured as followed:
In Section \ref{sec:strp:me} the meta model and its model elements are presented.
In Section \ref{sec:strp:e-scooter} an example of the Structural Perspective is given.
The strengths and limitations of the Structural Perspective are discussed in Section \ref{sec:strp:eval}.
A FAQ finalizes the description in Section \ref{sec:strp:faq}.

\section{Model elements}
\label{sec:strp:me}
\FloatBarrier

\begin{sidewaysfigure}[h]
	\centering
	\includegraphics[width=\linewidth]{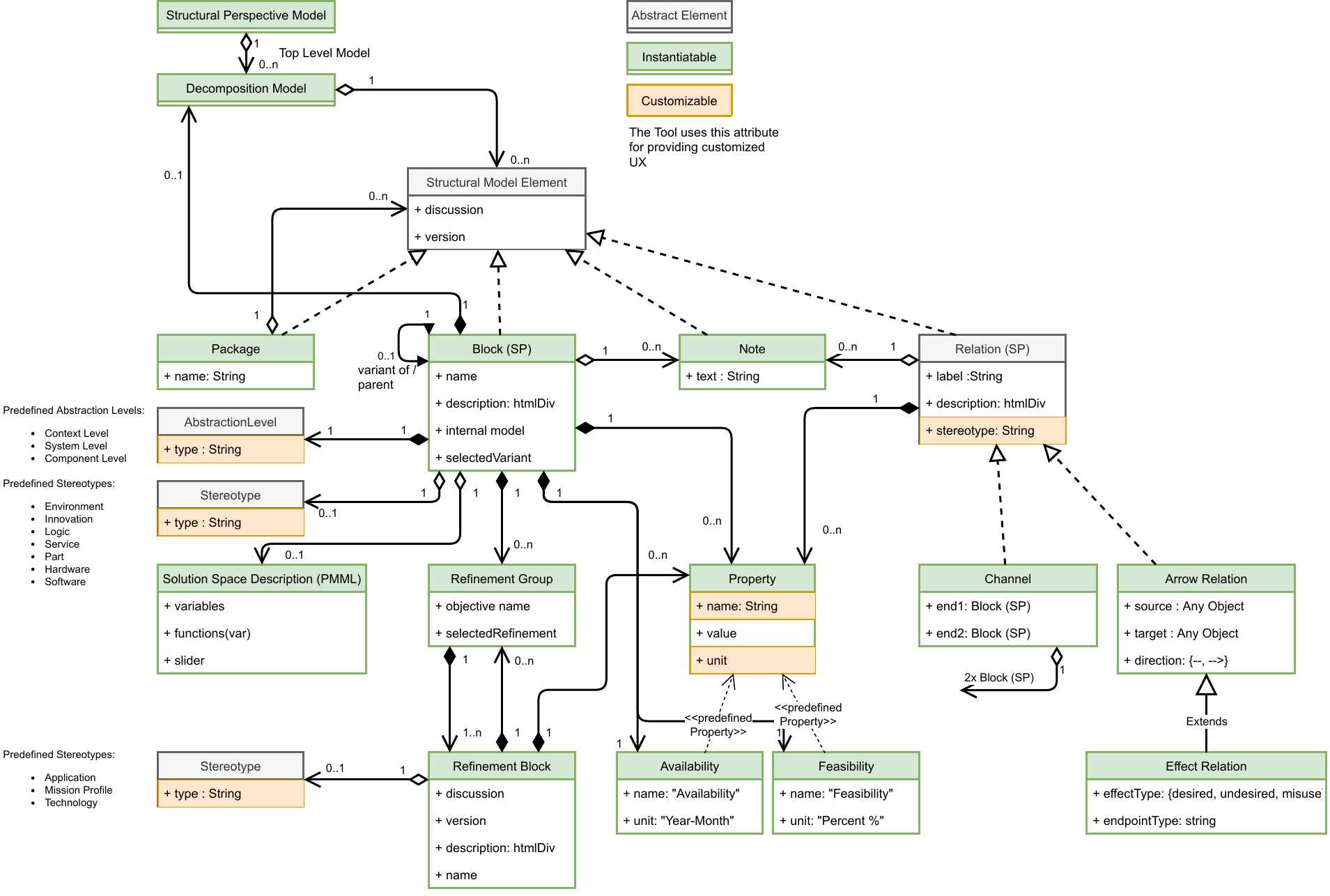}
	\caption{The model elements of the Structural Perspective.}
	\label{fig:strp:me}
\end{sidewaysfigure}

The meta model of the Structural Perspective (see Figure \ref{fig:strp:me}) builds on the \textit{Decomposition} and \textit{Refinement} concepts.
The meta model has the \textit{Structural Perspective Model} as the top level unit of the Structural Perspective.
The Structural Perspective Model contains a set of \textit{Decomposition Models}.
Decomposition Models represent the canvas fields with their sketchy system models known from system modeling tools like Cameo or Enterprise Architect.
The Structural Perspective can have multiple top level models, however it is recommended to only take one unless more are needed.

A Decomposition Model consists of \textit{Structural Model Elements}.
The Structural Model Elements include \textit{Blocks} (on the Structural Perspective) and \textit{Relations} between them, \textit{Packages} and \textit{Notes}.
Blocks (SP) represent solutions and own a bunch of attributes.
Among them are a \textit{name}, a \textit{description}, a \textit{discussion} chat, potentially an \textit{internal model}, a \textit{version}, a flag for the \textit{selected refinement variants}, an \textit{abstraction level} (either of type \textit{Context Level, System Level, Component Level} or a \textit{Custom Abstraction Level}), a \textit{Decomposition Model} and a reference to a possibly refined p\textit{arent block}, \textit{Notes}, a \textit{Stereotype} and possibly some \textit{Refinement Groups}.
Refinement Groups are the second important type of concept incorporated in the Structural Perspective model.
They allow to refine the Blocks by giving additional information and properties.
Each refinement is represented by a Refinement Block owning a set of properties.
Like Blocks (SP), Refinement Blocks can have a custom or a defined Stereotype, either of type \textit{Technology}, \textit{Mission Profile} or \textit{Application}.
These definitions go loosely hand in hand with the content of the Mission Profile standard (MPFO).
Additionally Blocks (SP) can own zero to any number of properties (defined by a name value and a unit).
Blocks (SP) have two predefined properties: \textit{Availability} and \textit{Feasibility} properties.
\textit{Solution Space Descriptions} (using PMML) can be added to blocks to allow multidimensional adjustments on the property variable values.
A detailed description of each attribute can be found in the Block (SP) description.

Two types of relations exist: \textit{Channels} and \textit{Arrow Relations}.
The abstract base class \textit{Relations (SP)} defines the basis of both.
It contains an \textit{HTMLDiv} for adding information, a \textit{version} number, a \textit{discussion} chat and a \textit{text} label.
Unlike relations on other Perspectives, these \textit{Relations (SP)} are rather complex and independent elements.
\textit{Channels} represent an information exchange between two Blocks (SP).
\textit{Arrow Relations} are used for any other (often less-complex) types of relation that shall be represented by an arrow.
The \textit{Effect} relation extends the arrow relation by an \textit{endpointType} and an \textit{effectType}.
Worth to note is that relation \textit{Decomposition} and \textit{Refinement} are not supported to keep the model simple.
Use the properties to specify additional information instead.
In a similar direction, analysis like \textit{Block Interface} - \textit{Channel} - \textit{Block Interface} consistency is not supported.
Such sophisticated analysis is rarely used in abstract innovation descriptions.
These are kept for the development phases.

\FloatBarrier

\fbox{
	\begin{minipage}{0.955 \textwidth}
		\large \textbf{Meta Model Base:}
		\normalsize \setstretch{0.8}
		\begin{itemize}
			\item Structural Perspective Model
			\item Decomposition Model
			\item Structural Model Element
			\item Relation (SP)
		\end{itemize}
	\end{minipage}
}

\rule{\textwidth}{1pt}
Meta Model Element:
\begin{center}
	\includegraphics{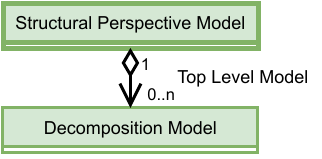}
\end{center}

Description:

\fcolorbox{gray!30!black}{gray!20!white}{
	\begin{minipage}{0.955 \textwidth}
		\large \textbf{Structural Perspective Model}\\
		\normalsize The \textit{Structural Perspective Model} is the diagram of the Structural Perspective of an innovation.
		It contains a \textit{Decomposition Model} which contains all model elements of the Structural Perspective.
	\end{minipage}
}

Example: A full Structural Perspective Model example is shown in Section \ref{sec:strp:e-scooter}.

\rule{\textwidth}{1pt}
Meta Model Element:
\begin{center}
	\includegraphics{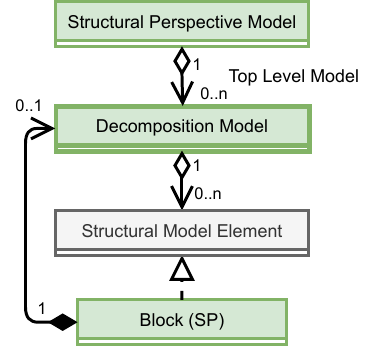}
\end{center}

Description:

\fcolorbox{gray!30!black}{gray!20!white}{
	\begin{minipage}{0.955 \textwidth}
		\large \textbf{Decomposition Model} \\
		\normalsize The \textit{Decomposition Model} represents one of the major concepts (\textit{Decomposition, Refinement, Variation}) and is fundamental for the modeling activities.
		The Decomposition Model can be thought of as the main canvas to draw the innovation on.
		The Decomposition Models consist of any number of \textit{Structural Model Elements}, which include \textit{Blocks} (of the Structural Perspective), \textit{relations} between them, as well as \textit{Notes} and \textit{Packages}.
		It is used for describing the top level model of the overall Structural Perspective.
		Each Block can have its own Decomposition Model and thus the Blocks with their Decomposition Models span up a hierarchy.
	\end{minipage}
}

Example:
\begin{center}
	\includegraphics{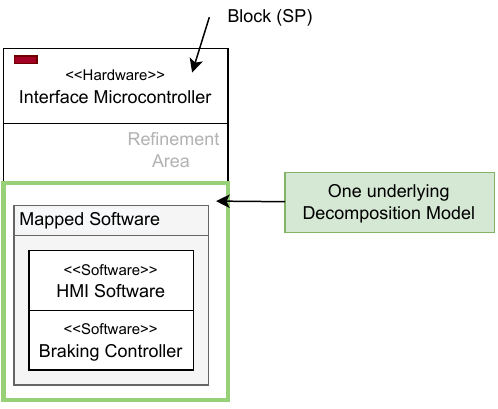}
\end{center}

\rule{\textwidth}{1pt}
Meta Model Element:
\begin{center}
	\includegraphics[width=\linewidth]{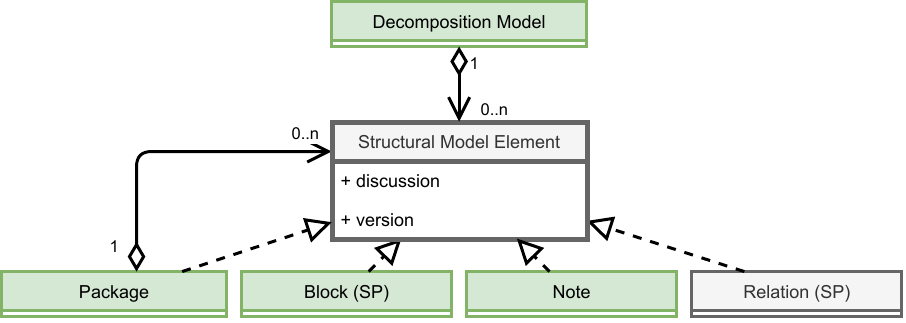}
\end{center}

Description:

\fcolorbox{gray!30!black}{gray!20!white}{
	\begin{minipage}{0.955 \textwidth}
		\large \textbf{Structural Model Element)} \\
		\normalsize The \textit{Structural Model Element} is the abstract object any model element of the \textit{Decomposition Model} derives: \textit{Package}, \textit{Block (SP)}, \textit{Note} and \textit{Relation (SP)}.
		It owns the following attributes:
		\begin{itemize}
			\item The \textit{discussion} attribute to take comments about this element.
			\item The \textit{version} attribute to support version control.
		\end{itemize}
	\end{minipage}
}

Example: An example is shown for each Structural Model Element under their respective description.

\rule{\textwidth}{1pt}
Meta Model Element:
\begin{center}
	\includegraphics{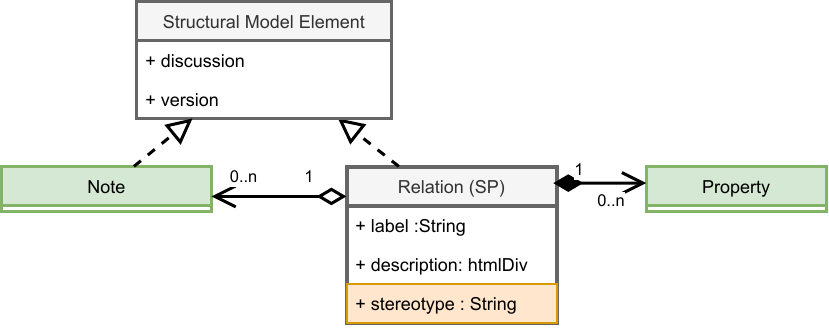}
\end{center}

Description:

\fcolorbox{gray!30!black}{gray!20!white}{
	\begin{minipage}{0.955 \textwidth}
		\large \textbf{Relation (SP)} \small \textbf{(Read: Relation on the Structural Perspective)} \\
		\normalsize The abstract\textit{ Relation (SP)} describes relations between \textit{Blocks (SP)}.
		Relations are implemented by \textit{Channels} and \textit{Arrow Relations}.
		In contrast to relations on the Functional Perspective or Quality Perspective, \textit{Relations (SP)} can own labels and the following attributes:
		\begin{itemize}
			\item A \textit{description} to solve the problem of lack of clarity of the relation by adding information.
			The description shall answer shortly \enquote{What shall the Relation represent?}, the reasoning behind the relation and its basic conditions to work.
			There is no template needed.
			Images or drafts provide valuable information.
			\item A \textit{stereotype} to define what type of relation it represents.
			The stereotype is not defined to be part of any set values, but can hold any string.
			This way, the stereotype can be used for describing any customized relation.
			\item A set of \textit{Properties} can be used to specify the relation and to form a basis for any consistency analysis.
			\item Additionally the derived attributes \textit{discussion} and \textit{version} from the \textit{Structural Model Element} as well as \textit{Notes} to enrich the relations comprehensibility.
		\end{itemize}
		Each of the implementing relations are described in more detail on its own.
	\end{minipage}
}

Example: An example is shown for each relation under their respective description.

\fbox{
	\begin{minipage}{0.955 \textwidth}
		\large \textbf{Model Elements:}\\
		Block related elements:
		\begin{itemize}
			\normalsize \setstretch{0.8}
			\item Block (SP)
			\item Package
			\item Note
			\item Refinement Group
			\item Refinement Block
			\item Solution Space Description
			\item Property
		\end{itemize}
		Relation Types:
		\begin{itemize}
			\normalsize \setstretch{0.8}
			\item Channel
			\item Arrow Relation
			\item Effect Relation
		\end{itemize}
	\end{minipage}
}

In the following the model elements are introduced.
First the seven Block relation elements are introduced and then the relations are presented.

\fbox{
	\begin{minipage}{0.955 \textwidth}
		\Large Block related elements
	\end{minipage}
}

\rule{\textwidth}{1pt}
Meta Model Element:
\begin{center}
	\includegraphics[width=\linewidth]{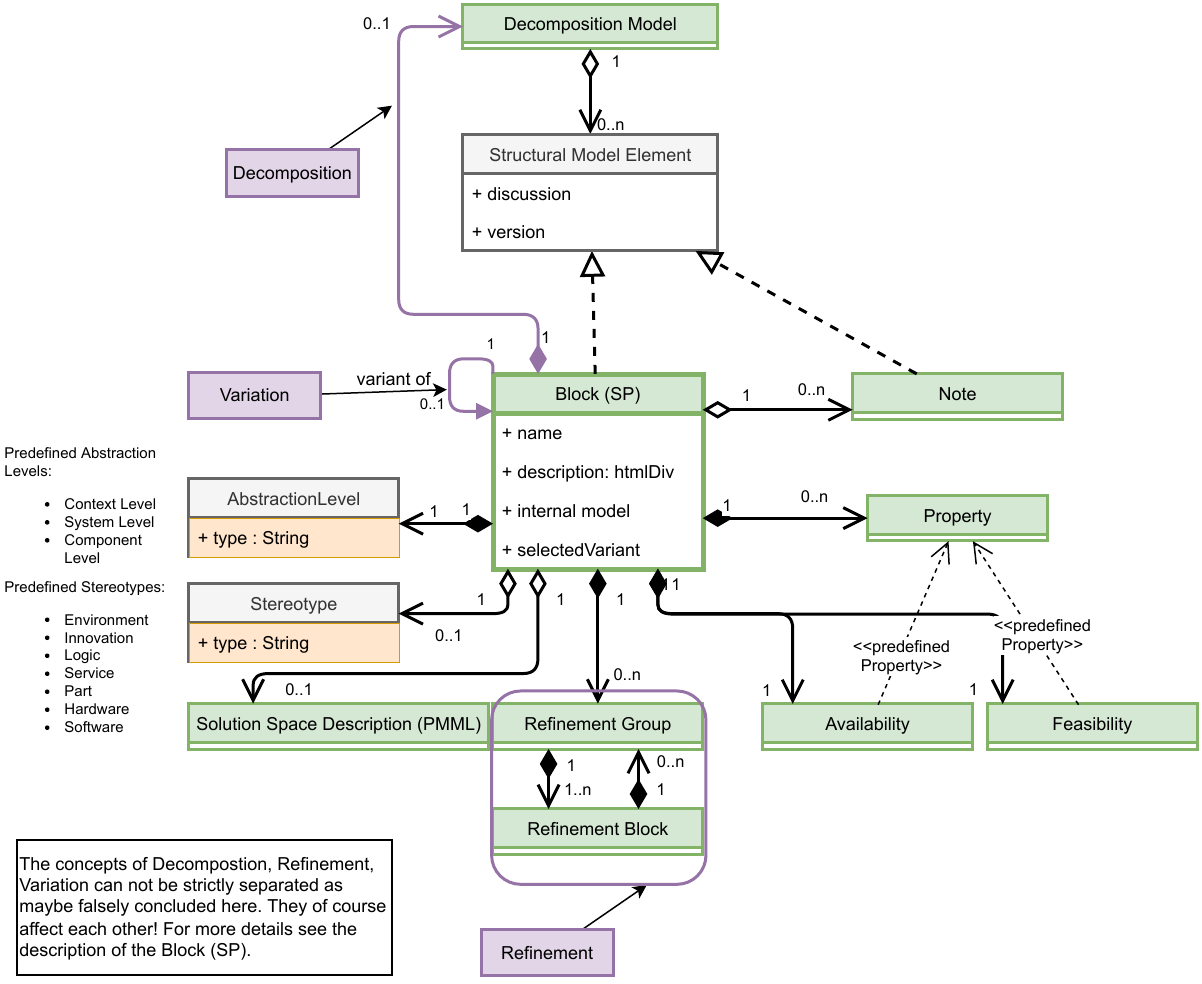}
\end{center}

Description:

\fcolorbox{gray!30!black}{gray!20!white}{
	\begin{minipage}{0.955 \textwidth}
		\large \textbf{Block (SP)} \small \textbf{(Read: Block on the Structural Perspective)} \\
		\normalsize The \textit{Block (SP)} is one of the main elements of the Structural Perspective.
		It is used to represent any system and component that is modeled and is thus the most complex element. It derives the following attributes of the \textit{Structural Model Element} definition:
		\begin{itemize}
			\item The \textit{discussion} attribute to take comments about this element.
			\item The \textit{version} attribute to support version control.
		\end{itemize}
		In addition to the Structural Model Element definition, it contains the following attributes:
		\begin{itemize}
			\item A \textit{name}.
		\end{itemize}
	\end{minipage}
}

\fcolorbox{gray!30!black}{gray!20!white}{
	\begin{minipage}{0.955 \textwidth}
		\large \textbf{Block (SP) continued} \small \textbf{(Read: Block on the Structural Perspective)} \\
		\begin{itemize}
			\item A \textit{description} of the Block to solve the problem of lack of clarity by adding information.
			The description shall answer shortly \enquote{What shall the Block represent?}, the reasoning behind the Block and its basic conditions to work.
			There is no template needed.
			Images or drafts provide valuable information.
			\item An \textit{abstraction level} to define the level of abstraction the Block represents.
			It can be either a predefined abstraction level (\textit{Context Level, System Level, Component Leve}l) or any other string.
			\item An optional \textit{Stereotype}.
			For any \textit{Block (SP)}, this could be any custom string or one of the predefined Stereotypes:
			\begin{itemize}
				\item \textit{Environment} for representing the environment of the innovation.
				\item \textit{Innovation} for representing the main focus in this model: the innovation.
				\item \textit{Logic} for representing a functional unit that is undefined if it is implemented as a physical unit, Hardware or Software.
				\item \textit{Service} for representing actions of executing some functionality requested by users.
				\item \textit{Part} for representing any physical unit including materials.
				\item \textit{Hardware} for representing electrical units that can execute algorithms.
				\item \textit{Software} or representing algorithms.
			\end{itemize}
			\item Optional \textit{Properties} to specify the Block in more detail and to form a basis for consistency analysis.
			Predefined properties are:
			\begin{itemize}
				\item An \textit{Availability} property for the estimation of the availability of the Block or in other words, \enquote{To which timestamp is the component available?}.
				\item A \textit{Feasibility} property for the estimation of the feasibility if the Block is available and implementable to the given \textit{Availability} timestamp.
			\end{itemize}
			\item An optional \textit{Solution Space Description} (in form of PMMLs) to represent important dependencies between the Blocks properties.
			The properties and the description form together the solution space.
			\item An optional \textit{internal model} providing more specification details to enhance the Block description.
			\item The attribute \textit{selected variant} represents if and which variant is chosen.
		\end{itemize}
		The Block (SP) supports the three main concepts (Decomposition, Refinement, Variation) in the form of more complex attributes:
		\begin{itemize}
			\item An optional \textit{Decomposition Model} to model how the Block is set up.
			This element is mainly used to build the system model hierarchy.
			By representing the concept of 'Decomposition' it is especially important.
			\item An optional set of disjoint \textit{Refinement Groups}, to model possible refinements of the Block's properties.
			The \textit{Refinement Groups} can contain several \textit{Refinement Blocks} to model varieties of the properties.
			The set of properties defined among all Refinement Groups and the block itself shall be disjoint!
		\end{itemize}
	\end{minipage}
}

\fcolorbox{gray!30!black}{gray!20!white}{
	\begin{minipage}{0.955 \textwidth}
		\large \textbf{Block (SP) continued} \small \textbf{(Read: Block on the Structural Perspective)} \\
		\begin{itemize}
			\item The \textit{variant} attribute defines a set of variants - which are itself Blocks (SP) too - to represent deviations from the Block.
			Meanwhile each variant sets its parent reference to the block.
			A variant can thus only be part of one parent block!
			Variants represent the implementation of the concept of 'Alternatives' for the Structural Perspective.
			Selecting a variant of a Block affects the Blocks attributes in the following way:
			\begin{itemize}
				\item \textit{Overwriting (but the Block's original value shall be still represented in the tool)}:
				\begin{itemize}
					\item The \textit{name} of the variant overwrites the Block's name.
					E.g: The variant name 'Comfort E-Scooter' overwrites the Block's name 'E-Scooter'.
					\item Same holds for the \textit{discussion, version, description, abstraction level} and \textit{Stereotype}.
				\end{itemize}
				\item \textit{Extended:}
				\begin{itemize}
					\item The \textit{properties} of variant extend the Block's properties.
					If the variant property name is the same as the property of the block, then the variant property overwrites the blocks property.
					E.g: Property 'Weight' is set for the 'E-Scooter' and the variant 'Comfort E-Scooter'.
					In this case, the weight property from the 'Comfort E-Scooter' overwrites the weight property from the block 'E-Scooter'.
					\item The \textit{Solution Space Description} (SSE) of the variant extends the blocks SSE, if and only if both SSEs have at maximum one common property.
					If both SSEs have more than one common property (e.g: 'weight' and 'speed') then only the SSE from the variant is considered (because there is typically no function satisfying both SSEs).
					\item The \textit{Decomposition Model} of the variant extends the Decomposition model of the block.
					There is nothing like overwriting here.
					If both - variant and block - have a block with the same name in their decomposition model, then they are both considered!
					Because of the strict variant / parent reference, variants can own relations between blocks of their decomposition model and any blocks of decomposition model of the parent block!
					This makes the selection of variants a pretty powerful tool to change things!
					\item The \textit{Refinement Groups} and \textit{Refinement Block} of the variant extends the Refinement Groups of the block if the Refinement Groups name are not the same.
					If any Refinement Group exists in both variant and block, then the variant Refinement Group overwrites the Refinement Group from the block.
				\end{itemize}
				\item \textit{Unaffected:}
				\begin{itemize}
					\item \textit{Internal models} of variants take precedence in the list.
					However, because Internal models are only referenced and not used in the modeling methodology IMoG, there is no clear overwriting or extending in place.
					The user shall consider both or just one based on their needs.
					\item The \textit{selected variant} stays of course relevant (otherwise it would be unclear which variant properties affect the block!).
					If the variant itself has this property set and its own attributes are affected by their own variants, then this selection among the variants of the variant stays relevant too!
				\end{itemize}
			\end{itemize}
		\end{itemize}
	\end{minipage}
}

Example:
\begin{center}
	\includegraphics[width=\linewidth]{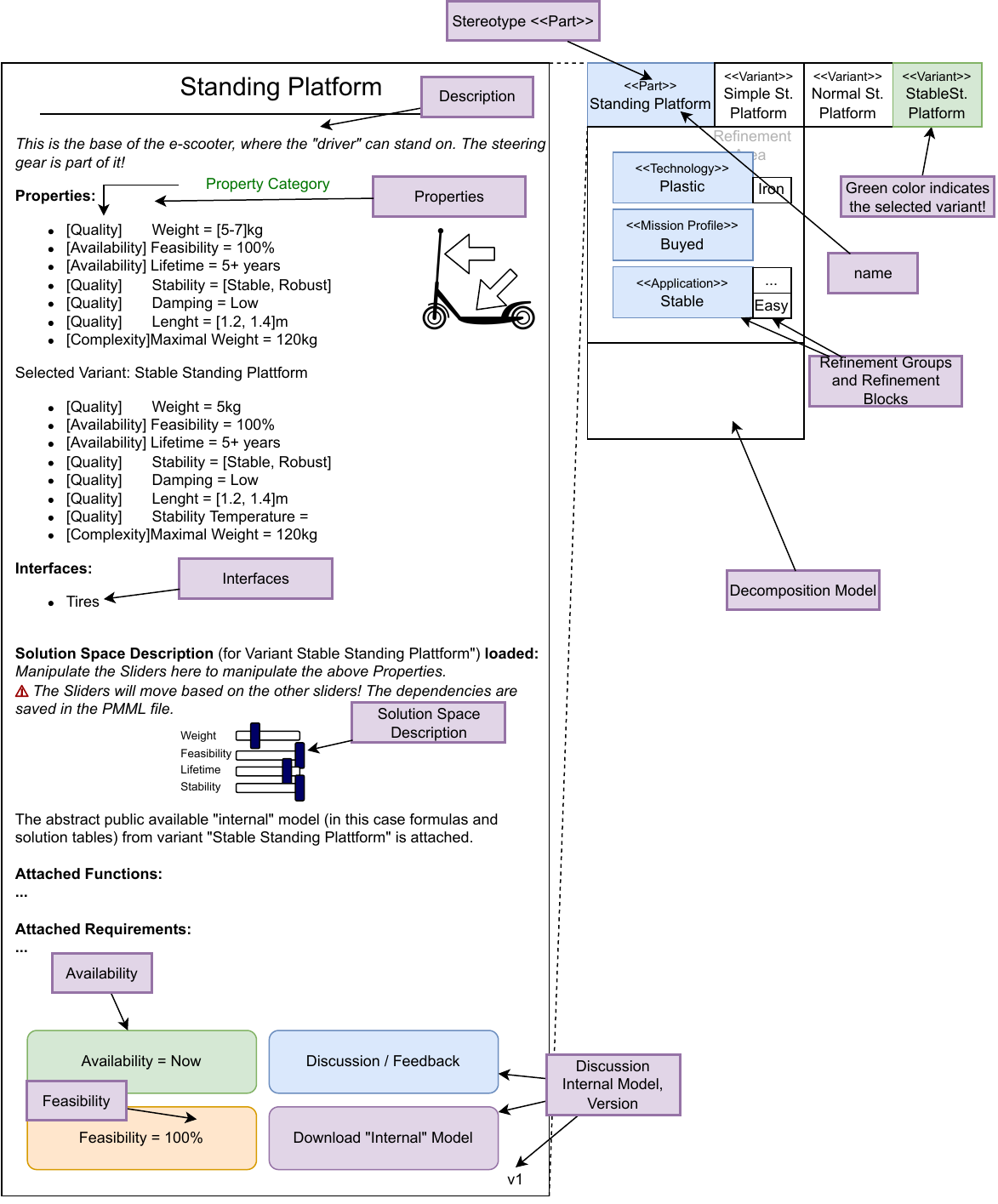}
\end{center}

\rule{\textwidth}{1pt}
Meta Model Element:
\begin{center}
	\includegraphics[width=\linewidth]{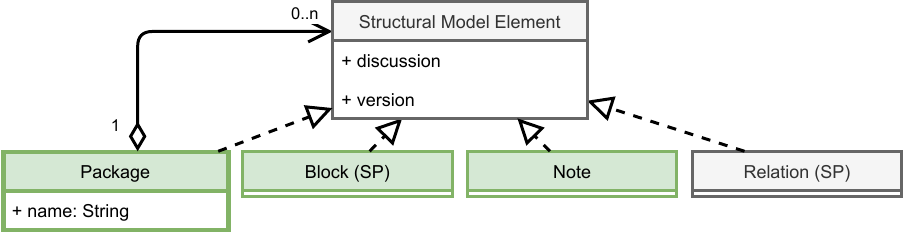}
\end{center}

Description:

\fcolorbox{gray!30!black}{gray!20!white}{
	\begin{minipage}{0.955 \textwidth}
		\large \textbf{Package} \\
		\normalsize The \textit{Package} describes an collection of Structural Model Elements to allow to create a hierarchy in the Decomposition Model and distinct between Packages.
		There is nothing special about Packages otherwise: They have a \textit{name} and a set of \textit{Structural Model Elements}.
	\end{minipage}
}

Example: A package named 'Mapped Software' with two blocks inside.
\begin{center}
	\includegraphics{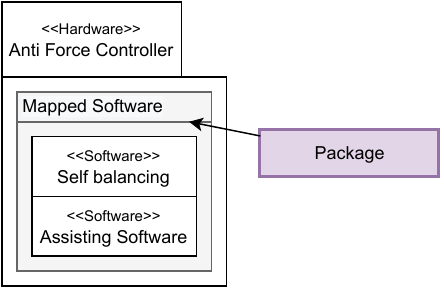}
\end{center}

\rule{\textwidth}{1pt}
Meta Model Element:
\begin{center}
	\includegraphics{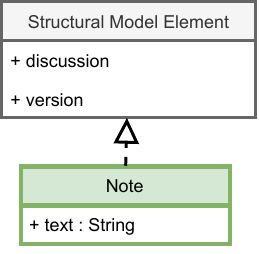}
\end{center}

Description:

\fcolorbox{gray!30!black}{gray!20!white}{
	\begin{minipage}{0.955 \textwidth}
		\large \textbf{Note} \\
		\normalsize The \textit{Note} can be used to add information to the model that can not or should not be modeled.
		Notes should be used sparsely!
	\end{minipage}
}

Example:
\begin{center}
	\includegraphics{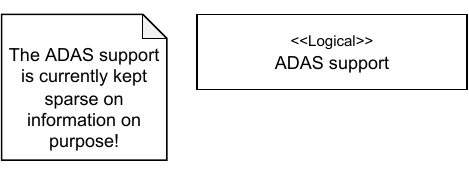}
\end{center}

\rule{\textwidth}{1pt}
Meta Model Element:
\begin{center}
	\includegraphics{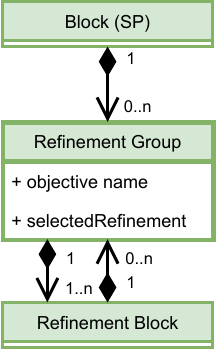}
\end{center}

Description:

\fcolorbox{gray!30!black}{gray!20!white}{
	\begin{minipage}{0.955 \textwidth}
		\large \textbf{Refinement Group} \\
		\normalsize The \textit{Refinement Group} represents a container for modeling variety of Refinements.
		It contains a non empty set of \textit{Refinement Blocks} and a marker (called \textit{selectedRefinement}), which of the \textit{Refinement Blocks} is currently selected.
		\textit{Refinement Groups} can not exist on their own.
		They must be included by a Block (SP) or a Refinement Block.
		The Blocks (SP) and Refinement Blocks may contain none or multiple Refinement Groups.
		The Refinement Blocks inside a container are recommended, but not restricted to have the same stereotype.
		It makes much sense to group only blocks with the same stereotype instead of mixing say \textit{Application} specifications with \textit{Technology} specifications.
		The selected Refinement Block will be used to overtake Properties to the Block (SP) specification or to the Refinement Block specification, which are then used for consistency analysis.
	\end{minipage}
}

Example: An example Refinement Area with one Refinement Group containing two <<technology>> stereotyped Blocks: 'Copper' and 'Iron'
\begin{center}
	\includegraphics{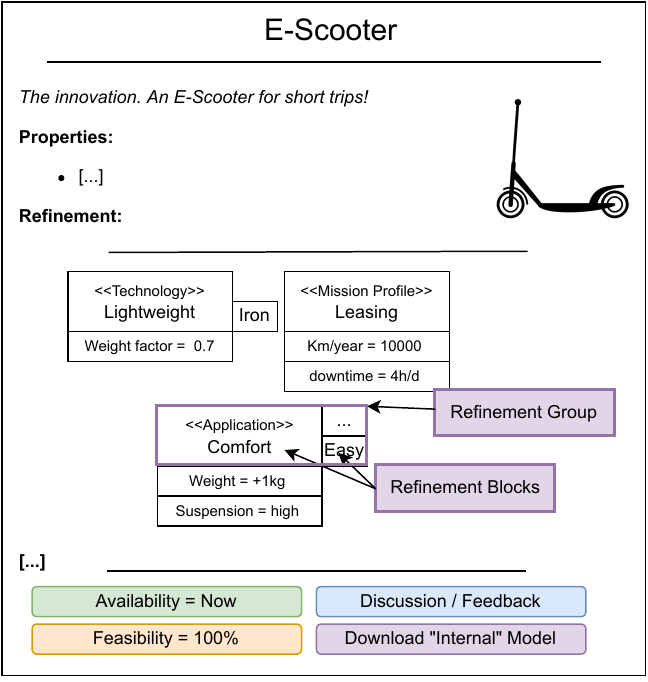}
\end{center}

\rule{\textwidth}{1pt}
Meta Model Element:
\begin{center}
	\includegraphics[width=\linewidth]{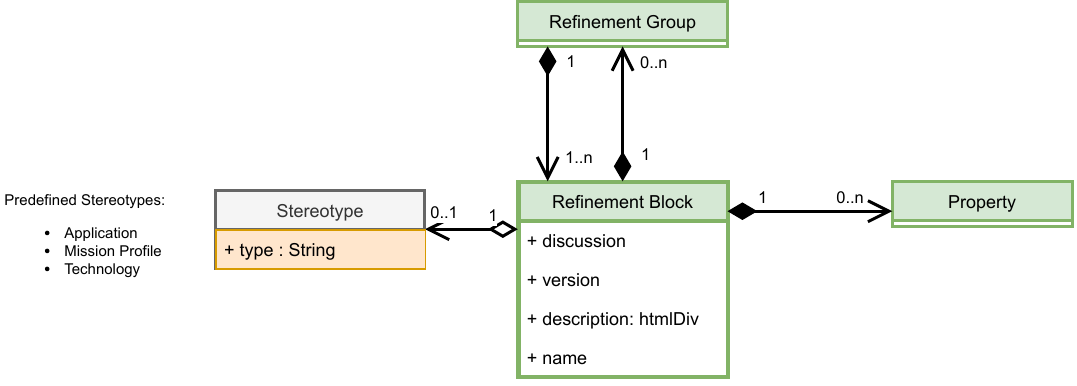}
\end{center}

Description:

\fcolorbox{gray!30!black}{gray!20!white}{
	\begin{minipage}{0.955 \textwidth}
		\large \textbf{Refinement Block} \\
		\normalsize The \textit{Refinement Block} represents 'Refinements' of Blocks (SP).
		Refinement Blocks represent the 'Refinement' concepts and are thus of special importance.
		Refinement Blocks are a lighter variant of the Block (SP) and own fewer common attributes:
		\begin{itemize}
			\item A \textit{discussion} attribute to take comments about this element.
			\item A \textit{version} attribute to support version control.
			\item A \textit{description} of the Block to solve the problem of lack of clarity by adding information.
			The description shall answer shortly \enquote{What shall the Block represent?}, the reasoning behind the Block and its basic conditions to work.
			There is no template needed. Images or drafts provide valuable information.
			\item A \textit{name}.
			\item Optional \textit{Properties} to specify the Refinement Block and to form a basis for consistency analysis.
		\end{itemize}
		In addition to the common attributes above, the Refinement Block may contain an optional \textit{Stereotype}: For any Refinement Block this could be a custom string or one of the predefined Stereotypes:
		\begin{itemize}
			\item \textit{Technology} for representing information about influencing factors of technology and materials of the innovation.
			\item \textit{Mission Profile} for representing environmental factors that influence the decision.
			It is especially used for restricting the environment to a special set of situations.
			\item \textit{Application} for representing specialties about how the system or component is used.
		\end{itemize}
		Additionally, Refinement Blocks can be further refined.
		This can be achieved by adding \textit{Refinement Groups} with Refinement Blocks to the block.
		Important to note is, that Refinement Blocks can not exist on their own and thus must be part of a Refinement Group.
		Refinement Groups are either owned by Blocks (SP) or Refinement Blocks.
		Thus it inevitable follows that any Refinement Block has on the highest level an Block (SP) as an owner.
	\end{minipage}
}

Example:
\begin{center}
	\includegraphics[width=\linewidth]{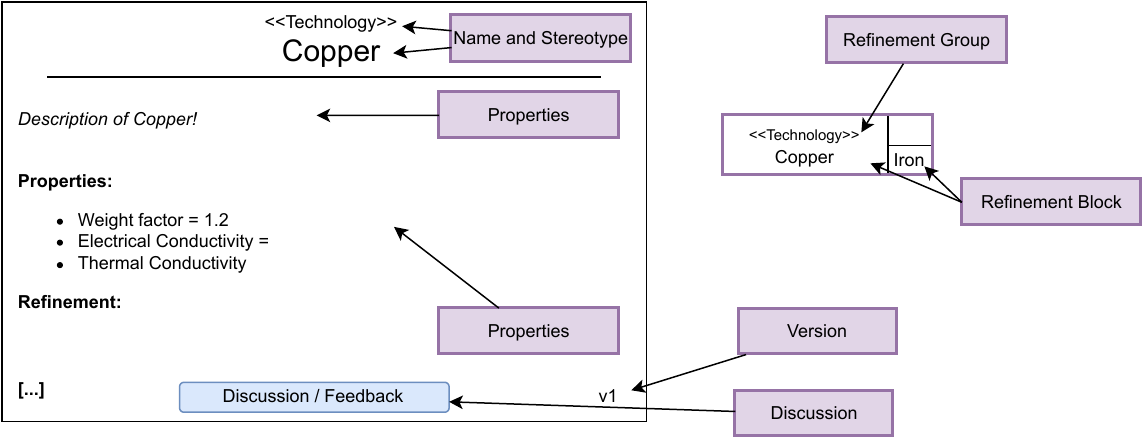}
\end{center}

\rule{\textwidth}{1pt}
Meta Model Element:
\begin{center}
	\includegraphics{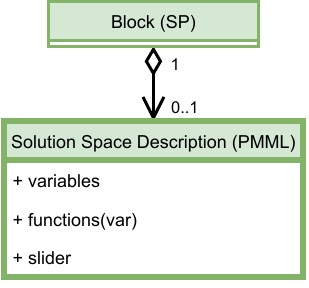}
\end{center}

Description:

\fcolorbox{gray!30!black}{gray!20!white}{
	\begin{minipage}{0.955 \textwidth}
		\large \textbf{Solution Space Description (PMML)} \\
		\normalsize The \textit{Solution Space Description} enables the user to add additional functions how the Block properties relate to each other.
		The format is PMML and it describes the relation of a set of input variables to a set of output variables as multivariate functions.
		The chosen values shall be manipulate able by a graphical slider representation.
	\end{minipage}
}

Example:
\begin{center}
	\includegraphics[width=\linewidth]{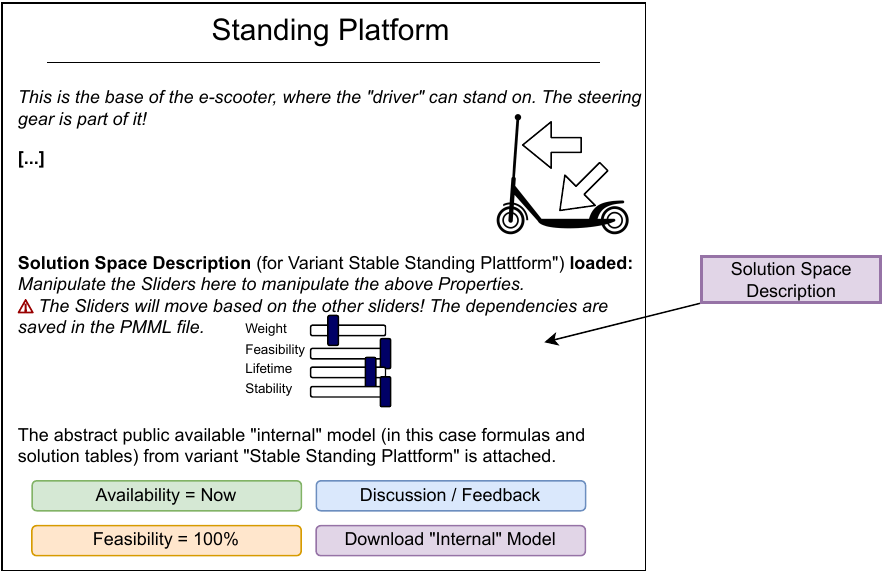}
\end{center}

\rule{\textwidth}{1pt}
Meta Model Element:
\begin{center}
	\includegraphics{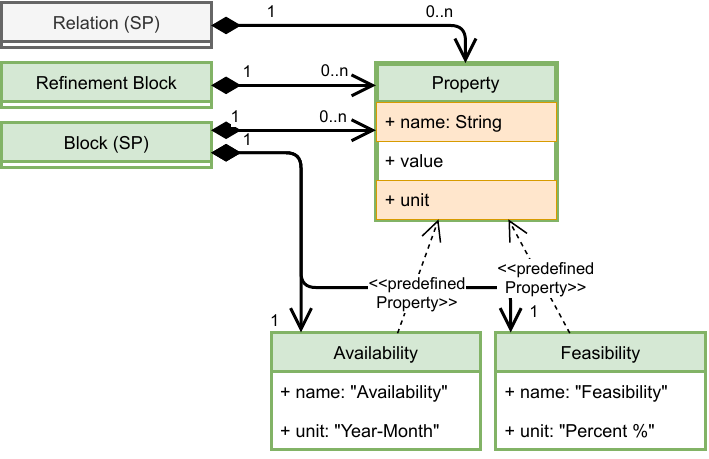}
\end{center}

Description:

\fcolorbox{gray!30!black}{gray!20!white}{
	\begin{minipage}{0.955 \textwidth}
		\large \textbf{Property} \\
		\normalsize \textit{Properties} are part of the specification of Blocks (SP), Refinement Blocks and Relations (SP).
		Properties are used to provide important information and build a basis for further analysis.
		Properties can not exist on their own and must have any of the above mentioned elements as an owner.
		Two properties are predefined in the meta model: The \textit{Availability} and the \textit{Feasibility} properties are defined for every Block (SP):
		\begin{itemize}
			\item The \textit{Availability} property is used for the estimation of the availability of the Block or in other words, \enquote{To which timestamp is the component available?}.
			\item The \textit{Feasibility} property is used for the estimation of the feasibility if the Block is available and implementable to the given \textit{Availability} timestamp.
		\end{itemize}
		Properties own a \textit{name}, a \textit{value} and a \textit{unit}.
		The \textit{name} and \textit{unit} are not preset to a specific set of values.
		They can be used for any customization to allow the user to add innovation specific properties.
		For any domain and innovation it is recommended to have a existing 'domain properties' set to build upon instead of starting with just the two predefined properties.
	\end{minipage}
}

Example:
\begin{center}
	\includegraphics{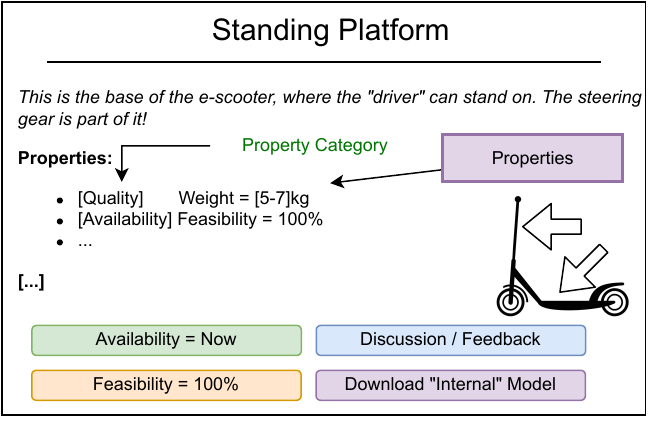}
\end{center}

\fbox{
	\begin{minipage}{0.955 \textwidth}
		\Large Relations
	\end{minipage}
}

\rule{\textwidth}{1pt}
Meta Model Element:
\begin{center}
	\includegraphics{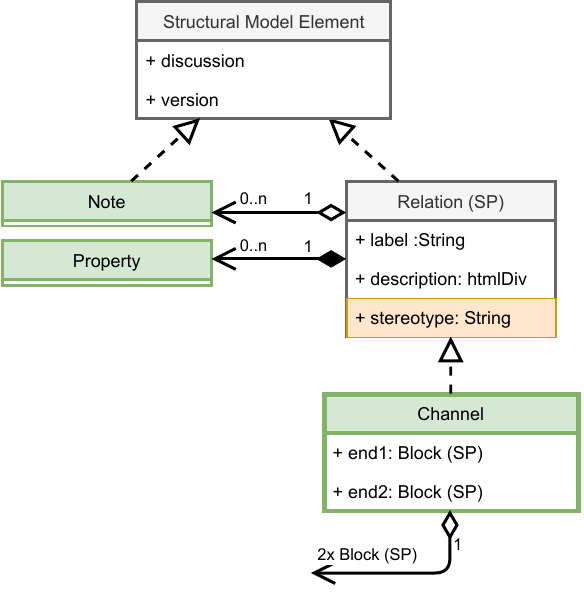}
\end{center}

Description:

\fcolorbox{gray!30!black}{gray!20!white}{
	\begin{minipage}{0.955 \textwidth}
		\large \textbf{Channel} \\
		\normalsize The \textit{Channel} represents a communication medium between Blocks (SP).
		Channels are used, whenever the communication between two Blocks (SP) plays a significant role and needs a specification.
		Channels can trigger communication based analysis.
		The definition of the Channel relation reflects that the meta model is not pure generic but slightly customized to the microelectronic domain.

		Channels implement the abstract Relation (SP).
		Channels thus own the following attributes:
		\begin{itemize}
			\item Two \textit{Block (SP)} endpoints.
			\item A \textit{label} and a \textit{description} to solve the problem of lack of clarity of the relation by adding information.
			The description shall answer shortly \enquote{What shall the Relation represent?}, the reasoning behind the relation and its basic conditions to work.
			There is no template needed.
			Images or drafts provide valuable information.
			\item A \textit{stereotype} to define what type of relation it represents.
			The stereotype is not defined to be part of any set values, but can hold any string.
			This way, the stereotype can be used for describing any customized relation.
			\item A set of \textit{Properties} can be used to specify the Relation and to form a basis for any consistency analysis.
			Additionally the derived attributes \textit{discussion} and \textit{version} from the \textit{Structural Model Element} as well as \textit{Notes} to enrich the relations comprehensibility.
		\end{itemize}
	\end{minipage}
}

Example:
\begin{center}
	\includegraphics{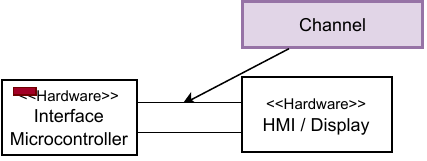}
\end{center}

\rule{\textwidth}{1pt}
Meta Model Element:
\begin{center}
	\includegraphics{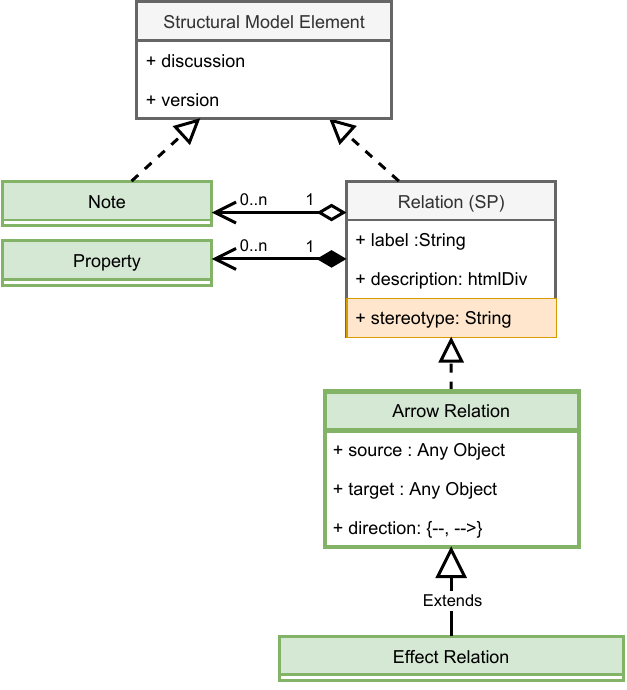}
\end{center}

Description:

\fcolorbox{gray!30!black}{gray!20!white}{
	\begin{minipage}{0.955 \textwidth}
		\large \textbf{Arrow Relation} \\
		\normalsize The \textit{Arrow Relation} represents a bidirectional relation between two elements.
		The Arrow Relation remains pretty generic.
		The Arrow Relation implements the abstract Relation (SP).
		Arrow Relations thus own the following attributes:
		\begin{itemize}
			\item A \textit{source} and a \textit{target} endpoint.
			\item A \textit{direction} either of the type \textit{unidirectional} or \textit{bidirectional}.
			\item A \textit{label} and a \textit{description} to solve the problem of lack of clarity of the relation by adding information.
			The description shall answer shortly \enquote{What shall the Relation represent?}, the reasoning behind the relation and its basic conditions to work.
			There is no template needed.
			Images or drafts provide valuable information.
			\item A \textit{stereotype} to define what type of relation it represents.
			The stereotype is not defined to be part of any set values, but can hold any string.
			This way, the stereotype can be used for describing any customized relation.
			\item A set of \textit{Properties} can be used to specify the relation and to form a basis for any consistency analysis.
			Additionally the derived attributes \textit{discussion} and \textit{version} from the \textit{Structural Model Element} as well as \textit{Notes} to enrich the relations comprehensibility.
		\end{itemize}
	\end{minipage}
}

Example:
\begin{center}
	\includegraphics{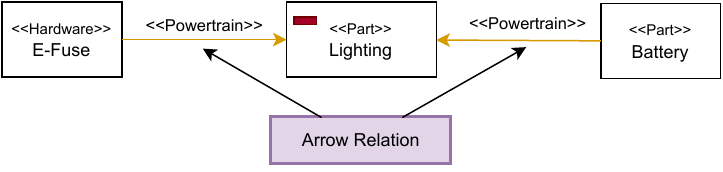}
\end{center}

\rule{\textwidth}{1pt}
Meta Model Element:
\begin{center}
	\includegraphics{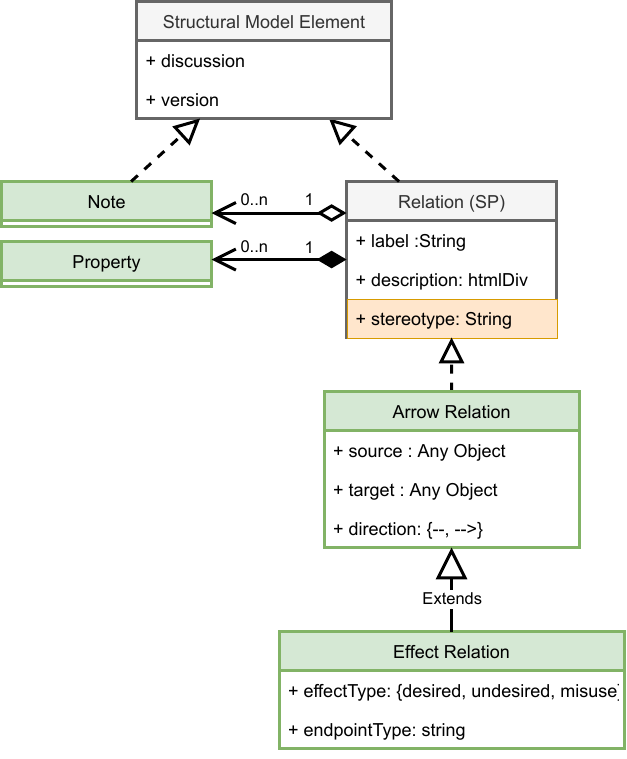}
\end{center}

Description:

\fcolorbox{gray!30!black}{gray!20!white}{
	\begin{minipage}{0.955 \textwidth}
		\large \textbf{Effect Relation} \\
		\normalsize The \textit{Effect Relation} represents an effect between two elements.
		The Effect Relation originates from the well known effect chain analysis and can be used in a similar manner.
		Special types of Effect Relations can be either implemented by customizing the \textit{label} attribute or by creating a new relation category by implementing \textit{Custom} relations.
		The Effect Relation implements the abstract \textit{Relation (SP)}.
		Effect Relations thus own the following attributes:
		\begin{itemize}
			\item An \textit{effectType}, which can be set to either \textit{desired}, \textit{undesired} or \textit{misuse}.
			\item A \textit{source} and a \textit{target} endpoint with the target endpoint having a \textit{endpointType} (e.g. 'thermal' or 'acoustic' or 'radiation').
			\item A \textit{direction} either of the type \textit{unidirectional} or \textit{bidirectional}.
			\item A \textit{label} and a \textit{description} to solve the problem of lack of clarity of the relation by adding information.
			The \textit{description} shall answer shortly \enquote{What shall the Relation represent?}, the reasoning behind the relation and its basic conditions to work.
			There is no template needed.
			Images or drafts provide valuable information.
			\item A \textit{stereotype} to define what type of relation it represents.
			The stereotype is not defined to be part of any set values, but can hold any string.
			This way, the stereotype can be used for describing any customized relation.
			\item A set of \textit{Properties} can be used to specify the relation and to form a basis for any consistency analysis.
			Additionally the derived attributes \textit{discussion} and \textit{version} from the \textit{Structural Model Element} well as \textit{Notes} to enrich the relations comprehensibility.
		\end{itemize}
	\end{minipage}
}

Example:
\begin{center}
	\includegraphics[width=\linewidth]{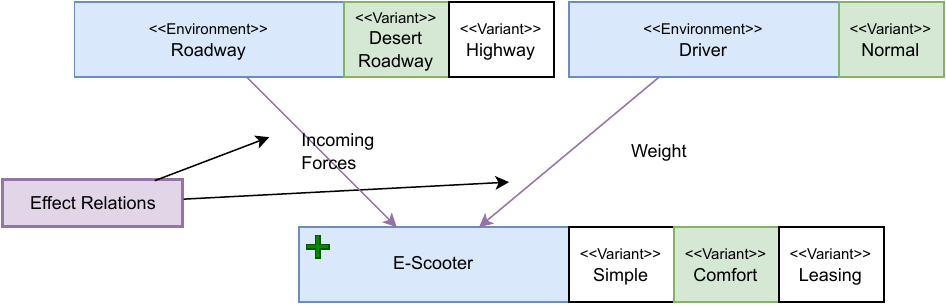}
\end{center}

\section{E-Scooter example}
\label{sec:strp:e-scooter}

\begin{figure}[!h]
	\centering
	\includegraphics[width=\linewidth]{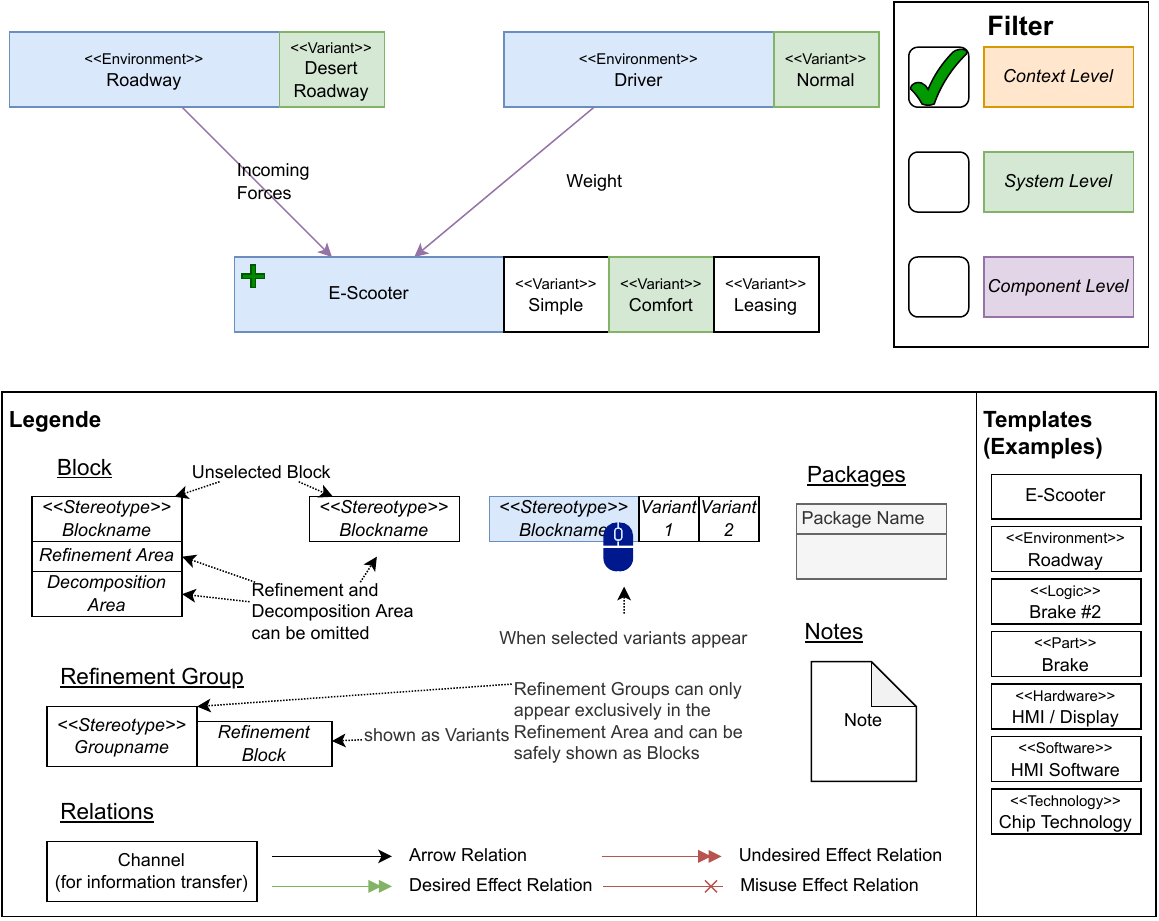}
	\caption{Top view: The solutions of the innovation \enquote{Providing mobility with an e-scooter} from a high level.}
	\label{fig:strp:example1}
\end{figure}

\begin{sidewaysfigure}[!h]
	\centering
	\includegraphics[width=\linewidth]{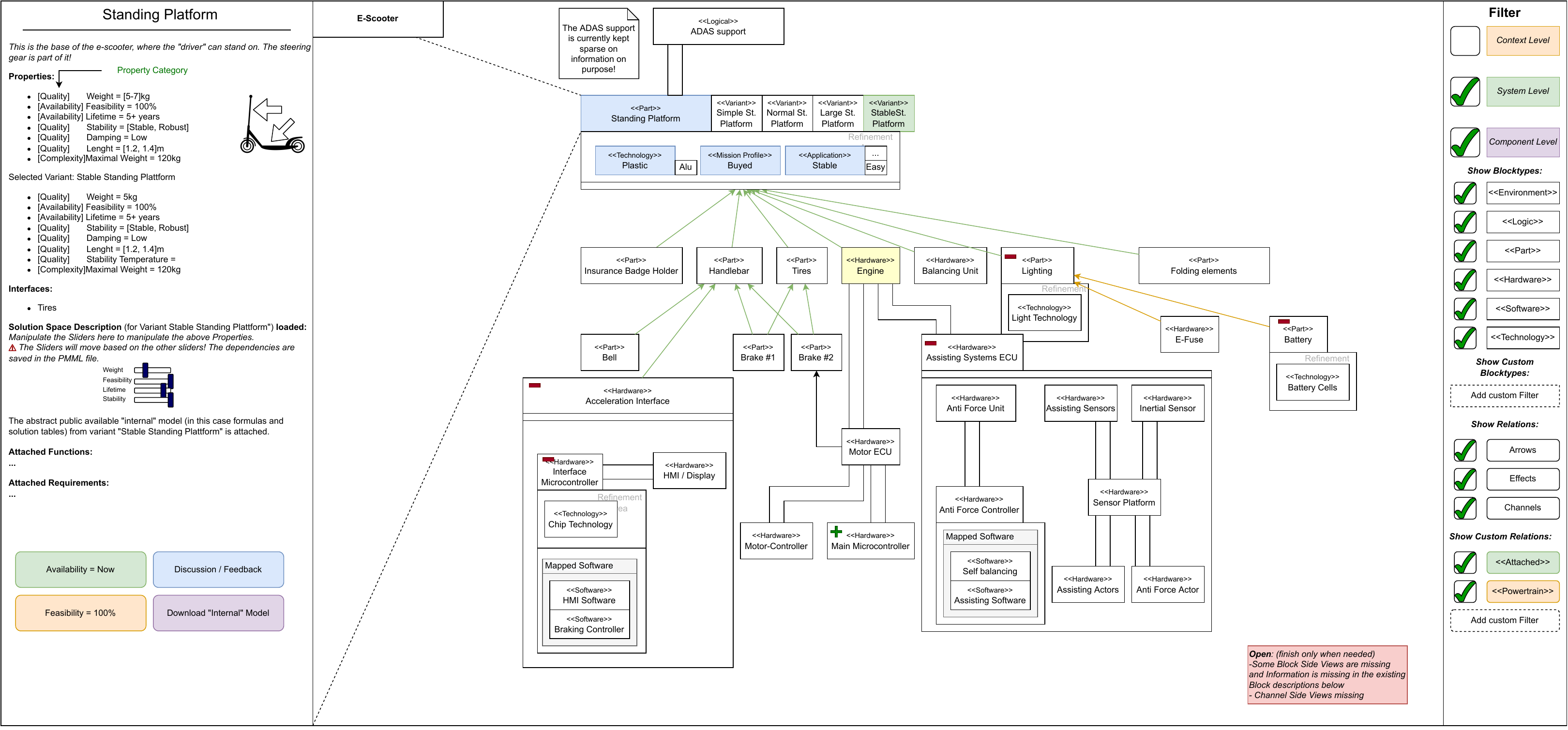}
	\caption{Block view: The solutions for the innovation of \enquote{Providing mobility with an e-scooter} of the solution-elements of the system level and the component level.}
	\label{fig:strp:example2}
\end{sidewaysfigure}

\begin{sidewaysfigure}[!h]
	\centering
	\includegraphics[width=\linewidth]{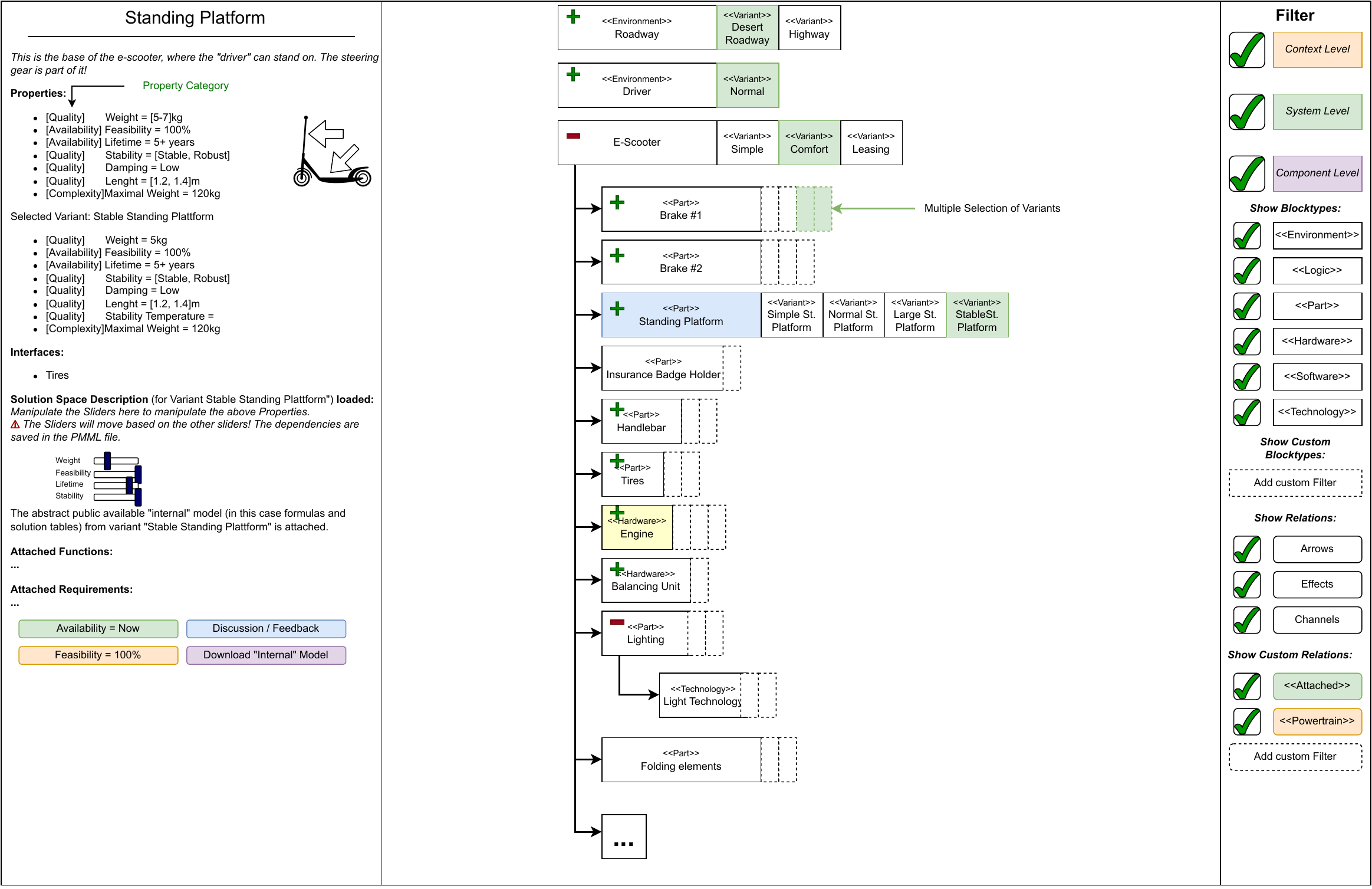}
	\caption{Tree View.}
	\label{fig:strp:example3}
\end{sidewaysfigure}

The example of the Structural Perspective comprises the innovation requirements of \enquote{Providing mobility with an e-scooter} in one view called Structural View.
The example is divided into three sub views:
\begin{itemize}
	\item The \textit{top view} (see Figure \ref{fig:strp:example1}) describes the solutions of the innovation \enquote{Providing mobility with an e-scooter} from a high level.
	\item The \textit{Block view} (see Figure \ref{fig:strp:example2}) shows – next to the context level elements from the top view – the solution-elements of the system level and the component level.
	Its purpose targets the exemplary representation of all model elements from the meta model.
	\item The \textit{Tree view} (see Figure \ref{fig:strp:example3}) represents the model as a tree.
	This view is useful when the model gets too large to handle visually or for searching purposes.
\end{itemize}

\section{Structural Perspective: Strengths and Limitations}
\label{sec:strp:eval}

The Structural Perspective was the most difficult Perspective to design, because of the vast amount of possibilities to model solutions.
The general concept to keep things abstract and simple remains the same for the Structural Perspective.
The main design decisions of the other perspectives also hold here:
\begin{itemize}
	\item The perspective provides also complex filtering mechanisms via abstraction levels and stereotypes.
	\item The concepts of \textit{Decomposition}, \textit{Refinement} and \textit{Variability} are all well supported.
	\item The Structural Perspective is constrain able via requirements.
	\item The Structurl Perspective is still not part of the development or design phases!
	\item The Structural Perspective bases on Model Based Systems Engineering.
\end{itemize}
The focus on  the Structural Perspective is also to go not too deep into the behavior to keep solutions abstract enough for innovation modeling.
In line with this abstract focus, the Structural Perspective also contains no complex elements like Ports for interface consistency checks.
Nonetheless, abstract communication modeling is supported via channels and effect chains.
The solution spaces itself can be well assessed via the use of Key Performance Indicators, which are simply chosen properties of Blocks and relations.
The Structural Perspective also supports properties for Blocks and relations and provides more than only 'structure'.

As limitations, there are only few elements supported to model domain specific specialties.
These domain specific elements should however be added based on the needs of the innovation.
There exists also the high risk of 'model explosion' making the model unmaintainable.
However, innovation modeling tends to be abstract and thus manageable.

Overall, the Structural Perspective seems to be in a good shape for innovation modeling.

\FloatBarrier

\section{Structural Perspective FAQ}
\label{sec:strp:faq}

\includegraphics[width=\linewidth]{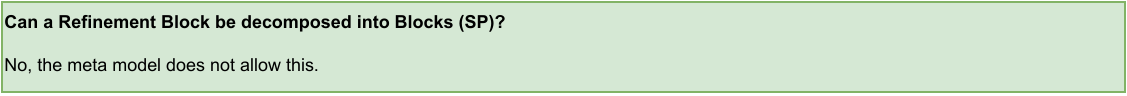}
\includegraphics[width=\linewidth]{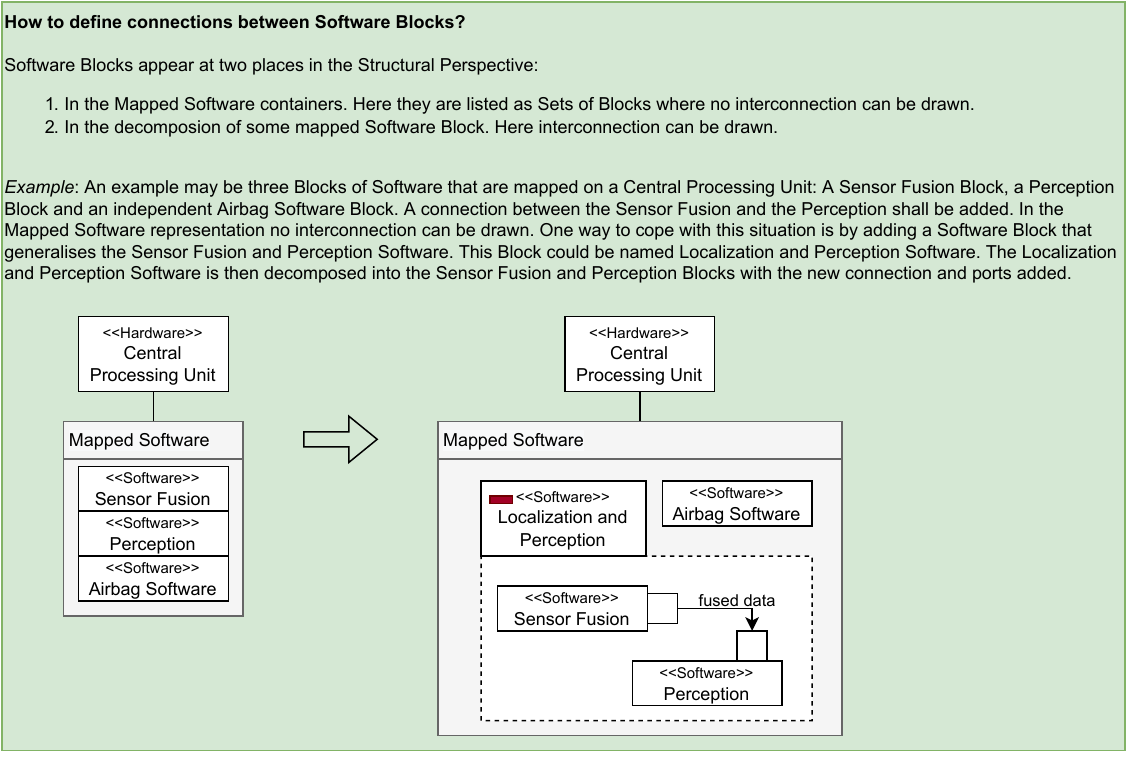}
\includegraphics[width=\linewidth]{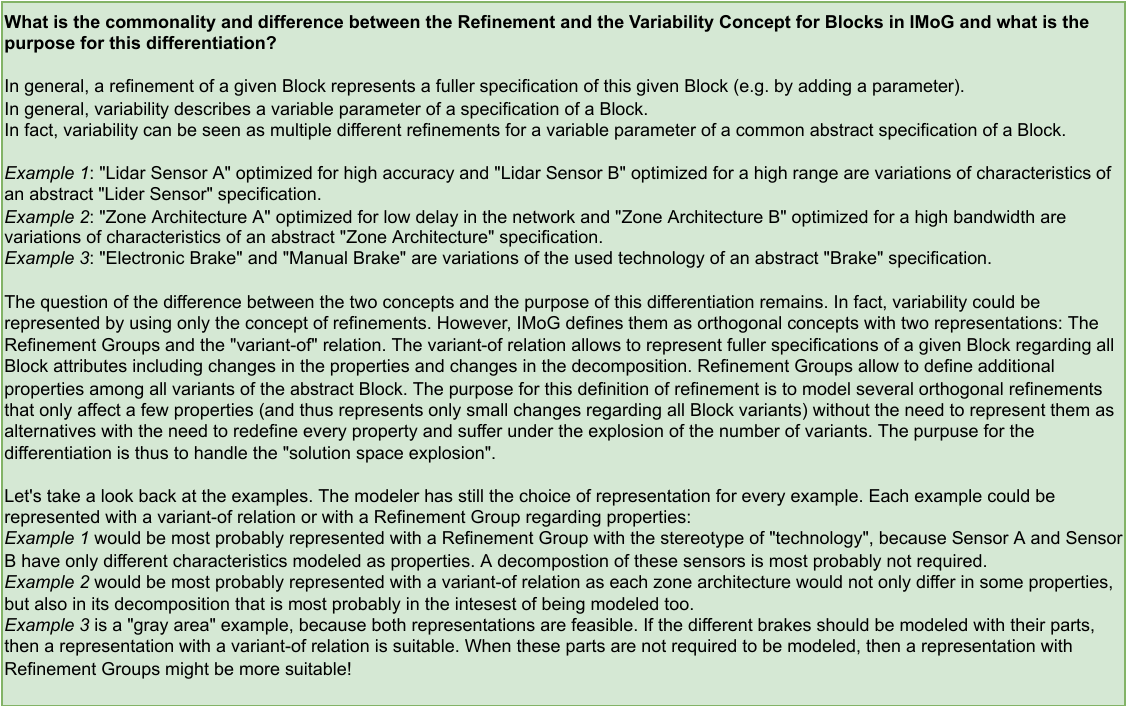}
\includegraphics[width=\linewidth]{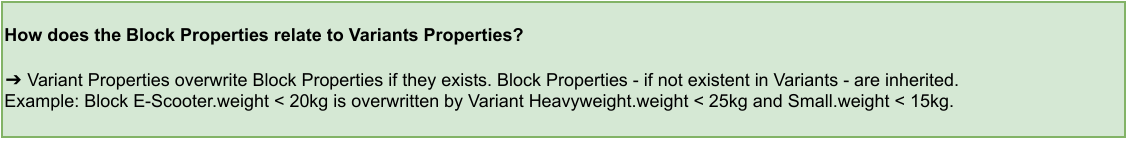}
\includegraphics[width=\linewidth]{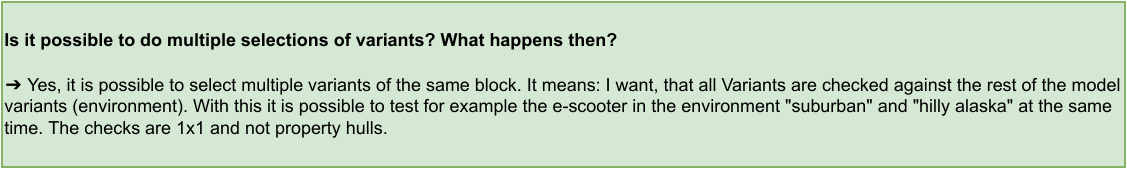}
\includegraphics[width=\linewidth]{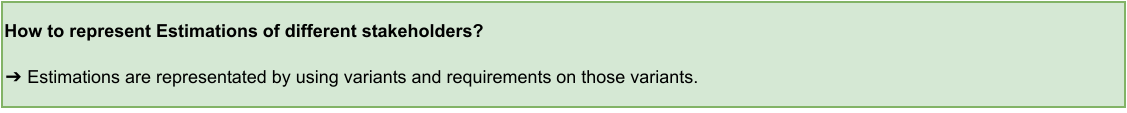}
\includegraphics[width=\linewidth]{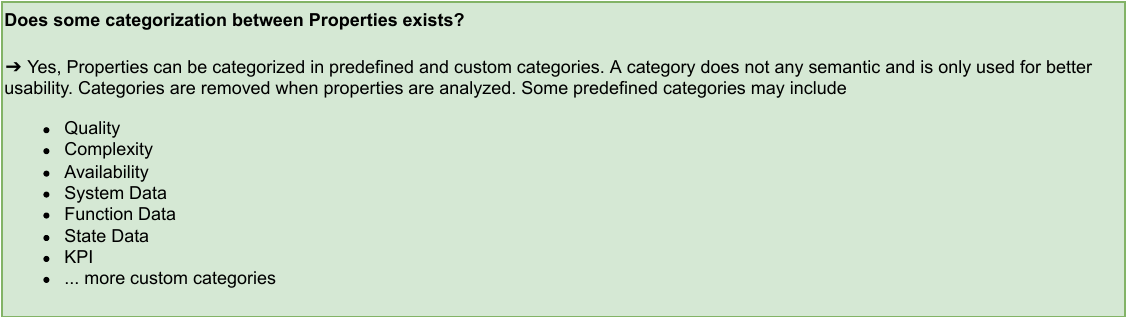}
\includegraphics[width=\linewidth]{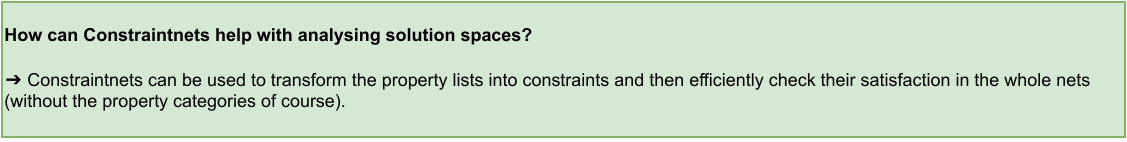}

%% file: content/knowledge_perspective.tex
\chapter{Domain Knowledge Perspective}
\label{chap:kp}

\begin{figure}[b!]
	\centering
	\begin{tikzpicture}
		\newcommand\scf{0.9} 
		\node[anchor=south west] {\includegraphics[width=\scf\linewidth]{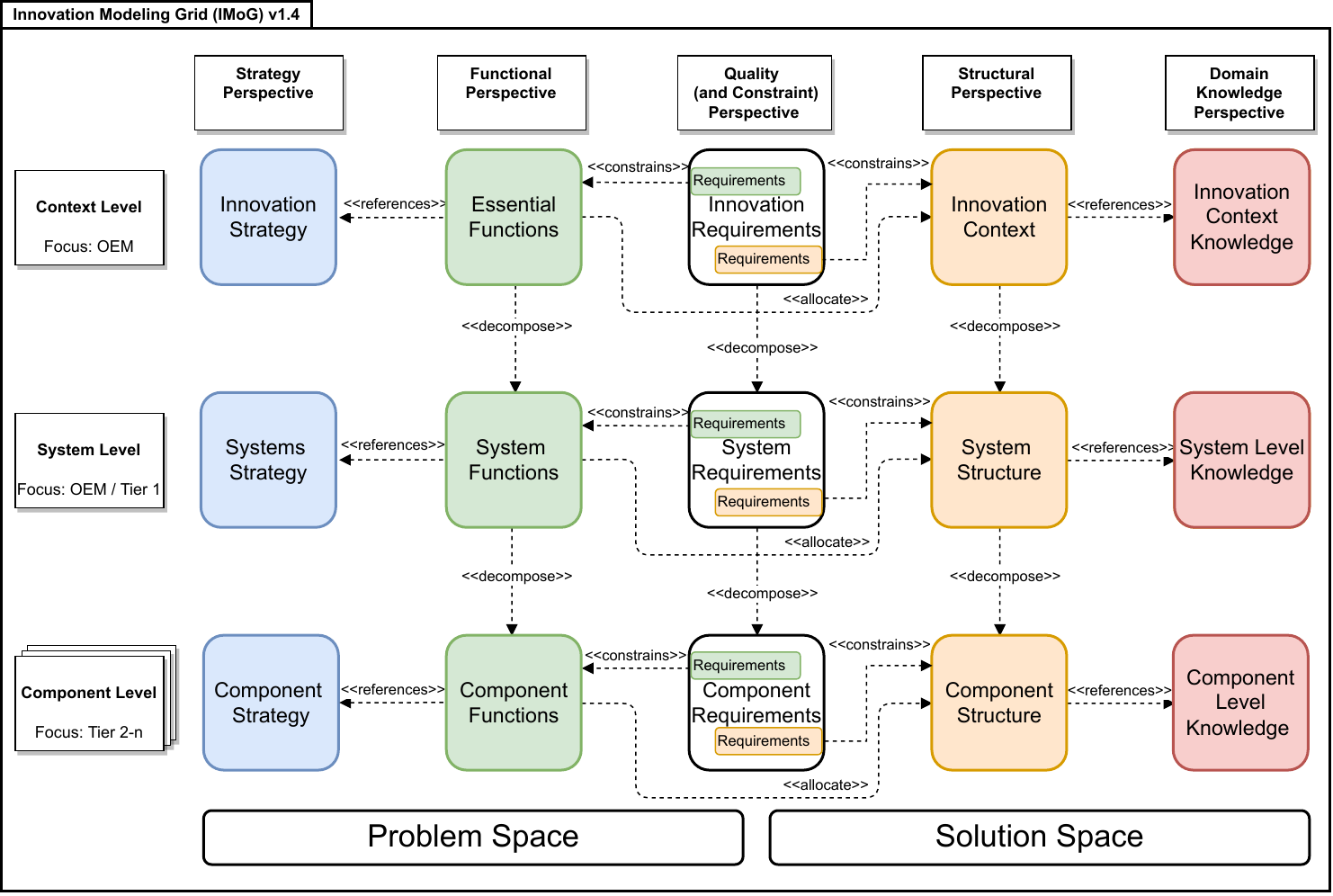}};
		\path[fill=white,opacity=0.5] (\scf*2.2,\scf*1.2) rectangle (\scf*4,\scf*9.5);
		\path[fill=white,opacity=0.5] (\scf*4.9,\scf*1.2) rectangle (\scf*6.7,\scf*9.5);
		\path[fill=white,opacity=0.5] (\scf*7.55,\scf*1.2) rectangle (\scf*9.4,\scf*9.5);
		\path[fill=white,opacity=0.5] (\scf*10.25,\scf*1.2) rectangle (\scf*12.05,\scf*9.5);
		\draw[ultra thick, blue, rounded corners] (\scf*12.9,\scf*1.2) rectangle (\scf*14.7,\scf*9.5);
	\end{tikzpicture}
	\caption{Location of the Domain Knowledge Perspective in IMoG}
	\label{fig:strp:imog}
\end{figure}

The Domain Knowledge Perspective is the fifth perspective in IMoG (see Figure \ref{fig:strp:imog}).
There has not been any more work done on defining the Domain Knowledge Perspective other than given in Chapter \ref{chap:methodology}.

%% file: content/roadmapping.tex

%% file: content/updating_the_roadmap.tex

%% file: content/inter_perspectives_relations.tex

%% file: content/tooling.tex
\chapter{Tooling Prototype}
\label{chap:tooling}

We created a tooling prototype for IMoG to evaluate our modeling methodology.
The scope of the prototype is the Functional Perspective.
The prototype was limited to this scope because the effort for consistent tooling between all perspectives within this project was too high.
A sophisticated tooling is thus left open for an industrial development after this project.
The Functional Perspective is based on the well-known Feature Models \cite{kang1990feature}.
As already mentioned in Chapter \ref{chap:fp} (Functional Perspective), we adjusted the meta model to our needs in the context of public committee-based road mapping.
This includes the differentiation of \enquote{Features} and \enquote{Functions} and the addition of a description, an abstraction level and various attributes to each block.
The abstraction level is primarily used for filtering purposes.
A \enquote{configuration} of the Functional Perspective is defined similarly to Feature Models.
However, configurations are currently not supported in the prototype.

We considered two approaches in achieving tooling support for the Functional Perspective:
\begin{enumerate}
	\item Translating or implementing IMoG into an existing modeling language like UML or SysML to take advantage of the existing tools.
	This approach is faster and eases the integration of IMoG into the internal modeling processes of the industry companies.
	\item Or by implementing a dedicated tooling prototype for IMoG by ourselves.
	This approach takes more effort, but brings a more elegant solution with a better learnability curve and a better user experience.
\end{enumerate}
For the purpose of an IMoG evaluation, we chose the latter approach, which promises a better user experience and less distortion in an evaluation.
Figure \ref{fig:tooling:comparison} underlines this reasoning by discussing what kind of a benefit we expect from using different kinds of tooling support for IMoG in a committee:

\begin{figure}[t]
	\centering
	\includegraphics[width=\linewidth]{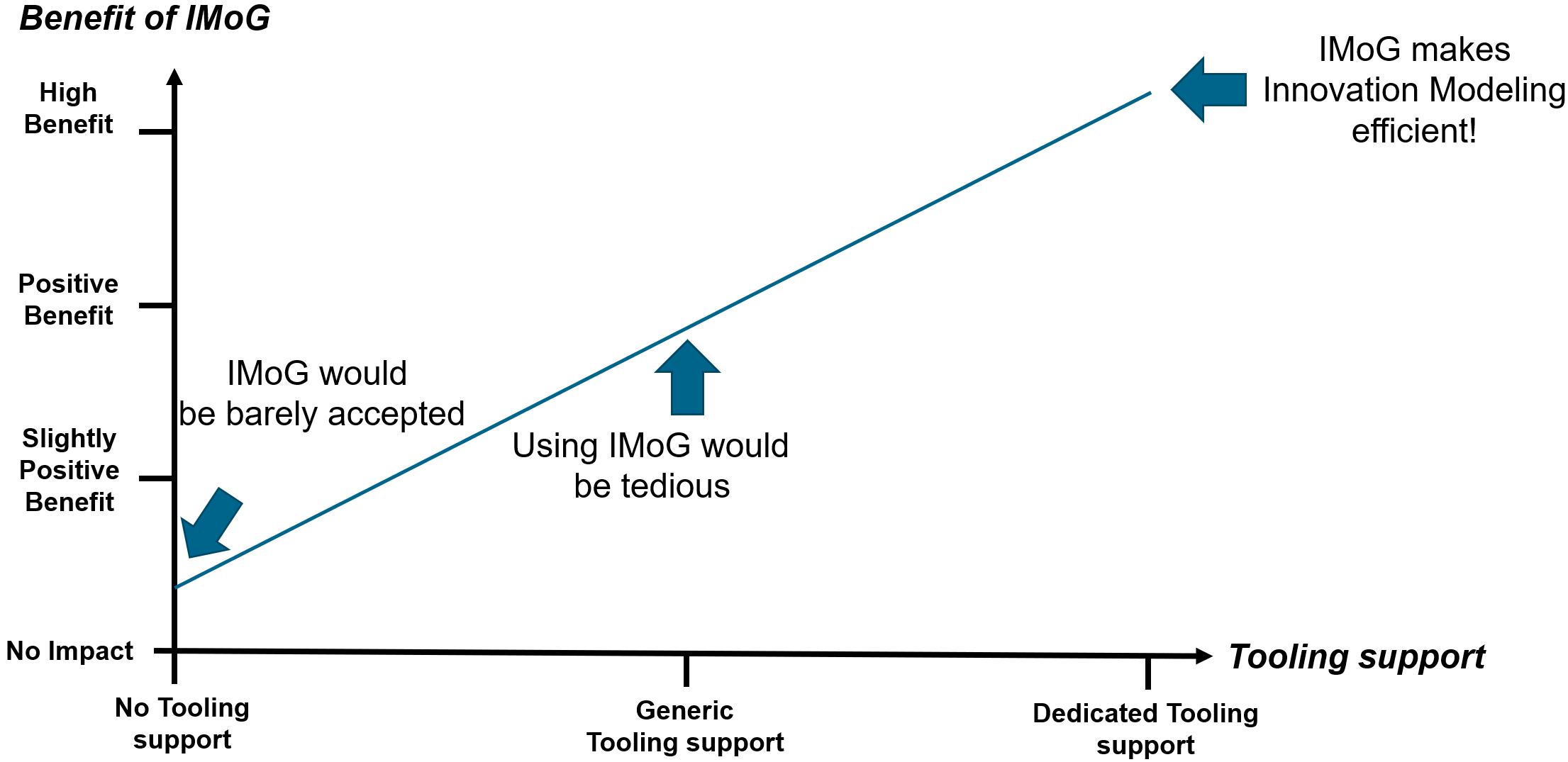}
	\caption{Comparison between the expected benefit of having no tooling support, generic tooling support with an existing tool and using a dedicated tool for IMoG.
		The horizontal axis represents the amount of tooling support in the committee, ranging from \enquote{No Tooling support} over a \enquote{Generic Tooling support} to a \enquote{Dedicated Tooling support}.
		The vertical axis represents the expected benefit of IMoG in a committee.
		In general, the more the tooling is designed around its application, the higher is the expected acceptance!}
	\label{fig:tooling:comparison}
\end{figure}

With \textit{no tooling support} used in the committee, the expectation is, that IMoG generates a huge modeling overhead.
The models would be drawn and the committee would need guidance on how and what to model.
An IMoG expert would be essentially required to create the models.
The strong separation of perspectives and abstraction levels would make it hard to remain efficient.
Thus, IMoG would be barely accepted in the committee.
Only the use of IMoG’s process would provide a guidance and would lead to a slightly positive benefit in innovation modeling.

When \textit{IMoG is translated to a generic modeling language} like UML or SysML and supported by a generic tooling, then IMoG would provide a positive benefit to the committee.
IMoG would be roughly supported by different types of diagrams for perspective and view separation, but the modeling elements would not perfectly fit and would be cumbersome to handle.
It would take the committee some time learning on how to handle the models.
Only an expert would be able to set up IMoG in the external tool, such that it would be usable by others.
All in all, the tooling would only have a moderate acceptance in the committee.

IMoG makes innovation modeling efficient when \textit{fully supported by a dedicated tooling approach}.
The dedicated tooling would support templates for perspectives and mapping views between different models.
Each modeling element would have a dedicated interface corresponding for its considered use.
There would be hardly any learning overhead, and the good user experience would be motivating.
The tooling would understand IMoG’s process and would request the necessary inputs from the user.
A new user would be guided through the tooling and would not have to understand IMoG to the degree of an expert to provide meaningful content.
IMoG together with its tooling would significantly reduce the time required for innovation modeling and would help to understand the innovation efficiently.
The tooling would have a high acceptance in the committee.

\section{Functional Perspective Prototype}
\label{sec:tooling:prototype}

The tooling prototype is publicly available under \url{https://genial.uni-ulm.de/imog-dev/} (We would like to thank the University of Ulm for providing their tool \textit{IRIS} \cite{iris,breckel2022domain} that we used as a basis for our IMoG prototype).

\begin{figure}[h]
	\resizebox{\textwidth}{!}{
		\begin{tikzpicture}
			\node {\includegraphics[scale=0.8]{"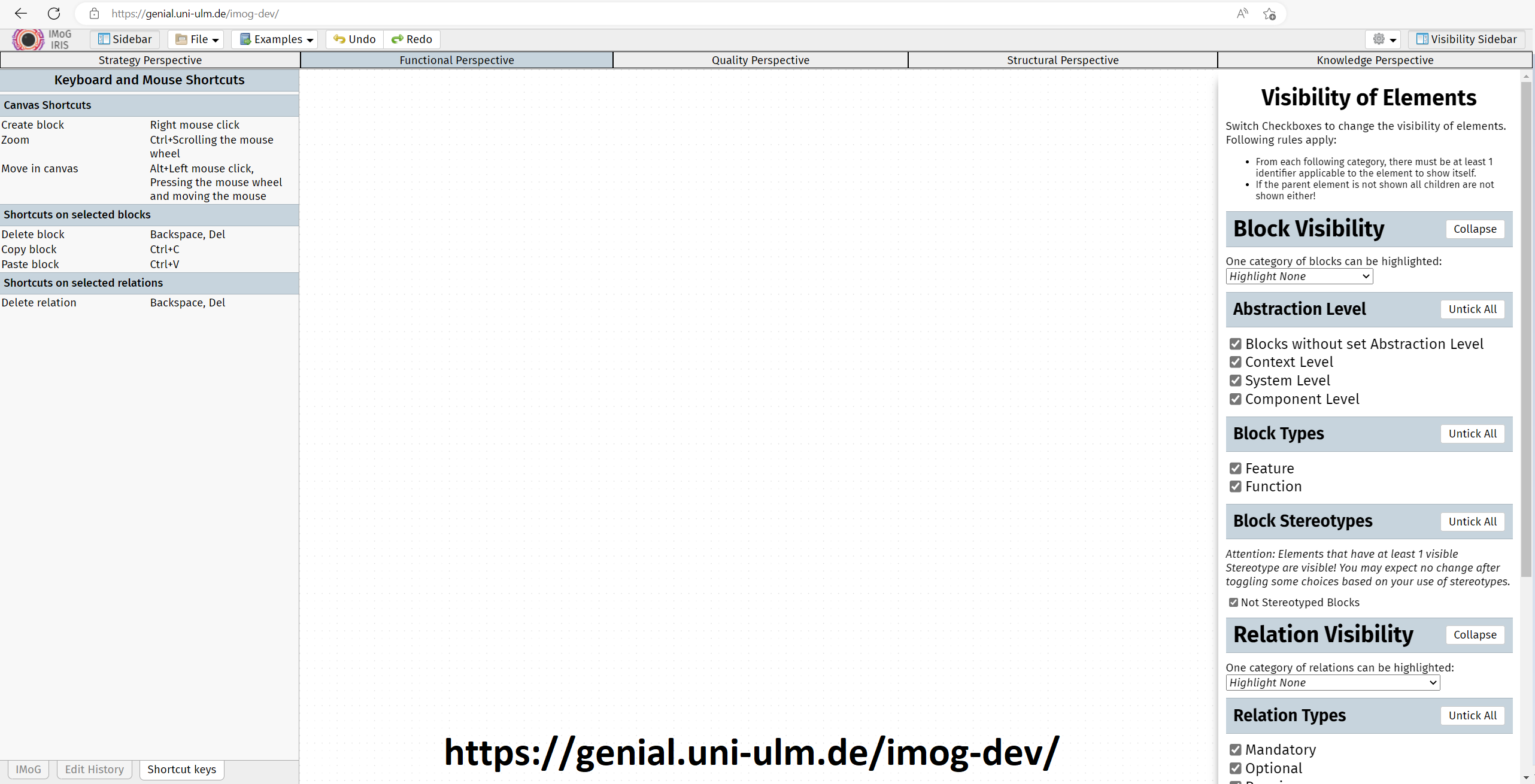"}};
			\begin{scope}[yshift=10]
				\node at (0,7) (persp) {\Huge \textbf{IMoG Perspectives}};
				\draw[-{latex},line width=1mm] (persp) -- (-15,8.7);
				\draw[-{latex},line width=1mm] (persp) -- (-8.5,8.5);
				\draw[-{latex},line width=1mm] (persp) -- (0,8.5);
				\draw[-{latex},line width=1mm] (persp) -- (8.5,8.5);
				\draw[-{latex},line width=1mm] (persp) -- (15,8.7);

				\node at (0,1) (medt) {\Huge \textbf{Model Editor}};
				\draw[-{latex},line width=1mm] (medt) -- (0,-1.4);

				\node at (9,7) (hide) {\Huge \textbf{Hide and}};
				\node at (9,6) () {\Huge \textbf{highlight elements}};
				\draw[-{latex},line width=1mm] (hide) -- (19,9.5);

				\node at (-17.5,-6.5) (sidebar) {\Huge \textbf{Sidebar including}};
				\node at (-18.5,-7.5) (info) {\Large \textbf{IMoG element Information}};
				\node at (-19,-8.25) (hist) {\Large \textbf{Model History}};
				\node at (-16.5,-9) (keys) {\Large \textbf{Shortcuts}};
				\draw[-{latex},line width=1mm] (-21,-7.9) -- (-21,-10.7);
				\draw[-{latex},line width=1mm] (hist) -- (-19,-10.7);
				\draw[-{latex},line width=1mm] (keys) -- (-16.5,-10.7);
			\end{scope}
		\end{tikzpicture}
	}
	\caption{IMoG tooling - a first glance after opening the prototype in the web browser.}
	\label{fig:tooling:imog-gui}
\end{figure}
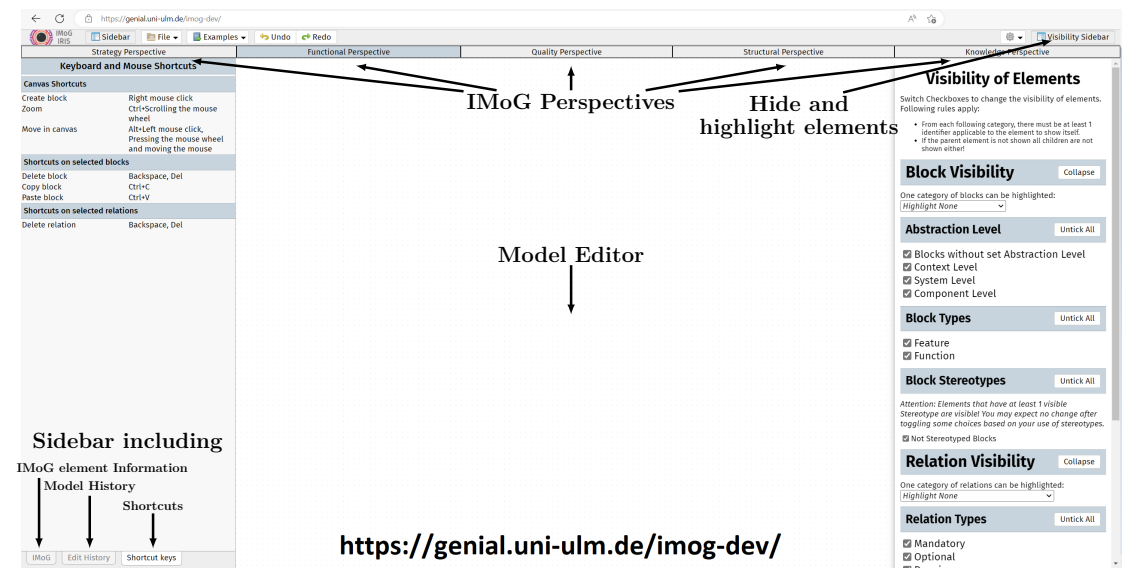

In the following, the model of the Functional Perspective of the e-scooter (see Chapter \ref{chap:fp}) is created in the prototype to present the features of the tooling.
Figure \ref{fig:tooling:imog-gui} shows the first view on the tool, when opened.
The first view is shortly described before the model of the e-scooter is created.
It contains a model editor in the center, one sidebar on the left and on the right side of the tool, an IMoG Perspectives toolbar and a menu toolbar.
The entry \enquote{File} in the menu bar can be used for creating new models, saving models and loading models, the entry \enquote{Examples} can be used for loading example models like the e-scooter model, the \enquote{Undo} and \enquote{Redo} buttons and the entry \enquote{Settings} can be used for changing the user interface (e.g., the grid size of the model editor).
The menu bar additionally includes toggles to open and close the \enquote{Sidebar} and the \enquote{Visibility Sidebar} on the left and right side of the editor.
The Sidebar on the left is used to display and manipulate information about selected model elements, for tracking of changes in the model history (Edit History) and for presenting the Keyboard and Mouse Shortcuts.
It is possible to view older model states through the model history.
The Visibility Sidebar on the right can be used to filter model elements and highlight them.
For example, it is possible to highlight all Function Blocks in a green color while hiding all mandatory relations.
The IMoG Perspectives toolbar allows to select the current perspective and view.
The Functional Perspective is, for example, selected in the image.
It is possible to swap to different perspectives in the prototype, but the model editor is disabled for the other Perspectives.
The model editor can be used to create the Functional Perspective model, like the e-scooter model.

\begin{figure}
	\begin{minipage}[b]{0.57\textwidth}
		\begin{subfigure}{\textwidth}
			\centering
			\includegraphics[width=\linewidth]{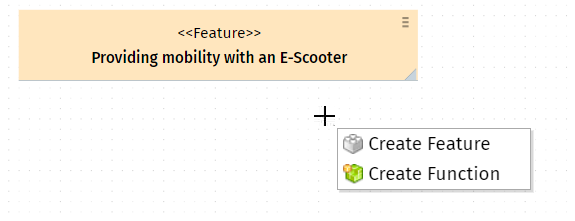}
			\caption{Context Menu: Create Features and Functions}
			\label{fig:tooling:imog-context-menu}
		\end{subfigure}
		\hfill
		\begin{subfigure}{\textwidth}
			\centering
			\includegraphics[width=0.8\linewidth]{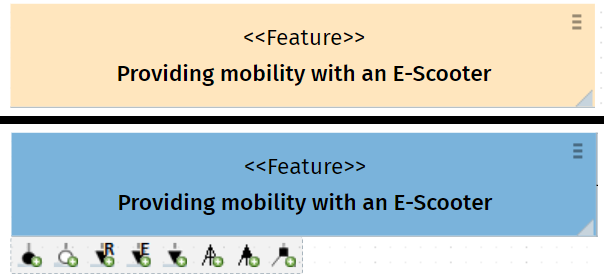}
			\caption{A Feature in both states: unselected (top) and selected (bottom)}
			\label{fig:tooling:imog-feature}
		\end{subfigure}
	\end{minipage}
	\hfill
	\begin{subfigure}{0.41\textwidth}
		\includegraphics[width=\linewidth]{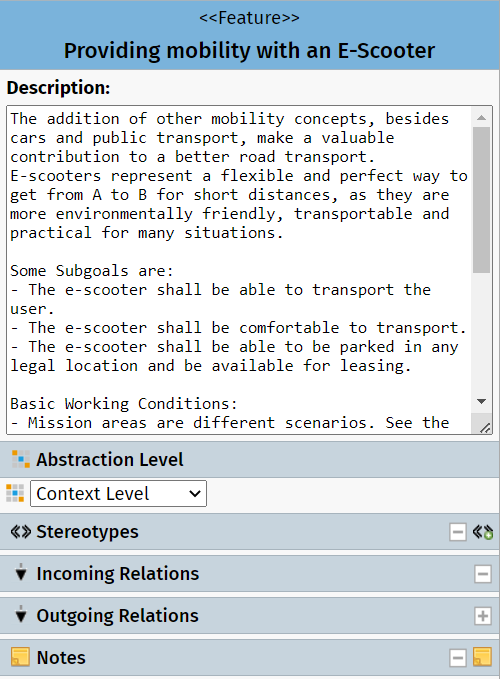}
		\caption{The description and information of the selected E-Scooter Feature}
		\label{fig:tooling:imog-description}
	\end{subfigure}
	\caption{The context menu, selected Features and the sidebar of the prototype.}
	\label{fig:tooling:parts}
\end{figure}

\begin{figure}
	\centering
	\includegraphics[width=\linewidth]{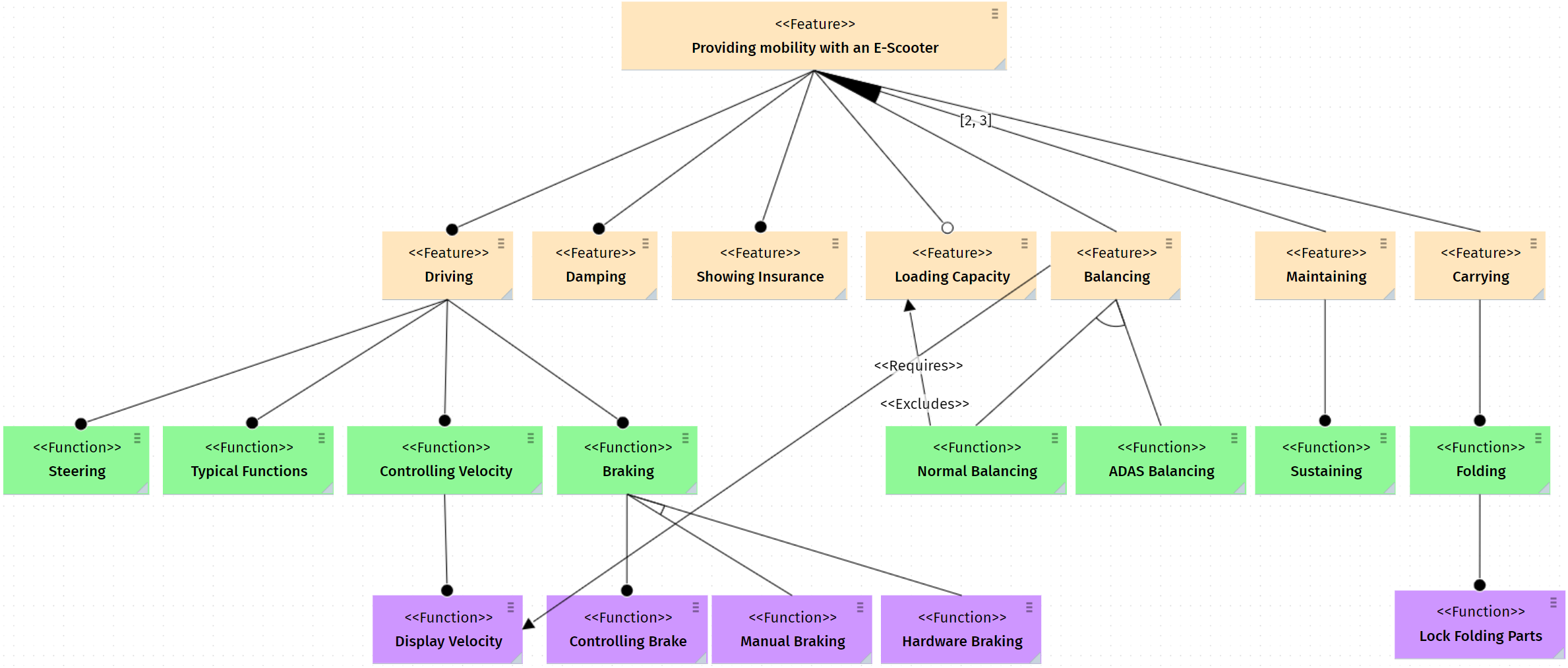}
	\caption{The model of the e-scooter in the prototype corresponding to the model presented on the Functional Perspective (see Chapter \ref{chap:fp}). It contains the root feature (yellow block), one layer of context level features (yellow blocks), one layer of system level functions (green blocks) and one level of component level functions (purple blocks) with the corresponding relations (mandatory, optional, constraint, $\ldots$).}
	\label{fig:tooling:e-scooter}
\end{figure}

The root feature of the e-scooter model can be created by using the context menu of the Model Editor (see Figure \ref{fig:tooling:imog-context-menu}) and by entering its name \enquote{Providing mobility with an E-Scooter}.
Each Feature and Function is represented by colored blocks.
Each block can be selected to operate with (see Figure \ref{fig:tooling:imog-feature}), like by renaming them, resizing them, changing their type and color or duplicate them.
When selected, the tab \enquote{IMoG} in the Sidebar will present further information about the block (see Figure \ref{fig:tooling:imog-description}).
This information can be manipulated, giving the blocks a description, changing their stereotype or setting their abstraction level.
After adding further Features and Functions to the model, the Features and Functions can be related with each other.
A relation can be created by selecting a block, choosing a relation from the relation toolbar under the block and then clicking on the target block.
All eight relations of the Functional Perspective can be used, changed and enriched with labels and information. These include the \enquote{Mandatory} relation, the \enquote{Optional} relation, the \enquote{Requires} constraint relation, the \enquote{Excludes} constraint relation, the \enquote{Custom} relation, the \enquote{Alternative} relation, the \enquote{Or} relation with a given cardinality and the \enquote{Multi directional Custom} relation.
Assuming the other Features, Functions and relations of the e-scooter model (see Chapter \ref{chap:fp}) are created, then the model of the e-scooter shall look similar to the model in Figure \ref{fig:tooling:e-scooter}.

\section{Tooling Evaluation}
\label{sec:tooling:eval}

We have conducted a user experience evaluation with two other researchers.
We gave them some predefined tasks to learn handling the user interface like creating some blocks, changing their properties, relating blocks with each other, saving and loading models and using the filtering features.
Afterwards, they had to redraw a model given on paper to identify their use of the tooling.
Furthermore, we asked them to open an example model and asked some more challenging questions about the model to identify if they understood what the model was meant to present.
Finally, we did a round of structural interviews with them regarding their user experience.
The prototype was overall evaluated as fast responding and easy-to-use.
We experienced the limitation of missing the other perspectives of IMoG in the prototype.
Nonetheless, our prototype demonstrated the large potential of a dedicated tooling approach for any IMoG related project by not bothering the user with cumbersome interactions.
Thus, we encourage the reader, other industry partners or committees to try out IMoG and its tooling prototype in their committee.

%% file: content/evaluation_results.tex
\chapter{Evaluation}
\label{chap:eval}

The initial evaluation in the original proposal of IMoG~\cite{Fakih2021} stated two strengths: the appropriate level of abstraction for modeling innovations and the examined ways through the matrix.
The examined ways include, for example, a top-down diagonal approach from the Context Level of the Strategy Perspective down to the Component Level of the Structural Perspective or a bottom-up approach from the ideas of the semiconductor suppliers back to the context of the car manufacturers.

The appropriate level of abstraction was confirmed and further underlined by the use cases where we have applied IMoG: IMoG helped us to adequately tackle the innovations.
We reconsidered our opinion regarding the mentioned ways through the matrix.
Instead of specifying several possible ways through IMoG, we think it is rather appropriate to follow the mentioned process for IMoG presented in section \ref{sec:imog:process}.
Furthermore, iterating between the problem space and solution space perspectives similar to the process defined in the twin peaks model \cite{nuseibeh2001weaving} is in our opinion the most appropriate approach.

The initial evaluation in the original proposal of IMoG~\cite{Fakih2021} identified three potential limitations based on an academic example of wireless charging: scalability, detailed behavioral models, and bridging to product level models.
The application of IMoG to the larger example of the e-scooter sheds more light onto these topics. 
Firstly, we did not encounter any issues regarding scalability in the use cases modeled here, which indicates that IMoG as such does not introduce unnecessary and unmanageable complexities.
Secondly, our use case here confirms the view that the absence of detailed behavioral models is actually a strength: Details are not required and should be left out in abstract innovation modeling.
Nonetheless, such detailed models should be possible to be attached to solution blocks whenever needed.
Finally, the bridge between an IMoG model to a product level model remains properly solvable:
Bridging the gap by referring IMoG's elements, using transformations of IMoG models to established system level development languages or by translating the IMoG model into a development focused framework (see Broy et al. in \cite{broy2009toward}) with adding the behavioral aspects to the designed framework are the recommended choices.

While applying the use cases we learned two more lessons:
Reordering the perspectives into the problem space and solution space made it easier to apply IMoG.
This distinction got added to the design principles of IMoG (see section \ref{sec:imog:overview}).
Another lesson was, that interpreting abstraction levels as filter functionality is better suited for the modeler than interpreting abstraction levels as a division into diagrams.
We examined that the division of an innovation model into several pieces would do more harm regarding its user experience and usefulness than it would help.

%% file: content/closing.tex
\chapter{Closing}
\label{chap:closing}

This technical document presented the Innovation Modeling Grid in detail.
This document is the successor of two publications on IMoG \cite{Fakih2021, klemp2023imog} and focuses on presenting all details of the methodology.
Beginning with the process and an overview, each perspective was presented in detail.
Afterwards the tooling and the evaluation was presented.

Overall, we think that IMoG has great potential to be really useful in committee driven innovation modeling.
Next to the applications of the e-scooter example and the application in a project of the GAIA-X family \cite{shakeri2023shaping}, IMoG is currently applied in the \enquote{Arbeitskreis Automotive} in a workshop series.
This document shows that much is already researched about IMoG, however IMoG still has some missing ends in parts of solution definition and tooling.
The model of IMoG can still improve and this improvement is enabled through getting some crucial feedback from applications like the workshop series.
Therefore, if one is interested in committee driven innovation modeling, we encourage to take a look at IMoG and tailor it to the needs of their committee.

%% file: content/appendix.tex
This appendix contains the following parts:
\begin{itemize}
	\item The glossary of IMoG
\end{itemize}



\section*{Supplementary material}

**Component:**
A Component is a Block on the Structure Perspective, that is follows the semantics of either (1), (2) or (3) [IMoG Definition]

Structural Perspective

(1) non-system level element (3.41) that is logically or technically separable and is comprised of more than one hardware part (3.71) or one or more software units (3.159) [ISO26262]

(2) A Component is a unit of composition with cantractually specified component interfaces and explicit context dependencies only; it can be deployed independently and is subject to composition by third parties.[Pohl]

*Note, that Components are restricted to Perspectives in IMoG, such that the context information are essential for the Components definition. The typical interpretation that everything is a component as often found in other model definitions just does not apply for IMoG.*

---

**Component Level:**
The Component Level describes the components of the system and parts in its atoms and is imposed over all perspectives. [IMoG Definition]

---

**Configuration:**
A feature configuration is a chosen set of features that represent an innovation where every choice is resolved. A feature configuration is permitted by a feature model if and only if it does not violate constraints imposed by the model. [Wikipedia - Feature Models]

*The IMoG Functional Perspective represents a Feature Model.*

---

**Constraint:**
See Requirement (Constraint)

Quality Perspective

---

**Context Level:**
The Context Level describes the innovation as a whole system embedded in its environment and is imposed over all perspectives. [IMoG Definition]

---

**Corporation Representative:**
The responsible person of a corporation to coordinate the corporation internal tasks to produce the needed imputs for the roadmap. [IMoG Definition]

---

**Decomposition and Aggregation:**
(1) Decomposition denotes the partitioning of an analysis element or design element (e.g., of a goal, a function, or a logical/technical component) into parts. [SPES]
(2) Abstracting a collection of units into a new unit is called aggregation. For example, school is an aggregation of students, teachers, etc. Refining an aggregation into its constituent units is called decomposition. [Kang90]

---

**Domain Expert:**
The domain expert represents a specialist of a particular discipline covering subdomains of development. The domain expert supports the innovation modeling and evaluates its influences and dependencies of certain domain elements on other domain elements [IMoG Definition]

---

**Feature:**
(1) A Feature is a noticeable or important part of an innovation. The Feature is an abstraction of functionalities that require system of systems solutions. The Feature functionality is described by the User Stories and decomposed into Functions. [IMoG Definition]
(2) A feature is a logical unit of behavior that is specified by a set of functional and quality requirements [Bosch]
(3) A Feature is an end-user visible characteristic of a system.[Pohl]

Functional Perspective

---

**Function:**
A task, action or activity that must be accomplished to achieve a desired outcome.
[AUTOSAR 2012]

Functional Perspective

---

**Functional Perspective:**
In the Functional Perspective the functional needs of the system and their breakdown into sub-functions is modeled. The set of all sub-functions then realize the top-level system functionality. [Architecture Modeling]
The Functional Perspective is part of the problem space and describes the features (like User Needs) and functionalities needed for the innovation without focusing on solutions. For details see the description of the Functional Perspective in its associated file. [IMoG Definition]

**Future Availability:**
The Future Availability represents the date when the Requirement or any other Block will be relevant. [IMoG Definition]

**Generalization:**
See Specialization and Generalization

**HTML Divs:**
The HyperText Markup Language (HTML) Content Division element (Div) element is the generic container for flow content. HTML is primarely used for representing the content of Websites. [Mozilla]

*Strategy Perspective*

**IMoG responsible Model Expert:**
The responsible person of creating and maintaining the IMoG model on the command of the committee members. The IMoG responsible Model Expert is also called IMoG Modeler. [IMoG Definition]

**Knowledge Perspective:**
The Knowledge Perspective targets the creation of a database of reuseable blocks of existing models and a knowledge base by transferring the knowledge of domain experts into an appropriate ontology. The Knowledge Perspective is part of the solution space and is used to refine the model with existing knowledge and constraints. The expert knowledge in the knowledge base may for example contain sensor characteristics and constraints from the road traffic regulations. [IMoG Definition]

**Mapping (between Views/between abstraction layers):**
A mapping between Views is a relationship between two models representing the Views. A mapping can exist between models of different Viewpoints or between models of the same View, but on adjacent abstraction layers. [SPES]

**Model:**
A model is an abstract representation of an existing reality or a reality to be created. Every model is created for a specific purpose of use. [SPES]

**OEM:**
The OEM (Original Equipment Manufacturer) is the manufacturer of the final product, which deals with the market launch of the vehicle. [Knauf]

**Perspective:**
A Perspective combines Views of different abstraction levels which are related to similar Viewpoints. Perspectives can be used to group and to structure Views of different disciplines in order to cope with the complex task of developinga system. [Architecture Modeling]

**Problem Space:**
(1) The Problem Space focuses on the description of the problem while ignoring technical informations. [IMoG Definition]

(2) The set of all valid system specifications in a domain (e.g. valid feature combinations) is referred to as the problem space and the set of all concrete systems in the domain is referred as to as the solution space. [Cz98]

(3) A lenghty example based explanation can be found under [Gupta19].

**Quality Perspective:**
The Quality Perspective represents the link between problem space and solution space and contains constraints and requirements to Features and Functions as well as to structural solutions. For more details see the description of the Quality Perspective in its associated file. [IMoG Definition]

**Refinement:**
A relationship that represents a fuller specification of something that has already been specified at a certain level of detail. [SysML]

**Requirement:**
(1) A requirement is:

*Quality Perspective*

    1. A need perceived by a stakeholder.
    2. A capability or property that a system shall have.
    3. A documented representation of a need, capability or property. [IREB]

(2) A statement that identifies a system, product, or process characteristic or constraint, which is unambiguous, clear, unique, consistent, stand-alone (not grouped), and verifiable, and is deemed necessary for stakeholder acceptability. [INCOSE Sys Eng Handbook V4]

**Requirement (Constraint):**
A requirement that limits the solution space beyond what is necessary for meeting the given functional requirements and quality requirements. [IREB]

Quality Perspective

**Requirement (Functional):**
A requirement concerning a result of behavior that shall be provided by a function of a system (or of a component or service). [IREB]

Quality Perspective

**Requirement (Non-Functional):**
A quality requirement or a constraint. Synonym: Extra-functional requirement [IREB]

Quality Perspective

**Requirement (Quality, Performance):**
A quality requirement is a requirement that pertains to a quality concern that is not covered by functional requirements.

A Performance requirement is describing a performance characteristic (timing, speed, volume, capacity, throughput...).
Performance requirements may be regarded as another category of non-functional requirements. In this glossary, performance requirements are considered to be a sub-category of quality requirements. [IREB]

Quality Perspective

**Requirements Engineer:**
The requirements engineer creates initial top-level requirements for the innovation and captures them uniformly (formally or in natural language). The requirements engineer leverages the expertise of the domain experts and system architects to uniformly refine the requirements in the system models. [IMoG Definition]

**Roadmap Manager:**
The roadmap manager monitors the innovation status, reports to top management on the feasibility of the innovation, surveys new technologies from other partners, and updates the roadmap. The roadmap manager investigates trends and innovations. During innovation modeling, the roadmap manager performs the initial tasks and writes the roadmap after consulting with the other domain experts, requirements engineers, and system architects. [IMoG Definition]

**Roadmap Manager of the Committee:**
The roadmap manager of the committee is responsible for the creation and maintenance of the roadmap. [IMoG Definition]

**Robustness:**
The degree to which a system or component can function correctly in the presence of invalid inputs or stressful environmental conditions. [ISO/IEC/IEEE 24765:2017]

**Specialization and Generalization:**
(1) Abstracting the commonalities among a collection of units into a new conceptual unit suppressing detailed differences is called generalization. Refining a generalized unit into a unit incorporating details is called specialization. For example, the conceptual entity "employee" is an abstraction of secretaries, managers, technical staffs, etc. [Kang90]

**Specification:**
(1) A set of requirements for a system or other element. [SysML]

(2) A systematically represented description of the properties of an entity (a system, a device, etc.) that satisfies given criteria. It may be about required properties (requirements specification) or implemented properties (e.g., a technical product specification). [IREP]

**Solution Space:**
(1) The Solution Space focuses on how to translate the problem into solution designs.  [IMoG Definition]

(2) The set of all valid system specifications in a domain (e.g. valid feature combinations) is referred to as the problem space and the set of all concrete systems in the domain is referred as to as the solution space. [Cz98]

(3) A lenghty example based explanation can be found under [Gupta19].

**SQL Query:**
A database query based on the formal language *Structured Query Language* compliant to the standard ISO/IEC 9075.

**Stakeholder:**
(1) Individuals, groups, and/or institutions which may be impacted by the system throughout its life cycle, including acquisition, development, production, deployment, operations, support, and disposal. [SysML]

(2) Individual or organization having a right, share, claim, or interest in a system or in its possession of characteristics that meet their needs and expectations [ISO/IEC/IEEE 15288:2015]

**Stereotype:**
A stereotype defines how an existing Block may be extended, and enables the use of platform or domain specific terminology or notation in place of, or in addition to, the ones used for the extended Block. [IMoG Definition inspired by SysML (replaced 'metaclass' with 'Block')]

**Strategy Perspective:**
The Strategy Perspective is expected to be the first considered Perspective of IMoG (part of the problem space) and targets the first phase of the creation and the capturing of one innovation. For details see the description of the Strategy Perspective in its associated file. [IMoG Definition]

**Structural Perspective:**
The Structural Perspective targets the modeling of the solution space by using the Requirements and Functions. It may include environment descriptions, system decompositions including software elements, hardware elements and a mapping between them, Mission Profiles and properties as well as sensor descriptions or abstract technologies with parameters, functions and key performance indicators. For more details see the description of the Structural Perspective in its associated file. [IMoG Definition]

**System Architect:**
The system architect has the role of an interdisciplinary expert who designs systems by using modeling techniques. The system architect has know-how in the area of software - hardware design. In innovation modeling, the system architect takes on the role of the innovation modeler and its decomposition into subsystems. [IMoG Definition]

**System Level:**
The System Level describes the system and its parts and is imposed over all perspectives. Decompositions and property consistency are of special importance on the System Level. [IMoG Definition]

**Tier 1:**
The Tier 1 suppliers develop system solutions that are tailored to the end product without major changes. [Knauf]

**Tier 2:**
The Tier 2 creates the components needed to be integrated into systems. This includes the production of semiconductors and microcontrollers. [inspired by Knauf]

**User Need:**
A prerequisite identified as necessary for a user, or a set of users, to achieve an intended outcome, implied or stated within a specific context of use. [ISO/IEC/IEEE 24765:2017]

**Variant (or Alternative):**
A variant is a representation of a variability object (a particular instance of a variable item of the real world or a variable property of such an item) within domain artefacts.[Pohl]

*For more information about the IMoG Model Element "Variant" see the section "Model Elements".*

**Variation Point:**
A Variation Point is a representation of a variable item of the real world or a variable property of such an item within domain artefact enriched by contextual information. [Pohl]

*For more information about the IMoG Model Element "Variation Point" see the section "Model Elements".*

**View:**
A View is a representation of a whole System under Development from the Perspective of a related set of concerns (based on [IEEE1471]). [SPES]

*In IMoG the "System under Development" can be substituted with an innovation.*

**Viewpoint:**
A Viewpoint is a specification of the conventions for constructing and using a View. Viewpoints comprise patterns or templates from which to develop individual Views by establishing the purpose and audience for a View and the techniques for its creation and analysis (based on [IEEE1471]). [SPES]

